\documentclass[letterpaper,11pt,leqno]{article}
\usepackage{paper}
\usepackage{array}
\hypersetup{pdftitle={Measuring Geopolitical Alignment and Economic Growth}}

\newcommand{\bib}{bibliography.bib}

\makeatletter
\newif\ifpaper@appendixcontents
\paper@appendixcontentsfalse

\AddToHookWithArguments{cmd/addcontentsline/after}[paper/appendixcontents]{%
  \ifpaper@appendixcontents
    \ifstrequal{#1}{toc}
      {\addtocontents{apc}{%
         \protect\contentsline{#2}{#3}{\thepage}{\@currentHref}%
         \protected@file@percent}}%
      {}%
  \fi
}

\newcommand{\appendixtableofcontents}{%
  \pdfbookmark[1]{Online Appendix: Table of Contents}{appendix.contents}%
  \section*{Online Appendix: Table of Contents}%
  \begingroup
    \small
    \singlespacing
    \setcounter{tocdepth}{2}%
    \@starttoc{apc}%
  \endgroup
  \clearpage
}

\newcommand{\startappendixcontents}{\paper@appendixcontentstrue}

\newcommand{\stopappendixcontents}{\paper@appendixcontentsfalse}
\makeatother

\begin{document}

\title{Measuring Geopolitical Alignment and Economic Growth\thanks{Tianyu Fan: Princeton University. Email: tianyufan.econ@gmail.com. Previously circulated as ``The Geopolitical Determinants of Economic Growth, 1960--2024.'' I am deeply indebted to Michael Peters, Pascual Restrepo, and Fabrizio Zilibotti for their invaluable guidance and unwavering support. I also thank Leah Boustan, Chris Clayton, Ethan Ilzetzki, Ruixue Jia, Sam Kortum, Fernando Leibovici, Hannes Malmberg, Andrea Presbitero (discussant), and Linchuan Xu (discussant) for insightful discussions. I am grateful for helpful comments from participants at the Yale Trade Workshop, the Workshop at the University of Hong Kong, the 2025 NEUDC Conference, the International Macro History Online Seminar Series, the NBER Conference on AI and Economic Measurement, the Third Kiel--G\"{o}ttingen--CEPR Conference on China in the Global Economy, the 2026 SED Annual Meeting in Athens, Peking University, and University of Minnesota. I thank Ruihan Cao, Zihao Cong, Kana Hirotomi, Krisztian Pinter, Yiyang Sun, and Haoyu Wang for excellent research assistance with the human audit. Replication data and code for the main results are available at \url{https://github.com/tianyufan-econ/research-replication/tree/main/geopolitical-growth}. Bilateral geopolitical alignment scores are available at \url{https://github.com/tianyufan-econ/global-geopolitics}.}}

\author{Tianyu Fan\\ \textsc{Princeton University}}

\date{\today\\ \href{https://www.tianyu-fan.com/files/FAN_Tianyu_Geopolitical_Growth.pdf}{(Click here for the most recent version)}}

\begin{titlepage}
\maketitle

\begin{center}
\begin{minipage}{0.88\textwidth}
We introduce an event-based measure of bilateral geopolitical alignment from 373,020 events across 193 countries over 1960–2024, compiled using large language models and validated with external benchmarks and a human audit. Aggregating alignment with major nations summarizes each country's external geopolitical environment. In local projections using within-country variation, a sustained one-standard-deviation improvement in this environment is followed by about 10 percent higher GDP per capita after 25 years. This relationship is similar across major-nation partners and historical periods, and is accompanied by greater stability, investment, trade openness, and productivity. Accounting exercises map observed realignments into substantial GDP differences. (\textit{JEL}: F50, F51, O11, O40, C80, P16)
\end{minipage}
\end{center}

\end{titlepage}

\section{Introduction}\label{s:introduction}

Research on the wealth of nations has focused on domestic determinants. Early empirical contributions by \citet{Barro1991-bi} and \citet{Mankiw1992-op} identified physical and human capital accumulation as key drivers of cross-country income differences. Subsequent research has examined technology, institutions, geography, culture, and financial systems as fundamental determinants, with \citet{Acemoglu2019-bo} providing causal evidence that democratization increases GDP per capita by approximately 20 percent in the long run. Yet large cross-country differences in growth and income remain unexplained \citep{Kremer2022-ha}.

Countries develop not only through these domestic factors, but also within an external environment. This paper studies one dimension of this environment: geopolitical alignment, the state of a country's diplomatic and strategic relationships with the major nations. Cooperative relations with major nations can open access to their markets, capital, and technology, while conflictual relations can bring sanctions, isolation, and security threats. The recent fragmentation of international relations makes understanding the economic consequences of these relationships especially pressing. We construct and validate a measure of geopolitical alignment and show that improvements in a country's external geopolitical environment are followed by large and persistent GDP gains: a sustained one-standard-deviation improvement is followed by about 10 percent higher GDP per capita after 25 years.

To characterize this external geopolitical environment, we introduce a new event-based measure of bilateral geopolitical alignment that captures both the timing and intensity of changes in relations between countries. We first use a large language model with web search to compile, for each country pair and year, the geopolitical events that significantly affect or indicate the state of their relationship. These events cover six issue areas: economic relations, diplomatic and political relations, security and defense, legal and territorial issues, multilateral governance, and other bilateral interactions. The model then codes each event following established conventions: it classifies the event using the Conflict and Mediation Event Observations (CAMEO) framework and assigns a Goldstein score, from $-10$ (maximum conflict) to $+10$ (maximum cooperation), that measures the conflictual or cooperative intensity of the event.\footnote{CAMEO is a framework for coding international political events into cooperation--conflict categories. See \citet{schrodt2012cameo} for CAMEO, \citet{goldstein1992} for the Goldstein scale, and Appendix~\ref{app_c:cameo} for our coding and scoring rules.} The resulting database contains 373,020 events in the bilateral relationships between each of the 193 United Nations member states and the 24 major nations from 1960 to 2024.

We validate the database against independent evidence. We first benchmark it against 4,670 documented bilateral developments in eight classes, ranging from war onsets and sanctions to new diplomatic ties and alliances. Our database records an event in the expected conflict or cooperation direction for 92 percent of them within each class's matching window, comparable to the rates of machine-coded databases such as ICEWS \citep{boschee2015icews}, GDELT \citep{leetaru2013gdelt}, and Phoenix \citep{althaus2020phoenix}, which record far more events. The key difference is that, by focusing on consequential events rather than averaging over large volumes of routine reports, our annual scores better capture the dynamics of international relations: relative to each pair's usual level, they fall more in years of documented conflict and rise more in years of documented cooperation. Finally, in a human audit of 200 dyad-years, trained coders first reconstruct the events independently with the model's codebook but without seeing its output. Coders' individual annual scores correlate with the model's at 0.80, compared with 0.76 between the two coders. The audit also finds only three fabricated records among the model's 856 events.

Building on this event database, we construct a \emph{bilateral geopolitical alignment score} for each country pair and year. The score is a weighted average of current and past event scores, with the weight on past events depreciating at 30 percent per year, so that it tracks how each relationship evolves over time. These scores capture well-known episodes in international relations, such as US--Soviet d\'{e}tente in the 1970s, the India--Pakistan wars of 1965 and 1971, and the deterioration of Greece--Turkey relations after the 1974 Cyprus invasion. At the country level, the \emph{geopolitical alignment index} averages each country's bilateral alignment scores with the 24 major nations, weighted by their shares of world GDP, and so summarizes the external geopolitical environment each country faces. Our measure spans six decades, from Cold War bipolarity through post-1990 globalization to the recent era of geopolitical fragmentation.

In our main application, improvements in the index are followed by GDP gains that emerge within a few years, peak after about a decade, and remain positive two decades later. The estimates come from local projections that compare each country with itself over time, absorb shocks common to its region in each year, and control for past GDP and alignment. A sustained one-standard-deviation improvement, about the size of the deterioration in Russia's relations through the Crimean crisis (2005--2015), is followed by about 10 percent (10 log points) higher GDP per capita after 25 years. Because the index summarizes a country's relations with all major nations, this estimate reflects the combined changes in access and risk that accompany more cooperative relations.

Alternative measures of alignment do not reproduce this relationship. Measures based on UN voting similarity, or built the same way from machine-coded event databases, show little or no detectable relationship with subsequent growth. Adding sanctions or formal alliances, one at a time, leaves our estimates nearly unchanged, while the sanctions and alliance coefficients shrink and are no longer statistically distinguishable from zero.

The relationship is similar across partners and historical periods: GDP gains follow improved relations with the United States, with other Western nations, and with other major nations, and appear both during and after the Cold War. It also holds across issue areas: GDP responses are positive for improvements in diplomatic and security relations, not only in economic relations. It holds as well when the index uses only formal acts that contemporaneous sources record regardless of later outcomes, such as agreements, sanctions, and uses of force, a result that weighs against look-ahead bias in the types of events the model selects and in how it scores them. The estimates are robust to alternative fixed effects and to controlling for the past paths of trade openness, domestic unrest, war exposure, institutions, and a broader set of economic fundamentals, factors that could drive both international relations and growth. Our long-run estimates are also robust to dynamic panel estimation and to an unsmoothed index (Section~\ref{s:add_robust}).

Shifts in relations with the major nations often coincide with domestic reforms and regime changes that may affect growth on their own. We address this concern in two ways. First, we use only the changes in relations around leadership transitions that follow deaths in office or close elections in the major nations. These estimates are positive and similar to or larger than the baseline magnitude, though less precise. Second, as a conservative check, we exclude variation from all years around major domestic reforms and regime changes. The estimated gain remains three-fifths of the baseline or more over most of the first two decades.

Improvements in the index are accompanied by changes in growth fundamentals, each with its own timing: domestic unrest falls and trade openness rises immediately, investment and productivity rise within a few years, and the capital stock builds over the following decade. This sequence is consistent with an early reduction in risk followed by gains from better access to markets and technology. We then revisit the democracy--growth relationship in light of geopolitical alignment. Democracy and geopolitical alignment are related but distinct: democratization is followed by closer relations with Western nations, and controlling for alignment lowers its estimated growth effect, most at short and medium horizons, while its long-run benefits persist.

Applying the estimated responses to observed changes in alignment, we find that each decade's realignment from the 1970s through the 1990s implied a positive median long-run GDP gain, peaking at about 4.6 log points for the 1980s, while fragmentation in the 2010s turned the median slightly negative, at $-1.2$ log points. Across countries, realignment implies illustrative growth differences of about $-31$ to $+31$ log points during and after the Cold War, and alignment in 2024 implies income differences of about $-29$ to $+12$ log points relative to the median.

\paragraph{Related Literature} We contribute to the measurement of geopolitical alignment, the study of growth determinants, and geoeconomics.

\textit{Measuring Bilateral Geopolitical Alignment}. Our primary contribution is a measure of bilateral geopolitical alignment built from consequential events. Existing measures differ in how they conceptualize alignment \citep{Voeten2026review}. The most widely used, UN General Assembly (UNGA) voting similarity \citep{Signorino1999-yb, Bailey2017-po}, treats alignment as agreement and captures countries' positions in multilateral forums rather than the state of their bilateral relations. Measures that treat alignment as a relationship, including strategic rivalries \citep{thompson2001identifying}, sanctions \citep{Felbermayr2021-ws}, military alliances \citep{leeds2002alliance}, and bilateral treaties \citep{broner2025hegemonic}, provide discrete classifications that record changes in a relationship's status but not its evolving intensity. Our measure follows the relational view and tracks the intensity of cooperation and conflict in each relationship continuously, at annual frequency, across six decades.

Our measure is closest to event data. Machine-coded event databases such as ICEWS, GDELT, and Phoenix also track bilateral interactions continuously, but they extract events automatically from large volumes of news reports. We instead compile only consequential events and validate them against independent evidence. Our construction draws on the text-as-data framework \citep{gentzkow2019text} and its applications to economic and political risks \citep{Baker2016-ni, Hassan2019-rf, Caldara2022-fy, Hassan2024-tx}, as well as the growing use of large language models for data construction in economics \citep{clayton2025geoeconomic, Dell2025-qj, asirvatham2026gpt}. As \citet{Ludwig2026-llm} emphasize, such measurement requires a validation sample, and our external benchmarks and human audit serve that role.

\textit{Economic Growth Determinants}. We contribute to the literature on cross-country growth determinants \citep{Levine1992-wi, barro1996determinants, Durlauf2005-yc}. Existing research has identified fundamental drivers internal to countries---institutions \citep{Acemoglu2001-io, Dell2010-go}, human capital \citep{Barro1991-bi, Mankiw1992-op}, culture \citep{Nunn2008-zl, Tabellini2010-ib}, and geography \citep{gallup1999geography, Dell2012-sa}---as well as external economic integration through trade \citep{Frankel1999-tp, Alcala2004-bl} and cross-border technology diffusion \citep{Parente1994-bt, Coe1995-rd}. Yet the role of countries' external political environment remains underexplored. We measure this environment and, building on the empirical framework of \citet{Acemoglu2019-bo}, use local projections \citep{Jorda2005-te} to estimate its dynamic relationship with growth.

\textit{Geoeconomics}. Our research bridges classical economic statecraft \citep{hirschman1945national, baldwin1985economic} with contemporary geoeconomic analysis \citep{clayton2024theory, clayton2023framework, NBERw32638, Kleinman2024-yl, Gopinath2025-kw, liu2025international}. This literature has studied economic coercion and the fragmentation of trade and investment, but evidence linking geopolitical relations to aggregate growth remains limited. Our measure places aid \citep{Alesina2000-ad, Nunn2014-food}, sanctions \citep{Morgan2023-oq}, and wars \citep{Martin2008-vf, FederleEtAl2026PriceOfWar} on a unified cooperation--conflict scale. A companion paper \citep{fan2025barriers} extends the database to all country pairs to study how geopolitical relations shape trade globalization.

\paragraph{Road Map} Section~\ref{s:measure_geo_relation} introduces and validates our event-based measure of geopolitical alignment. Section~\ref{s:dyn_growth_effects} estimates the dynamic relationship between the index and GDP per capita, assesses its robustness, and discusses its interpretation. Section~\ref{s:correlates} examines accompanying changes in growth fundamentals, the relationship with democracy, and the implied GDP magnitudes. Section~\ref{s:add_robust} reports additional robustness checks and comparisons with alternative measures. Section~\ref{s:conclusion} concludes.

\section{Measuring Geopolitical Alignment and the External Environment}\label{s:measure_geo_relation}

We use a large language model with web search to compile and score 373,020 geopolitical events from 1960 to 2024 involving at least one of 24 major nations.\footnote{In the companion paper, \citet{fan2025barriers}, we extend the database to all bilateral pairs among the 193 UN member states over 1950--2024, yielding 833,485 events with complete CAMEO and Goldstein coding.} These events cover all 193 UN member states through their relations with the 24 major nations, denoted $\mathcal{N}$.\footnote{The 24 major nations are Argentina, Australia, Belgium, Brazil, Canada, Switzerland, People's Republic of China, Germany, Denmark, Spain, France, United Kingdom, Indonesia, India, Italy, Japan, Republic of Korea, Mexico, Netherlands, Poland, Russian Federation (Soviet Union), Saudi Arabia, Turkey, and United States. Appendix \ref{app_a:major_nations} discusses these choices. Collectively, these 24 nations account for 82--90 percent of global GDP over the study period.} Using these events, we construct bilateral geopolitical alignment scores that vary by country pair and year, then aggregate them into a country-level geopolitical alignment index. The resulting scores track well-known historical episodes, agree with externally documented events and a blind human audit, and co-move with bilateral political and economic activity such as trade, aid, and sanctions (Section~\ref{ss:data_validation}).

\subsection{A Global Database of Geopolitical Events, 1960--2024} \label{ss:llm}

Bilateral geopolitical events---diplomatic agreements, economic sanctions, military actions, and territorial disputes---are prominent features of the historical record, extensively documented across news archives, diplomatic records, government publications, and scholarly databases. Unlike granular economic transactions or social interactions that may leave limited historical traces, these events are salient and widely corroborated across independent sources, making them well-suited for systematic extraction by large language models (LLMs) trained on such corpora. We use Gemini 2.5 Pro, an LLM with web search, to collect, verify, and analyze bilateral geopolitical events. Figure~\ref{fig:geopolitical_analysis} illustrates our LLM-based analysis procedure. Appendices~\ref{app_a:geo_events} and~\ref{app_c:llm} provide the complete analytical framework and prompt specifications.

\begin{figure}[ht]
\centering
\begin{tikzpicture}[
  startstop/.style={rectangle, rounded corners, minimum height=0.6cm, text width=1.8cm, text centered, draw=black, fill=red!10, font=\scriptsize},
  process/.style={rectangle, minimum height=0.6cm, text width=1.8cm, text centered, draw=black, fill=blue!10, font=\scriptsize},
  arrow/.style={thick, -Stealth},
  every node/.style={inner sep=1pt}
]

\node[startstop] (start) at (0,0) {Input \{countries, year\}};
\node[process, right=0.4cm of start] (verify) {Verify Political Entities};
\node[process, right=0.4cm of verify] (search) {Compile Geopolitical Events};
\node[process, right=0.4cm of search] (cameo) {Assign CAMEO Codes};
\node[process, right=0.4cm of cameo] (goldstein) {Estimate Goldstein Score};

\draw[arrow] (start) -- (verify);
\draw[arrow] (verify) -- (search);
\draw[arrow] (search) -- (cameo);
\draw[arrow] (cameo) -- (goldstein);

\end{tikzpicture}
\caption{LLM Geopolitical Event Analysis Procedure}
\label{fig:geopolitical_analysis}
\note{For each country pair and year, the model verifies the political entities (for example, the Soviet Union for Russia before 1992), compiles the year's bilateral events, assigns each event its CAMEO codes, and scores it on the Goldstein scale from $-10$ to $+10$. Appendix~\ref{app_c:llm} reports the prompt.}
\end{figure}

The prompt instructs the LLM to perform four sequential tasks: (i) verify the historical political entities for each country pair and year, accounting for state succession (e.g., Soviet Union to Russian Federation); (ii) search its knowledge base and the internet to identify bilateral geopolitical events from authoritative sources, spanning six event categories: economic relations, diplomatic and political relations, security and defense, legal and territorial issues, multilateral governance, and other bilateral interactions (Table~\ref{tab:event_categories} in Appendix~\ref{app_a:geo_events} lists the complete taxonomy); (iii) classify each event using the CAMEO framework; and (iv) assign a Goldstein score ranging from $-10$ (maximum conflict) to $+10$ (maximum cooperation) based on event intensity (Appendix~\ref{app_c:prompt} adds an overall relationship assessment that the scores do not use).\footnote{Appendix~\ref{app_c:cameo} describes the coding conventions and anchored cases that the LLM uses with the CAMEO and Goldstein frameworks to calibrate each score to the event's context and severity.} The design thus separates two stages: an ex ante taxonomy fixed in the prompt determines \textit{which} events enter the database, while the CAMEO and Goldstein frameworks guide \textit{how} selected events are classified and scored. This separation matters for the validation below: what distinguishes the resulting database from machine-coded event databases is the first stage, the selection of consequential bilateral events.

\begin{table}[ht]
\centering
\caption{US--Russia Bilateral Geopolitical Events in 2022: LLM Analysis Results}
\label{tab:us_russia_2022}
\footnotesize
\begin{tabular}{@{}p{2.5cm}p{5.6cm}p{1.3cm}p{1.75cm}p{1.45cm}@{}}
\toprule
\raggedright\textbf{Event Name} & \raggedright\textbf{Event Description} & \raggedright\textbf{CAMEO Class.} & \raggedright\textbf{Category} & \textbf{Goldstein Score} \\
\midrule

\raggedright US Military Assistance to Ukraine &
\raggedright US committed billions in security assistance through Presidential Drawdown Authority and Ukraine Security Assistance Initiative &
\raggedright Material Conflict (17-170) &
\raggedright Security \& Defense &
$-8.0$ \\
\addlinespace[0.3em]

\raggedright Sweeping Sanctions on Russia &
\raggedright Following Russia's invasion of Ukraine, US coordinated with G7 and EU to impose extensive sanctions &
\raggedright Material Conflict (16-163) &
\raggedright Economic &
$-7.5$ \\
\addlinespace[0.3em]

\raggedright US Leads Russia's Suspension from UN Human Rights Council &
\raggedright Following reports of civilian deaths in Bucha, US led successful campaign to suspend Russia from UN Human Rights Council by 93--24 vote &
\raggedright Material Conflict (16-166) &
\raggedright Multilateral &
$-6.5$ \\
\addlinespace[0.3em]

\raggedright Biden Labels Putin a ``War Criminal'' &
\raggedright President Biden publicly called President Putin a ``war criminal'' for actions during the Ukraine invasion, a label the Kremlin called ``unacceptable and unforgivable'' &
\raggedright Verbal Conflict (11-112) &
\raggedright Diplomatic \& Political &
$-5.5$ \\
\addlinespace[0.3em]

\raggedright US Leads Diplomatic Condemnation at UN &
\raggedright US led effort resulting in UN General Assembly Resolution ES-11/1, which passed 141--5 deploring Russia's aggression and demanding immediate withdrawal from Ukraine &
\raggedright Verbal Conflict (11-113) &
\raggedright Multilateral &
$-4.5$ \\
\addlinespace[0.3em]

\raggedright Griner--Bout Prisoner Exchange &
\raggedright US and Russia conducted prisoner swap in Abu Dhabi, exchanging Russian arms dealer Viktor Bout for American basketball player Brittney Griner after months of negotiations &
\raggedright Material Coop. (08-084) &
\raggedright Legal \& Territorial Issues &
$+7.0$ \\

\bottomrule
\end{tabular}
\note{The table lists all six events the model records for the US--Russia pair in 2022, ordered from most conflictual to most cooperative. Event names and descriptions are abridged from the model output. The CAMEO column gives the quad class with the root and event codes (Appendix~\ref{app_c:cameo}), and the category column gives the event category (Table~\ref{tab:event_categories}). Goldstein scores range from $-10$ (most conflictual) to $+10$ (most cooperative).}
\end{table}

{\clubpenalty=10000
Table~\ref{tab:us_russia_2022} shows the coded US--Russia bilateral events for 2022, the year Russia launched its full-scale war on Ukraine. Five of the six events represent conflict, with Goldstein scores ranging from $-4.5$ (UN General Assembly condemnation) to $-8.0$ (US military assistance to Ukraine). Events that involve third countries are scored by their implications for the focal pair: US military assistance to Ukraine, for example, enters as a conflictual US action toward Russia. The sole cooperative event---the Griner--Bout prisoner exchange---scores $+7.0$. Table~\ref{tab:us_soviet_1972} in Appendix~\ref{app_a:geo_events} provides a contrasting example from the 1972 d\'{e}tente period, when all nine US--Soviet events are cooperative with scores ranging from $+5.0$ to $+9.0$.\par}

\paragraph{Coverage and Descriptive Statistics}\label{ss:coverage}
\begin{table}[ht] 
\centering 
\caption{Geopolitical Events Summary by Decade, 1960--2024} 
\label{tab:geopolitical_short_summary} 
\footnotesize 
\begin{tabular}{lcccccccc} 
\toprule 
& 1960s & 1970s & 1980s & 1990s & 2000s & 2010s & 2020s & Total \\ 
\midrule 
Cooperation Events & 22,011 & 26,872 & 29,657 & 40,537 & 59,230 & 82,878 & 48,545 & 309,730 \\
Conflict Events & 7,278 & 7,353 & 9,390 & 7,989 & 9,649 & 13,297 & 8,334 & 63,290 \\
\addlinespace[0.1cm] 
Mean Goldstein Score & 2.86 & 3.28 & 2.92 & 3.98 & 4.09 & 3.88 & 3.73 & 3.67 \\ 
\bottomrule 
\end{tabular} 
\begin{tablenotes} 
\footnotesize 
\item \textit{Notes:} The table covers the 373,020 events of 1960--2024 in country pairs involving at least one of the 24 major nations. Cooperation events include verbal and material cooperation (CAMEO classes 1--2). Conflict events include verbal and material conflict (CAMEO classes 3--4). Goldstein scores range from $-10$ (most conflictual) to $+10$ (most cooperative). The 2020s column covers 2020--2024.
\end{tablenotes} 
\end{table}

Table~\ref{tab:geopolitical_short_summary} summarizes the 373,020 events of 1960--2024 by decade. Cooperation accounts for 75--79 percent of events in the 1960s--1980s and 84--86 percent in each subsequent decade. Mean Goldstein scores rise from 2.86--3.28 in the first three decades to 3.98--4.09 in the 1990s and 2000s, then decline to 3.88 in the 2010s, as lower-scored verbal cooperation gains share, and to 3.73 over 2020--2024, as the conflict share also rises. Appendix~\ref{app_a:geo_events_stats} provides detailed statistics on event classification and temporal patterns.\footnote{The event record becomes denser over time: the share of major-nation dyad-years with at least one recorded event rises from 32 percent in the 1950s to 80 percent in the 2020s. Appendix~\ref{app_a:coverage_diagnostics} documents this coverage gradient. Section~\ref{ss:data_validation} benchmarks detection against externally documented events.}

\subsection{Bilateral Geopolitical Alignment Scores} \label{ss:measure_geo_relations}

\paragraph{Dynamic Bilateral Geopolitical Alignment} 
Let $\left\{s^{n}_{ij,t}\right\}_{n=1}^{\tilde{N}_{ij,t}}$ denote the Goldstein scores for the set of events between countries $i$ and $j$ in year $t$, where $\tilde{N}_{ij,t}$ is the total number of events. The annual event score, normalized to the $[-1,+1]$ interval, is:

\begin{equation*}
\tilde{S}_{ij,t}=\frac{1}{\tilde{N}_{ij,t}} \sum_{n=1}^{\tilde{N}_{ij,t}} s_{ij,t}^n / 10
\end{equation*}

To capture both the immediate impact of political events and the institutional memory that characterizes international relationships, we then construct a dynamic geopolitical alignment score for each country pair as a weighted moving average:
\begin{align}
    S_{ij,t}&=\left(1-\phi_{ij,t}\right) \cdot S_{ij,t-1}+\phi_{ij,t} \cdot \tilde{S}_{ij,t} \label{eq:dynamic_score}\\
    \phi_{ij,t} &= \tilde{N}_{ij,t}/N_{ij,t},\quad N_{ij,t} = \left(1 - \delta \right)N_{ij,t-1} + \tilde{N}_{ij,t} \nonumber
\end{align}
where $\phi_{ij,t}$ represents the updating weight, $N_{ij,t}$ is the effective cumulative number of events, and $\delta$ is the depreciation rate for past observations. We set $\delta=0.3$, which gives the weight on past events a half-life of about two years and allows the measure to capture the persistence consistent with a typical four-year electoral cycle.\footnote{When no events are observed ($\tilde{N}_{ij,t} = 0$), we let both the effective event count and the score decay toward zero: $N_{ij,t} = (1-\delta)N_{ij,t-1}$ and $S_{ij,t} = (1-\delta)S_{ij,t-1}$. For the initial year of each bilateral pair, $S_{ij,t_0} = \tilde{S}_{ij,t_0}$ and $N_{ij,t_0} = \tilde{N}_{ij,t_0}$. Section~\ref{ss:irf_geo_shocks} considers $\delta = 1$ (an unsmoothed index). Holding the score unchanged in years without events instead gives a similar GDP path and a 25-year sustained-path coefficient of 10.0 log points (Appendix Figure~\ref{fig:carry_forward}). Because the local projections include lags of geopolitical alignment, they use current alignment variation after conditioning on its lagged history.} The dynamic geopolitical alignment score is normalized to range from $-1$ (maximum conflict) to $+1$ (maximum cooperation). Throughout the paper, we refer to $S_{ij,t}$ as the \textit{bilateral geopolitical alignment score} and to the GDP-weighted aggregate $p_{ct}$ (defined below) as the \textit{geopolitical alignment index}. Appendix~\ref{app_a:dyn_geo_relations} examines the dynamic properties of both measures.

\begin{figure}[htb]
    \centering
    \includegraphics[width = \linewidth]{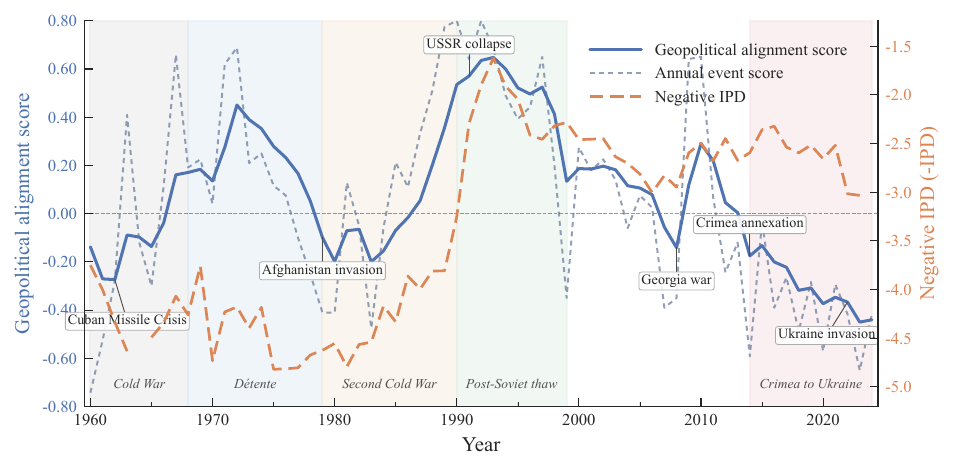}
    \caption{Geopolitical Alignment Scores Between the United States and Russia (Soviet Union)}
    \label{fig:geo_score_USA_RUS}
    \note{Time-series comparison of bilateral geopolitical alignment measures between the United States and Russia, 1960--2024. The blue solid line shows our dynamic geopolitical alignment score (left axis), the light blue dashed line shows the annual (unsmoothed) event score, and the orange dashed line displays the negative Ideal Point Distance ($-$IPD) from UN voting data (right axis). Shaded regions indicate five geopolitical eras: Cold War (1960--1968), D\'{e}tente (1968--1979), Second Cold War (1979--1990), Post-Soviet thaw (1990--1999), and Crimea to Ukraine (2014--2024). Key geopolitical events are annotated.}
\end{figure}

\begin{figure}[htb]
    \centering
    \begin{subfigure}[b]{0.48\textwidth}
        \includegraphics[width=\textwidth]{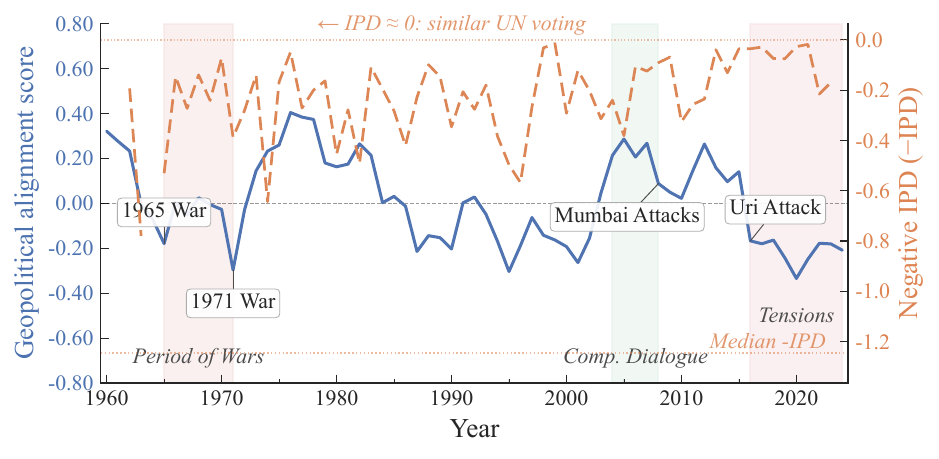}
        \caption{India and Pakistan}
        \label{fig:geo_score_IND_PAK_main}
    \end{subfigure}
    \hfill
    \begin{subfigure}[b]{0.48\textwidth}
        \includegraphics[width=\textwidth]{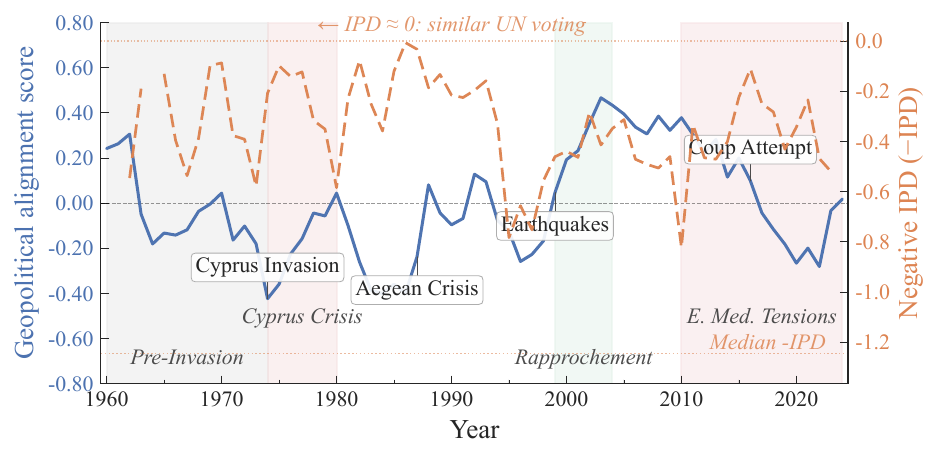}
        \caption{Greece and Turkey}
        \label{fig:geo_score_GRC_TUR_main}
    \end{subfigure}
    \caption{Geopolitical Alignment Scores: India--Pakistan and Greece--Turkey}
    \label{fig:geo_score_main_cases}
    \note{Time-series comparison of bilateral geopolitical alignment measures for India--Pakistan (panel~(a)) and Greece--Turkey (panel~(b)), 1960--2024. Blue solid lines show our dynamic geopolitical alignment score (left axis). Orange dashed lines display the negative Ideal Point Distance ($-$IPD) from UN voting data (right axis). The dotted horizontal line marks the pooled sample median of negative IPD. Shaded regions highlight key periods of conflict and cooperation. Boxed labels mark major bilateral events.}
\end{figure}

Figure~\ref{fig:geo_score_USA_RUS} shows the dynamic score for a familiar bilateral relationship. The measure tracks major historical episodes: the Cuban Missile Crisis marking peak Cold War tensions, the d\'{e}tente period (1968--1979), deterioration following the Soviet invasion of Afghanistan, improvement beginning with Gorbachev's reforms in the mid-1980s, the post-Soviet peak in 1993, and subsequent decline culminating in the Crimean crisis. The 2014--2024 period records the lowest values in the sample. The Ideal Point Distance (IPD) measure from UN voting data (orange dashed line) is less responsive to these bilateral shifts: it changes little during d\'{e}tente and after the Crimean crisis despite the sustained deterioration in bilateral relations. Figure~\ref{fig:geo_score_main_cases} provides two further historical illustrations involving non-superpower relationships.

Panel~(a) of Figure~\ref{fig:geo_score_main_cases} traces India--Pakistan relations through six decades of recurrent conflict. Our measure tracks the wars of 1965 and 1971, the Mumbai terrorist attacks (2008), and the Uri attack and subsequent surgical strikes (2016), with scores falling from $+0.32$ in 1960 to $-0.30$ during the 1971 war. The measure captures repeated cycles of conflict and cautious engagement: scores recover to $+0.40$ in the late 1970s, rise again during the Composite Dialogue period (2004--2008) to approximately $+0.30$, and collapse after the Mumbai attacks. The IPD measure, by contrast, shows these persistent rivals as increasingly aligned because both countries vote similarly on postcolonial issues at the UN.

Panel~(b) shows a similar pattern in Greece--Turkey relations: our measure registers the sharp deterioration following the 1974 Cyprus invasion, the 1987 Aegean crisis, the brief rapprochement after the 1999 earthquakes, and the sustained decline during Eastern Mediterranean tensions (2010--2024). The IPD measure shows these NATO allies as consistently close throughout, reflecting their similar voting patterns within the Western bloc despite bilateral disputes over Cyprus, the Aegean, and maritime boundaries. In both cases, negative IPD remains above the pooled sample median (dotted line): similar UN voting positions can coexist with substantial bilateral conflict.\footnote{\citet{fan2025barriers} show formally that UNGA voting similarity correlates with the low-dimensional bloc structure of our bilateral alignment scores---the distance over the first three principal components---but misses the higher-dimensional bilateral variation associated with economic outcomes.} Appendix~\ref{app_a:case_validation} provides additional historical illustrations for eight bilateral relationships, both those involving major nations (Figure~\ref{fig:appendix_major_powers}) and regional rivalries (Figure~\ref{fig:appendix_regional}). Geopolitical maps appear in Appendix~\ref{app_a:maps}.

\begin{figure}[ht]
    \centering
    \includegraphics[width = \linewidth]{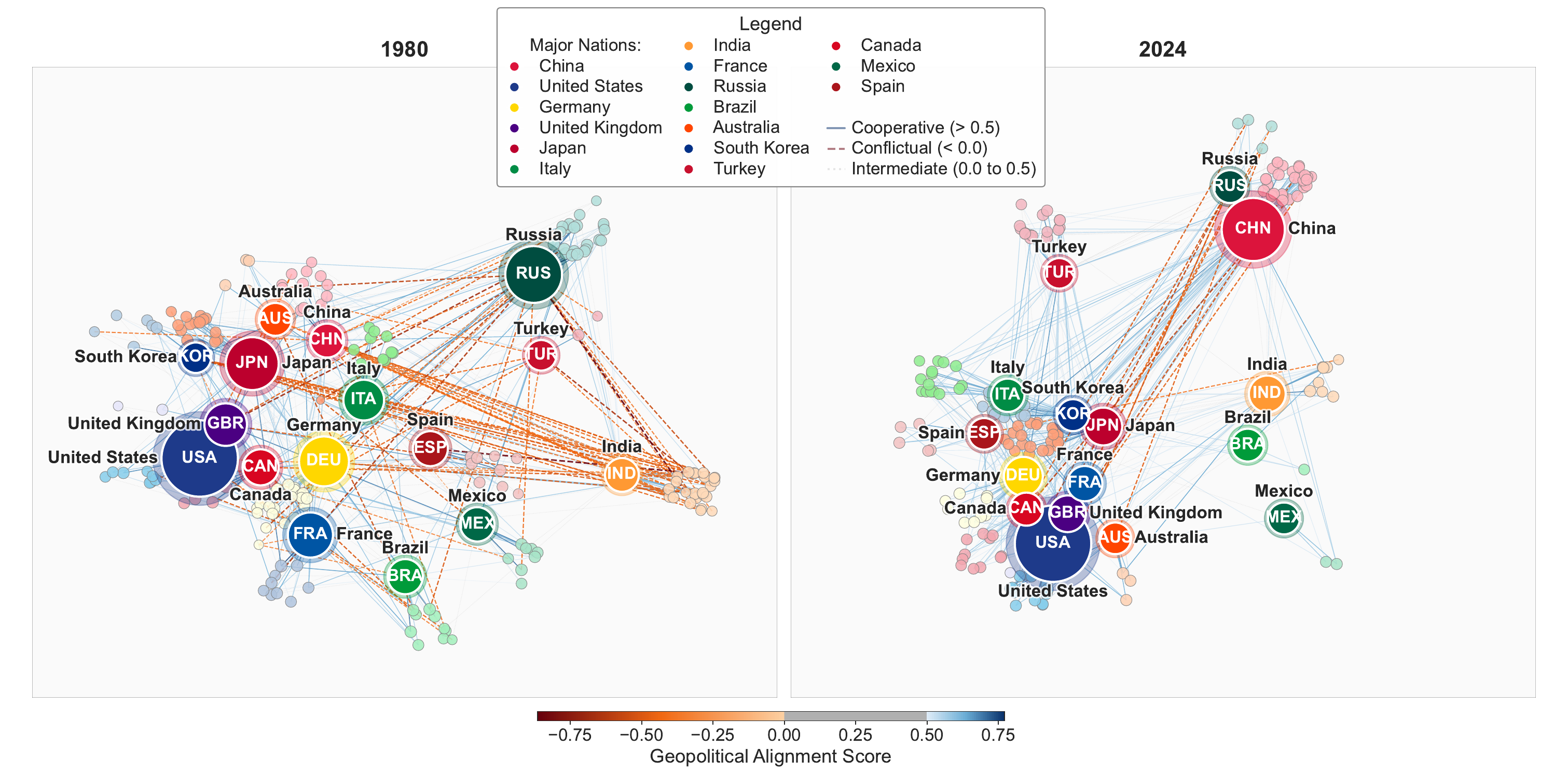}
    \caption{Geopolitical Alignment Topology, 1980 and 2024}
    \label{fig:topology_1980_2024}
    \note{Positions of 16 selected major nations come from multidimensional scaling of correlation distances between their bilateral-alignment profiles with all 193 countries. Closer positions indicate more similar global relationship profiles. Major-nation circle sizes scale with GDP shares. Smaller circles denote other countries, colored by the major nation with which they have the highest alignment score and positioned using their major-nation alignment profiles. Links show each smaller country's three largest major-nation relationships in absolute value: blue solid lines indicate scores above 0.5, red dashed lines scores below 0, and gray dotted lines scores from 0 to 0.5.}
\end{figure}

We also use the cross-sectional matrix of bilateral geopolitical alignment scores, $[S_{ij,t}]$, to characterize the topology of the international system. Figure~\ref{fig:topology_1980_2024} compares this topology in 1980 and 2024. In 1980, the United States and the Soviet Union anchor distinct clusters. China lies between them but closer to the Western cluster, consistent with the Sino-American rapprochement following Nixon's opening and shared opposition to Soviet expansion.

{\widowpenalty=10000
By 2024, this bipolar structure has given way to a more fragmented system. The United States, United Kingdom, Germany, France, Japan, Australia, and South Korea form a tightly integrated cluster, while China and Russia lie apart from it. India occupies an intermediate position, and ``connector'' states such as Turkey, Brazil, and Mexico bridge the clusters.\par}

\subsection{Validation} \label{ss:data_validation}\label{ss:score_validation}

Beyond the case studies in Section~\ref{ss:measure_geo_relations}, we assess the measure in five ways. We benchmark the event record against independently documented bilateral developments, compare it with machine-coded event databases, test stability across models and pipelines, conduct an independent human audit, and examine associations with bilateral economic outcomes. Section~\ref{ss:alt_geo_measures} separately compares alternative measures within the growth application.

\paragraph{Validation against Documented Events}
A first concern is that uneven historical source coverage may cause the database to miss developments in earlier periods or involving poorer countries \citep{fariss2014respect}. We benchmark the database, including the 1950s records that initialize the scores, against 4,670 bilateral developments over 1950--2023 from eight externally compiled event classes.\footnote{The four conflict classes are interstate war onsets, militarized interstate disputes involving force short of war, sanction impositions, and diplomatic severances. The four cooperative classes are sanction removals, diplomatic establishments, regional trade-agreement entries, and alliance formations. An event of the expected type has the expected event category or CAMEO root (a broad event-family match in Appendix~\ref{app_a:anchoring}).} Within the class-specific matching windows, our database records some event for 97.6 percent of these developments, an event in the expected conflict or cooperation direction for 92.1 percent, and an event of the expected type for 82.6 percent. Matching on event type does not vary systematically with country income and rises only mildly over time, with the increase concentrated in two of the eight event classes (Appendix Figure~\ref{fig:benchmark_match_gradients}). Matching also does not vary with the country's growth over the following decade, suggesting that the model does not record documented developments selectively by subsequent growth (Appendix Table~\ref{tab:benchmark_hindsight}).

We then examine whether the recorded intensity of bilateral relations moves with these documented developments. We regress the unsmoothed annual event score jointly on the eight documented-event indicators, absorbing dyad and country-by-year fixed effects. All eight conditional annual-score differences have the expected sign and seven are statistically distinguishable from zero, with the largest negative difference for war onsets and the largest positive difference for alliance formations. Appendix~\ref{app_a:anchoring} reports the detailed benchmark analysis.

\paragraph{Comparison with Machine-Coded Event Databases}
We next compare our database with three machine-coded event databases: ICEWS \citep{boschee2015icews}, GDELT \citep{leetaru2013gdelt}, and the Cline Center Historical Phoenix database \citep{althaus2020phoenix}. These databases automatically extract events from large volumes of news reports and code them using the CAMEO framework. The comparison separates two questions: whether each event database records independently documented developments and how its annual score differs during documented conflict and cooperation years. Appendix~\ref{app_a:event_comparison} reports the detailed comparison.

\begin{table}[ht]
\centering
\caption{Detection and Conditional Annual-Score Differences across Event Databases}
\label{tab:external_validation}
\footnotesize
\setlength{\tabcolsep}{4pt}
\begin{tabular}{lrrrr}
\toprule
 & Our database & ICEWS & GDELT & Phoenix \\
\midrule
\multicolumn{5}{l}{\textit{Panel A. Documented events, 1995--2018 ($N=2{,}038$)}} \\
Mean events in documented-event year & 3.0 & 371.7 & 4,083.0 & 61.3 \\
Any event recorded & 0.982 & 0.907 & 0.980 & 0.844 \\
Event in expected direction & 0.917 & 0.869 & 0.972 & 0.794 \\
\addlinespace
\multicolumn{5}{l}{\textit{Panel B. Common observed dyad-years ($N=34{,}412$)}} \\
Documented conflict & $-1.467$ (0.179) & $-0.475$ (0.078) & $-0.120$ (0.054) & $-0.295$ (0.084) \\
Documented cooperation & $+0.475$ (0.112) & $+0.194$ (0.060) & $+0.033$ (0.051) & $-0.133$ (0.096) \\
\bottomrule
\end{tabular}
\note{Panel~A uses the 2,038 documented developments over 1995--2018 that all four databases cover after actor harmonization; detection uses the matching windows in Appendix Table~\ref{tab:benchmark_design}, and median events per documented-event year are 3, 34, 288.5, and 6, respectively. Panel~B regresses each database's unsmoothed annual mean Goldstein score, in raw points, on indicators for any documented conflict and any documented cooperation, with dyad and country-by-year fixed effects, on identical dyad-years; standard errors, clustered by dyad, are in parentheses. Appendix~\ref{app_a:event_comparison} reports the specification and complete results.}
\end{table}

Table~\ref{tab:external_validation} shows that, in the common 1995--2018 sample, our database records some event around externally documented developments as often as the other event databases, while containing only a fraction of their event volume. The key distinction lies in the sensitivity of the annual score to documented events: on identical dyad-years and under a common fixed-effects specification, our database exhibits the largest negative conditional difference during documented conflict and the largest positive difference during documented cooperation. These results suggest that selecting a few consequential events, instead of averaging over many routine reports, produces a more differentiated annual conflict--cooperation signal. Section~\ref{ss:alt_geo_measures} shows that these databases capture little growth-relevant variation under the same construction.

\paragraph{Reliability and Replicability}
We also test whether annual event scores are stable across models and repeated runs. Cross-model comparisons of Gemini 2.5 Pro, GPT~5.4, Claude Opus~4.6, and Claude Sonnet~4.6 on 300~randomly sampled dyad-years yield a pooled correlation of 0.88. Repeated runs of each model give a pooled within-model correlation of 0.93.

A separate exercise changes both the LLM and the prompt structure. An alternative LLM pipeline constructed primarily with Gemini 2.5 Flash (preview) and an earlier prompt produces a country-year index that correlates at 0.87 with the baseline. It also yields very similar GDP coefficient paths (Appendix~\ref{app_a:model_robustness}).

\paragraph{Human Audit}
Cross-model agreement does not establish accuracy because models may share errors in interpreting and scoring events. We therefore conduct a staged human audit of a stratified sample of 200 eventful dyad-years, with two coder assessments per case. Six trained research assistants from Peking University first reconstruct and score bilateral events without seeing model output or using generative AI. Like the model, they work on one dyad-year at a time and see only the country pair, the year, and the codebook the model follows. After locking these records, they review and score the events described by the model while its classifications and scores remain concealed. Neither the model nor the coders are given economic data, and both search only for events with geopolitical consequences for the pair. Because both work from the current historical record, we test for look-ahead bias using formal acts only (Section~\ref{ss:decomposition}).

Blind human and model annual scores correlate at 0.80, compared with 0.76 between human coders. For events that coders identified independently and that match a model event, Goldstein scores correlate at 0.92. After review, correlations are 0.92 for recoded model events and 0.89 for annual scores. The audit identifies three fabricated records among the 856 model-generated events, less than 0.5 percent. The audit thus supports the annual conflict--cooperation signal and points to event selection as an important source of the remaining differences. Appendix~\ref{app_a:human_audit} reports the detailed audit design and results.

\paragraph{Bilateral Outcomes}
Finally, the bilateral alignment score co-varies with bilateral economic outcomes. In standard gravity specifications, more cooperative relations are associated with greater bilateral trade, more official development assistance, and a lower probability of sanctions. These patterns persist in specifications with dyad fixed effects, which exploit within-pair changes over time, and, with smaller coefficients, after rebuilding the score without events directly linked to each outcome. We interpret them as evidence on what the score captures, not as estimates of causal channels. Appendix~\ref{app_a:bilateral_outcomes} reports the full results, including migration.

\subsection{The External Geopolitical Environment} \label{ss:external_env}

In international relations, ``alignment'' is a relational and directional concept: countries align \textit{with} particular partners and \textit{against} others. Our dyadic score respects this usage: it measures the cooperative-versus-conflictual state of one bilateral relationship. The country-level aggregate we now construct is a different object. It does not record which side a country has chosen. It summarizes the overall state of a country's relations with the major nations, what we call the \textit{external geopolitical environment} facing that country. The political form of an improvement in this environment differs across international systems: under Cold War bipolarity it often reflected side-switching or normalization with the dominant economic bloc, while in the post-Cold War system it more often reflects broad-based cooperative engagement with multiple poles. The index does not identify a universal foreign-policy strategy. It places these distinct political changes on a common scale according to the resulting state of relations with major nations.

We aggregate bilateral geopolitical alignment scores $S_{cj,t}$ into a country-level index using GDP weights:
\begin{equation}
p_{ct} = \sum_{j\in \mathcal{N}\setminus\{c\}} S_{cj,t} \times \text{GDP share}_{jt} \label{eq:country_geo}
\end{equation}

Here, $\text{GDP share}_{jt}$ denotes country $j$'s share of world nominal GDP in current US dollars. We interpret nominal GDP as a proxy for geopolitical influence. The resulting \emph{geopolitical alignment index} $p_{ct}$ is our primary measure of a country's external geopolitical environment. Higher values of $p_{ct}$ indicate more cooperative relations with major nations and hence a more favorable external environment. Lower values indicate more conflictual relations.

Why should this environment matter for growth? More cooperative relations with major nations may expand a country's \textit{access} to export markets and imported inputs, capital and technology, aid and official finance, and the security guarantees and diplomatic legitimacy that underpin long-horizon investment. More conflictual relations may increase \textit{risk} and exclusion through sanctions and embargoes, technology and financial restrictions, military threats, diplomatic isolation, and the uncertainty that accompanies all of these. The index summarizes bilateral conditions on both margins, not a single policy intervention. As a check on the conflict side of the index, Appendix~\ref{app_a:gpr_comparison} shows that the GDP-weighted conflict component of the underlying events co-moves with the Geopolitical Risk Index of \citet{Caldara2022-fy}, particularly during major military confrontations.

\paragraph{The Global Distribution, 1960--2024}\label{ss:landscape_geo_relations}
The cross-country distribution of the index evolves through three periods: Cold War bipolarity (1960--1991), post-Cold War convergence (1992--2009), and contemporary fragmentation (2010--2024). These periods provide descriptive context for the within-country changes analyzed in Section~\ref{s:dyn_growth_effects}.

\begin{figure}[ht]
    \centering
    \includegraphics[width = \linewidth]{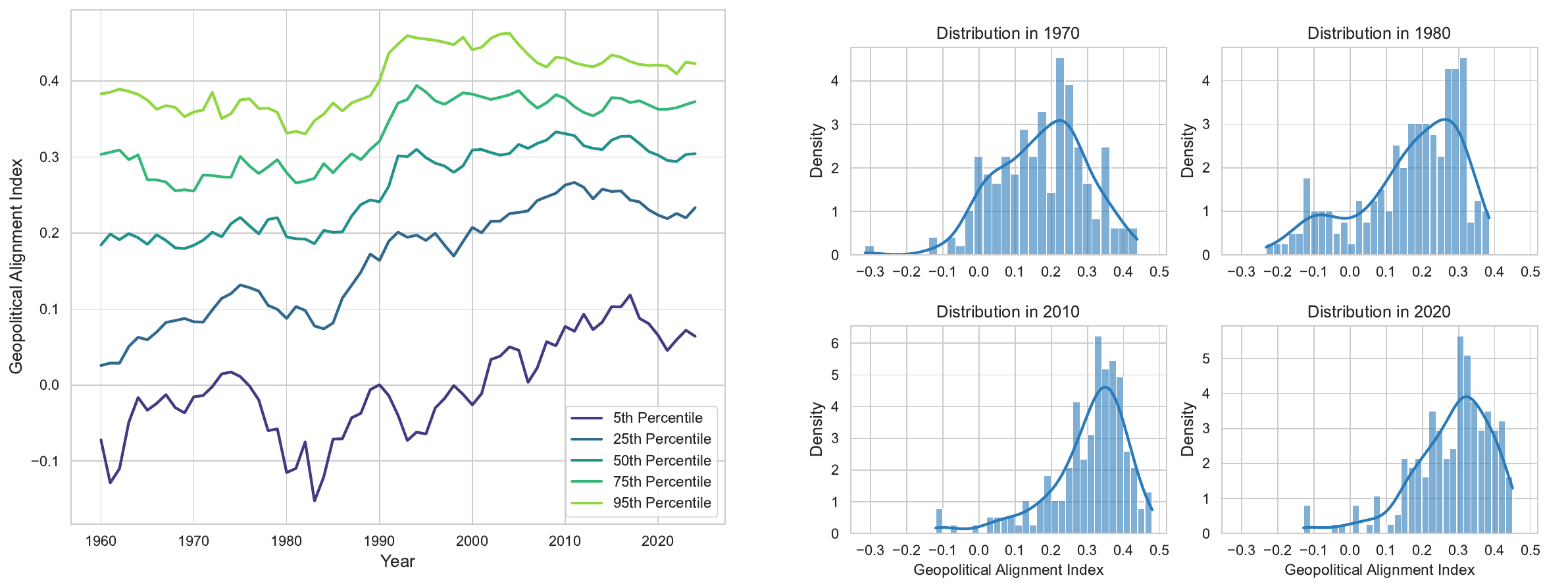}
    \caption{Evolution of the Distribution of the Geopolitical Alignment Index}
    \label{fig:evolution_distribution}
    \note{The left panel shows the 5th, 25th, 50th, 75th, and 95th percentiles of the country-level geopolitical alignment index from 1960 to 2024. The right panels show histograms and kernel-density estimates for 1970, 1980, 2010, and 2020, highlighting the shift from Cold War bipolarity through post-Cold War convergence to contemporary fragmentation.}
\end{figure}

Figure~\ref{fig:evolution_distribution} shows these shifts. During the Cold War era (1960--1991), the 5th percentile fluctuated between $-0.15$ and 0.02, reflecting the volatile position of countries caught in superpower competition, while upper percentiles remained relatively stable. The 1970 and 1980 distributions are left-skewed---with a primary peak around 0.2 and considerable mass extending into negative territory---capturing the division into opposing geopolitical camps.

During 1992--2009, the distribution moved toward broad convergence. By 2010, the 5th percentile had improved from $-0.04$ to 0.08, the median had risen from 0.30 to 0.33, and the distribution was concentrated tightly around 0.33 with minimal mass below 0.0. Over 2010--2024, the distribution widened modestly: the 5th percentile declined from 0.08 to 0.06, variance increased slightly, and the 2020 distribution shows a flatter peak with a left tail extending below 0.0.\footnote{The 5th--95th percentile spread narrowed from roughly 0.51 (ranging from $-0.13$ to 0.39) during peak Cold War to 0.35 (0.08 to 0.43) by 2010, before widening again to 0.36 (0.06 to 0.42) by 2024.} The measure is constructed without reference to these periods.\footnote{Summary statistics by decade, including means and standard deviations, appear in Table~\ref{tab:geo_relation_summary_stats} (Appendix~\ref{app_a:geo_relations}).}

\begin{table}[!htb]
\centering
\caption{Historical Realignments on the Index Scale}
\label{tab:episodes_1sd}
\footnotesize
\setlength{\tabcolsep}{4pt}
\resizebox{\linewidth}{!}{%
\begin{tabular}{lclc}
\toprule
Benchmark or episode & $\Delta$ index (SD) & Benchmark or episode & $\Delta$ index (SD) \\
\midrule
Interquartile shift (25th$\to$75th) & 0.186 (1.32) & Iraq: Gulf War and sanctions (1989--94) & $-0.58$ ($-4.1$) \\
South Africa: apartheid ends (1986--96) & $+0.69$ ($+4.9$) & Vietnam: normalization with the US (1990--2000) & $+0.19$ ($+1.4$) \\
Iran: Islamic Revolution (1977--83) & $-0.48$ ($-3.4$) & Russia: deterioration through Crimea (2005--15) & $-0.14$ ($-1.0$) \\
Poland: post-communist transition (1988--98) & $+0.27$ ($+1.9$) & Indonesia: post-Suharto reorientation (1996--2004) & $+0.13$ ($+1.0$) \\
China: reform and opening (1976--86) & $+0.15$ ($+1.1$) & Egypt: Camp David-era realignment (1973--81) & $+0.05$ ($+0.3$) \\
\bottomrule
\end{tabular}}
\note{$\Delta$ is the end-year minus start-year value for each selected window. Parentheses report the change in standard deviations, using the full-sample pooled $\sigma=0.141$. The interquartile benchmark is the shift from the 25th percentile (0.154) to the 75th percentile (0.340). Episode labels provide historical context, and examples are chosen for recognizability rather than as a ranking.}
\end{table}

To benchmark magnitudes, Table~\ref{tab:episodes_1sd} reports changes between selected historical endpoints in index units and standard deviations. One pooled-sample standard deviation is 0.141. An interquartile move is 0.186 units, or 1.32 standard deviations. Russia's 2005--2015 deterioration and Indonesia's 1996--2004 improvement are each about one standard deviation, while the changes in South Africa, Iran, and Iraq span several standard deviations.

\section{Geopolitical Alignment and Economic Growth: Dynamic Estimates} \label{s:dyn_growth_effects}

Improvements in a country's external geopolitical environment, measured by $p_{ct}$, are followed by a hump-shaped rise in GDP per capita that peaks about a decade later. This section presents these local-projection estimates and their robustness, and probes the causal interpretation of the dynamic relationship.

\subsection{Economic Data and Sample}

We use a panel of 193 countries from 1960 to 2024, of which 184 have GDP data from the World Development Indicators. The main estimates use a long-coverage sample of 148 countries with at least 40 years of GDP data. We report the full available unbalanced panel alongside the baseline. The primary outcome variable is log GDP per capita in constant US dollars from the World Development Indicators, which provides broad country coverage. We build upon the dataset constructed by \citet{Acemoglu2019-bo}, expanding coverage and incorporating additional variables from the Penn World Table \citep{Feenstra2015-vy} and \citet{Acemoglu2025-lv}.\footnote{Table~\ref{tab:data_description} in Appendix~\ref{app_b:econ_data} provides detailed information on country coverage and data availability for each variable.}

We also use growth determinants from several categories: augmented Solow fundamentals (physical capital, investment rates, population growth, education), institutional measures (democracy indices, governance quality), trade openness, and additional correlates identified in the growth literature. These variables allow us to examine whether changes in the external geopolitical environment are accompanied by changes in growth fundamentals.

\paragraph{Distinctiveness from Economic Fundamentals}
Table~\ref{tab:geo_vs_gdp_fundamentals} in Appendix~\ref{app_b:fundamentals} reports within-year cross-sectional correlations between the geopolitical alignment index and standard economic fundamentals. These correlations are small and unstable across decades, typically below 0.2 in absolute value. In contrast, human capital, capital stock, and investment exhibit strong and stable correlations with GDP per capita.

\subsection{Empirical Specification and Main Estimates} \label{ss:baseline}

We estimate how GDP per capita evolves at subsequent horizons following changes in the geopolitical alignment index. For each horizon $h$, we estimate the following local projection (LP) \citep{Jorda2005-te}:
\begin{equation}
y_{c,t+h} = \alpha_h^{\text{LP}} p_{ct} + \gamma_h^{\prime} \mathbf{x}_{ct} + \varepsilon_{c,t+h}, \quad h = 0, 1, \ldots, H
\label{eq:lp_spec}
\end{equation}
where $y_{c,t+h}$ is log GDP per capita ($\times$ 100), $p_{ct}$ is the geopolitical alignment index, and $\mathbf{x}_{ct}$ is a vector of controls. The coefficient $\alpha_h^{\text{LP}}$ is the horizon-specific population linear-projection coefficient on current alignment. Throughout, we report $\alpha_h^{\text{LP}}$ and all other coefficients on the index per one standard deviation of the index (0.141), so that each summarizes the conditional change in future GDP per capita associated with a one-standard-deviation increase in the index, holding fixed the included controls.\footnote{Coefficients are in log points (GDP per capita $\times$ 100) per standard deviation of the index, the pooled standard deviation across all 12,545 country-years. Estimation uses the index in its natural units, and the reported coefficients and standard errors are multiplied by 0.141.} The control vector is
\[
\mathbf{x}_{ct} = \left\{ \{y_{c,t-\ell}, p_{c,t-\ell}\}_{\ell=1}^{J}, \eta_c, \eta_{r(c)t}, \boldsymbol{m}_{ct} \right\},
\]
where $y_{c,t-\ell}$ and $p_{c,t-\ell}$ are lagged values of GDP and geopolitical alignment, respectively, $\eta_c$ denotes country fixed effects, $\eta_{r(c)t}$ represents region-year fixed effects, and $\boldsymbol{m}_{ct}$ includes time-varying country-specific controls employed in robustness checks. In our baseline specification, we include four lags ($J = 4$) of both GDP and geopolitical alignment, following \citet{Acemoglu2019-bo}, to capture growth dynamics and support inference in the presence of serial correlation \citep{Montiel-Olea2021-oo}.

\begin{figure}[ht]
    \centering
    \begin{subfigure}[b]{0.48\textwidth}
        \includegraphics[width=\textwidth]{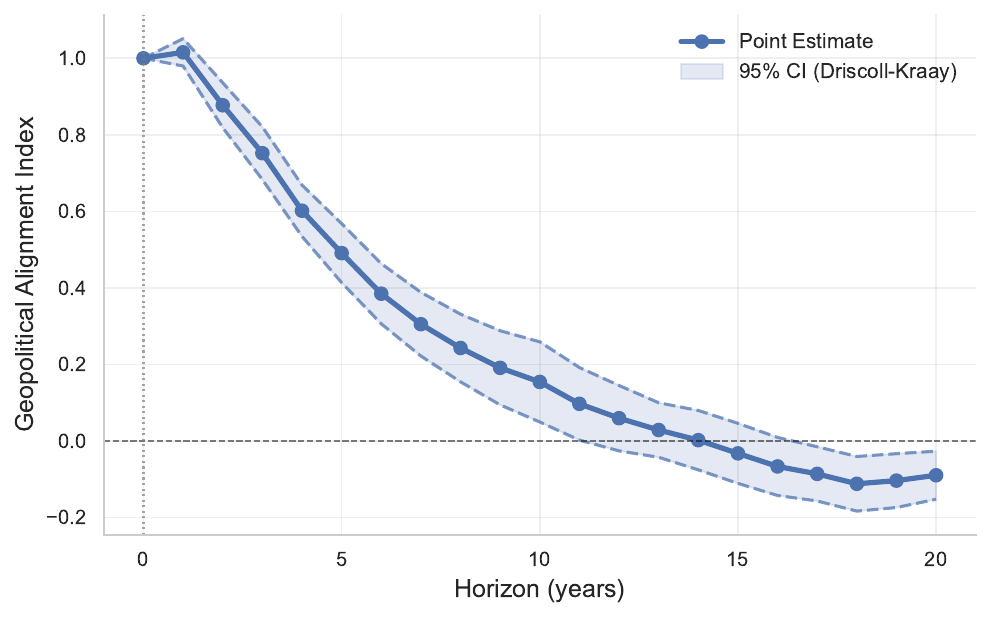}
        \caption{Persistence of Geopolitical Alignment}
        \label{fig:irf_geo_relation}
    \end{subfigure}
    \hfill
    \begin{subfigure}[b]{0.48\textwidth}
        \includegraphics[width=\textwidth]{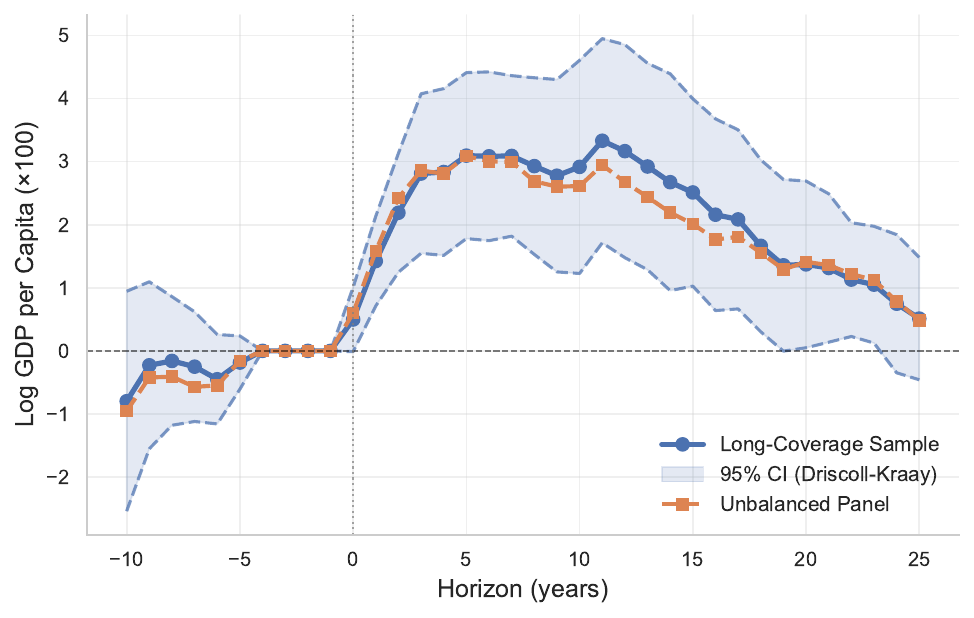}
        \caption{GDP per Capita}
        \label{fig:irf_gdp_pc}
    \end{subfigure}
    \caption{Dynamic Local-Projection Estimates}
    \label{fig:baseline_irf}
    \note{Panel (a) reports coefficients from projecting the geopolitical alignment index at $t+h$ on its current value, conditional on the baseline controls. The coefficient at $h=0$ equals one by construction. Panel (b) reports coefficients from projecting log GDP per capita ($\times 100$) at $t+h$ on current geopolitical alignment, per one-standard-deviation change in the index. The solid blue line uses the long-coverage sample of 148 countries with at least 40 years of GDP data. The dashed orange line uses the full unbalanced panel. Both panels include four lags of GDP and geopolitical alignment, country fixed effects, and region-year fixed effects. Panel (a) spans $h=0,\ldots,20$. Panel (b) spans $h=-10,\ldots,25$, with horizons $h=-10,\ldots,-5$ providing timing diagnostics. Coefficients at $h=-4,\ldots,-1$ are zero by construction because the corresponding lags of GDP enter the controls. Blue shaded areas show 95\% confidence intervals based on Driscoll--Kraay standard errors \citep{Driscoll1998-oh}.}
\end{figure}

Country fixed effects $\eta_c$ account for time-invariant heterogeneity across countries, so the coefficient is estimated from within-country variation over time \citep{Durlauf2005-yc}. Region-year fixed effects $\eta_{r(c)t}$ control for regional and global shocks, including shared temporal dynamics such as regional conflicts, financial crises, and globalization waves.\footnote{Our estimates are robust to using only year fixed effects, as shown in Section~\ref{ss:robustness}.}
Section~\ref{ss:interpretation} discusses the assumptions under which these conditional dynamic responses admit a causal interpretation.
Figure~\ref{fig:baseline_irf} reports horizon-specific coefficients from local projections
of future geopolitical alignment and GDP per capita
on current alignment.

\paragraph{Dynamics of Geopolitical Alignment and Economic Growth}

To characterize the conditional dynamics of the index, we estimate the same local-projection specification with future geopolitical alignment, $p_{c,t+h}$, as the dependent variable. Panel (a) of Figure~\ref{fig:baseline_irf} shows that alignment is persistent but mean-reverting: the coefficient declines from one at $h=0$ to 0.60 at year 4 and 0.15 at year 10. It reaches zero by year 14 and turns slightly negative thereafter.\footnote{Appendix~\ref{app_a:dyn_geo_relations} compares this persistence with that of the unsmoothed index and of the bilateral scores.} Accordingly, the estimated GDP dynamics reflect the response both to the initial improvement in alignment and to the index's subsequent changes.

From $h=0$ onward, the estimated GDP response in panel (b) is positive and hump-shaped. The coefficient rises from 0.5 log points at $h=0$ to 3.3 log points in year 11, then declines to 1.4 log points by year 20. Estimates from the 148-country long-coverage sample and the full unbalanced panel closely overlap. Subsequent specifications use the long-coverage sample unless otherwise noted. The response does not depend on shifts in partner GDP shares: fixing partner weights over time yields similar estimates, as do regime-fixed weights, equal weights, and weights based on military capability across the 24 major nations (Appendix Figure~\ref{fig:weighting_robustness}).\footnote{Appendix~\ref{app_b:add_baseline} also reports bootstrap-based inference and robustness to alternative lag specifications, and Appendix Figure~\ref{fig:fwl_binscatter} shows that the binned residual means are approximately linear over the observed range.}

Before impact, the coefficients are small and not individually distinguishable from zero, which provides one timing diagnostic. A reverse projection likewise shows no systematic relationship between current GDP and subsequent geopolitical alignment: no coefficient after impact is distinguishable from zero (Appendix Figure~\ref{fig:robustness_checks_main}, panel~(b)).

\paragraph{Transitory and Sustained Alignment Paths}

The conditional responses in Figure~\ref{fig:baseline_irf} combine a current change in geopolitical alignment with the subsequent dynamics of the index. To isolate the response associated with a purely transitory change, we follow \citet{Sims1986-mt} and \citet{bilal2024macroeconomic} and construct a counterfactual path that raises the index by one standard deviation on impact and returns the change to zero immediately thereafter. Cumulating the transitory coefficients gives the corresponding mapping for an increase maintained at one standard deviation at all subsequent horizons.\footnote{The mapping assumes linearity and that the response to a sequence of unanticipated innovations equals the response to the same alignment path anticipated at time zero. It adds no identifying variation beyond the baseline local projections. The constructed paths retain the same conditional interpretation. Appendix~\ref{app_b:irf_transitory_persistent} details the construction and assumptions.}

\begin{figure}[ht]
    \centering
    \begin{subfigure}[b]{0.48\textwidth}
        \includegraphics[width=\textwidth]{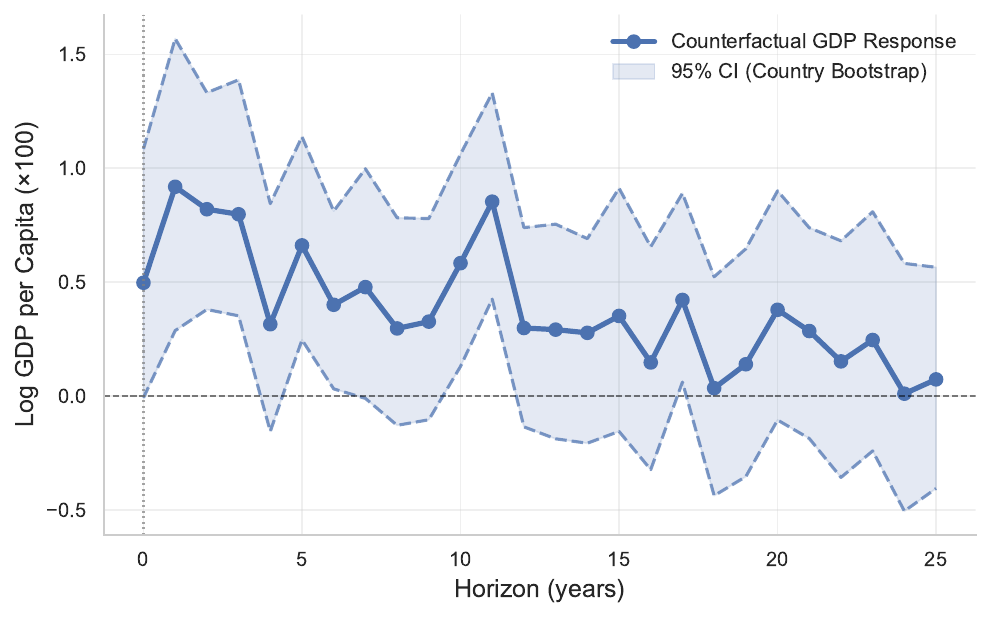}
        \caption{Transitory Alignment Path}
        \label{fig:irf_gdp_transitory}
    \end{subfigure}
    \hfill
    \begin{subfigure}[b]{0.48\textwidth}
        \includegraphics[width=\textwidth]{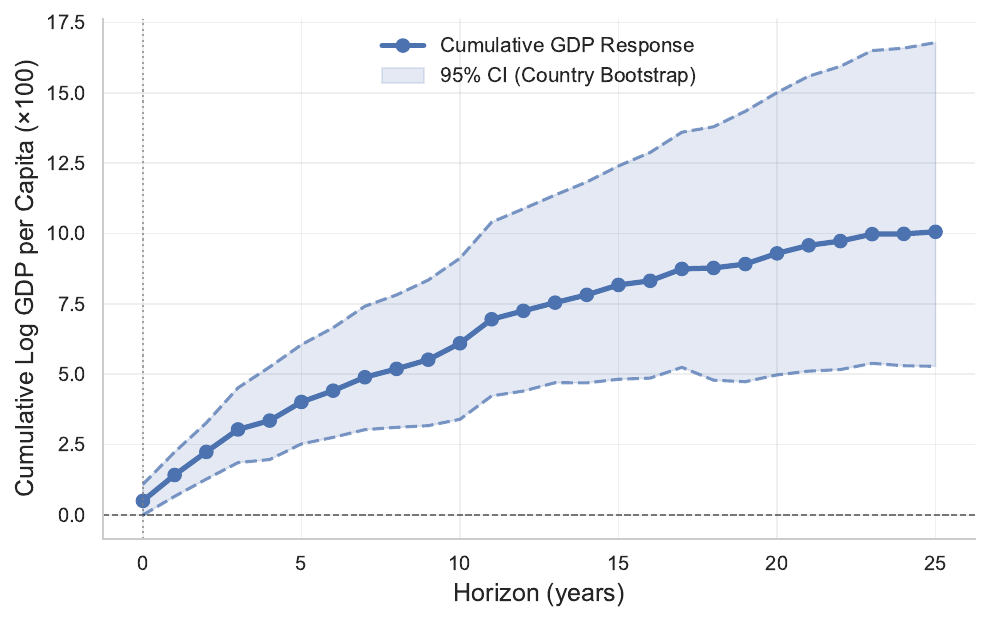}
        \caption{Sustained Alignment Path}
        \label{fig:irf_gdp_permanent}
    \end{subfigure}
    \caption{GDP Responses Associated with Transitory and Sustained Alignment Paths}
    \label{fig:permanent_shocks}
    \note{Panel (a) reports the response associated with a one-standard-deviation change in alignment on impact that returns to zero thereafter. Panel (b) reports the cumulative response associated with maintaining a one-standard-deviation change at each subsequent horizon. The outcome is log GDP per capita ($\times 100$). Baseline specification as in Figure~\ref{fig:baseline_irf}. Shaded areas represent 95\% confidence intervals from 500 country-block bootstrap iterations.}
\end{figure}

Figure~\ref{fig:permanent_shocks} maps the estimated conditional responses into transitory and sustained counterfactual paths. Under the transitory path in panel (a), the GDP coefficient is 0.5 log points on impact and 0.9 log points after one year. It remains positive but fluctuates over subsequent horizons. The wide confidence intervals indicate substantial uncertainty about the medium- and long-run path.

Panel (b) shows the cumulative response associated with the sustained path: the implied GDP coefficient rises steadily, reaching about 10 log points after 25 years. The trajectory continues to rise after year 20, though at a diminishing rate. Long-horizon local-projection estimates are imprecise. Section~\ref{ss:dynamic_panel} presents a complementary analysis using dynamic panel methods that impose an autoregressive structure to trace the mapping beyond the local-projection horizon.

\paragraph{Economic Magnitude}

The index has a standard deviation of 0.141 and ranges from $-0.43$ (hostile relations) to 0.55 (strong cooperation), with a median of 0.26. A sustained one-standard-deviation improvement corresponds to about 10 log points after 25 years (95 percent bootstrap interval: 5.3 to 16.8).\footnote{A delta-method interval clustered by country and by year gives 4.1 to 16.0 log points. A split-panel jackknife correction for fixed-effects bias \citep{Dhaene2015-spj} gives a 25-year coefficient of 10.2 log points.} An interquartile improvement, from 0.154 to 0.340 (1.32 standard deviations), implies about 13 log points. Table~\ref{tab:episodes_1sd} compares these magnitudes with observed realignments: Russia's deterioration through the Crimean crisis (2005--2015) is about one standard deviation and Vietnam's normalization with the United States (1990--2000) is 1.4.

\subsection{Robustness and Diagnostics} \label{ss:robustness}

\subsubsection{Partner Groups and Historical Periods} \label{ss:sources_of_variation}

To interpret what the index captures and to assess whether the growth relationship is concentrated in alignment with a single bloc, we decompose the index into relations with the United States, other Western nations, and other major nations. We then enter the three components jointly in the baseline local-projection specification.\footnote{Let $p_{ct}^{\mathrm{US}}$, $p_{ct}^{\mathrm{OW}}$, and $p_{ct}^{\mathrm{OM}}$ denote the GDP-weighted contributions from the United States, other Western nations, and other major nations. By construction, $p_{ct}=p_{ct}^{\mathrm{US}}+p_{ct}^{\mathrm{OW}}+p_{ct}^{\mathrm{OM}}$. At each horizon, we estimate
\[
y_{c,t+h}=\alpha_h^{\mathrm{US}}p_{ct}^{\mathrm{US}}+\alpha_h^{\mathrm{OW}}p_{ct}^{\mathrm{OW}}+\alpha_h^{\mathrm{OM}}p_{ct}^{\mathrm{OM}}+\boldsymbol{\gamma}_h^{\prime}\mathbf{x}_{ct}+\eta_c+\eta_{r(c)t}+\varepsilon_{c,t+h},
\]
where $\mathbf{x}_{ct}$ contains four lags of GDP per capita and four lags of each component, and $\eta_c$ and $\eta_{r(c)t}$ are country and region-year fixed effects. We report $h=0,\ldots,10$.} We also estimate the baseline specification separately for 1960--1989 and 1990--2019 to assess whether the relationship appears across two distinct international systems.

\begin{figure}[ht]
    \centering
    \begin{subfigure}[b]{0.48\textwidth}
        \includegraphics[width=\textwidth]{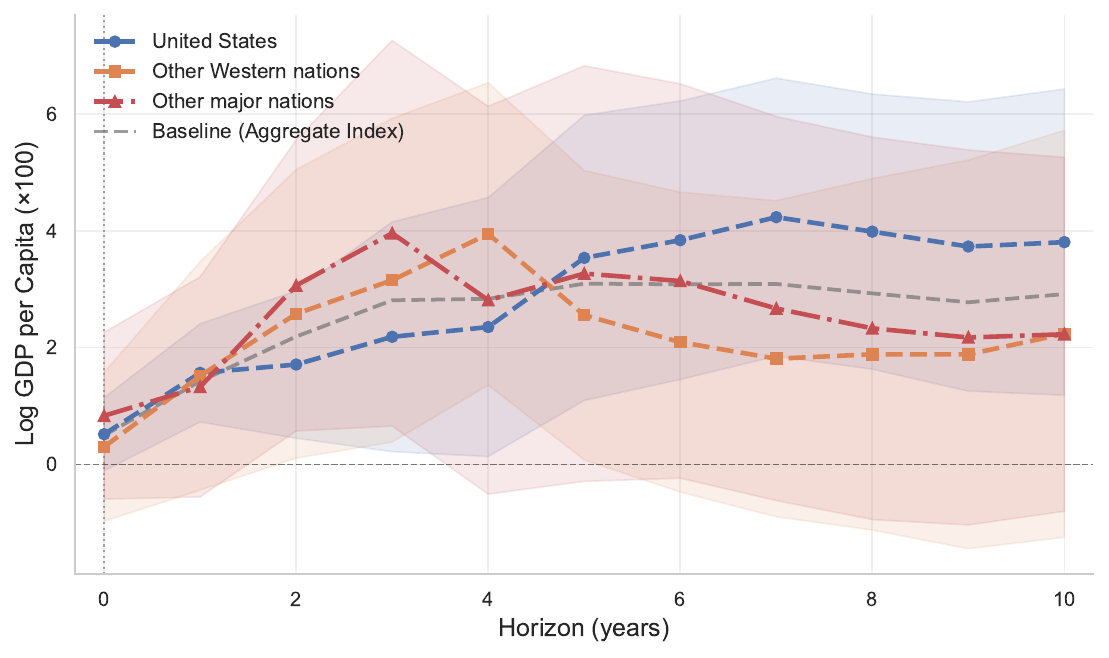}
        \caption{Partner Groups}
        \label{fig:symmetry_partner}
    \end{subfigure}
    \hfill
    \begin{subfigure}[b]{0.48\textwidth}
        \includegraphics[width=\textwidth]{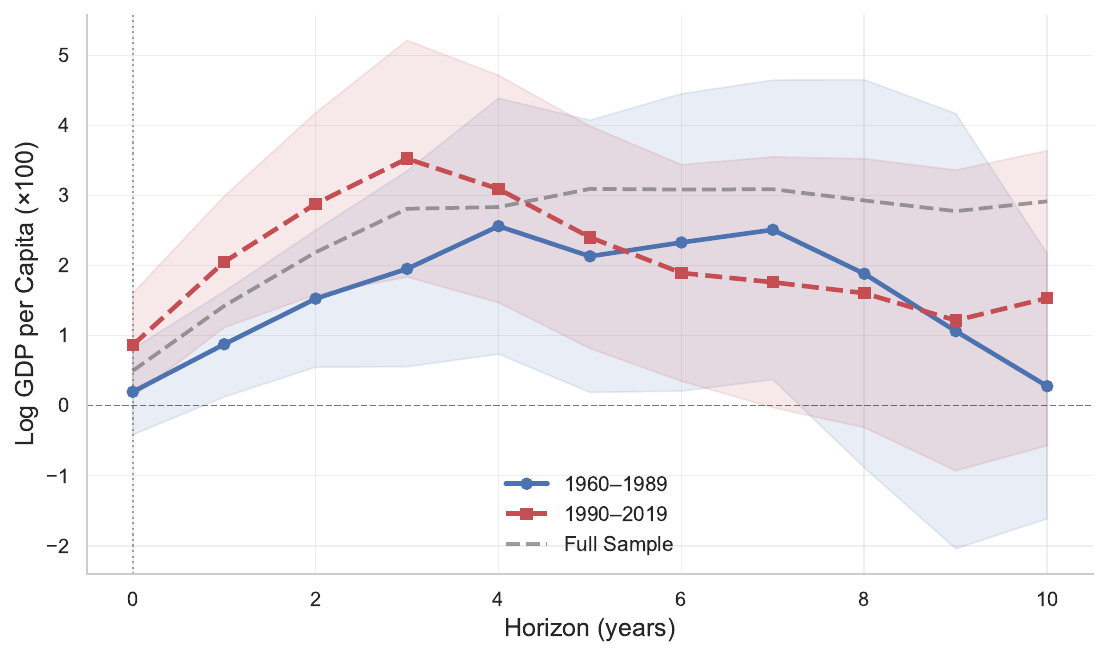}
        \caption{Historical Periods}
        \label{fig:symmetry_temporal}
    \end{subfigure}
    \caption{Growth Responses by Partner Group and Historical Period}
    \label{fig:symmetry}
    \note{Panel~(a) jointly estimates the three additive partner-group components: the United States, other Western nations (Australia, Belgium, Canada, Denmark, France, Germany, Italy, the Netherlands, Poland, Spain, Switzerland, and the United Kingdom), and other major nations (Argentina, Brazil, China, India, Indonesia, Japan, Mexico, the Republic of Korea, Russia, Saudi Arabia, and Turkey). Group membership is fixed over the sample period. The gray dashed line is the baseline response to the aggregate index, shown only as a reference. It is not the sum of the three component-specific coefficients. Panel~(b) estimates the baseline specification separately for 1960--1989 and 1990--2019 and shows the full-sample estimate for reference. All series show conditional local-projection coefficients for log GDP per capita ($\times 100$) over horizons $h=0,\ldots,10$, per one standard deviation of the aggregate index. Shaded bands are 95\% confidence intervals based on Driscoll--Kraay standard errors.}
\end{figure}

Panel~(a) of Figure~\ref{fig:symmetry} shows broadly similar positive, hump-shaped point-estimate paths across the three partner groups. The component estimates are less precise than the aggregate response and their confidence bands overlap, but all three coefficients peak at roughly 4 log points per standard deviation of the aggregate index within the first decade, with the US path peaking later than the other two. Each component is the group's GDP-weighted contribution to the aggregate index, so similar paths support the index's construction, in which relations matter in proportion to partner GDP. After partialling out the baseline fixed effects and lagged controls, the US and other-major-nations components have a conditional correlation of 0.04 (shared $R^2=0.002$), so the two components vary largely independently. The positive point-estimate paths indicate that the growth relationship is not confined to one hegemon or ideological bloc.

Panel~(b) shows positive short- and medium-horizon coefficients in both historical periods. The two paths rise with the full-sample estimate over the first few years and peak between 2.6 and 3.5 log points. They attenuate thereafter, while precision declines as longer horizons progressively reduce the usable observations within each 30-year subperiod. The relationship therefore appears in two structurally different international systems, even though geopolitical improvement took different political forms in each.\footnote{Both periods contain similar variation in the index. In the long-coverage estimation sample at $h=5$, the index has residualized standard deviations of 0.045 in 1960--1991 and 0.037 in 1992--2024 after the baseline fixed effects and lagged controls. Panel~(b) uses 1960--1989 and 1990--2019 to accommodate forward horizons. Raw within-country variation is also similar under that split.}

\subsubsection{Decomposing the Index: Event Domains and Formal Acts} \label{ss:decomposition}

A natural concern is that the aggregate growth relationship may be concentrated in events with direct economic content, including trade agreements, sanctions, and investment treaties, which may themselves respond to growth conditions. Using the six event categories defined ex ante in the LLM prompt and held fixed across the sample (Section~\ref{ss:llm}), we construct three components of the geopolitical alignment index: \textit{economic relations} (category A), \textit{diplomatic and political relations} (category B), and \textit{security, defense, and territorial relations} (categories C and D).\footnote{Category A covers trade policy, financial relations, economic coercion, strategic sectors, and economic integration. Category B covers formal diplomacy, high-level interactions, public diplomacy, and cultural exchanges. Categories C and D cover military cooperation, security incidents, intelligence, legal actions, territorial and maritime disputes, and the movement of people. The remaining categories, multilateral governance (E) and other events (F), account for 20 percent of events and are omitted. Including them as a fourth component does not affect the results. See Table~\ref{tab:event_categories} for subcategory details.} We aggregate each component using the same GDP-weighted procedure as the baseline index in equation~\eqref{eq:country_geo}. Their correlations with the aggregate index are 0.73, 0.78, and 0.62, while their pairwise cross-correlations range from 0.30 to 0.48. The components therefore share meaningful but incomplete variation.

\begin{figure}[ht]
    \centering
    \begin{subfigure}[b]{0.48\textwidth}
        \includegraphics[width=\textwidth]{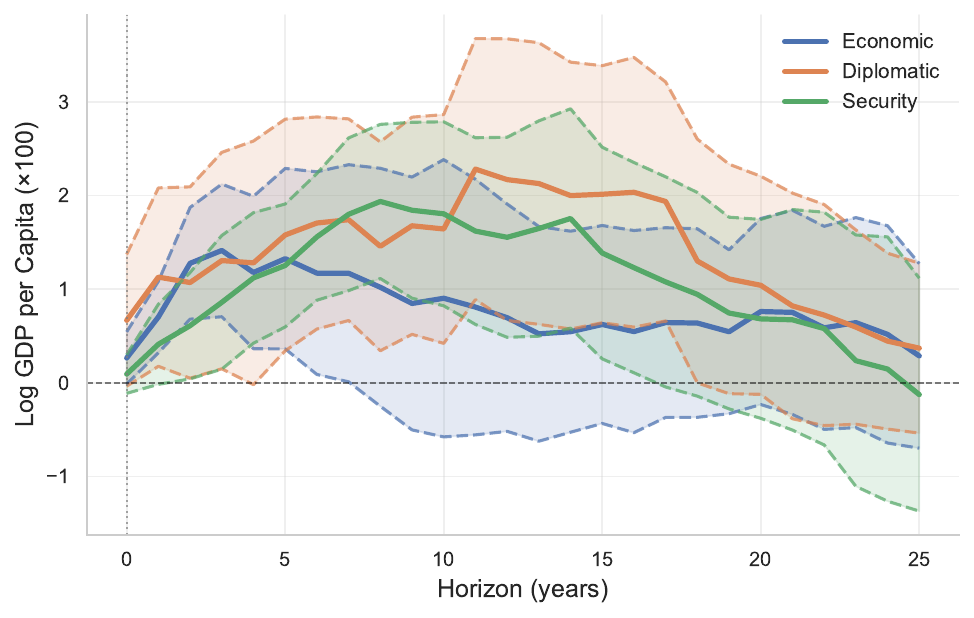}
        \caption{Event Domains, Joint Specification}
        \label{fig:component_horserace}
    \end{subfigure}
    \hfill
    \begin{subfigure}[b]{0.48\textwidth}
        \includegraphics[width=\textwidth]{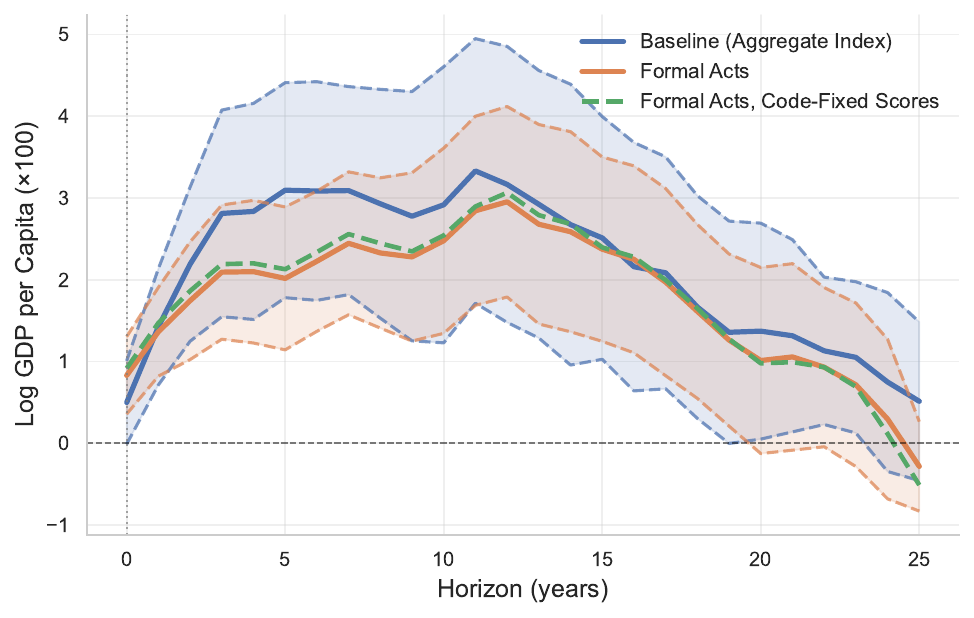}
        \caption{Formal Acts}
        \label{fig:formal_acts}
    \end{subfigure}
    \caption{Decomposing Geopolitical Alignment: Event Domains and Formal Acts}
    \label{fig:component_decomposition}
    \note{Panel (a) estimates a multivariate extension of equation~\eqref{eq:lp_spec} with the three event-domain components entered simultaneously, each with four lags. Components: Economic (category A), Diplomatic (category B), Security (categories C and D). Panel (b) compares the baseline with an index built only from formal acts (solid) and with the same acts scored at the median Goldstein value of their CAMEO code (dashed). Both panels report conditional local-projection coefficients for log GDP per capita ($\times 100$) on the long-coverage sample, with baseline fixed effects and Driscoll--Kraay standard errors as in Figure~\ref{fig:baseline_irf}. Shaded areas represent 95\% confidence intervals. Coefficients are per one standard deviation of the aggregate index.}
\end{figure}

The aggregate growth relationship is not confined to events classified as economic. When all three components enter jointly, each with four lags, in a multivariate extension of equation~\eqref{eq:lp_spec} (panel~(a) of Figure~\ref{fig:component_decomposition}), the coefficients on all three components are positive and statistically distinguishable from zero over portions of the horizon. The economic coefficient peaks early, at approximately 1.4 log points within the first few years, then attenuates. The diplomatic coefficient is larger and more persistent, reaching 1.5--2.3 log points between horizons 5 and 12, while the security and territorial coefficient remains near 1.8 log points around horizon 10.

A second concern is look-ahead bias. The model and the human coders work from the current historical record, so they could select or score events with knowledge of what followed, making countries that later grew look better aligned in hindsight. Such bias would operate mainly through discretionary judgments about which consultations, statements, or cooperation count as consequential and how intense they were. We therefore rebuild the index from formal acts only: agreements and diplomatic recognition, sanctions and their removal, diplomatic breaks and expulsions, uses of force, and state visits.\footnote{The formal-acts set comprises CAMEO codes 042--043 (visits), 054 (diplomatic recognition), 057 (formal agreements), 081 and 085 (easing of sanctions), 161--163 and 166 (breaking relations, halting aid, imposing sanctions, and expulsions or withdrawals), 172 and 174 (administrative sanctions and expulsions of individuals), and 190--204 (military force and mass violence).} These acts are dated public decisions that contemporaneous sources record regardless of later outcomes. They account for 35 percent of events. Apart from the event set, the index is constructed as the baseline, and its correlation with the baseline is 0.81.

Panel~(b) of Figure~\ref{fig:component_decomposition} shows that the formal-acts index reproduces the baseline path. Its coefficient lies within the baseline 95 percent confidence interval at every horizon from 0 to 15 and peaks at 3.0 log points in year 12, compared with 3.3 log points in year 11 for the baseline. The model also assigns each event an intensity score, which could carry hindsight even when the event itself is a formal act. For the dashed line, we score each act at the median Goldstein value of its CAMEO code, so that its score no longer depends on the model's judgment of intensity. Doing so changes the path by at most 0.2 log points. Together, the two paths weigh against look-ahead bias in the types of events selected or in their scores as the source of the relationship. As Section~\ref{ss:data_validation} shows, whether the database records externally documented events also does not depend on the subsequent growth of the countries involved.

\subsubsection{Additional Controls and Fixed Effects} \label{ss:add_controls}

\paragraph{Alternative Fixed Effects Specifications}

\begin{figure}[ht]
    \centering
    \begin{subfigure}[b]{0.48\textwidth}
        \includegraphics[width=\textwidth]{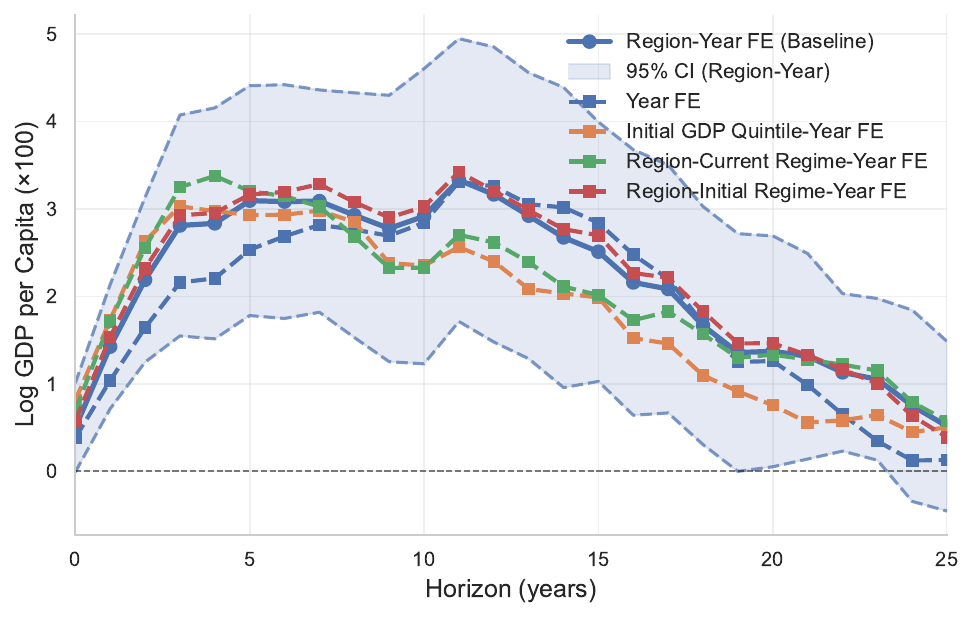}
        \caption{Alternative Fixed Effects}
        \label{fig:fe_robustness}
    \end{subfigure}
    \hfill
    \begin{subfigure}[b]{0.48\textwidth}
        \includegraphics[width=\textwidth]{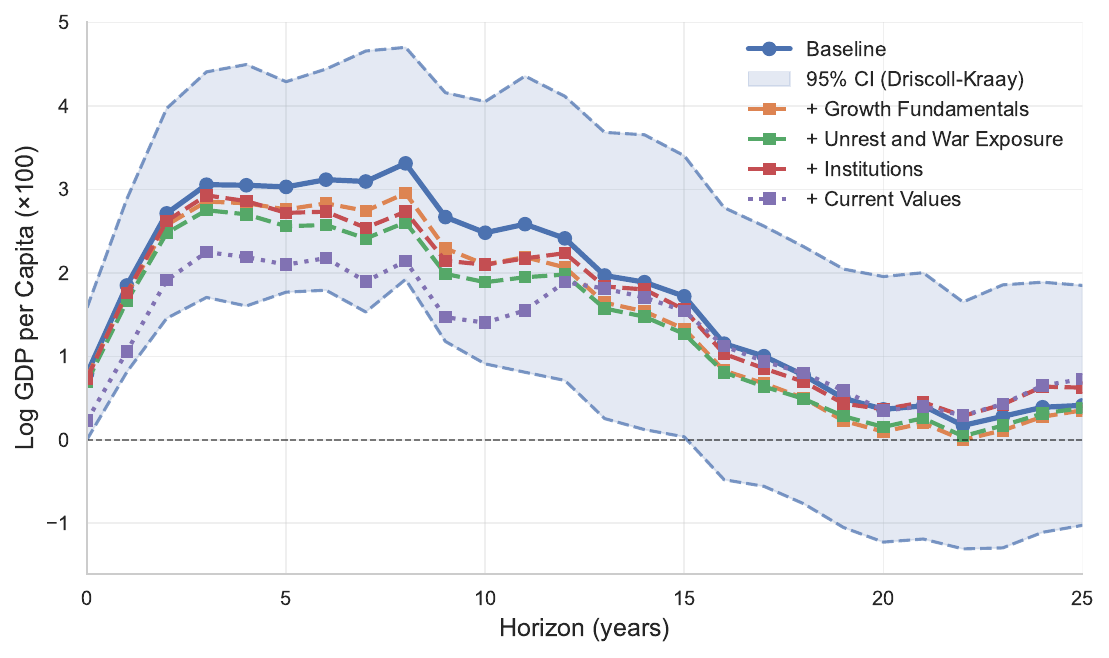}
        \caption{Additional Controls}
        \label{fig:control_robustness}
    \end{subfigure}
    \caption{Robustness to Alternative Specifications}
    \label{fig:robustness_specs}
    \note{Panel~(a) compares conditional coefficient paths under alternative fixed-effects specifications. The baseline (solid blue with 95\% Driscoll--Kraay confidence interval) uses region-year fixed effects. Dashed overlays show year fixed effects only, initial GDP quintile-year, region-initial regime-year, and region-current regime-year fixed effects (regime = democracy status). Panel~(b) cumulatively adds four lags of growth fundamentals (trade openness, log population, age-structure shares, investment share, log capital stock, and log human capital index), domestic unrest and five casualty-based war exposure measures, and five Varieties of Democracy (V-Dem) institutional quality indices. The final overlay adds the current values of all these controls. All paths in panel~(b) use identical observations from 98 countries. Both panels use the baseline lag structure as in Figure~\ref{fig:baseline_irf}. Coefficients are for log GDP per capita ($\times 100$) per one-standard-deviation change in the index. In panel~(a), the baseline and year fixed effects use the 148-country long-coverage sample, and the other three paths use 122, 144, and 144 countries. In panel~(b), the shaded area is the 95\% Driscoll--Kraay confidence interval for the baseline.}
\end{figure}

Panel (a) of Figure~\ref{fig:robustness_specs} shows similar coefficient paths under four alternative fixed-effects structures: year fixed effects only, initial GDP quintile-year effects,\footnote{Following \citet{Acemoglu2019-bo}, we rank countries using Angus Maddison's GDP estimates for 1960, which are available for 145 countries in our panel, to maximize sample coverage.} region-initial regime-year effects (where regime indicates a country's democracy status), and region-current regime-year effects. Across specifications, the paths closely track the baseline and peak between about 3.0 and 3.4 log points. The modest attenuation under region-current regime-year effects, which allow region-year shocks to differ by current regime status, is consistent with some overlap between regime status and the variation summarized by the index. Section~\ref{ss:geo_democracy} examines this relationship.

\paragraph{Time-Varying Economic and Political Controls}

Time-varying factors, including growth fundamentals, domestic instability, wars, and institutional change, may confound the estimated relationship if they correlate with geopolitical changes. We probe this concern by progressively adding controls, which enter as lags because contemporaneous controls may themselves respond to alignment. To limit compositional differences, all estimates use identical observations from the 98 countries with complete data on all controls.

Figure~\ref{fig:robustness_specs}, panel~(b), shows that the coefficient paths remain close as controls are cumulatively added, with peaks of about 3 log points in every lag-only specification. Adding four lags of standard growth fundamentals, including trade openness, investment, capital, and human capital, produces limited changes. Four lags of domestic unrest\footnote{The unrest indicator equals one if any domestic conflict event (assassinations, strikes, guerrilla warfare, government crises, purges, riots, revolutions, or anti-government demonstrations) is recorded in the Cross-National Time-Series Data Archive. See Appendix~\ref{app_b:econ_data} for details.} and lagged casualty-based war exposure\footnote{We include four lags of five war variables from \citet{FederleEtAl2026PriceOfWar}: domestic war-site onset, domestic casualty intensity (scaled by prewar population), trade-weighted foreign war exposure, proximity-weighted foreign war exposure, and the count of active foreign war sites.} likewise leave the path close to the common-sample baseline. Adding four lags of five V-Dem institutional quality indices (electoral, liberal, participatory, deliberative, and egalitarian democracy) \citep{Coppedge2024-vdem}, following \citet{Acemoglu2025-lv}, produces little further change.

The final specification adds the current values of all controls to their lags. The estimates attenuate to 2.1 log points at year 5 and 1.4 at year 10 but remain statistically distinguishable from zero at most horizons. Because current values may respond to alignment or lie on its transmission path, they are potentially bad controls, and we report this specification only as an overcontrol sensitivity. These results suggest that geopolitical alignment carries information about future growth beyond a broad set of observed economic and political conditions.

\subsubsection{Placebo and Timing Diagnostics} \label{ss:placebo}

We conduct two placebo exercises that ask whether the estimated GDP response depends on the realized country-specific assignment and timing of changes in geopolitical alignment. Both exercises re-estimate the full baseline local projection in each placebo draw, preserving the specification's dynamics and fixed-effects structure.

\paragraph{Within-Region-Year Reassignment}

The first exercise randomly reassigns contemporaneous alignment values across countries within each region-year cell. The reassignment preserves the index's within-cell distribution while breaking its observed country assignment. Because the baseline already includes region-year fixed effects, the exercise does not test common regional dynamics. Instead, it asks whether the estimated response depends on the realized pairing between countries' GDP paths and alignment values within a region-year cell.

\begin{figure}[htbp]
    \centering
    \begin{subfigure}[b]{0.48\textwidth}
        \includegraphics[width=\textwidth]{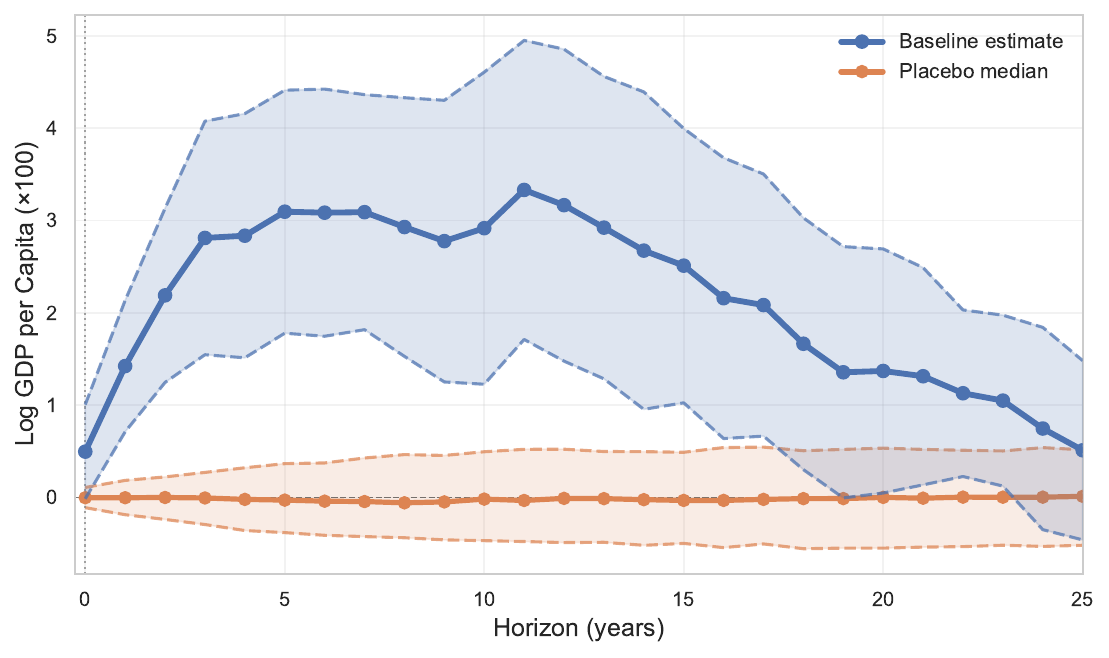}
        \caption{Within-Region-Year Reassignment}
        \label{fig:placebo_irf_A}
    \end{subfigure}
    \hfill
    \begin{subfigure}[b]{0.48\textwidth}
        \includegraphics[width=\textwidth]{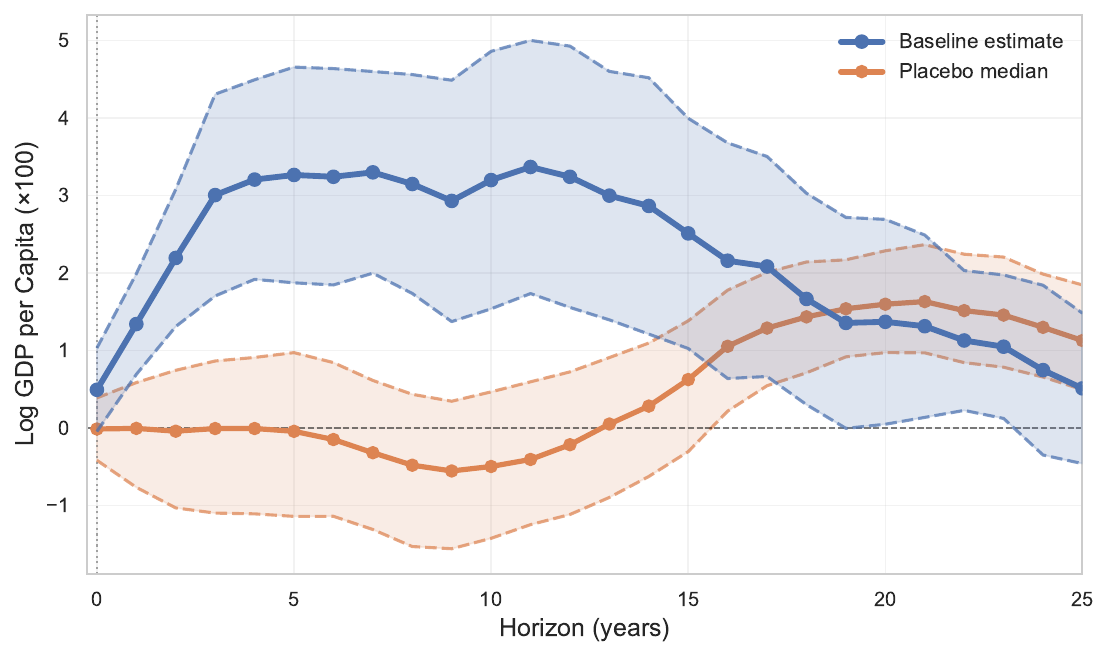}
        \caption{Future-Year Timing Placebo}
        \label{fig:placebo_irf_B}
    \end{subfigure}
    \caption{Placebo Tests for the Conditional Dynamic GDP Response}
    \label{fig:placebo_irf}
    \note{Blue: baseline estimates and 95\% Driscoll--Kraay CIs (in panel~(b), estimated on 1960--2009 base years). Orange: placebo median and 5th--95th percentile range across 500 draws. Panel (a) reassigns alignment within region-year cells. Panel (b) replaces it with values 8--15 years ahead. Each draw re-estimates the baseline specification. The outcome is log GDP per capita ($\times 100$).}
\end{figure}

Panel (a) of Figure~\ref{fig:placebo_irf} shows that the placebo median remains flat near zero across all horizons and the placebo band stays tight, while the realized response lies well outside the placebo band throughout the main response window. This contrast shows that the realized within-cell pairing of alignment and GDP paths matters for the estimated response. It does not, however, distinguish the alignment relationship from time-varying omitted determinants that covary with actual alignment, because the reassignment breaks both relationships.

\paragraph{Future-Year Timing Placebo}

The second exercise replaces each country's contemporaneous alignment value with its value 8 to 15 years in the future, with the lead drawn randomly for each country. The shift keeps each country's own alignment series but breaks its contemporaneous timing with GDP. The exercise therefore asks whether the baseline path depends on the temporal match between geopolitical changes and subsequent GDP dynamics.

In panel (b) of Figure~\ref{fig:placebo_irf}, the placebo median stays near zero through horizon 5, dips to about $-0.6$ log points around horizon 9, and then rises to the actual late-horizon path once the projection window reaches the years from which the future alignment values are drawn. This delayed emergence is a positive control: the design is sensitive to timing. The exercise does not distinguish geopolitical changes from omitted factors that move contemporaneously with them, and the long shift is less informative about confounding that operates over shorter intervals.

Averaged over horizons 0 to 10, the realized response exceeds all 500 placebo draws in both exercises (Appendix~\ref{app_b:placebo}).

\subsection{Robustness to Bundled Domestic Transformations} \label{ss:bundled}
\enlargethispage{1pt}

Changes in a country's external geopolitical environment often coincide with major domestic transformations, including democratization, regime change, post-socialist transition, post-conflict reconstruction, and state collapse. These episodes may affect growth through channels beyond geopolitical alignment. We address this concern in two ways. The first absorbs transformation-period observations with separate country-year fixed effects. The second uses bilateral shifts around leadership transitions in major nations, anchoring the timing to events outside the focal country. Neither exercise eliminates all omitted-variable concerns, but together they reduce reliance on the variation most directly exposed to bundled domestic transformations.

\subsubsection{Excluding Domestic Transformation Episodes}

We identify domestic transformation periods using a historical-institutional codebook. The codebook evaluates candidate episodes according to whether they mark a recognized turning point, involve broad changes in domestic political or economic institutions, span multiple institutional margins, and plausibly affect the country's long-run economic trajectory. Geopolitical realignment itself is not a classification criterion. The resulting classification contains 329 episodes across 172 countries and designates 11.8 percent of country-years as transformation periods. Examples include China's reform and opening (1978--1984), Chile's regime change (1973--1976), and South Africa's democratization (1990--1994). Appendix~\ref{app_b:transformation_episodes} describes the criteria, and Table~\ref{tab:transformation_periods} reports summary statistics and additional examples.

We re-estimate the baseline local projection after adding a separate country-year fixed effect for every transformation-period observation, thereby removing these observations from the variation used in estimation. As an alternative classification check, we repeat the exercise using V-Dem's \textit{Regimes of the World} measure, absorbing covered observations within three years before or after a change in regime type.\footnote{Appendix~\ref{app_b:transformation_episodes} explains why we absorb transformation-period observations rather than add fixed effects for all country-era cells.}

\begin{figure}[ht]
    \centering
    \begin{subfigure}[b]{0.48\textwidth}
        \includegraphics[width=\textwidth]{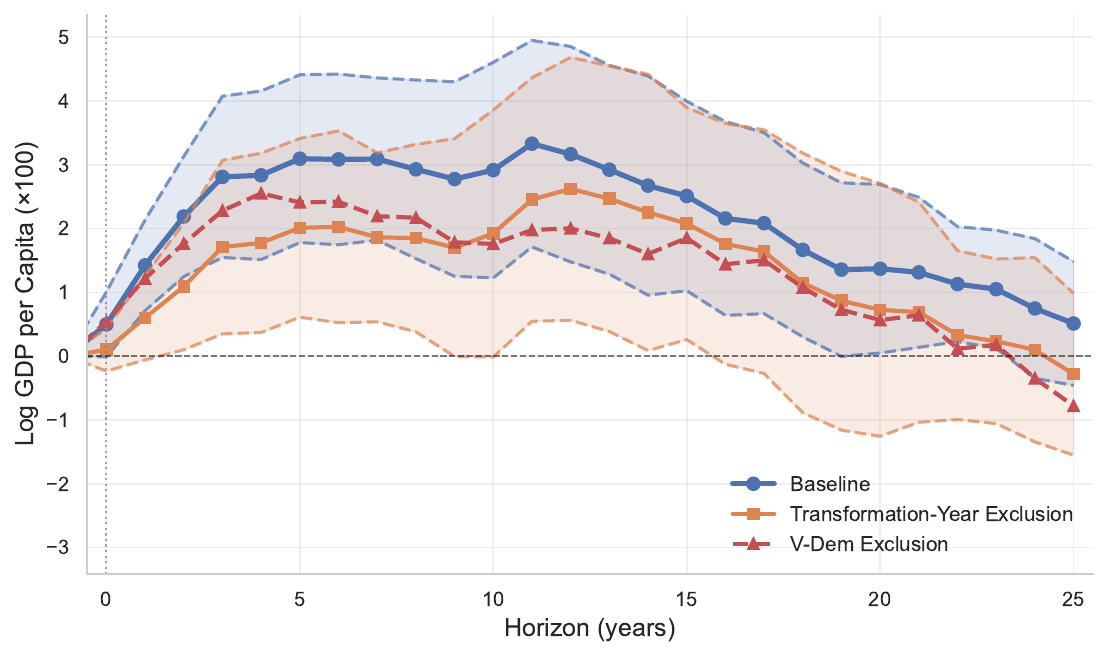}
        \caption{Excluding Transformation Years}
        \label{fig:stable_era_exclusion}
    \end{subfigure}
    \hfill
    \begin{subfigure}[b]{0.48\textwidth}
        \includegraphics[width=\textwidth]{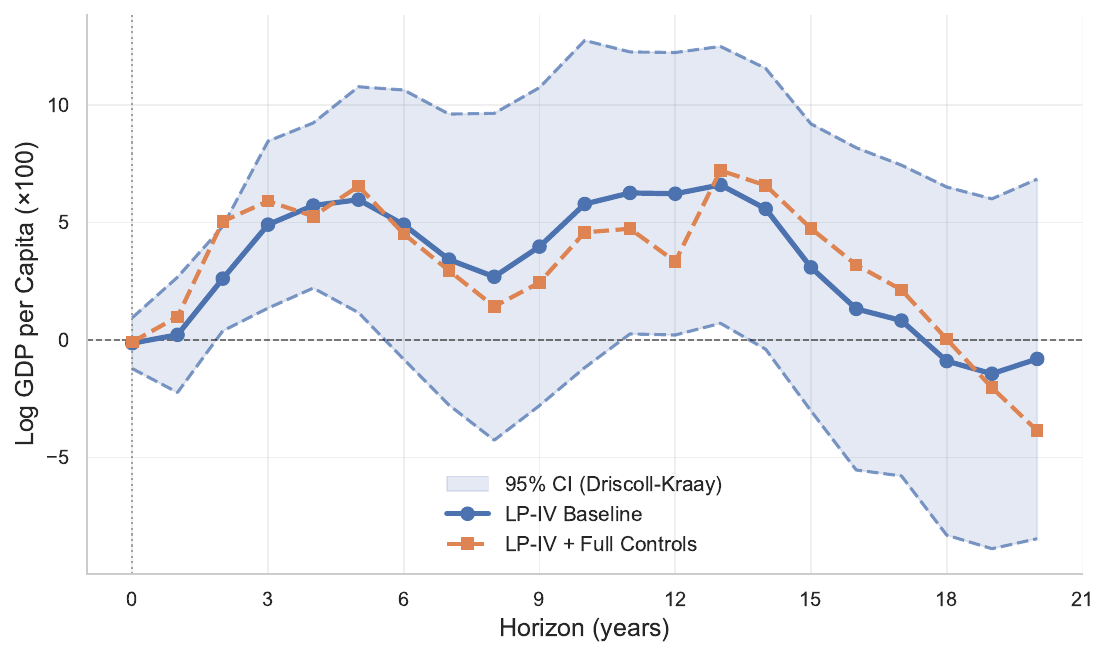}
        \caption{Leadership Transitions in Major Nations}
        \label{fig:leader_transition_iv}
    \end{subfigure}
    \caption{Robustness to Bundled Domestic Transformations}
    \label{fig:bundled_transformations}
    \note{Panel~(a) compares baseline conditional local-projection coefficients with estimates that absorb country-years classified as domestic transformations. Panel~(b) reports LP-IV estimates instrumenting alignment with bilateral shifts around leadership changes in major nations. The baseline and transformation-year paths use 148 countries. The V-Dem path uses covered observations, and the full-controls IV uses a 109-country common sample. The V-Dem path absorbs observations within three years before or after a change in regime type in V-Dem's \textit{Regimes of the World} measure. The full-controls IV adds four lags of trade openness, domestic unrest, war exposure, and V-Dem institutional indices. Shaded bands report 95\% Driscoll--Kraay intervals for the baseline and transformation-year paths in panel~(a) and the baseline IV in panel~(b). Other paths are point estimates. The outcome is log GDP per capita ($\times 100$).}
\end{figure}

Panel~(a) of Figure~\ref{fig:bundled_transformations} shows that both exclusion rules attenuate the estimated GDP path but preserve its positive medium-run profile. At year 5, the historical transformation exclusion yields 2.0 log points, compared with 3.1 in the 148-country baseline. The V-Dem exclusion gives 2.4 log points on its covered-observation sample.\footnote{Because V-Dem coverage changes the estimation sample, Appendix~\ref{app_b:transformation_episodes} reports a fixed 124-country comparison: 3.6 log points in the corresponding baseline and 2.5 under the V-Dem exclusion.} Both exclusion estimates are distinguishable from zero and remain positive through year 15, although precision weakens at later horizons. Applying the same one-standard-deviation, 25-year sustained-path mapping within each estimation sample gives 6.3 log points under the historical exclusion and 5.9 under the V-Dem exclusion, compared with 10.1 in the baseline. The historical exclusion therefore preserves about three-fifths of the baseline magnitude, and the V-Dem mapping is similar on its covered sample.

\subsubsection{Leadership Transitions in Major Nations}

The second exercise uses bilateral alignment shifts around 67 leadership transitions across 21 major nations: 25 deaths in office from the Archigos database \citep{Goemans2009-archigos} and 42 close-election turnovers with vote margins below 10 percentage points \citep{marx2024elections}. For a transition in major nation $j$ in year $t$, we construct
\begin{equation*}
    z_{ct}^{\mathrm{LC}} = \left(S_{cj,t+1} - S_{cj,t-1}\right) \times \text{GDP share}_{j,t-1},
\end{equation*}
summing across transitions when more than one occurs in the same year. We instrument the index with this GDP-weighted bilateral shift in the baseline local projection and add four lags of the instrument.\footnote{Following \citet{Jorda2005-te} and \citet{Plagborg-Moller2021-hi}, at each horizon the estimator combines a reduced-form projection of $y_{c,t+h}$ on $z_{ct}^{\mathrm{LC}}$ with a first-stage projection of $p_{ct}$ on $z_{ct}^{\mathrm{LC}}$. Both use the baseline fixed effects and lag controls, augmented with four lags of the instrument. The LP-IV response is $\widehat{\alpha}_h^{\mathrm{LP\text{-}IV}}=\widehat{\alpha}_h^{\mathrm{RF}}/\widehat{\alpha}^{\mathrm{FS}}$.}

Panel~(b) of Figure~\ref{fig:bundled_transformations} shows positive paths that are less precise than the baseline. The baseline IV path reaches 4--6 log points at several horizons in the first decade and rises again around years 11--14, and its 95 percent intervals exclude zero in years 2--5 and 11--13. The full-controls specification displays a similar pattern. The wider confidence intervals reflect the limited number of underlying transitions. Because the instrument uses the realized bilateral change around each transition, not the transition alone, it may also capture other bilateral developments in the short event window. A narrower window, $S_{cj,t}-S_{cj,t-1}$, gives a stronger first stage ($F=53.7$, against 16.9) and 3.1 log points at year 5, close to the baseline. Appendix~\ref{app_b:leader_iv} reports the event-type split and alternative bilateral-shift windows.

\subsection{Interpretation of the Growth Estimates} \label{ss:interpretation}

The estimates above are conditional dynamic responses. They admit a causal interpretation if, after conditioning on country and region-year fixed effects and on past GDP and alignment, countries whose alignment changes would otherwise have followed similar GDP paths. That is, current changes in alignment must be unrelated to other shocks that affect GDP now or later. Formally, the maintained condition is as follows.
\begin{assumption}
\label{assumption_1}
The error $\varepsilon_{c,t+h}$, the effect on GDP of all structural shocks other than the current change in alignment, satisfies $\mathbb{E} \left[\varepsilon_{c,t+h} \mid p_{ct}, \left\{ \mathbf{x}_{c\tau}\right\}_{t_{0} \leq \tau \leq t} \right] = 0$ for all $\left\{ \mathbf{x}_{c\tau}\right\}_{t_{0} \leq \tau \leq t}$, all countries $c$, and all $t \geq t_0$.
\end{assumption}

Under Assumption~\ref{assumption_1} and the linear specification, the LP coefficient $\alpha_h^{\text{LP}}$ identifies the impulse response at horizon $h$, for any value $p_{c0}$ of the index:
\begin{equation}
\alpha_h^{\mathrm{LP}}
=
\mathcal{R}_{p \rightarrow y}(h)
\equiv
\mathbb{E}\!\left[y_{c,t+h}\mid p_{ct}=p_{c0}+1,\mathbf{x}_{ct}\right]
-
\mathbb{E}\!\left[y_{c,t+h}\mid p_{ct}=p_{c0},\mathbf{x}_{ct}\right],
\qquad h=0,1,\ldots,H.
\label{eq:impulse_response_def}
\end{equation}
The impulse response traces the dynamic effect on GDP per capita of a one-unit improvement in current geopolitical alignment together with the subsequent evolution of the index \citep{Jorda2005-te}. The reported coefficients scale the response in equation~\eqref{eq:impulse_response_def} to a one-standard-deviation improvement, $0.141\,\mathcal{R}_{p \rightarrow y}(h)$.

What does this response capture? Because the index summarizes a country's relations with all major nations rather than a single policy, the causal object is broad: a change in the access and risk conditions described in Section~\ref{ss:external_env}. The accompanying changes documented in Section~\ref{pwt_gdp} fit both margins. Because the index is a broad summary, the estimates cannot separate these margins or isolate the effect of any single treaty, sanction, or partner.

No single diagnostic verifies Assumption~\ref{assumption_1}. The exercises in Sections~\ref{ss:baseline}--\ref{ss:bundled} each target a distinct threat. Conditioning on past GDP and alignment addresses reverse causality from past growth: negative-horizon coefficients are not individually distinguishable from zero, and a reverse projection shows no systematic relationship between current GDP and subsequent alignment. The component and partner-group estimates show that the relationship is not confined to economic events or to relations with one bloc, and the formal-acts index indicates that it does not depend on discretionary event types or on the model's intensity scores. Alternative fixed effects and lagged growth fundamentals, conflict, and institutional controls address observed pre-existing conditions, and adding their current values attenuates but does not remove the relationship. Placebo reassignment and future-year shifts address mechanical artifacts and timing, and the transformation exclusions and the shifts around major nations' leadership transitions address domestic reforms and regime changes.

These exercises do not rule out every unobserved contemporaneous shock that moves alignment and future GDP together, and the timing diagnostics have limited power against confounders that accumulate gradually. The main remaining concern is a contemporaneous shock, at home or in a major partner, that changes both bilateral relations and future growth. Subject to that limitation, the evidence is consistent with a causal interpretation of the conditional dynamic response.

\section{Channels and Implications} \label{s:correlates}

This section interprets the dynamic relationship between GDP and the external geopolitical environment. We first compare the relationship across output measures and examine accompanying changes in candidate growth fundamentals. We next study the interaction between democratic institutions and geopolitical reorientation. We conclude with illustrative accounting exercises that map observed realignments into implied GDP differences under maintained assumptions.

\subsection{Output Measures and Growth Fundamentals} \label{pwt_gdp}

\begin{figure}[ht]
    \centering
    \begin{subfigure}[b]{0.48\textwidth}
        \includegraphics[width=\textwidth]{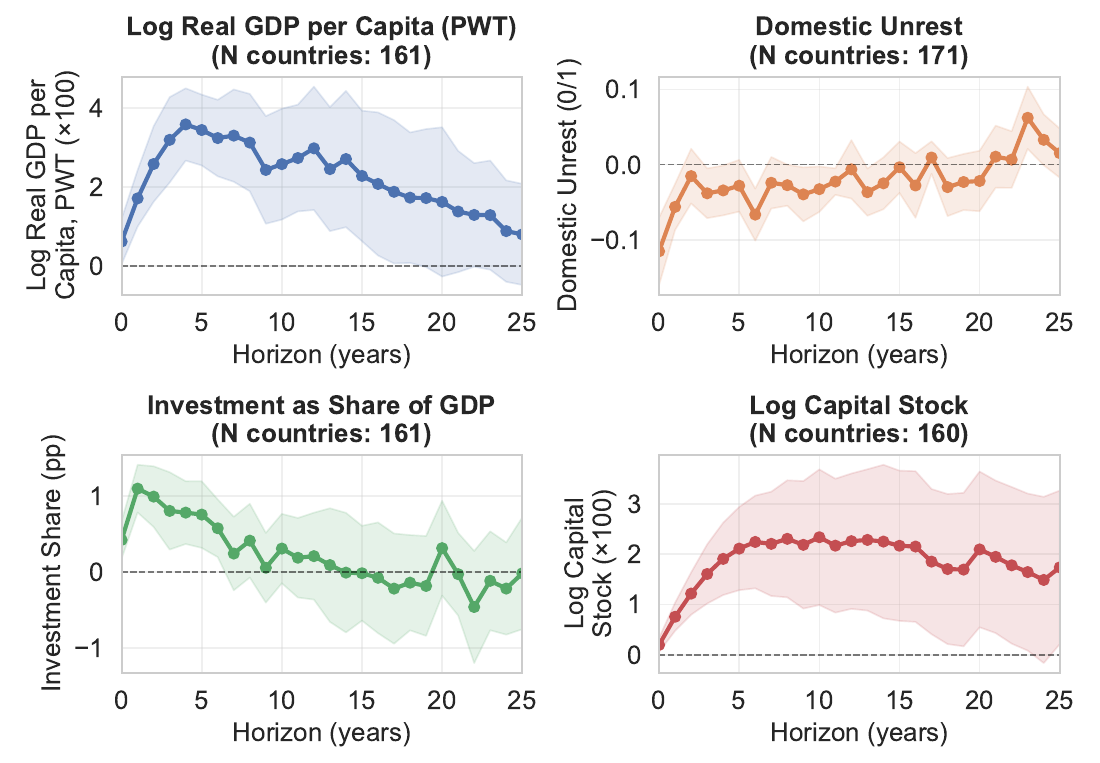}
        \caption{Output, Stability, and Capital Accumulation}
    \end{subfigure}
    \hfill
    \begin{subfigure}[b]{0.48\textwidth}
        \includegraphics[width=\textwidth]{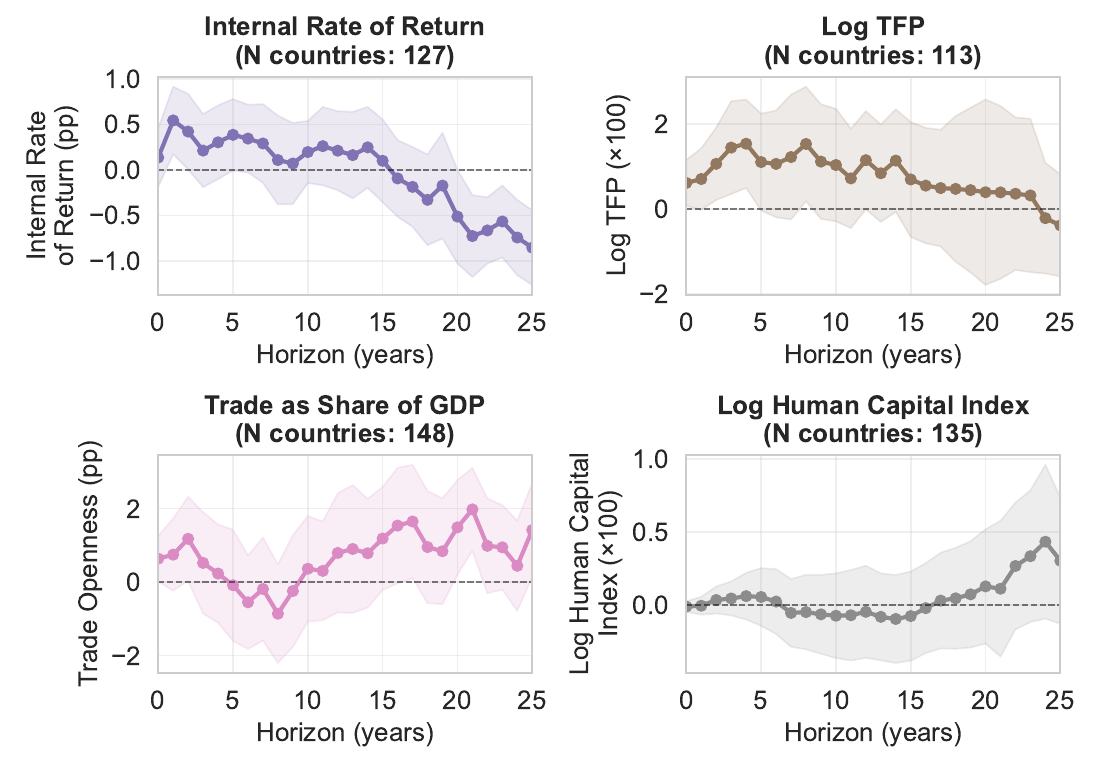}
        \caption{Growth Fundamentals}
    \end{subfigure}
    \caption{Growth Correlates Following Changes in the Geopolitical Alignment Index}
    \label{fig:growth_correlates}
    \note{This figure reports horizon-specific coefficients on the country-level geopolitical alignment index from local projections of the listed outcomes. Panel (a) shows real GDP per capita from the Penn World Table, domestic unrest, investment share, and capital stock. Panel (b) shows internal rate of return, total factor productivity, trade openness, and human capital. All specifications follow equation~\eqref{eq:correlates} with baseline fixed effects and lag structure as in Figure~\ref{fig:baseline_irf}, adding four lags of the dependent variable. Each outcome uses an outcome-specific country set, with counts shown in parentheses. Country counts are constant across horizons except for trade, for which 147 of 148 countries contribute at horizons 18--24. Coefficients are per one-standard-deviation change in the index. Log outcomes are multiplied by 100. Investment share, trade openness, and the internal rate of return are in percentage points. Domestic unrest is a binary indicator. Shaded areas represent 95\% confidence intervals based on Driscoll--Kraay standard errors.}
\end{figure}

\paragraph{Alternative Output Measures} Our baseline uses GDP per capita from the World Development Indicators to maximize country and temporal coverage. As an alternative, panel~(a) of Figure~\ref{fig:growth_correlates} uses real GDP per capita from the Penn World Table. The response peaks at about 3.6 log points around year 4 and remains above 2 log points through year 15.

\paragraph{Growth Fundamentals} To examine whether changes in the index are accompanied by changes in candidate growth fundamentals, we estimate the following reduced-form responses:
\begin{equation}
    m_{c,t+h} = \alpha_h^{m} p_{ct} + \sum_{\ell=1}^4 \beta_{h\ell} y_{c,t-\ell} + \sum_{\ell=1}^4 \gamma_{h\ell} p_{c,t-\ell} + \sum_{\ell=1}^4 \lambda_{h\ell} m_{c,t-\ell} + \eta_c + \eta_{r(c)t} + \varepsilon^m_{c,t+h}
    \label{eq:correlates}
\end{equation}
where $m_{c,t+h}$ denotes the outcome at horizon $h$. Each specification adds four lags of the outcome to the baseline fixed effects and lagged controls.

Figure~\ref{fig:growth_correlates} shows distinct timing across outcomes. Domestic unrest, our measure of domestic political instability, falls sharply on impact and then moves back toward zero. The fall in unrest speaks to the risk margin described in Section~\ref{ss:external_env}: political stability can sustain domestic production and reduce the political and sovereign risks that deter investment and capital accumulation. Consistent with this interpretation, the internal rate of return rises during the first two post-impact years, investment rises by about 1 percentage point within the first few years, and the capital stock reaches about 2.2 log points around years 8--12. At long horizons, after the initial alignment improvement has dissipated, the internal rate of return turns negative.

Total factor productivity rises to about 1.5 log points by year 4 and remains above 1 log point through year 10. This response is consistent with the access margin described in Section~\ref{ss:external_env} and with the roles of technology adoption and international R\&D spillovers emphasized by \citet{Parente1994-bt} and \citet{Coe1995-rd}. Trade openness, defined as exports plus imports of goods and services relative to GDP, rises by about 0.6 percentage points on impact and, after imprecise intermediate estimates, reaches 1.4--2.0 percentage points at several horizons after year 15.\footnote{\citet{fan2025barriers} study the dynamic relationship between trade flows and bilateral alignment in a dyad-level gravity framework and find that bilateral trade rises in the same year as an improvement in alignment.} Human-capital estimates remain near zero until late in the horizon, and no post-impact coefficient is individually distinguishable from zero.

Appendix~\ref{app_b:other_growth_correlates} reports positive point-estimate paths for the tax-to-GDP ratio and primary enrollment, and GDP-like paths for consumption and domestic absorption per capita. It finds no systematic response in market reform, secondary enrollment, employment, or the labor share.

The sequence of lower domestic political instability and higher investment and productivity, followed by capital accumulation, helps interpret the aggregate GDP response. Trade openness rises at short and long horizons. Because the outcomes are estimated separately, their causal ordering and roles in mediating the GDP response remain unresolved.

\subsection{Democracy and Geopolitical Alignment} \label{ss:geo_democracy}

Domestic political institutions and the external political environment are distinct dimensions of development, but they need not evolve independently. Institutions are a central determinant of development \citep{Acemoglu2001-io}, and \citet{Acemoglu2019-bo} document substantial long-run GDP gains from democratization. Internationally, democratic institutions are associated with preferential foreign aid \citep{Alesina2000-ad} and more stable relations among democracies \citep{Maoz1993-bi}. We first examine how democratic status is associated with bilateral reorientation across major nations, then compare the GDP responses to democracy and geopolitical alignment when both enter jointly.

\subsubsection{Democracy and Bilateral Geopolitical Reorientation}

Using the binary democracy indicator from \citet{Acemoglu2019-bo} (henceforth ANRR), we estimate
\begin{equation}
    S_{cj,t+h} = \beta_h^j D_{ct}^{\text{ANRR}} + \sum_{\ell=1}^4 \gamma_{h\ell}^j S_{cj,t-\ell} + \eta_c + \eta_t + \varepsilon_{cj,t+h}
    \label{eq:dem_bilateral}
\end{equation}
where $S_{cj,t}$ is country $c$'s bilateral alignment score with major nation $j$, and $D_{ct}^{\text{ANRR}}$ is its democratic status. The specification exploits within-country changes in democratic status after conditioning on four lags of the bilateral score, country fixed effects, and year fixed effects. Because $D_{ct}^{\text{ANRR}}$ records democratic status rather than a separately coded transition, $\beta_h^j$ traces its dynamic conditional association with bilateral relations at each horizon.

\begin{figure}[ht]
    \centering
    \begin{subfigure}[b]{\textwidth}
        \includegraphics[width=\textwidth]{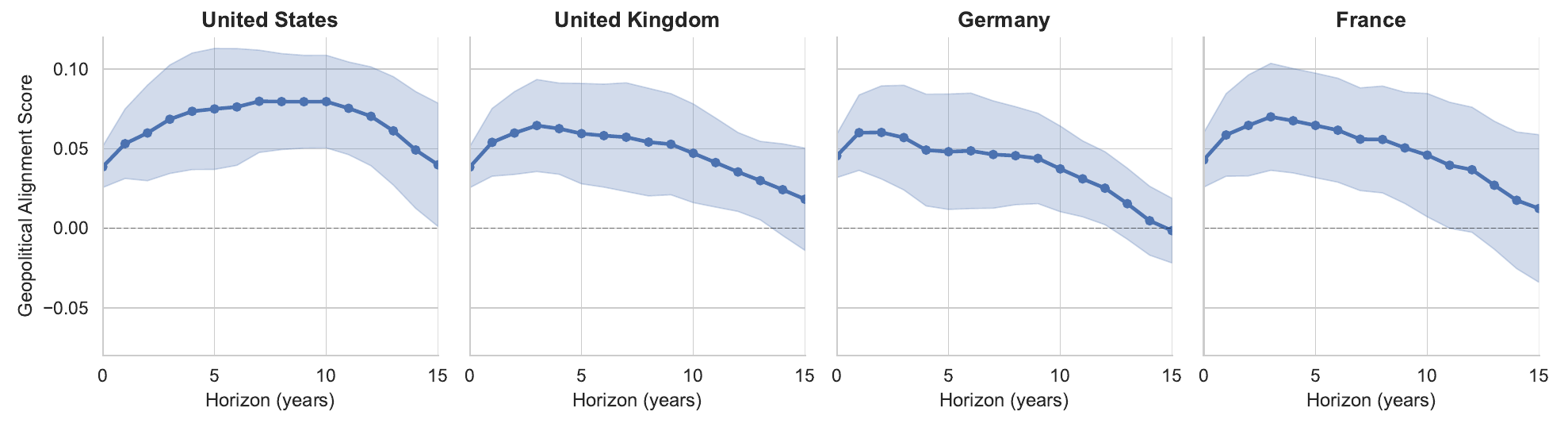}
        \caption{Selected Western Democracies}
    \end{subfigure}

    \vspace{0.5em}
    \begin{subfigure}[b]{\textwidth}
        \includegraphics[width=\textwidth]{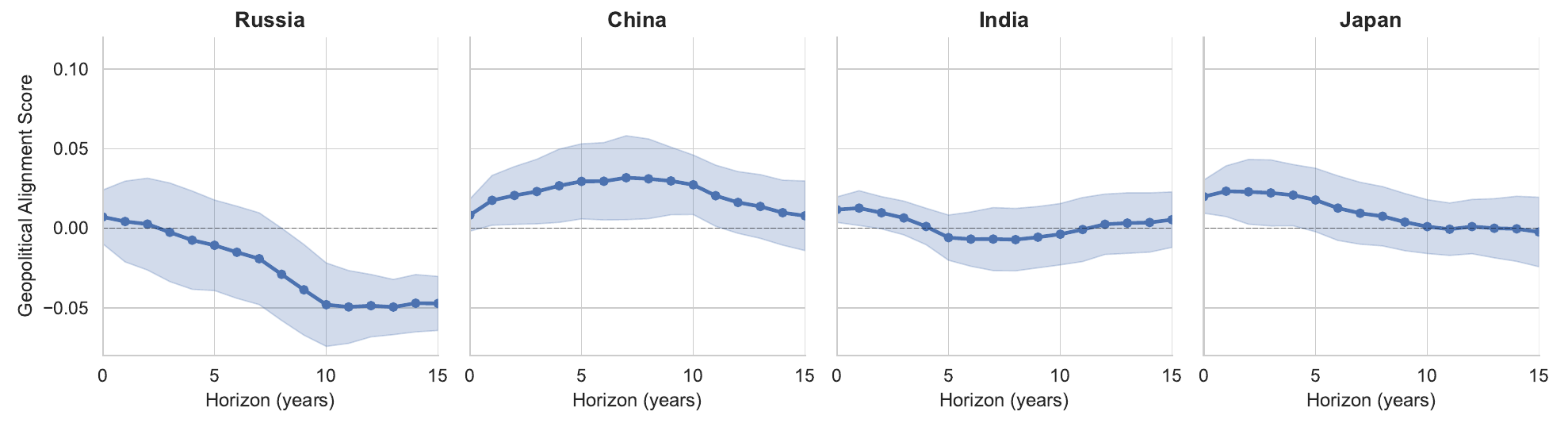}
        \caption{Other Major Nations}
    \end{subfigure}
    \caption{Democracy and Bilateral Geopolitical Reorientation}
    \label{fig:dem_bilateral_effects}
    \note{This figure reports horizon-specific coefficients on the ANRR democracy indicator from equation~\eqref{eq:dem_bilateral} over horizons 0--15. Panel (a) covers the United States, United Kingdom, Germany, and France. Panel (b) covers Russia, China, India, and Japan. The dependent variable is the bilateral geopolitical alignment score with the named major nation. The sample covers the ANRR democracy data (182 countries, 1960--2010). All specifications include four lags of the bilateral score, country fixed effects, and year fixed effects. Shaded areas are pointwise 95\% confidence intervals based on Driscoll--Kraay standard errors.}
\end{figure}

Panel~(a) of Figure~\ref{fig:dem_bilateral_effects} shows positive coefficients for all four Western democracies. The US coefficient rises from about 0.04 on impact to 0.08 around years 7--10 and remains positive through year 15. The coefficients for the United Kingdom, Germany, and France peak between 0.06 and 0.07 within the first five years before attenuating. Their common direction is consistent with closer bilateral ties among countries that share democratic institutions.

Panel~(b) shows heterogeneous associations with the other major nations. The Russian coefficient is initially near zero and declines to approximately $-0.05$ after year 10. The Chinese coefficient is positive at medium horizons, reaching about 0.03, while the Indian and Japanese coefficients are smaller and less persistent. The panels therefore show a geopolitical reorientation toward Western democracies rather than a uniform improvement across all bilateral relations.

\subsubsection{Democracy, Geopolitical Alignment, and GDP}

We next ask whether democratic status and the index contain the same information about subsequent GDP by estimating
\begin{equation*}
    y_{c,t+h} = \alpha_h^{\text{Geo}} p_{ct} + \alpha_h^{\text{Dem}} D_{ct}^{\text{ANRR}} + \boldsymbol{\gamma}_h^{\prime} \mathbf{x}_{ct} + \eta_c + \eta_t + \varepsilon_{c,t+h}
\end{equation*}
where both the geopolitical alignment index ($p_{ct}$) and democratic status enter jointly, and $\mathbf{x}_{ct}$ contains four lags of GDP per capita and geopolitical alignment. Following ANRR, we use year rather than region-year fixed effects, retaining the variation associated with regional democratization waves.\footnote{Regional democratization waves, including Latin America in the 1980s, Eastern Europe in the 1990s, and the Arab Spring in the 2010s, induce strong regional correlation in democratic transitions. Region-year fixed effects absorb this variation and may therefore attenuate the democracy coefficients. Under our baseline country and region-year fixed effects, however, adding democratic status leaves the GDP coefficient path of the index nearly unchanged (Appendix Figure~\ref{fig:democracy_geo_appendix}, panel~(b)).}

\begin{figure}[ht]
    \centering
    \begin{subfigure}[b]{0.48\textwidth}
        \includegraphics[width=\textwidth]{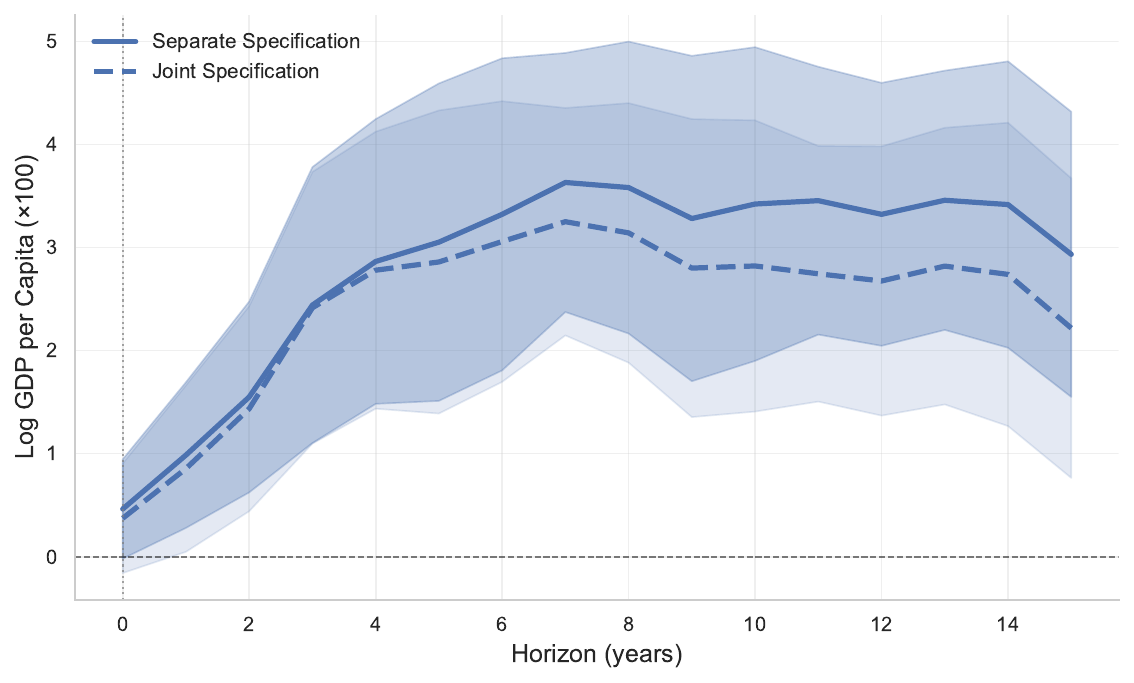}
        \caption{Geopolitical Alignment}
    \end{subfigure}
    \hfill
    \begin{subfigure}[b]{0.48\textwidth}
        \includegraphics[width=\textwidth]{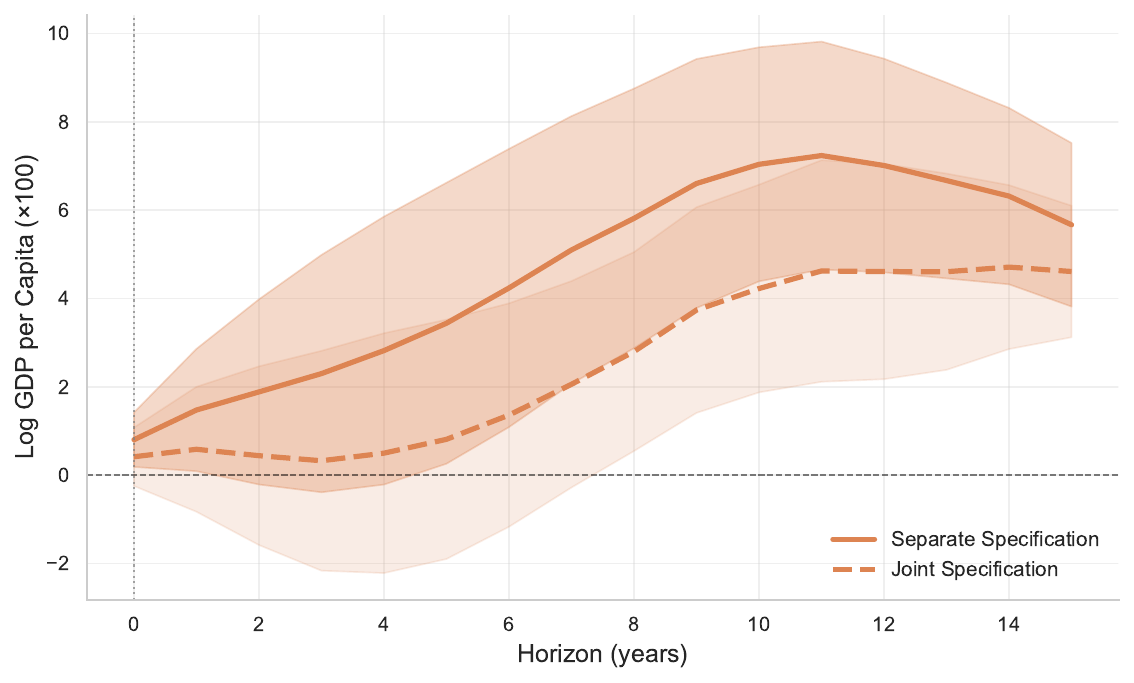}
        \caption{Democracy}
    \end{subfigure}
    \caption{Democracy, Geopolitical Alignment, and GDP}
    \label{fig:horse_race}
    \note{This figure compares horizon-specific GDP coefficient paths from separate and joint local projections using the original ANRR GDP series. All specifications include four lags of GDP, country fixed effects, and year fixed effects. In panel~(a), both specifications include four lags of geopolitical alignment, and the joint specification adds democratic status. In panel~(b), the separate specification follows ANRR, while the joint specification adds current geopolitical alignment and its four lags. The outcome is log GDP per capita ($\times 100$). Alignment coefficients correspond to a one-standard-deviation index change, while democracy coefficients correspond to a zero-to-one difference. Shaded areas are pointwise 95\% confidence intervals based on Driscoll--Kraay standard errors.}
\end{figure}

Panel~(a) of Figure~\ref{fig:horse_race} shows that the index retains a positive, hump-shaped GDP path when democratic status is added. The separate and joint coefficients peak at 3.6 and 3.3 log points at year~7. The joint path is moderately lower thereafter but remains positive through year~15. The external geopolitical environment therefore contains growth-related variation not subsumed by political regime type.

Panel~(b) shows that adding alignment lowers the democracy coefficient most at short and medium horizons. The separate democracy coefficient rises from 0.8 log points on impact to 7.2 log points at year~11, then declines to 5.7 log points by year~15. Adding current geopolitical alignment and its four lags lowers the coefficient throughout: the joint path reaches 4.7 log points at year~14 and is 4.6 log points at year~15.\footnote{Appendix Figure~\ref{fig:democracy_shock_decomposition} maps the democracy coefficients into transitory and sustained paths. The separate sustained democracy path reaches 19.1 log points after two decades, compared with 13.1 log points when alignment is included. The separate estimate closely matches the approximately 20 percent long-run GDP effect estimated by \citet{Acemoglu2019-bo}.}

Together with the bilateral reorientation in Figure~\ref{fig:dem_bilateral_effects}, these estimates show that the early democracy--GDP relationship overlaps with changes in geopolitical alignment, while a distinct democracy relationship emerges more gradually, consistent with slower-moving domestic institutional forces. The coefficient changes measure conditional overlap rather than a mediation share.

\subsection{Geopolitical Growth Accounting} \label{s:geo_growth_acct}

We use the estimated response paths to map observed changes in geopolitical alignment into implied GDP contributions. The first exercise applies the paths to within-country realignments. The second maps cross-country alignment gaps into GDP differences under a homogeneity assumption and is illustrative rather than separately identified.

\subsubsection{Decadal Realignments and Implied GDP Contributions}

For each country and calendar period, we compute implied within-period and 25-year GDP contributions from the transitory and sustained response paths.\footnote{Specifically, for within-period contributions, we calculate $\sum_{t=0}^{9} \alpha^{\text{transitory}}_{9-t} (p_{c,\tau+t} - p_{c,\tau})/\sigma_p$, where $\sigma_p = 0.141$ is the standard deviation of the index, so that index changes enter in the standard-deviation units of the reported coefficients, $\alpha^{\text{transitory}}_h$ is the transitory-path coefficient at horizon $h$, $p_{c,\tau+t} - p_{c,\tau}$ is the cumulative change in the geopolitical alignment index from the start of the decade, and $\tau$ denotes the first year of the decade. For 25-year contributions, we use $\alpha^{\text{sustained}}_{25} \times (p_{c,\tau+9} - p_{c,\tau})/\sigma_p$, where $\alpha^{\text{sustained}}_{25}$ is the sustained-path coefficient at the 25-year horizon. For 2020--2024, the analogous calculation uses the available five-year interval.}

\begin{figure}[ht]
    \centering
    \begin{subfigure}[b]{0.48\textwidth}
        \includegraphics[width=\textwidth]{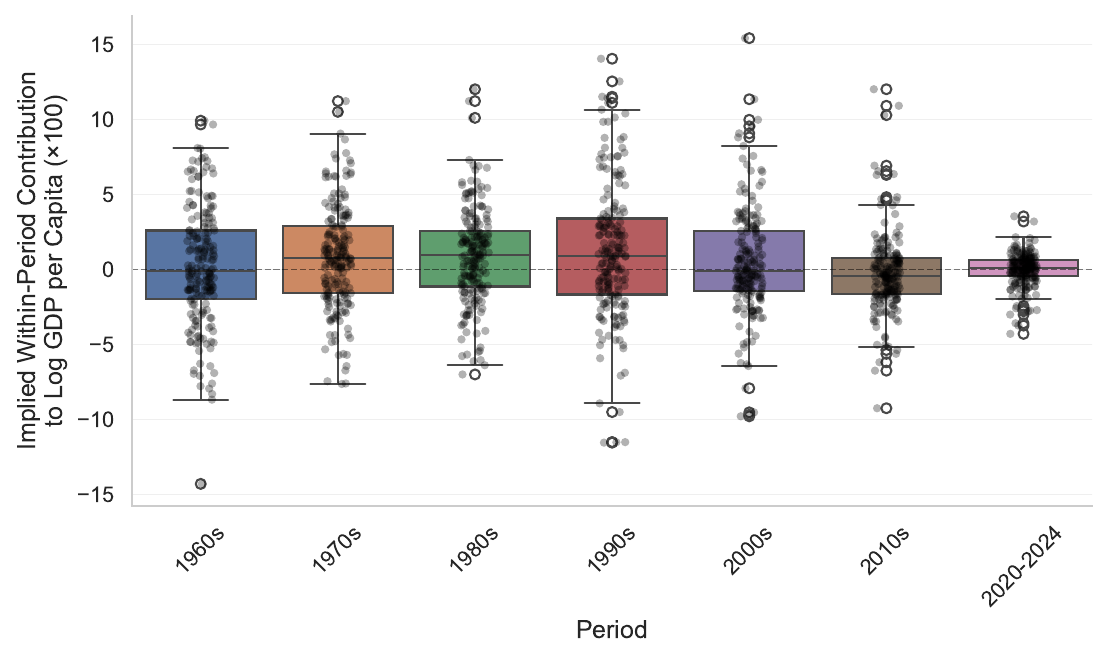}
        \caption{Within-Period Contribution}
        \label{fig:contemp_effects}
    \end{subfigure}
    \hfill
    \begin{subfigure}[b]{0.48\textwidth}
        \includegraphics[width=\textwidth]{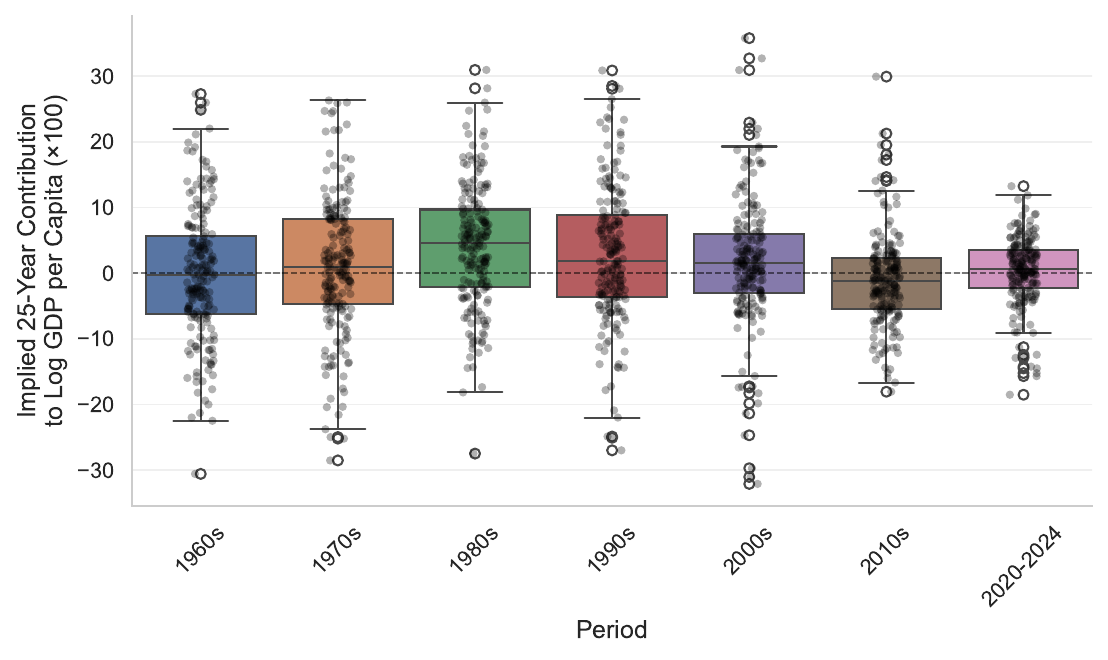}
        \caption{25-Year Contribution}
        \label{fig:longrun_effects}
    \end{subfigure}
    \caption{Implied GDP Contributions of Geopolitical Realignments by Period}
    \label{fig:growth_accounting_temporal}
    \note{This figure reports the distribution across countries of implied GDP contributions associated with observed within-period changes in geopolitical alignment. Panel (a) applies the transitory response path to changes realized during each period. Panel (b) applies the 25-year sustained-path coefficient to the period-end change. Values are log GDP per capita multiplied by 100. Boxes show interquartile ranges and whiskers extend to 1.5 times the interquartile range. Gray dots show individual countries, and open circles mark values beyond the whiskers. The dashed line marks zero. Periods cover calendar decades from the 1960s through the 2010s and 2020--2024, the last spanning only five years. Each period includes all 193 countries in the panel, since the implied contributions use only the alignment index.}
\end{figure}

Figure~\ref{fig:growth_accounting_temporal} shows positive median accounting contributions in the 1970s through the 1990s. The within-period median is 0.7--0.9 log points across those decades, while the 25-year median reaches 4.6 log points in the 1980s and remains 1.5--1.8 log points in the 1990s and 2000s. The within-period median is near zero in the 2000s, and both medians turn negative in the 2010s. This reversal gives economic scale to the distributional shift documented in Section~\ref{ss:landscape_geo_relations}: post-Cold War convergence maps into positive median contributions, while fragmentation after 2010 maps into a slightly negative median. The 2020--2024 period shows a partial median recovery but continued wide dispersion across countries.

\subsubsection{Illustrative Cross-Country GDP Contributions}

To provide cross-country scale, we apply the estimated within-country response paths to each country's deviation in geopolitical alignment from the annual cross-country median.\footnote{Formally, the geopolitical contribution to country $c$'s GDP in year $t$ is: $\Delta y_{c,t}^{\text{geo}} = \sum_{s=t-25}^{t} \alpha^{\text{transitory}}_{t-s} \times (p_{c,s} - p_{s}^{\text{median}})/\sigma_p$, where $\sigma_p = 0.141$ is the standard deviation of the index, $\alpha^{\text{transitory}}_h$ is the transitory-path coefficient at horizon $h$ and $p_s^{\text{median}}$ is the cross-country median of the geopolitical alignment index in year $s$. Deviations before 1960 are set to their 1960 values. Panel~(a) plots the within-period change $\Delta y_{c,\text{end}}^{\text{geo}} - \Delta y_{c,\text{start}}^{\text{geo}}$ for each period. Panel~(b) applies the 25-year sustained-path coefficient to cross-country differences in 2024: $(p_{c,2024} - p_{2024}^{\text{median}})/\sigma_p \times \alpha^{\text{sustained}}_{25}$.}

\begin{figure}[ht]
    \centering
    \begin{subfigure}[b]{0.48\textwidth}
        \includegraphics[width=\textwidth]{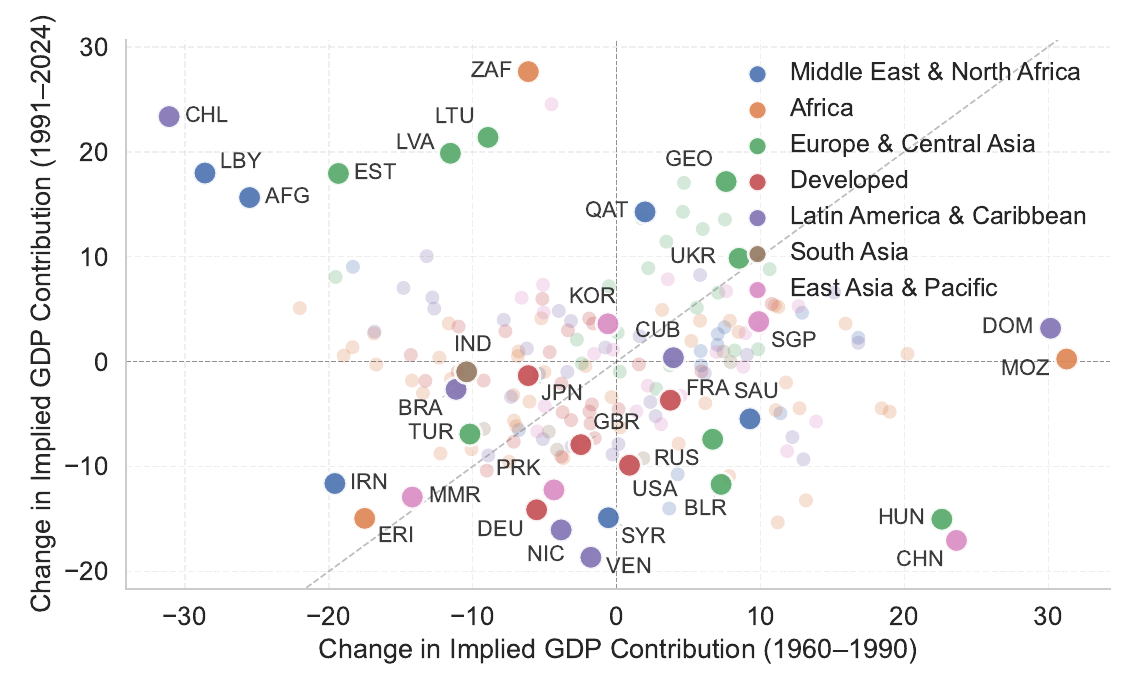}
        \caption{Changes in Implied GDP Contributions}
        \label{fig:geo_growth_scatter}
    \end{subfigure}
    \hfill
    \begin{subfigure}[b]{0.48\textwidth}
        \includegraphics[width=\textwidth]{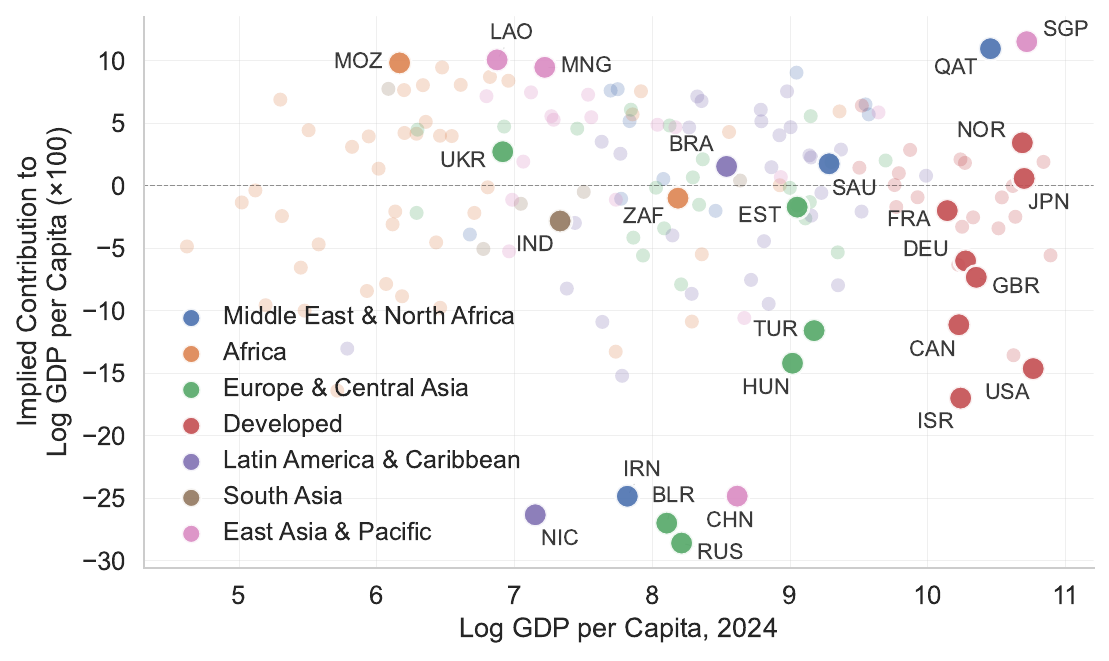}
        \caption{Implied GDP Contributions, 2024}
        \label{fig:geo_2024}
    \end{subfigure}
    \caption{Illustrative Cross-Country Accounting for Geopolitical Alignment}
    \label{fig:geo_cross_country}
    \note{This figure reports illustrative GDP accounting contributions obtained by applying the estimated response paths to countries' deviations in geopolitical alignment from the annual cross-country median. Panel~(a) plots the change in each country's implied contribution during 1960--1990 on the horizontal axis and during 1991--2024 on the vertical axis. Panel~(b) plots each country's implied contribution in 2024 against log GDP per capita, among countries with 2024 GDP data, with deviations measured from the 2024 median among these countries. Accounting contributions are measured as log GDP per capita multiplied by 100. Colors indicate regions. Horizontal dashed lines mark zero in both panels. Panel~(a) also includes a vertical zero line and a diagonal 45-degree line.}
\end{figure}

Panel~(a) of Figure~\ref{fig:geo_cross_country} compares changes in implied contributions over 1960--1990 and 1991--2024. Values span approximately $-31$ to $+31$ log points and are dispersed across all four quadrants. Chile shows the largest reversal, from about $-31$ log points in the first period to $+23$ in the second, and South Africa's contribution rises by about 28 log points in the second period. China and Hungary move in the opposite direction, and Venezuela and Nicaragua also move sharply toward more negative contributions in the second period. The limited concentration around the 45-degree line indicates that relative accounting contributions differ substantially across periods.

Panel~(b) reports the 2024 cross-country mapping. Implied contributions range from approximately $-29$ to $+12$ log points, with a longer negative than positive tail. Russia, Belarus, Nicaragua, Iran, and China lie at the lower end. Israel and Canada also have sizable negative contributions. Levels for the 24 major nations are mechanically lower because their index omits their own share of world GDP. Singapore lies at the upper end, followed by Qatar, Laos, and Mozambique.

There is no evident monotonic relationship between GDP per capita and the implied contribution. The range shows the scale implied by cross-country differences in the external geopolitical environment under the maintained homogeneity assumption. It is not a causal decomposition of cross-country income differences.

\section{Additional Robustness and Alternative Measures} \label{s:add_robust}

Dynamic panel estimates and an unsmoothed index give 25-year sustained-path coefficients of 10.5 and 10.7 log points, close to the baseline 10.1, while indices built from machine-coded event databases or UN voting do not reproduce the GDP relationship. Sanctions and formal alliances leave the alignment path nearly unchanged.

\subsection{Dynamic and Construction Robustness} \label{ss:dynamic_panel}

\paragraph{Dynamic Panel Estimates}

We complement the local-projection estimates with a dynamic panel model. Following the preferred specification in \citet{Acemoglu2019-bo}, we regress log GDP per capita on current geopolitical alignment, four lags of GDP and alignment, and the baseline country and region-year fixed effects (Appendix~\ref{app_b:dynamic_panel}). Under stable finite-order linear dynamics, this model and the local projections recover the same population response paths \citep{Plagborg-Moller2021-hi}, and the dynamic panel can improve precision by imposing the autoregressive structure \citep{olea2024double}. The response to a transitory alignment change rises from 0.5 log points on impact to 0.9 after one year and then declines while remaining positive. Cumulated into a sustained one-standard-deviation path, it reaches 10.5 log points at year 25, with a 95 percent confidence interval of 5.4 to 16.7, close to the local-projection estimate of 10.1 and its interval of 5.3 to 16.8 (Appendix Figure~\ref{fig:ar_irf}). The dynamic panel mainly extends the horizon by imposing parametric dynamics.

\paragraph{Unsmoothed Index} \label{ss:irf_geo_shocks}

Our baseline index smooths bilateral event scores to emphasize persistent relations. To test whether smoothing itself accounts for the GDP dynamics, we replace the index with the unsmoothed GDP-weighted average of bilateral event scores $\tilde{S}_{ct}$, equivalent to setting $\delta=1$ in equation~\eqref{eq:dynamic_score}.\footnote{We define $\tilde{S}_{ct}=\sum_{j\in\mathcal{N}\setminus\{c\}}\tilde{S}_{cj,t}\times\text{GDP share}_{jt}$, with $\tilde{S}_{cj,t}=0$ in dyad-years without events, and substitute it for $p_{ct}$ in the baseline local projection. The specification includes four lags of GDP and $\tilde{S}_{ct}$, country fixed effects, and region-year fixed effects.}

Appendix Figure~\ref{fig:event_based_irf} shows that the GDP coefficient remains positive when we use the unsmoothed index but is smaller than for the baseline index: it peaks at 1.2 log points in year 11, compared with 3.3 log points for the smoothed index. This attenuation is consistent with the rapid mean reversion of annual event scores: 58 percent of an initial change dissipates within one year. When the event-score path is sustained, however, the year-25 coefficient reaches 10.7 log points, close to the baseline estimate of 10.1. Smoothing therefore primarily filters transitory event variation. Appendix~\ref{app_b:irf_events} reports the supporting persistence and transitory-path results.

\subsection{Alternative Data Sources and Bilateral Benchmarks} \label{ss:alt_geo_measures}

Machine-coded event databases provide alternative event streams, UNGA ideal points summarize multilateral voting positions, and sanctions and formal alliances capture narrower bilateral margins.

\paragraph{Machine-Coded Event Databases} \label{ss:corpus_growth}

We reconstruct country-level indices from ICEWS, GDELT, and Phoenix using the same aggregation and dynamic recursion as for the baseline index, then standardize each index within the relevant overlap window. Appendix~\ref{app_b:measurement_alternatives} shows that our geopolitical alignment index produces a positive, hump-shaped GDP coefficient path in all three windows. The ICEWS and GDELT paths are statistically indistinguishable from zero, while the Phoenix path is small and fades after the first few years. Because the indices are standardized within each window, pure differences in scale cannot account for the contrast. Together with the measurement comparison in Section~\ref{ss:data_validation}, this pattern is consistent with the selection of consequential events as the source of the difference.

\paragraph{UN General Assembly Voting} \label{ss:UNGA_votes}

UNGA voting similarity is widely used to measure foreign-policy affinity \citep{Signorino1999-yb}. It locates countries in a common multilateral voting space, a different object from bilateral relationship intensity. We use the negative Ideal Point Distance (IPD) from \citet{Bailey2017-po}, so higher values indicate closer voting positions.

Panel~(a) of Figure~\ref{fig:unga_results} in Appendix~\ref{app_b:UNGA_votes} reports our primary comparison, negative IPD to the United States. Under the baseline country and region-year fixed effects, the GDP coefficient path is statistically indistinguishable from zero throughout the horizon. Panel~(b) replaces the United States reference with a GDP-weighted average of negative IPD to the 24 major nations. Because ideal-point distances locate countries in a common multilateral space, their GDP-weighted average is not a direct counterpart to our aggregation of bilateral relations. We nonetheless report it as a broader check that the US comparison is not driving the result. Its coefficient path is also statistically indistinguishable from zero.

These results and the case comparisons in Section~\ref{ss:measure_geo_relations} indicate that IPD captures broad geopolitical affinity but not the same time-varying bilateral information as the geopolitical alignment index.

\paragraph{Sanctions and Formal Alliances} \label{ss:sanctions}

Sanctions capture a salient but narrow margin of bilateral coercion. Using the Global Sanctions Database \citep{felbermayr2020global,yalcin2025global}, we aggregate active sanctions imposed by the 24 major nations, weighted by their GDP shares, and enter this exposure jointly with geopolitical alignment. Figure~\ref{fig:sanctions_horse_race} in Appendix~\ref{app_b:sanctions} shows that adding sanctions leaves the alignment coefficient path nearly unchanged. The sanctions coefficient is negative when entered separately and reaches its trough in year~3. At that horizon, it attenuates from $-8.2$ to $-5.6$ log points in the joint specification, a reduction of 31 percent, and no joint sanctions coefficient is statistically distinguishable from zero. This attenuation reflects the different timing of the two measures: the alignment index records the diplomatic deterioration surrounding a sanction and its announcement or lifting, while the sanctions measure tracks active enforcement.

Formal alliances capture a narrow, slow-moving margin of institutionalized security cooperation. Using the Alliance Treaty Obligations and Provisions dataset \citep{leeds2002alliance}, we aggregate active alliances with the same major nations and enter the resulting exposure jointly with geopolitical alignment. Figure~\ref{fig:alliance_horse_race} in Appendix~\ref{app_b:alliances} again shows little change in the alignment coefficient path. The alliance coefficients are positive and build through the first decade. In year~9, the estimate declines from 17.8 to 15.3 log points when alignment is included, a 14 percent attenuation, and the joint estimates are statistically indistinguishable from zero. Here the alignment index records diplomatic cooperation surrounding alliance formation and maintenance, whereas the alliance variable is a persistent treaty stock. The stability of the alignment path in both exercises shows that neither narrow bilateral margin accounts for the broader alignment--GDP relationship.

\section{Conclusion}\label{s:conclusion}

This paper introduces and validates an event-based measure of bilateral geopolitical alignment and uses it to document a large, persistent relationship between improvements in a country's external geopolitical environment and its subsequent economic growth. Our results carry a broader message for the growth literature: the traditional separation between ``domestic'' and ``international'' determinants of growth obscures important interactions between domestic factors and the external political environment. Improvements in geopolitical alignment are accompanied by lower domestic unrest and higher investment, capital, productivity, and trade openness, each with its own timing.

Two limitations remain. First, because the estimates exploit within-country changes, they speak to changes in geopolitical relations, not to the level consequences of sustained cooperation or isolation, and their causal reading requires that these changes be unrelated to other shocks to future growth. Second, the measure relies on LLM-compiled event records. Documented-event benchmarks and the human audit support its conflict--cooperation signal, but the selection of events remains an important source of measurement error.

The implications go beyond the general case for openness. Our accounting exercises suggest that the gains from cooperative relations are neither automatic nor permanent: fragmentation after 2010 turned the median implied contribution slightly negative, after positive contributions in the preceding decades. Because GDP gains follow improved relations with Western and other major nations alike, the relationship is not tied to alignment with any single bloc. As the post-Cold War order gives way to renewed strategic competition, the external political environment is an increasingly important dimension of economic development.

\bibliography{\bib}

\newpage
\appendix

\numberwithin{equation}{section}
\numberwithin{table}{section}
\numberwithin{figure}{section}
\makeatletter
\@addtoreset{theorem}{section}
\@addtoreset{proposition}{section}
\@addtoreset{lemma}{section}
\@addtoreset{corollary}{section}
\@addtoreset{definition}{section}
\@addtoreset{assumption}{section}
\@addtoreset{remark}{section}
\renewcommand{\thetheorem}{\thesection.\arabic{theorem}}
\renewcommand{\theproposition}{\thesection.\arabic{proposition}}
\renewcommand{\thelemma}{\thesection.\arabic{lemma}}
\renewcommand{\thecorollary}{\thesection.\arabic{corollary}}
\renewcommand{\thedefinition}{\thesection.\arabic{definition}}
\renewcommand{\theassumption}{\thesection.\arabic{assumption}}
\renewcommand{\theremark}{\thesection.\arabic{remark}}
\makeatother

\appendixtableofcontents
\startappendixcontents
\section{Event-based Measure of Geopolitical Alignment}\label{app_a:geo_relation_measures}

Appendix~\ref{app_a:geo_relation_measures} documents the event database and the alignment measures in the order of Section~\ref{s:measure_geo_relation}: the reference set (A.1), event compilation and statistics (A.2), dynamics of the scores (A.3), case studies and maps (A.4--A.5), the validation exercises (A.6--A.10), and index statistics and the comparison with the Geopolitical Risk Index (A.11--A.12).

\subsection{Major Nations} \label{app_a:major_nations}
\begin{figure}[ht]
    \centering
    \includegraphics[width=\linewidth]{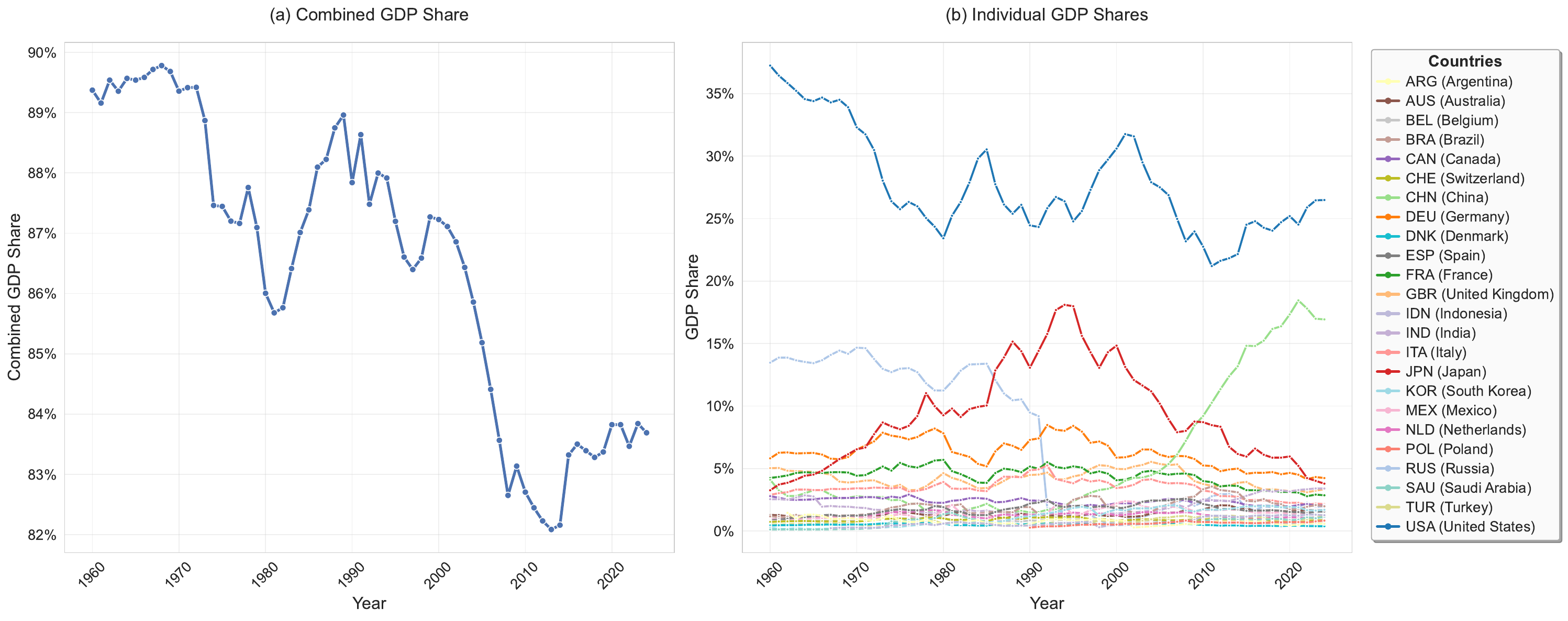}
    \caption{GDP Shares of Major Nations}
    \label{fig:gdp_shares}
    \note{Panel~(a) shows the combined GDP share of 24 major nations (Argentina, Australia, Belgium, Brazil, Canada, Switzerland, People's Republic of China, Germany, Denmark, Spain, France, United Kingdom, Indonesia, India, Italy, Japan, Republic of Korea, Mexico, Netherlands, Poland, Russian Federation, Saudi Arabia, Turkey, and United States) relative to global GDP over time. Panel~(b) shows the individual GDP share of each nation. Data are sourced from the World Development Indicators, with Soviet Union GDP before 1992 equal to US nominal GDP times the Soviet--US real GDP ratio at purchasing power parity from the Maddison Project \citep{bolt2025maddison}.}
\end{figure}

The reference set consists of 24 major nations: the G20 member countries other than South Africa, together with Belgium, Denmark, the Netherlands, Poland, Spain, and Switzerland. Twenty-three of them rank among the 24 largest economies by average share of world GDP over 1960--2024, and together they account for 82--90 percent of world GDP over the sample period. Figure~\ref{fig:gdp_shares} presents the time series of aggregate and individual GDP shares for these 24 major nations.\footnote{GDP data are nominal GDP in current USD and sourced from the World Development Indicators. For the Soviet Union before 1992, GDP is US nominal GDP multiplied by the ratio of Soviet to US real GDP at purchasing power parity from the Maddison Project, so the Soviet weight is not valued at market exchange rates.} 

\FloatBarrier

\subsection{Geopolitical Event Data: Methodology and Descriptive Statistics} \label{app_a:geo_events}

The all-pairs event compilation contains 833,485 events with complete CAMEO and Goldstein coding over 1950--2024. Unless otherwise noted, the descriptive statistics below concern the 373,020 major-involving events used in this paper over 1960--2024. Observations from the 1950s initialize the dynamic bilateral scores. Appendix~\ref{app_c:llm} summarizes the prompt, and the full production prompt is released with the replication package.

\subsubsection{Event Compilation and Analysis} \label{app_a:geo_events_analy}

This subsection provides additional detail on the event compilation and analysis procedure described in Section~\ref{ss:llm}.

\paragraph{Entity Verification and Historical Accuracy}
Our framework begins by verifying the political entities corresponding to each country pair for the target year. This step handles state succession, such as the dissolution of the Soviet Union and emergence of the Russian Federation on December 26, 1991. The prompt directs the model to official, international-organization, news-archive, and academic sources. For years when specified entities did not exist, the framework identifies the primary political entities controlling the relevant territories or populations and applies them consistently throughout the analysis. For example, the Baltic states' relations before 1991 are recorded through the Soviet Union, and Germany before 1990 combines the two German states.

\paragraph{Event Identification}
The LLM searches across multiple dimensions of bilateral relations to identify major geopolitical events that significantly affect or strongly indicate the state of the bilateral relationship. Table~\ref{tab:event_categories} summarizes the six dimensions and 25 subcategories that define the scope of geopolitical events in our analysis, with representative examples for each. The prompt instructs the model to check events against reliable sources and to exclude unverified ones. The prompt prioritizes events with demonstrable effects on the relationship's trajectory, with particular attention to political actions carried out through economic means and major developments concerning trade or security agreements. Appendix~\ref{app_c:events} details the selection criteria.

\begin{table}[p]
\centering
\caption{Categories of Bilateral Geopolitical Events}
\label{tab:event_categories}
\resizebox{\linewidth}{!}{
\begin{tabular}{>{\raggedright\arraybackslash}p{2.8cm}>{\raggedright\arraybackslash}p{7.0cm}>{\raggedright\arraybackslash}p{2.8cm}>{\raggedright\arraybackslash}p{7.0cm}}
\toprule
\textbf{Subcategory} & \textbf{Representative Events} & \textbf{Subcategory} & \textbf{Representative Events} \\
\midrule
\multicolumn{2}{l}{\textit{A. Economic Relations}} & \multicolumn{2}{l}{\textit{D. Legal, Territorial \& Movement}} \\
\addlinespace[0.1cm]
A1. Trade Policy
& Tariff imposition/removal; non-tariff barriers; FTA negotiations, signing, ratification, withdrawal
& D1. Legal Actions
& ICJ/WTO disputes; international arbitration; extradition; law enforcement cooperation \\
A2. Financial Relations
& Asset freezes; SWIFT exclusions; capital market denial; currency swaps; FDI restrictions; investment treaties
& D2. Territorial \& Maritime
& EEZ/continental shelf disputes; freedom of navigation operations; border demarcation; sovereignty claims \\
A3. Economic Coercion
& Trade embargoes; sectoral sanctions; Entity List designations; export controls; foreign aid; debt relief
& D3. Movement of People
& Visa regime changes; travel bans; guest worker programs; refugee/asylum policies \\
A4. Strategic Sectors
& Energy supply agreements; pipeline projects; nuclear cooperation; 5G bans; rare earth controls
& & \\
A5. Integration
& BRI/B3W infrastructure; port concessions; friend-shoring; supply chain arrangements; customs unions
& \multicolumn{2}{l}{\textit{E. Multilateral \& Global Governance}} \\
\addlinespace[0.1cm]
A6. Other Economic
& Regulatory harassment; boycott campaigns; government procurement restrictions; gray zone pressures
& E1. Int'l Organizations
& UNSC confrontations; GA coalition building; regional bloc membership changes \\
& & E2. Global Issues
& Climate cooperation; pandemic coordination; vaccine diplomacy; human rights \\
\addlinespace[0.15cm]
\multicolumn{2}{l}{\textit{B. Diplomatic \& Political Relations}} & \multicolumn{2}{l}{\textit{F. Other Significant Events}} \\
\addlinespace[0.1cm]
B1. Formal Diplomacy
& Embassy/consulate openings and closures; ambassador recalls; staff expulsions; d\'{e}marches
& F1. Historical \& Symbolic
& Apologies for historical wrongs; memorial visits; monument disputes \\
B2. High-Level Interactions
& Presidential visits; bilateral summits; summit boycotts; ministerial meetings; strategic dialogues
& F2. Humanitarian
& Disaster aid offers/rejections; joint rescue operations; evacuation cooperation \\
B3. Public Diplomacy
& Policy speeches; parliamentary resolutions; propaganda campaigns
& F3. Sports \& Events
& Olympic boycotts; joint hosting; World Expo participation \\
B4. Cultural \& Educational
& Cultural agreements/boycotts; scholarship programs; university partnerships; student visa policies
& F4. Technology \& Space
& Joint space missions; tech theft accusations; research terminations \\
& & F5. Environment
& Transboundary river/pollution disputes; conservation conflicts \\
\addlinespace[0.15cm]
\multicolumn{2}{l}{\textit{C. Security \& Defense}}
& F6. Communications
& Journalist expulsions; broadcasting restrictions; information warfare \\
\addlinespace[0.1cm]
C1. Military Cooperation
& Alliances; defense pacts; base arrangements; arms sales/embargoes; joint exercises; military aid
& F7. Other
& Emerging forms of bilateral interaction \\
C2. Security Incidents
& Border skirmishes; airspace/maritime violations; naval encounters; hybrid warfare
& & \\
C3. Intelligence \& Cyber
& Espionage revelations; intel officer expulsions; intel-sharing pacts; state-sponsored cyber attacks
& & \\
\bottomrule
\end{tabular}
}
\note{The production prompt assigns each event one primary subcategory (\texttt{event\_category}). The table lists all 25 subcategories with representative events. Appendix~\ref{app_c:events} describes the main ones.}
\end{table}

\paragraph{CAMEO Classification and Goldstein Scoring}
Each identified event is classified using the Conflict and Mediation Event Observations (CAMEO) framework, which categorizes international political actions along two dimensions: cooperation versus conflict and verbal versus material. The two dimensions define four quadrant classes. Within them, two-digit root codes identify general categories and three-digit event codes identify specific actions. The prompt supplies definitions, coding conventions, and anchored examples for this classification. After selecting a CAMEO code, the LLM assigns a Goldstein scale score ranging from $-10.0$ (maximum conflict) to $+10.0$ (maximum cooperation), using intensity bands and action-based anchors while allowing the event's specific bilateral context and severity to inform the final value.

Economic events are those in subcategories A1--A6 of Table~\ref{tab:event_categories}. Appendix~\ref{app_c:cameo} summarizes the CAMEO classification and Goldstein scoring rules.

\begin{table}[!t]
\centering
\caption{US--Soviet Union Bilateral Geopolitical Events in 1972: LLM Analysis Results}
\label{tab:us_soviet_1972}
\footnotesize
\resizebox{0.9 \linewidth}{!}{
\begin{tabular}{@{}>{\raggedright\arraybackslash}p{2.5cm}>{\raggedright\arraybackslash}p{5.6cm}>{\raggedright\arraybackslash}p{1.3cm}>{\raggedright\arraybackslash}p{1.75cm}>{\raggedright\arraybackslash}p{1.45cm}@{}}
\toprule
\textbf{Event Name} & \textbf{Event Description} & \textbf{CAMEO Class.} & \textbf{Category} & \textbf{Goldstein Score} \\
\midrule

Declaration of Basic Principles & 
Joint declaration of twelve principles guiding bilateral relations, including peaceful coexistence and renouncing unilateral advantage & 
Verbal Coop. (01-019) & 
Diplomatic \& Political &
$+5.0$ \\
\addlinespace[0.3em]

US-Soviet Grain Deal & 
Soviet purchase of 19 million metric tons of American grain, including nearly a quarter of the US wheat harvest & 
Material Coop. (06-061) & 
Economic &
$+6.0$ \\
\addlinespace[0.3em]

Moscow Summit & 
President Nixon's historic visit to Moscow, the first official US presidential visit to the USSR & 
Verbal Coop. (04-042) & 
Diplomatic \& Political &
$+6.0$ \\
\addlinespace[0.3em]

Environmental Protection Agreement & 
Bilateral agreement establishing cooperation on 11 areas including air and water pollution and nature preservation & 
Verbal Coop. (05-057) &
Other Significant Events &
$+6.5$ \\
\addlinespace[0.3em]

Biological Weapons Convention & 
US and USSR signed the BWC prohibiting development and stockpiling of biological weapons & 
Verbal Coop. (05-057) &
Multilateral &
$+7.0$ \\
\addlinespace[0.3em]

US-Soviet Trade Agreement & 
Comprehensive agreement providing reciprocal MFN tariff treatment, official trade offices, and government credits & 
Verbal Coop. (05-057) &
Economic &
$+7.0$ \\
\addlinespace[0.3em]

Apollo-Soyuz Agreement & 
Agreement for joint US-Soviet space mission, leading to the 1975 orbital rendezvous of American astronauts and Soviet cosmonauts & 
Verbal Coop. (05-057) &
Other Significant Events &
$+7.5$ \\
\addlinespace[0.3em]

Incidents at Sea Agreement & 
Protocols to prevent accidents between US and Soviet navies, including rules against collisions and simulated attacks & 
Verbal Coop. (05-057) &
Security \& Defense &
$+8.0$ \\
\addlinespace[0.3em]

SALT I Accords & 
ABM Treaty limiting defensive systems and Interim Agreement freezing ICBM/SLBM launchers & 
Verbal Coop. (05-057) &
Security \& Defense &
$+9.0$ \\

\bottomrule
\end{tabular}
}
\note{The table lists all nine events the model records for the US--Soviet Union pair in 1972, ordered by Goldstein score. For years before 1992 the model analyzes the Soviet Union in place of Russia (Appendix~\ref{app_c:prompt}). Event names and descriptions are abridged from the model output. Columns are defined as in Table~\ref{tab:us_russia_2022}.}
\end{table}

Tables~\ref{tab:us_russia_2022} and~\ref{tab:us_soviet_1972} illustrate the range of scores: 2022 US--Russia events are dominated by conflict scores ($-4.5$ to $-8.0$), while the 1972 détente period features exclusively cooperative events ($+5.0$ to $+9.0$).

\subsubsection{Statistics of Geopolitical Events} \label{app_a:geo_events_stats}

\begin{table}[ht]
\centering
\caption{Summary Statistics of Geopolitical Events by Decade, 1960--2024}
\label{tab:geopolitical_events_summary}
\footnotesize
\begin{tabular}{lcccccccc}
\toprule
 & 1960s & 1970s & 1980s & 1990s & 2000s & 2010s & 2020s & Total \\
\midrule
\multicolumn{9}{l}{\textbf{CAMEO Event Classification}} \\
\addlinespace[0.1cm]
\quad Verbal Cooperation & 14,092 & 17,523 & 18,393 & 24,878 & 36,418 & 53,698 & 31,555 & 196,557 \\
\quad Material Cooperation & 7,919 & 9,349 & 11,264 & 15,659 & 22,812 & 29,180 & 16,990 & 113,173 \\
\quad Verbal Conflict & 4,729 & 4,643 & 6,012 & 4,816 & 6,480 & 8,587 & 5,235 & 40,502 \\
\quad Material Conflict & 2,549 & 2,710 & 3,378 & 3,173 & 3,169 & 4,710 & 3,099 & 22,788 \\
\addlinespace[0.2cm]
\multicolumn{9}{l}{\textbf{Goldstein Scale Statistics}} \\
\addlinespace[0.1cm]
\quad Mean & 2.86 & 3.28 & 2.92 & 3.98 & 4.09 & 3.88 & 3.73 & 3.67 \\
\quad Std. Dev. & 4.87 & 4.62 & 4.74 & 4.27 & 3.87 & 3.78 & 3.88 & 4.18 \\
\quad Median & 5.00 & 5.00 & 4.50 & 6.00 & 5.00 & 4.50 & 4.50 & 5.00 \\
\addlinespace[0.2cm]
\multicolumn{9}{l}{\textbf{Event Category Classification}} \\
\addlinespace[0.1cm]
\quad Economic Relations (A) & 6{,}296 & 9{,}045 & 10{,}288 & 14{,}067 & 20{,}364 & 26{,}366 & 14{,}847 & 101{,}273 \\
\quad Diplomatic \& Political (B) & 11{,}559 & 12{,}515 & 14{,}082 & 16{,}058 & 25{,}591 & 41{,}086 & 23{,}529 & 144{,}420 \\
\quad Security \& Defense (C) & 3{,}635 & 3{,}205 & 3{,}879 & 4{,}361 & 5{,}507 & 8{,}343 & 5{,}175 & 34{,}105 \\
\quad Legal \& Territorial (D) & 1{,}588 & 1{,}889 & 1{,}614 & 2{,}226 & 3{,}565 & 4{,}736 & 2{,}443 & 18{,}061 \\
\quad Multilateral \& Governance (E) & 5{,}239 & 6{,}212 & 6{,}624 & 9{,}765 & 10{,}780 & 11{,}268 & 7{,}459 & 57{,}347 \\
\quad Other Significant Events (F) & 972 & 1{,}359 & 2{,}560 & 2{,}049 & 3{,}072 & 4{,}376 & 3{,}426 & 17{,}814 \\
\addlinespace[0.2cm]
\multicolumn{9}{l}{\textbf{Summary}} \\
\addlinespace[0.1cm]
\quad Total Events & 29,289 & 34,225 & 39,047 & 48,526 & 68,879 & 96,175 & 56,879 & 373,020 \\
\bottomrule
\end{tabular}
\note{CAMEO classifications follow the Conflict and Mediation Event Observations framework. The Goldstein scale ranges from $-10$ (most conflictual) to $+10$ (most cooperative). Event categories (A)--(F) and their subcategories are those of the LLM prompt (Table~\ref{tab:event_categories}). All entries represent event counts except Goldstein scale statistics. Each event is assigned to exactly one category, so category totals sum to the overall total. The table covers the 373,020 events of 1960--2024 in country pairs involving at least one of the 24 major nations; the 2020s column covers 2020--2024.}
\end{table}

\begin{figure}[ht]
\centering
\includegraphics[width=\textwidth]{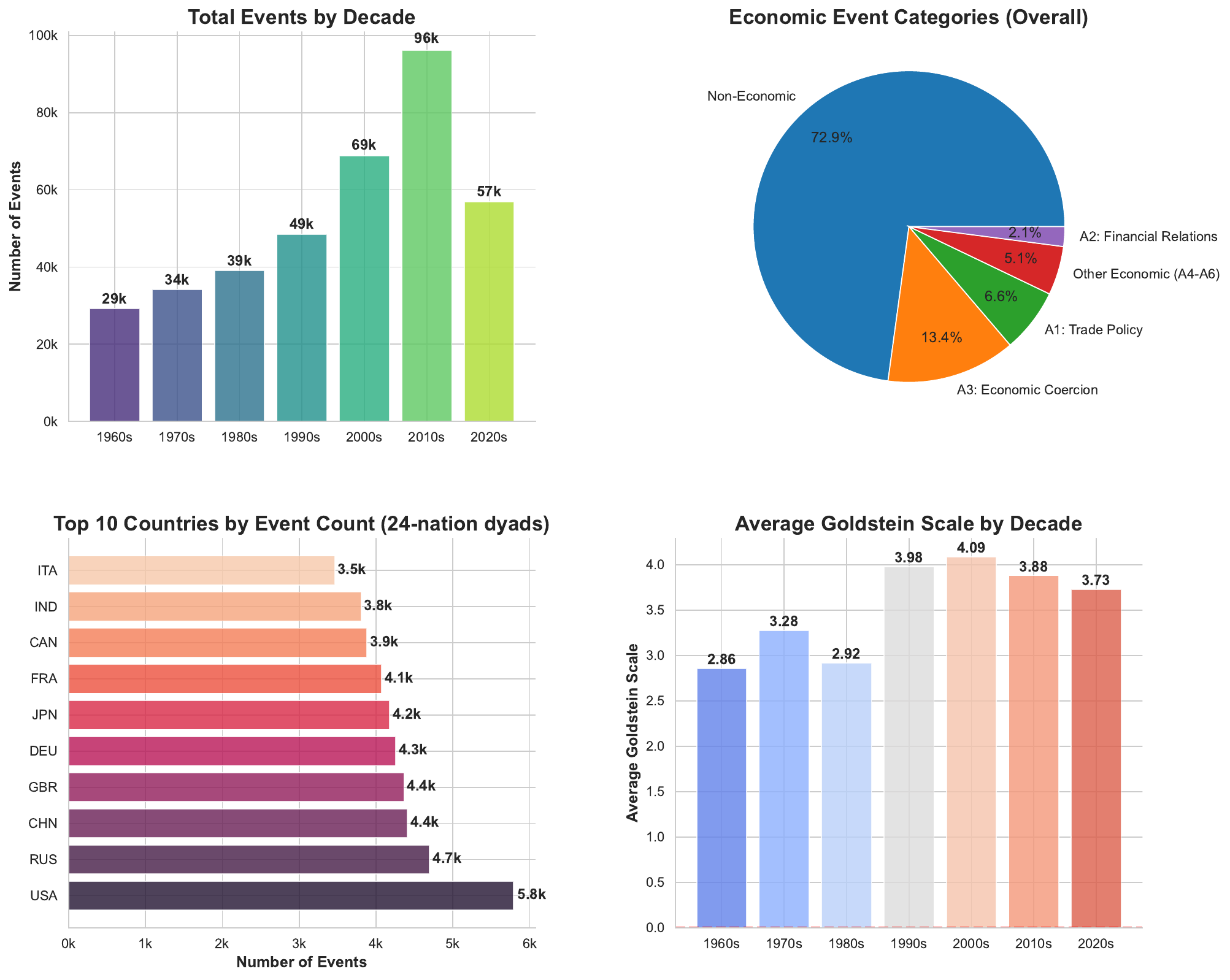}
\caption{Geopolitical Events Summary (1960--2024)}
\label{fig:geopolitical_events_summary}
\note{The upper panels show total events by decade (the 2020s cover 2020--2024) and the shares of events by economic category, where categories A1--A6 of Table~\ref{tab:event_categories} are economic and all other events non-economic. The lower-left panel counts each country's events in dyads between two of the 24 major nations. The lower-right panel shows the average Goldstein score by decade.}
\end{figure}

Table~\ref{tab:geopolitical_events_summary} and Figure~\ref{fig:geopolitical_events_summary} show three patterns. First, recorded events are mostly cooperative: 83 percent fall in the verbal or material cooperation classes. Second, total event volume more than tripled from 29,289 in the 1960s to 96,175 in the 2010s, driven primarily by expanding cooperative interactions. Third, mean Goldstein scores rose from Cold War levels (2.86--3.28) to a post-Cold War peak (3.98--4.09) before declining modestly from 3.88 in the 2010s to 3.73 in the 2020s, consistent with the shifts in the distribution of alignment in Section~\ref{ss:landscape_geo_relations}. Across the six event categories, diplomatic and political interactions (category B) account for 39 percent of all events, followed by economic relations (A, 27 percent) and multilateral governance (E, 15 percent). Security and defense events (C) constitute 9 percent, while legal/territorial (D) and other events (F) account for 5 percent each.

\paragraph{Coverage}\label{app_a:coverage_diagnostics}
The event record becomes denser over time and is uneven across dyads. The share of major-involving dyad-years with at least one recorded event rises from 32 percent in the 1950s to 80 percent in the 2020s, reflecting both the intensification of bilateral interactions and the greater availability of digitized source material for recent decades in LLM training corpora. Overall, 46 percent of major-involving dyad-years contain no recorded event, density is lowest for the poorest partners, and the most active fifth of dyads accounts for 42 percent of all events. Because the annual event scores average event-level Goldstein scores within each dyad-year, they capture the average intensity of recorded events rather than raw event volume, although the dynamic score moves toward neutral in years without recorded events. Appendix~\ref{app_a:anchoring} benchmarks the event record and intensity against externally documented events by decade and by partner income.

\FloatBarrier

\subsection{Dynamics of Geopolitical Alignment} \label{app_a:dyn_geo_relations}

This section estimates the persistence of the geopolitical alignment index and of the bilateral geopolitical alignment scores, comparing each with its unsmoothed counterpart. Persistence matters because the GDP coefficients in Section~\ref{ss:baseline} reflect both the initial change in alignment and the index's subsequent dynamics.

\paragraph{Methodology}
For the geopolitical alignment index, we estimate:
\begin{equation*}
    p_{c,t+h} = \alpha_h^{\text{Geo. Align.}} p_{ct} + \boldsymbol{\gamma}_h^{\prime} \mathbf{x}_{ct} + \varepsilon_{c,t+h}, \quad h = 0, 1, \ldots, 30
\end{equation*}
where $p_{ct}$ denotes either our dynamic geopolitical alignment index or the unsmoothed index, $\mathbf{x}_{ct}$ includes four lags of GDP and the respective geopolitical alignment measure, and we control for country and region-year fixed effects.

For bilateral geopolitical alignment scores, we estimate:
\begin{equation*}
    S_{ij,t+h} = \alpha_h^{\text{Score}} S_{ij,t} + \boldsymbol{\gamma}_h^{\prime} \mathbf{x}_{ij,t} + \varepsilon_{ij,t+h}, \quad h = 0, 1, \ldots, 30
\end{equation*}
where $S_{ij,t}$ stands for either the bilateral geopolitical alignment score or the annual event score $\tilde{S}_{ij,t}$, $\mathbf{x}_{ij,t}$ contains four lags of the score variable, and we include country-pair and year fixed effects with Driscoll--Kraay standard errors.

\begin{figure}[ht]
    \centering
    \begin{subfigure}[b]{0.49\textwidth}
        \centering
        \includegraphics[width=\textwidth]{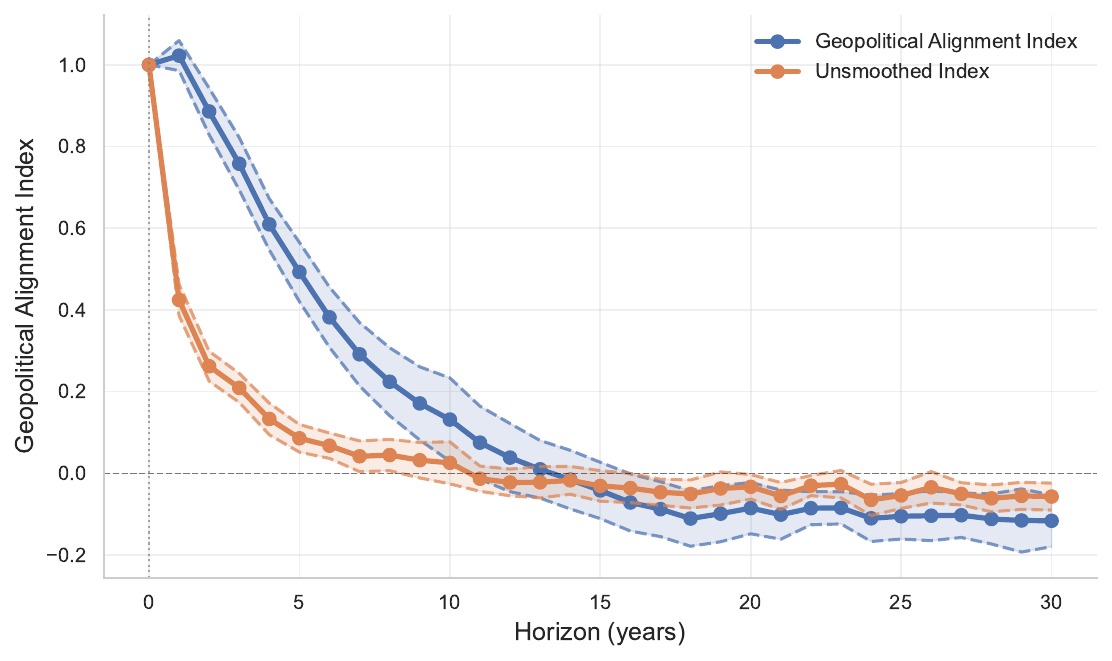}
        \caption{Country-Level Geopolitical Alignment Index}
        \label{fig:irf_geo_relation_compare}
    \end{subfigure}
    \hfill
    \begin{subfigure}[b]{0.49\textwidth}
        \centering
        \includegraphics[width=\textwidth]{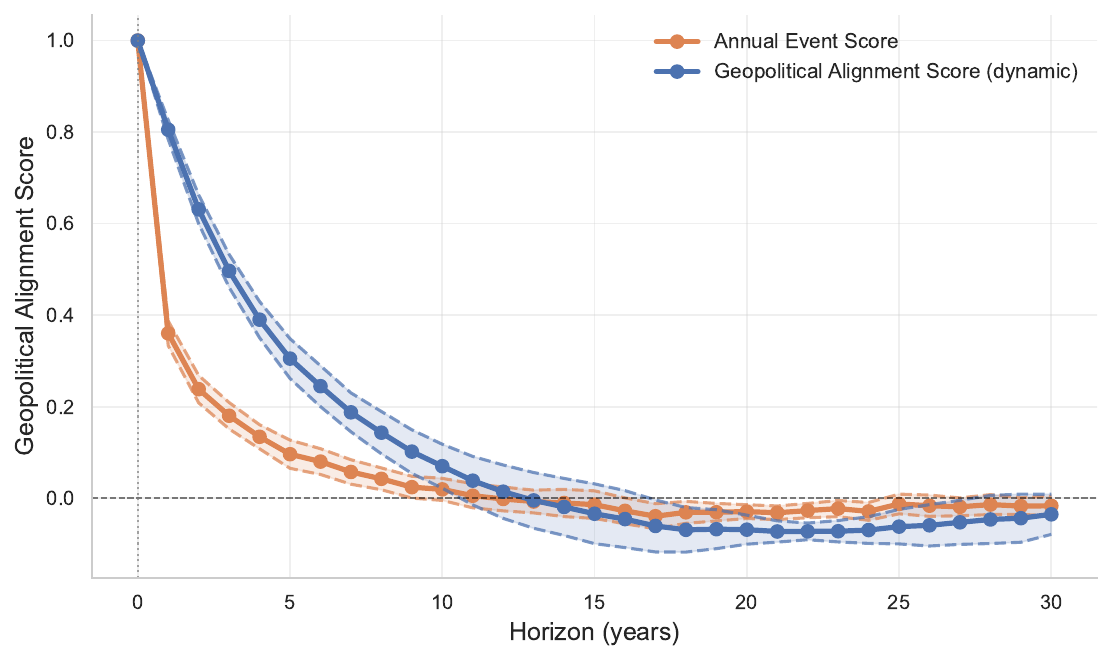}
        \caption{Bilateral Geopolitical Alignment Scores}
        \label{fig:irf_geo_score}
    \end{subfigure}
    \caption{Persistence of Geopolitical Alignment Measures}
    \label{fig:irf_geo_dynamics}
    \note{Coefficients from projecting each measure at $t+h$ on its current value. Panel (a): Geopolitical alignment index with dynamic measure (blue) and unsmoothed index (orange). Panel (b): Bilateral geopolitical alignment scores with dynamic score (blue) and annual event score (orange). Shaded areas are 95\% confidence intervals based on Driscoll--Kraay standard errors. Panel (a) controls for four lags of GDP and of the respective measure, country fixed effects, and region-year fixed effects. Panel (b) controls for four lags of the respective score, country-pair fixed effects, and year fixed effects. Panel~(a) uses the full unbalanced country panel. Panel~(b) uses all country pairs involving at least one of the 24 major nations, 1960--2024.}
\end{figure}

\paragraph{Results}
Panel (a) of Figure~\ref{fig:irf_geo_dynamics} shows that the dynamic geopolitical alignment index is persistent, with a half-life of about five years, while the unsmoothed index mean-reverts much faster (see also Figure~\ref{fig:event_dynamics}). The dynamic measure turns modestly negative after about year 15, as in panel~(a) of Figure~\ref{fig:baseline_irf}.

Panel (b) shows similar patterns at the bilateral level. The annual event score mean-reverts quickly, with a half-life under one year. In contrast, the dynamic score is more persistent, with coefficients remaining significant through year 10. This persistence comes from the decay parameter $\delta = 0.3$, which gives the weight on past events a half-life of about two years, consistent with a typical four-year electoral cycle (Section~\ref{ss:measure_geo_relations}).

\FloatBarrier

\subsection{Additional Case Validation for Geopolitical Alignment Scores} \label{app_a:case_validation}

This section provides additional validation of our bilateral geopolitical alignment scores beyond the cases presented in the main text (Figures~\ref{fig:geo_score_USA_RUS} and \ref{fig:geo_score_main_cases}). We examine eight bilateral relationships spanning different regions and relationship types: great-power rivalries, ideological revolutions, war-to-normalization transitions, sectarian conflicts, and diplomatic reversals.

\begin{figure}[ht]
    \centering
    \includegraphics[width=\linewidth]{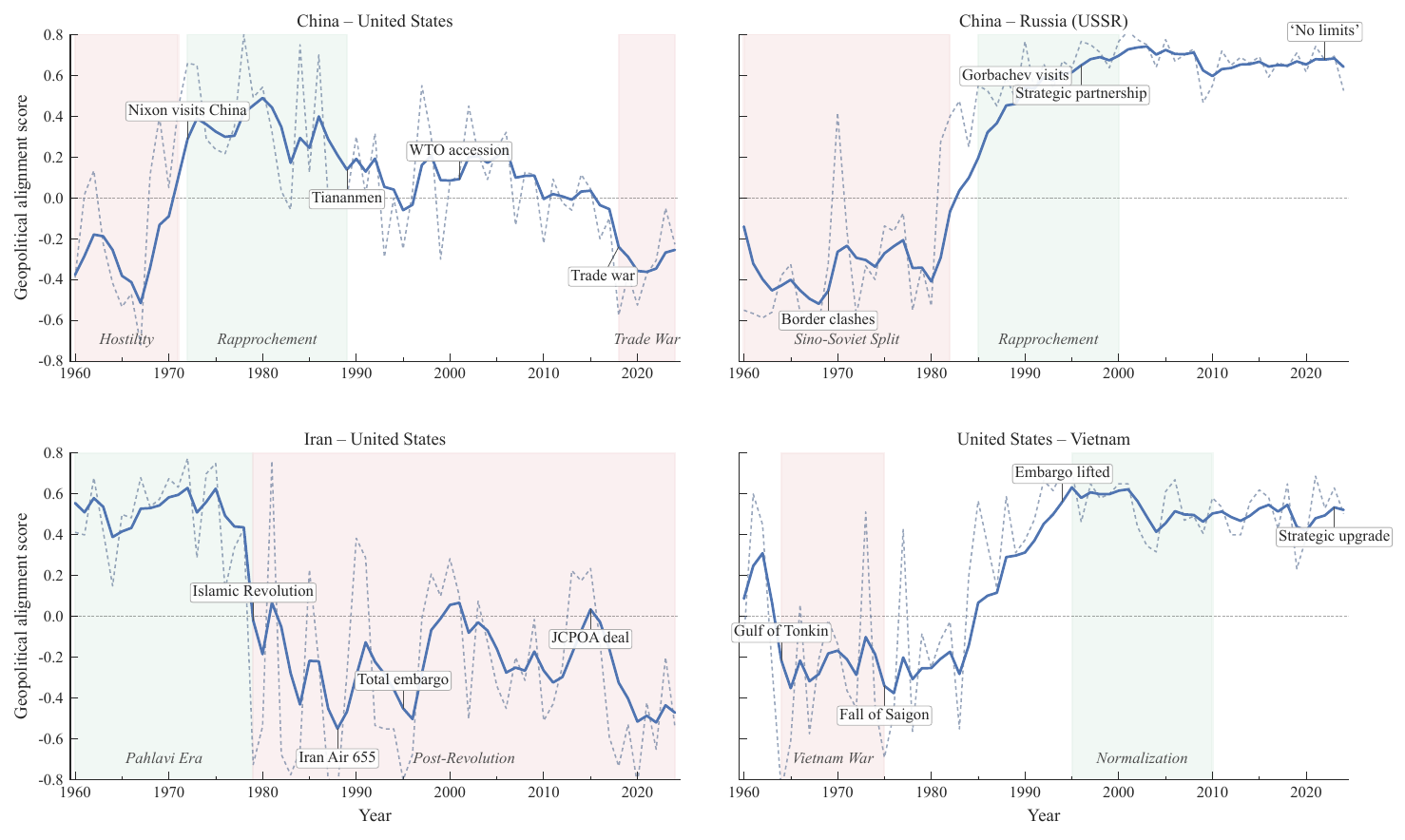}
    \caption{Geopolitical Alignment Scores: Relationships Involving Major Nations}
    \label{fig:appendix_major_powers}
    \note{Dynamic geopolitical alignment scores for four bilateral relationships involving major nations, 1960--2024. In each panel, the blue solid line shows the dynamic alignment score and the gray dashed line the annual (unsmoothed) event score. Shaded regions mark key historical periods, and boxed labels mark major bilateral events. Panels~(a)--(d) run left to right and top to bottom: China--United States, China--Russia, Iran--United States, and United States--Vietnam.}
\end{figure}

Figure~\ref{fig:appendix_major_powers} presents four relationships involving major nations. The China--US trajectory (panel~(a)) traces five distinct phases: Cold War-era hostility in the 1960s with scores near $-0.4$, Nixon's 1972 opening (scores rising to about $+0.4$), a post-Tiananmen deterioration around 1989, a period of WTO-era cooperation in the early 2000s, and the sharp trade-war decline after 2018 that returns scores to levels comparable to the 1960s. The China--Russia case (panel~(b)) moves from alliance through the Sino-Soviet split and 1969 border clashes (scores near $-0.5$) to Gorbachev's 1989 visit, the strategic partnership of the late 1990s, and the contemporary ``no limits'' partnership with scores near $+0.7$. Iran--US relations (panel~(c)) break sharply at the 1979 Islamic Revolution, with scores collapsing from approximately $+0.5$ during the Pahlavi era to sustained hostility. Scores fall below $-0.4$ in several periods, punctuated by the Iran Air 655 incident (1988), the total embargo (1995), and the brief JCPOA-related recovery around 2015. The US--Vietnam case (panel~(d)) demonstrates a complete war-to-normalization transition: scores fall from moderate levels in the early 1960s through the Gulf of Tonkin escalation and the Fall of Saigon to deep hostility, before recovering steadily from the mid-1980s, reaching cooperative levels around the 1994 embargo removal and remaining there through the 2023 strategic upgrade. Before 1976, the database records US relations with South Vietnam through 1963 and mainly with North Vietnam from 1964, so the early decline also reflects this change in counterpart.

\begin{figure}[ht]
    \centering
    \includegraphics[width=\linewidth]{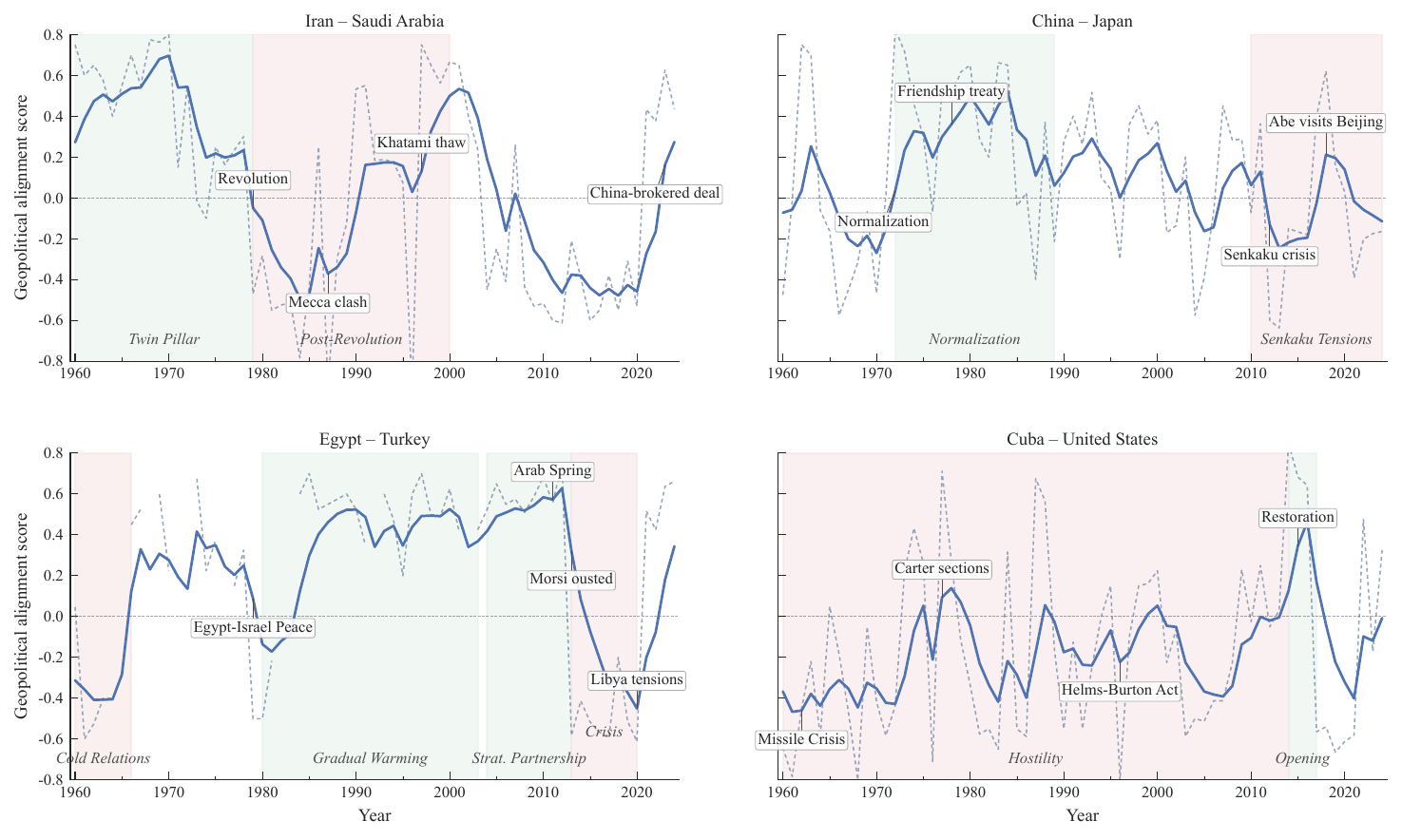}
    \caption{Geopolitical Alignment Scores: Regional Rivalries and Bilateral Dynamics}
    \label{fig:appendix_regional}
    \note{Dynamic geopolitical alignment scores for four regional bilateral relationships, 1960--2024. In each panel, the blue solid line shows the dynamic alignment score and the gray dashed line the annual (unsmoothed) event score. Shaded regions mark key historical periods, and boxed labels mark major bilateral events. Panels~(a)--(d) run left to right and top to bottom: Iran--Saudi Arabia, China--Japan, Egypt--Turkey, and Cuba--United States.}
\end{figure}

Figure~\ref{fig:appendix_regional} presents four regional relationships. The Iran--Saudi Arabia case (panel~(a)) tracks the Twin Pillar era of cooperation under the Shah (scores peaking near $+0.7$ in 1970), the sharp post-revolutionary deterioration including the 1987 Mecca clash (scores dropping to approximately $-0.4$), the Khatami-era thaw that temporarily restored scores above $+0.5$, and the subsequent deterioration followed by the 2023 China-brokered rapprochement. China--Japan relations (panel~(b)) show the post-war normalization around 1972 and the friendship treaty period of the late 1970s (scores reaching about $+0.5$ by the early 1980s), followed by gradual erosion and the sharp deterioration after the 2012 Senkaku/Diaoyu crisis, with a modest recovery around Abe's 2018 visit to Beijing. The Egypt--Turkey case (panel~(c)) shows cold relations in the early 1960s (scores near $-0.4$), mostly positive scores from 1966 apart from a dip in the early 1980s, a strategic partnership peak during the Arab Spring and Morsi presidency in 2011--2012 (scores reaching approximately $+0.6$), and the collapse after Morsi's ouster in 2013 with continuing tensions over Libya. Finally, Cuba--US relations (panel~(d)) show persistent hostility following the Castro revolution and the 1962 Missile Crisis, interrupted briefly by the Carter-era interest sections in the late 1970s, tightened further by the 1996 Helms-Burton Act, and punctuated by the sharp Obama-era restoration around 2015 before subsequent reversal.

\FloatBarrier

\subsection{Maps of Bilateral Geopolitical Alignment with the US and Its Rivals} \label{app_a:maps}

This section provides geographic visualization of geopolitical alignment, comparing Cold War patterns (1980) with the contemporary pattern (2024) for both the United States and its principal rivals.

\begin{figure}[ht]
    \centering
    \begin{subfigure}[b]{0.48\textwidth}
        \includegraphics[width=\textwidth]{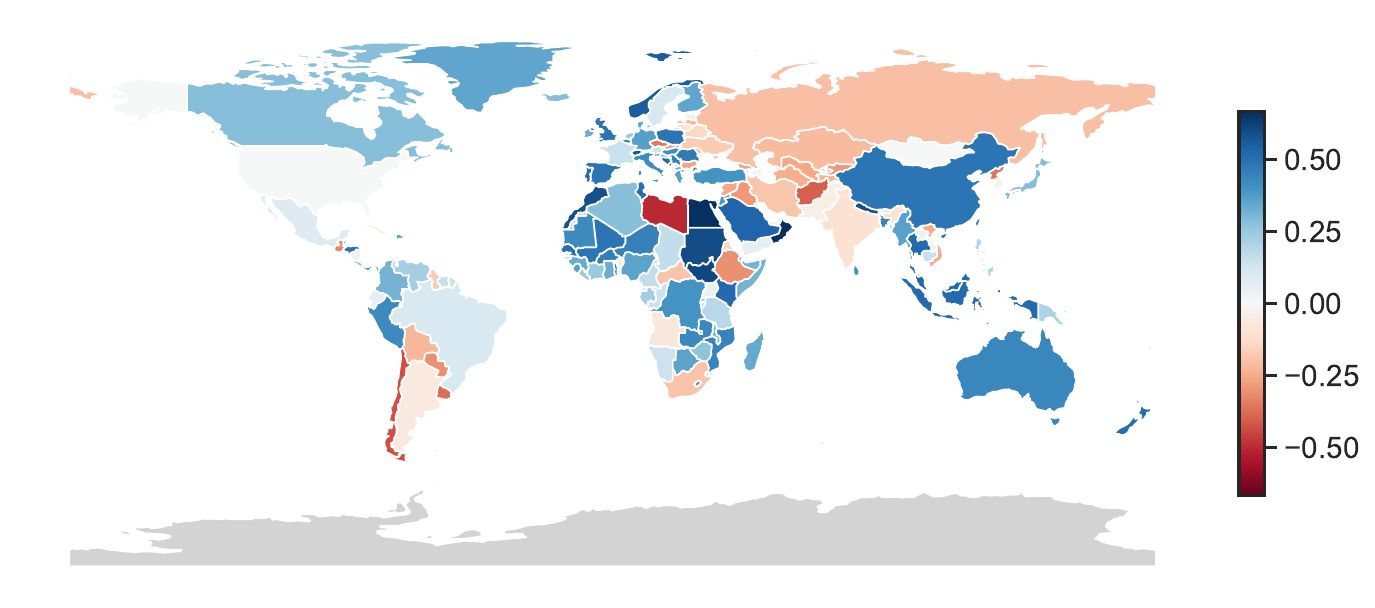}
        \caption{United States, 1980}
        \label{fig:geo_usa_1980}
    \end{subfigure}
    \hfill
    \begin{subfigure}[b]{0.48\textwidth}
        \includegraphics[width=\textwidth]{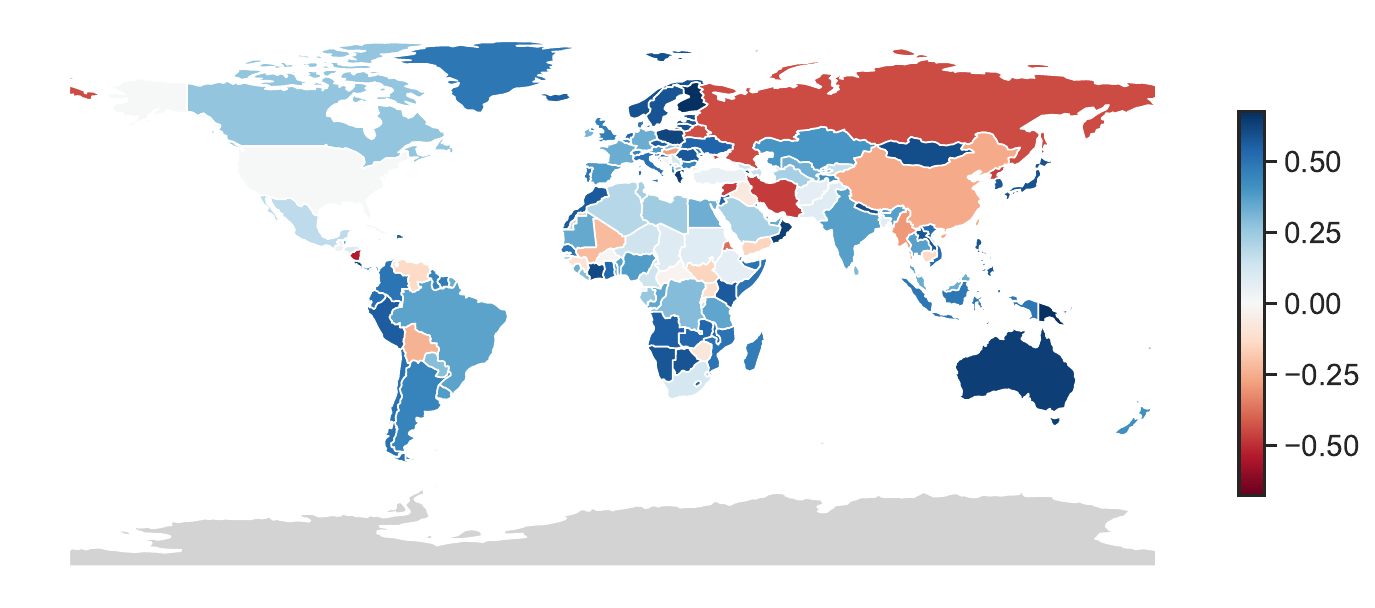}
        \caption{United States, 2024}
        \label{fig:geo_usa_2024}
    \end{subfigure}
    \caption{Map of Geopolitical Alignment with the United States}
    \label{fig:geo_usa_comparison}
    \note{World maps showing bilateral geopolitical alignment scores with the United States in 1980 and 2024. Blue indicates positive and red negative alignment.}
\end{figure}

Figure~\ref{fig:geo_usa_comparison} shows the shift from the Cold War pattern of 1980, with strong relations across NATO and Pacific allies and negative relations with the Soviet Union and parts of Eastern Europe, to a 2024 pattern in which core Western alliances remain intact, Russia appears in red and China in lighter red, and much of Latin America, Africa, and Southeast Asia has moderate or neutral relations.

\begin{figure}[ht]
    \centering
    \begin{subfigure}[b]{0.48\textwidth}
        \includegraphics[width=\textwidth]{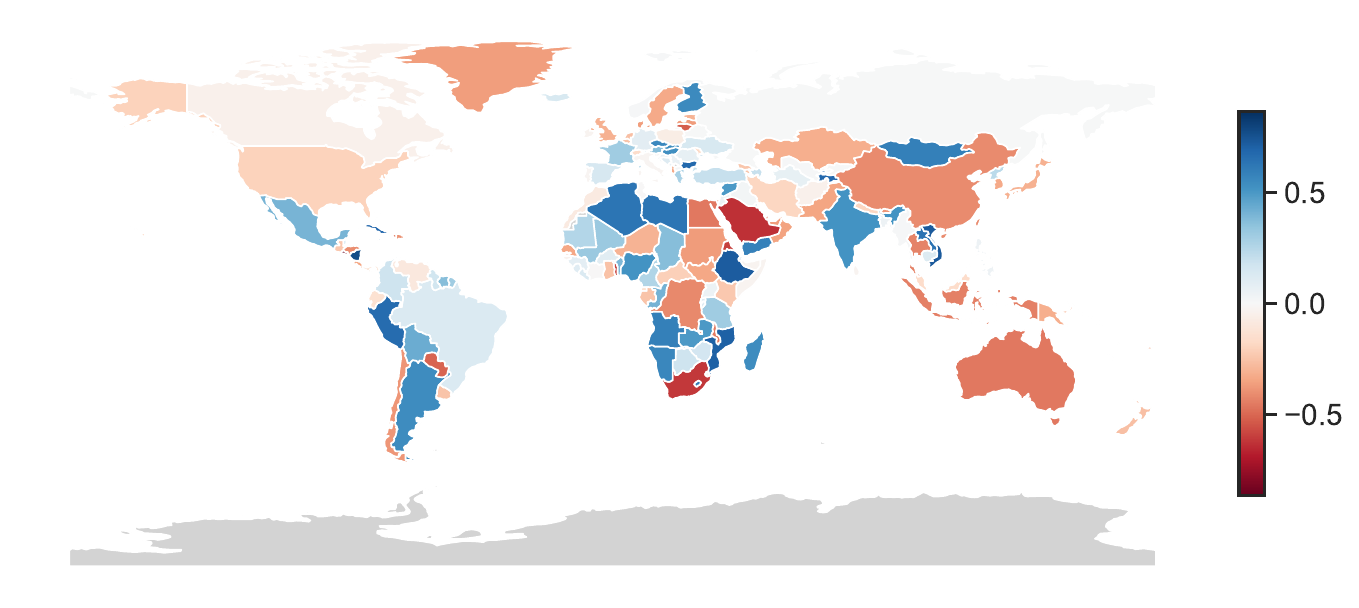}
        \caption{Soviet Union, 1980}
        \label{fig:geo_rus_1980}
    \end{subfigure}
    \hfill
    \begin{subfigure}[b]{0.48\textwidth}
        \includegraphics[width=\textwidth]{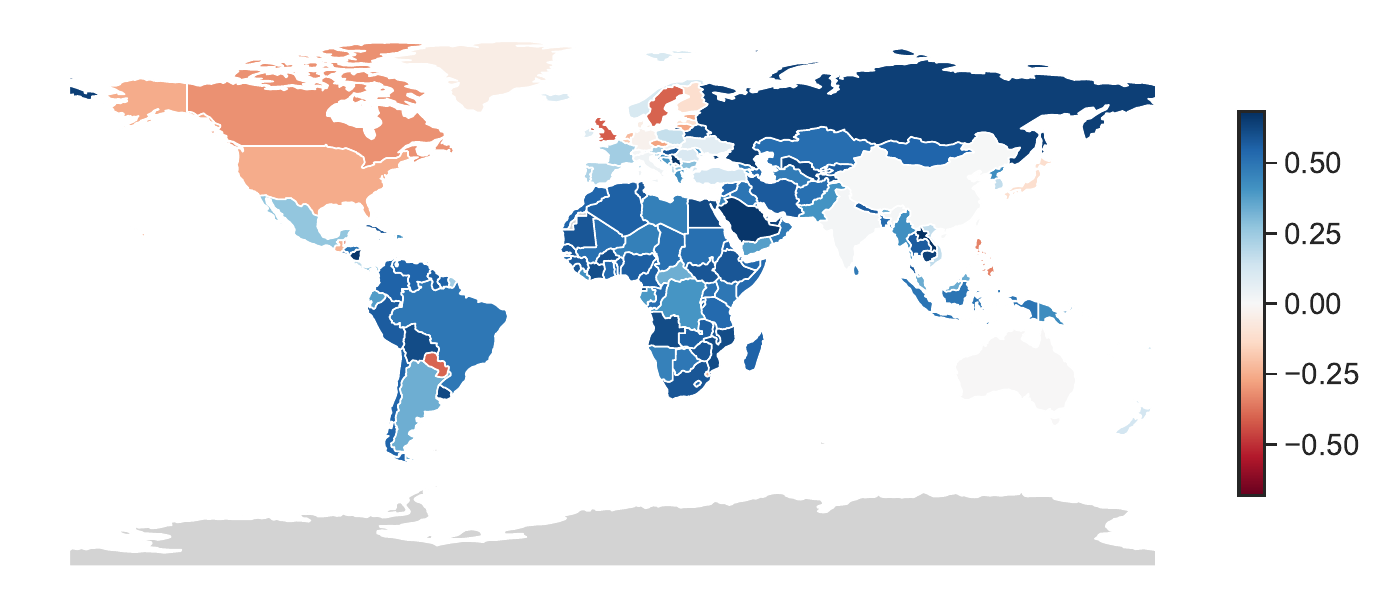}
        \caption{China, 2024}
        \label{fig:geo_chn_2024}
    \end{subfigure}
    \caption{Map of Geopolitical Alignment: Soviet Union (1980) vs. China (2024)}
    \label{fig:rival_comparison}
    \note{World maps showing bilateral geopolitical alignment scores with the principal US rival in 1980 (Soviet Union) and 2024 (China). Blue indicates positive and red negative alignment, as in Figure~\ref{fig:geo_usa_comparison}.}
\end{figure}

Figure~\ref{fig:rival_comparison} contrasts the geographic reach of America's principal rivals. The Soviet Union's 1980 positive relations were concentrated in Eastern Europe, Cuba, and select African and Middle Eastern allies, with China hostile (reflecting the Sino-Soviet split). China's 2024 map shows a different pattern: positive relations extend broadly across Sub-Saharan Africa, Central Asia, and Latin America, though China's relations with the United States, Canada, the Philippines, and several European countries are negative. The contrast between the opposing core blocs of the 1980 US and Soviet maps and the overlapping US and China maps of 2024, where most countries maintain positive relations with both, is consistent with the shift from Cold War bipolarity to the more complex distributional patterns documented in Figure~\ref{fig:evolution_distribution}.

\FloatBarrier

\FloatBarrier
\subsection{Validation against Externally Documented Events} \label{app_a:anchoring}

This section evaluates the event database against 4,670 bilateral developments independently dated by established external sources. The benchmark covers eight mutually exclusive conflict and cooperative event classes involving at least one of the 24 major nations between 1950 and 2023. Table~\ref{tab:benchmark_design} summarizes the event definitions, sources, matching windows, and class sizes.

\begin{table}[!htbp]
\centering
\caption{Construction of the Externally Documented Benchmark}
\label{tab:benchmark_design}
\scriptsize
\setlength{\tabcolsep}{3.5pt}
\begin{tabular}{@{}p{0.22\textwidth}p{0.29\textwidth}p{0.23\textwidth}cr@{}}
\toprule
Documented event class & Definition and date & External source & \shortstack{Matching\\window} & \shortstack{Documented\\events} \\
\midrule
\multicolumn{5}{@{}l}{\textit{Conflict events}} \\
War onset & First year of an interstate war & COW Dyadic MID 4.03 & $[-1,+1]$ & 51 \\
Use-of-force MID, excluding wars & First year of a nonwar militarized dispute involving force & COW Dyadic MID 4.03 & $[-1,+1]$ & 619 \\
Sanction imposition & First year of a unilateral sanction spell & Global Sanctions Database v4 & $[-1,+1]$ & 1,326 \\
Diplomatic severance & First census wave with no exchange following mutual resident representation & COW Diplomatic Exchange 2006.1 & $[-4,+1]$ & 28 \\
\addlinespace
\multicolumn{5}{@{}l}{\textit{Cooperative events}} \\
Sanction removal & First year after a unilateral sanction spell ends, $\textit{end}+1$ & Global Sanctions Database v4 & $[-1,+1]$ & 705 \\
Diplomatic establishment & First census wave with mutual resident representation following no exchange & COW Diplomatic Exchange 2006.1 & $[-4,+1]$ & 505 \\
Trade-agreement entry & Year that a regional trade agreement enters into force & CEPII Gravity V202211 & $[-1,+1]$ & 1,199 \\
Alliance formation & Year that a bilateral alliance enters into force & ATOP 5.1 & $[-1,+1]$ & 237 \\
\midrule
All classes & & & & 4,670 \\
\bottomrule
\end{tabular}
\note{The sample contains mutually exclusive bilateral developments involving at least one of the 24 major nations between 1950 and 2023. COW = Correlates of War; MID = Militarized Interstate Disputes. Sources: \citet{maoz2019dyadic}, \citet{felbermayr2020global}, \citet{yalcin2025global}, \citet{bayer2006diplomatic}, \citet{conte2022cepii}, and \citet{leeds2002alliance}. Zero in the matching-window column denotes the documented year. When a use-of-force dispute overlaps a war in the same dyad-year, only the war onset is retained: of 46 such disputes, 45 are represented by war onsets, and one, KOR--VNM in 1965, is excluded because common country codes cannot distinguish North from South Vietnam.}
\end{table}

\paragraph{Detection and Annual-Score Differences}
For each documented event, \textit{any event recorded} indicates that the database contains an event for the dyad within the matching window. \textit{Event in the expected direction} additionally requires conflict for the first four classes or cooperation for the final four. A \textit{broad event-family match} further requires either the expected substantive category or the expected CAMEO root. This criterion permits related actions within the same bilateral development. It does not require the recorded text to name the specific documented episode.

To examine the annual score, let $g_{ij,t}$ denote the mean Goldstein score across all events recorded for dyad $ij$ in year $t$, before smoothing and expressed in raw Goldstein points. Equivalently, $g_{ij,t}=10\tilde S_{ij,t}$, where $\tilde S_{ij,t}$ is the annual event score defined in Section~\ref{ss:measure_geo_relations}. Let $B^k_{ij,t}$ equal one when a documented event of class $k$ occurs in that dyad-year. We estimate
\begin{equation}
g_{ij,t}
=
\sum_{k=1}^{8}\beta_k B^k_{ij,t}
+\mu_{ij}
+\lambda_{i,t}
+\lambda_{j,t}
+\varepsilon_{ij,t},
\label{eq:benchmark_event_association}
\end{equation}
where all eight documented-event indicators enter jointly, $\mu_{ij}$ are dyad fixed effects, and $\lambda_{i,t}$ and $\lambda_{j,t}$ are country-by-year fixed effects for the two endpoints. Standard errors are clustered by dyad. Because $g_{ij,t}$ is defined only when the dyad records at least one event, the estimation sample consists of dyad-years with recorded events. The coefficient $\beta_k$ is the descriptive conditional difference in the annual mean Goldstein score associated with a documented event of class $k$, not a causal effect.

Table~\ref{tab:anchor_ladder} reports the detection rates and annual-score differences for the full sample, before 1990, and from 1990 onward. Across the 4,670 documented events, the database records some event for 97.6 percent, an event in the expected direction for 92.1 percent, and a broad event-family match for 82.6 percent. All eight full-sample annual-score differences have the expected direction, and seven are statistically distinguishable from zero at the 5 percent level. The expected directions are preserved before and after 1990, although the absolute differences are smaller after 1990 in seven of the eight classes.

\begin{table}[!t]
\begin{center}
\captionof{table}{Detection and Conditional Annual-Score Differences}
\label{tab:anchor_ladder}
\scriptsize
\setlength{\tabcolsep}{2.0pt}
\begin{tabular*}{\textwidth}{@{\extracolsep{\fill}}lcccccc@{}}
\toprule
& \multicolumn{3}{c}{Detection rate} & \multicolumn{3}{c}{Conditional annual-score difference (SE)} \\
\cmidrule(lr){2-4}\cmidrule(l){5-7}
Documented event class & \shortstack{Any event\\recorded} & \shortstack{Expected\\direction} & \shortstack{Broad event-family\\match} & Full sample & Before 1990 & 1990 onward \\
\midrule
\multicolumn{7}{@{}l}{\textit{Conflict events}} \\
War onset & 1.00 & 1.00 & 1.00 & $-5.02$ (0.62) & $-5.75$ (0.67) & $-1.77$ (0.97) \\
Use-of-force MID, excluding wars & 0.99 & 0.92 & 0.78 & $-2.03$ (0.22) & $-2.21$ (0.31) & $-0.85$ (0.27) \\
Sanction imposition & 1.00 & 0.89 & 0.73 & $-2.15$ (0.13) & $-2.53$ (0.33) & $-1.76$ (0.13) \\
Diplomatic severance & 0.96 & 0.86 & 0.79 & $-1.15$ (1.08) & $-1.39$ (1.77) & $-1.15$ (1.25) \\
\addlinespace
\multicolumn{7}{@{}l}{\textit{Cooperative events}} \\
Sanction removal & 1.00 & 0.94 & 0.80 & $+0.39$ (0.17) & $+0.01$ (0.43) & $+0.66$ (0.17) \\
Diplomatic establishment & 0.93 & 0.92 & 0.88 & $+0.59$ (0.18) & $+0.68$ (0.23) & $+0.46$ (0.26) \\
Trade-agreement entry & 0.95 & 0.93 & 0.91 & $+0.58$ (0.09) & $+1.52$ (0.30) & $+0.36$ (0.08) \\
Alliance formation & 1.00 & 1.00 & 0.99 & $+1.82$ (0.19) & $+2.77$ (0.33) & $+0.85$ (0.21) \\
\midrule
All classes & 0.98 & 0.92 & 0.83 & & & \\
\bottomrule
\end{tabular*}
\note{The detection columns use the matching windows in Table~\ref{tab:benchmark_design}. ``Expected direction'' requires conflict for the first four classes and cooperation for the final four. A broad event-family match further requires either the expected substantive category or the expected CAMEO root. The final three columns report coefficients from equation~\eqref{eq:benchmark_event_association}, estimated over the full panel and separately before 1990 and from 1990 onward. All eight documented-event indicators enter jointly in every regression. The three estimation samples contain 174,104, 70,158, and 103,946 dyad-years, respectively. Standard errors in parentheses are clustered by dyad.}
\end{center}

\end{table}

\paragraph{Variation in Matching over Time and Partner Income}
We examine whether broad event-family matching varies systematically with calendar time or partner income. For documented event $a$, we estimate separate linear probability models,
\begin{equation}
D^F_a=\alpha_{k(a)}+\theta Z_a+u_a,
\label{eq:benchmark_match_gradient}
\end{equation}
where $D^F_a$ indicates a broad event-family match, $\alpha_{k(a)}$ are documented-event-class fixed effects, and $Z_a$ is either decades since 1950 or the 1950s income quintile of the non-major partner. Standard errors are clustered by dyad. The income specification excludes major--major events and events without a mapped income quintile.

\begin{figure}[!tbp]
    \centering
    \includegraphics[width=0.94\linewidth]{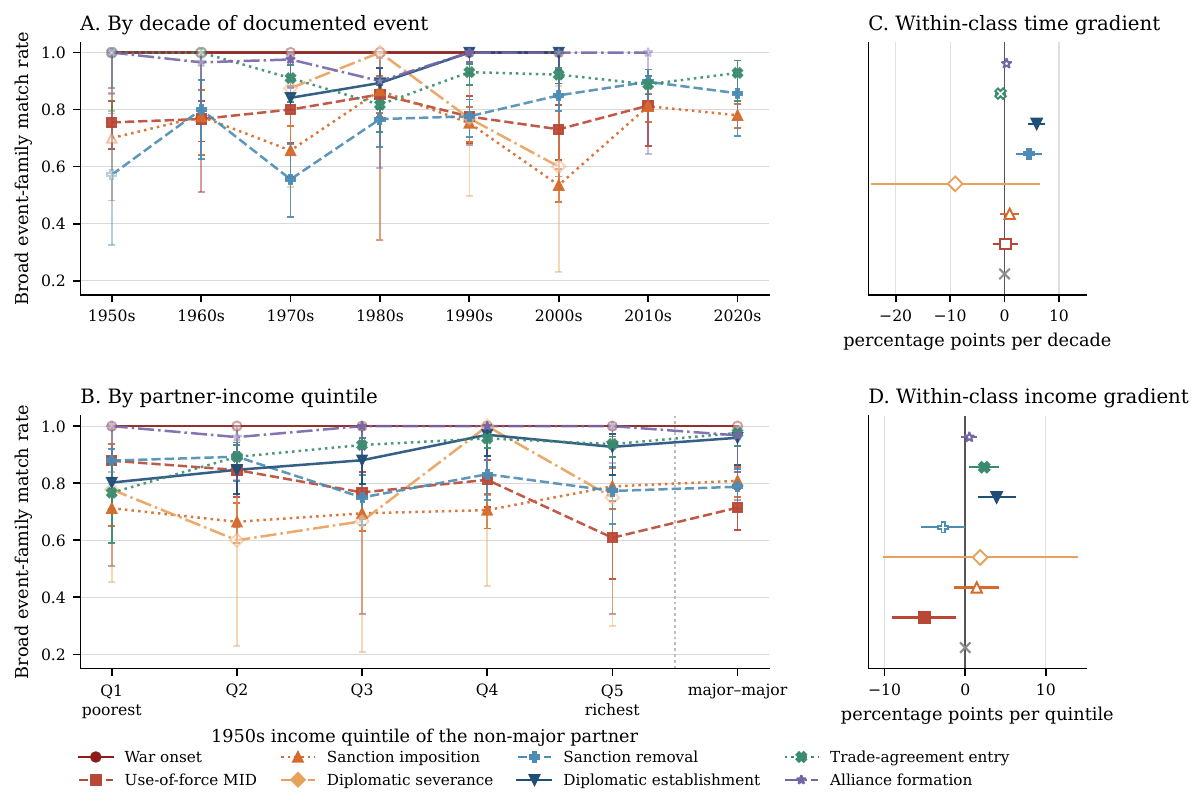}
    \captionof{figure}{Broad Event-Family Match Rates by Time and Partner Income}
    \label{fig:benchmark_match_gradients}
    \note{Panels~A and B report class-specific broad event-family match rates with pointwise 95\% Wilson confidence intervals. Panels~C and D report the corresponding within-class slopes with 95\% confidence intervals based on standard errors clustered by dyad. Cells based on fewer than 30 documented events are shown with open or faded markers. A gray cross denotes a class with a constant outcome, for which the slope is not estimable. Panels~B and D exclude events without a mapped 1950s income quintile. Panel~B shows major--major events separately.}
\end{figure}

Figure~\ref{fig:benchmark_match_gradients} reports the class-specific match rates and corresponding within-class gradients. The pooled income gradient is $+0.39$ percentage points per quintile, with a 95 percent confidence interval of $[-0.80,+1.58]$. The estimate does not indicate systematically lower matching for poorer partners. The pooled match rate rises by $1.16$ percentage points per decade, with a 95 percent confidence interval of $[+0.36,+1.96]$. This increase is concentrated in sanction removals and diplomatic establishments: excluding these two classes reduces the estimate to $+0.24$ percentage points per decade, with an interval of $[-0.66,+1.14]$. War and diplomatic severance are too sparse to establish class-specific stability across decades.

\paragraph{Variation in Matching over Subsequent Growth} We also test whether the database records documented developments selectively by the non-major country's subsequent growth. For documented event $a$ involving non-major country $c$ in year $t$, we estimate, separately for conflict and cooperative events,
\begin{equation}
D_a=\alpha_{k(a)}+\delta_{d(t)}+\theta\, G_{c,t}+\gamma X_{c,t}+\mu_c+u_a,
\label{eq:benchmark_hindsight}
\end{equation}
where $D_a$ indicates an expected-direction or broad event-family match, $G_{c,t}$ is the growth of log GDP per capita over $t$ to $t+10$ relative to the region-year mean, $X_{c,t}$ is log GDP per capita in year $t$, $\delta_{d(t)}$ are decade fixed effects, and $\mu_c$ are country fixed effects. Table~\ref{tab:benchmark_hindsight} shows that $\theta$ is small and statistically indistinguishable from zero for both matching measures and both event types.

\begin{table}[!htbp]
\centering
\caption{Documented-Event Matching and Subsequent Growth}
\label{tab:benchmark_hindsight}
\footnotesize
\begin{tabular}{lcc}
\toprule
 & Conflict events & Cooperative events \\
\midrule
Expected-direction match & $-0.5$ $(1.5)$ & $-0.4$ $(1.0)$ \\
Broad event-family match & $-0.5$ $(2.4)$ & $+1.8$ $(1.7)$ \\
\midrule
Documented events & 844 & 1,678 \\
Countries & 135 & 151 \\
\bottomrule
\end{tabular}
\note{Entries are estimates of $\theta$ in equation~\eqref{eq:benchmark_hindsight}, in percentage points per one standard deviation of the non-major country's growth in log GDP per capita over the following ten years relative to its region-year mean, with standard errors clustered by dyad in parentheses. Each regression includes documented-event-class, decade, and country fixed effects and log GDP per capita in the documented year. The sample contains documented events with exactly one major nation and observed growth over the following decade (events through 2014).}
\end{table}

\subsection{Comparison with Machine-Coded Event Databases} \label{app_a:event_comparison}

We next compare our annual event score with the corresponding scores from the Integrated Crisis Early Warning System \citep[ICEWS;][]{boschee2015icews}, the Global Database of Events, Language, and Tone \citep[GDELT;][]{leetaru2013gdelt}, and the Cline Center Historical Phoenix database \citep{althaus2020phoenix}. We harmonize actors to common country codes and construct undirected dyad-year measures under common aggregation rules. For each database, the annual measure is the mean Goldstein score in raw Goldstein points. None of the measures in this section is dynamically smoothed.

\paragraph{Event Volume and Detection}
Table~\ref{tab:volume_detection} compares event volume and the two detection outcomes. GDELT records a median of 288.5 events in a documented-event dyad-year, compared with three in our database, yet the two databases record some event within the matching window at nearly identical rates, 98.0 and 98.2 percent respectively. GDELT records an event in the expected direction more often, 97.2 versus 91.7 percent. ICEWS and Phoenix also record substantially more events than our database but have lower pooled rates on both outcomes. Event volume therefore does not by itself determine whether a database records the benchmark dyad or its expected conflict or cooperation direction.

\begin{center}
\begin{minipage}{\textwidth}
\centering
\captionof{table}{Event Volume and Detection across Event Databases, 1995--2018}
\label{tab:volume_detection}
\footnotesize
\setlength{\tabcolsep}{5pt}
\begin{tabular}{lrrrrr}
\toprule
 & \shortstack{Documented\\events} & Our database & ICEWS & GDELT & Phoenix \\
\midrule
\multicolumn{6}{l}{\textit{Panel A. Pooled volume and detection}} \\
Median events in documented-event year & & 3 & 34 & 288.5 & 6 \\
Mean events in documented-event year & & 3.0 & 371.7 & 4,083.0 & 61.3 \\
Any event recorded within matching window & 2,038 & 0.982 & 0.907 & 0.980 & 0.844 \\
Event in expected direction within matching window & 2,038 & 0.917 & 0.869 & 0.972 & 0.794 \\
\addlinespace
\multicolumn{6}{l}{\textit{Panel B. Event in expected direction, by documented-event class}} \\
War onset & 7 & 1.000 & 1.000 & 1.000 & 1.000 \\
Use-of-force MID, excluding wars & 167 & 0.922 & 1.000 & 1.000 & 0.982 \\
Sanction imposition & 561 & 0.836 & 0.900 & 0.984 & 0.816 \\
Diplomatic severance & 13 & 0.769 & 0.615 & 0.769 & 0.385 \\
Sanction removal & 399 & 0.962 & 0.952 & 0.995 & 0.905 \\
Diplomatic establishment & 79 & 1.000 & 0.962 & 1.000 & 0.937 \\
Trade-agreement entry & 772 & 0.940 & 0.764 & 0.944 & 0.666 \\
Alliance formation & 40 & 1.000 & 0.975 & 1.000 & 0.875 \\
\midrule
All classes & 2,038 & 0.917 & 0.869 & 0.972 & 0.794 \\
\bottomrule
\end{tabular}
\note{The sample contains the 2,038 documented events over 1995--2018 that all four databases can support after actor harmonization. Volume is measured in the documented-event year; detection uses the class-specific matching window.}
\end{minipage}
\end{center}

We estimate the class-adjusted linear probability model in equation~\eqref{eq:benchmark_match_gradient} separately for each event database and detection outcome. Per decade, any-event detection changes by $-0.19$ $[-0.96,+0.58]$ percentage points for our database, $-4.45$ $[-6.59,-2.32]$ for ICEWS, $+0.10$ $[-0.96,+1.17]$ for GDELT, and $-9.06$ $[-11.83,-6.30]$ for Phoenix, where brackets report 95 percent confidence intervals. Expected-direction detection changes by $+1.95$ $[+0.24,+3.66]$, $-4.48$ $[-6.93,-2.03]$, $-0.72$ $[-2.04,+0.60]$, and $-10.23$ $[-13.36,-7.09]$, respectively. Across partner income, Figure~\ref{fig:benchmark_corpus_gradients} shows that our database and GDELT have no statistically distinguishable gradients, whereas ICEWS and Phoenix detect richer-country dyads more often.

\begin{center}
\begin{minipage}{\textwidth}
    \centering
    \includegraphics[width=0.94\linewidth]{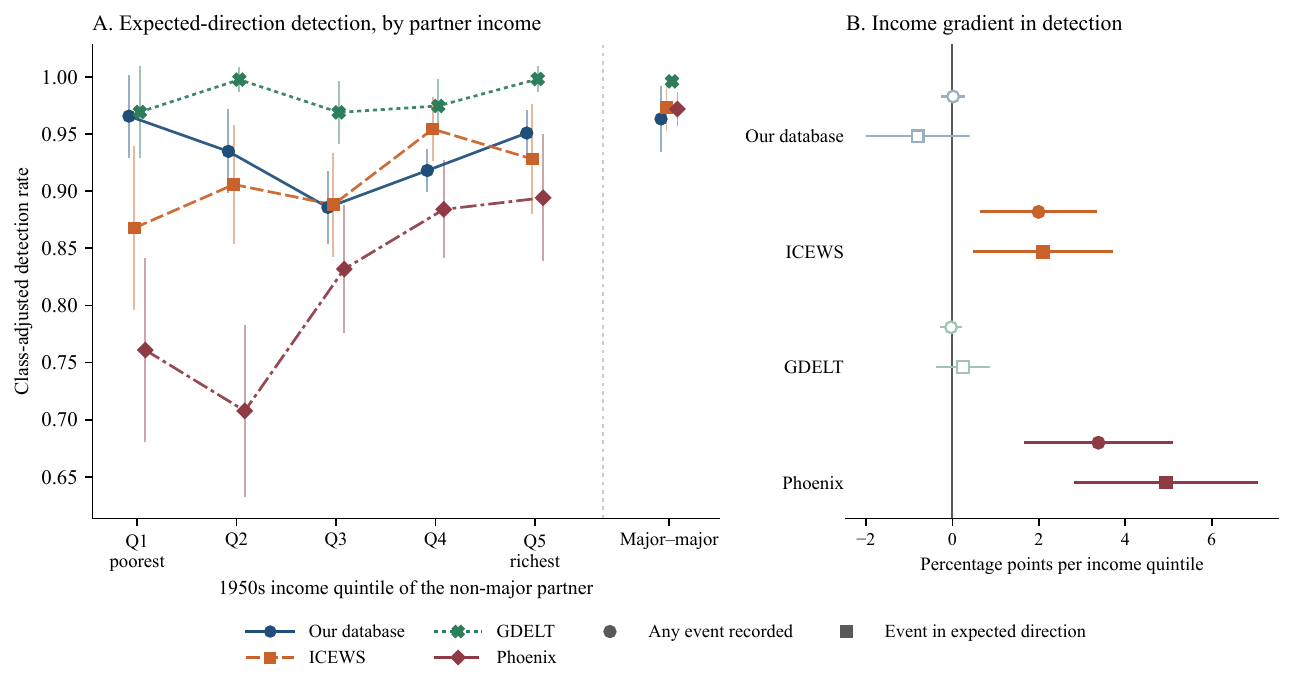}
    \captionof{figure}{Event Detection by Partner Income across Databases}
    \label{fig:benchmark_corpus_gradients}
    \note{Panel~A reports expected-direction detection rates, adjusted for documented-event class, by the 1950s income quintile of the non-major partner. Major--major events are shown separately. Panel~B reports percentage-point changes per income quintile from class-adjusted linear probability models for any-event and expected-direction detection. Filled markers indicate intervals that exclude zero. Bars in both panels are 95\% confidence intervals based on standard errors clustered by dyad. Of the 2,038 documented events in the common sample, 1,405 enter the quintile sample, 238 are major--major events shown separately, and the remaining 395 lack a mapped partner-income quintile and are excluded.}
\end{minipage}
\end{center}

\paragraph{Conditional Associations on Common Dyad-Year Samples}
To compare the annual measures on a common specification, let $C_{ij,t}$ indicate that at least one of the four documented conflict classes occurs in dyad $ij$ in year $t$, and let $K_{ij,t}$ indicate that at least one of the four documented cooperative classes occurs. For each annual measure $m$, we estimate
\begin{equation}
z^m_{ij,t}
=
\beta^m_C C_{ij,t}
+\beta^m_K K_{ij,t}
+\mu^m_{ij}
+\lambda^m_{i,t}
+\lambda^m_{j,t}
+\varepsilon^m_{ij,t},
\label{eq:benchmark_measure_association}
\end{equation}
where $z^m_{ij,t}$ is measure $m$ for dyad $ij$ in year $t$ and both documented-event indicators enter jointly. The specification absorbs dyad fixed effects and country-by-year fixed effects for both endpoints, and standard errors are clustered by dyad. The coefficients report conditional differences in measure $m$ during documented conflict and cooperation years. They are contemporaneous descriptive associations, not impulse responses or causal effects.

\begin{table}[!t]\begin{center}
\begin{minipage}{\textwidth}
\centering
\captionof{table}{Conditional Associations of Annual Event Scores with Documented Conflict and Cooperation}
\label{tab:benchmark_measure_associations}
\footnotesize
\setlength{\tabcolsep}{4pt}
\begin{tabular}{lrrrr}
\toprule
 & Our database & ICEWS & GDELT & Phoenix \\
\midrule
\multicolumn{5}{l}{\textit{Panel A. Common observed dyad-years}} \\
Documented conflict & $-1.467$ (0.179) & $-0.475$ (0.078) & $-0.120$ (0.054) & $-0.295$ (0.084) \\
Documented cooperation & $+0.475$ (0.112) & $+0.194$ (0.060) & $+0.033$ (0.051) & $-0.133$ (0.096) \\
Dyad-years & 34,412 & 34,412 & 34,412 & 34,412 \\
Documented conflict events & 627 & 627 & 627 & 627 \\
Documented cooperative events & 789 & 789 & 789 & 789 \\
\addlinespace
\multicolumn{5}{l}{\textit{Panel B. Each database's observed dyad-years}} \\
Documented conflict & $-1.388$ (0.155) & $-0.561$ (0.085) & $-0.144$ (0.060) & $-0.260$ (0.090) \\
Documented cooperation & $+0.403$ (0.085) & $+0.116$ (0.061) & $+0.064$ (0.057) & $-0.086$ (0.096) \\
Dyad-years & 71,538 & 59,036 & 81,141 & 43,009 \\
Documented conflict events & 726 & 691 & 722 & 643 \\
Documented cooperative events & 1,181 & 1,019 & 1,204 & 848 \\
\bottomrule
\end{tabular}
\note{Each column reports one regression of the annual event score in levels on jointly entered indicators for any documented conflict event and any documented cooperative event, as specified in equation~\eqref{eq:benchmark_measure_association}. All regressions absorb dyad fixed effects and country-by-year fixed effects for both endpoints; standard errors in parentheses are clustered by dyad. Panel~A uses identical observations for all four databases. Panel~B uses each database's complete observed panel over 1995--2018. All scores are in raw Goldstein points on the $[-10,10]$ scale.}
\end{minipage}
\end{center}
\end{table}

Panel~A of Table~\ref{tab:benchmark_measure_associations} uses the 34,412 dyad-years observed by all four databases that remain after fixed-effect singletons are dropped. Documented conflict is associated with annual differences of $-1.467$ Goldstein points in our database, $-0.475$ in ICEWS, $-0.120$ in GDELT, and $-0.295$ in Phoenix. Documented cooperation is associated with differences of $+0.475$, $+0.194$, $+0.033$, and $-0.133$, respectively. The cooperation coefficients for our database and ICEWS are statistically distinguishable from zero. Scaled by each database's standard deviation of annual scores in this sample, the conflict differences are $-0.45$, $-0.26$, $-0.08$, and $-0.12$ and the cooperation differences are $0.15$, $0.11$, $0.02$, and $-0.05$, so the ordering does not reflect differences in score dispersion alone. Panel~B uses each database's complete observed panel and preserves the relative magnitudes, although the ICEWS cooperation coefficient is not statistically distinguishable from zero at the 5 percent level.

The annual-score regressions do not establish that event selection causes the differences across databases. Appendix~\ref{app_b:measurement_alternatives} compares the smoothed indices built from each database and their relationships with economic growth.

\FloatBarrier

\FloatBarrier

\subsection{Model Robustness and Replicability} \label{app_a:model_robustness}

LLM-based measures may depend on the specific model or vary across independent runs. We evaluate both dimensions using four models from three families: Gemini, GPT, and Claude. Cross-model comparisons use Gemini 2.5 Pro as the production baseline and GPT 5.4, Claude Opus 4.6, and Claude Sonnet 4.6 as alternatives. Repeated-run tests use all four models.

\begin{figure}[ht]
    \centering
    \includegraphics[width=\linewidth]{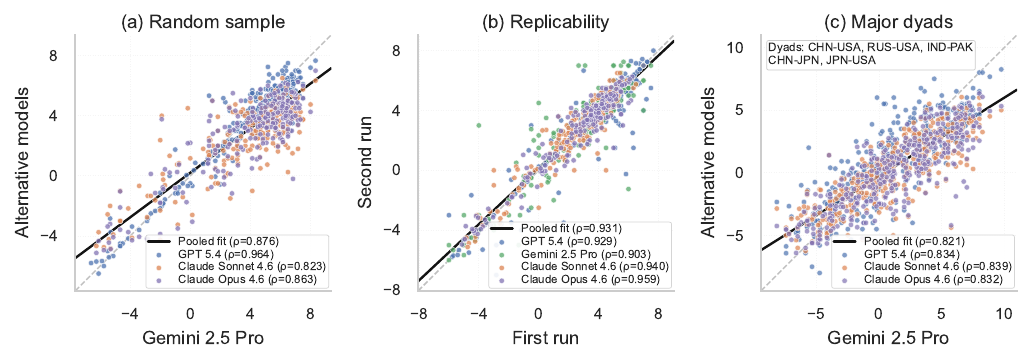}
    \caption{Model Robustness and Replicability of Annual Event Scores}
    \label{fig:model_robustness}
    \note{Each point represents the mean Goldstein score for a dyad-year observation. Panel~(a) compares alternative-model scores with the Gemini 2.5 Pro production baseline for 300 randomly selected dyad-years. Panel~(b) compares two runs on the same inputs within each model. The Gemini series uses Gemini 2.5 Pro. Panel~(c) compares alternative models with the Gemini 2.5 Pro baseline for five high-salience bilateral relationships (CHN--USA, RUS--USA, IND--PAK, CHN--JPN, JPN--USA) across all years. The solid black line is the pooled OLS fit and the dashed gray line the 45-degree line. $\rho$ is Pearson correlation. Exact model identifiers and provider-specific run settings are documented in the replication package.}
\end{figure}

Figure~\ref{fig:model_robustness} reports three complementary exercises. Panel~(a) compares scores from alternative LLMs with the Gemini 2.5 Pro baseline for 300 randomly selected dyad-years. Scores cluster along the 45-degree line, with pairwise correlations exceeding 0.82 (pooled $\rho = 0.88$). Panel~(b) tests within-model replicability. The pooled correlation is 0.93. Model-specific correlations are 0.96 for Opus 4.6, 0.94 for Sonnet 4.6, 0.93 for GPT 5.4, and 0.90 for Gemini 2.5 Pro. Because scores vary across runs and served models can change, we treat the released event database, rather than a model endpoint, as the reproducible artifact. Panel~(c) compares scores for five high-salience bilateral relationships across all available years, a more demanding test because their dense and shifting event records require greater interpretive judgment. The pooled correlation is 0.82. Together, the exercises show that the common signal is large relative to model- and run-specific variation.

\paragraph{Full-Pipeline Robustness}

The preceding exercises hold the prompt structure fixed and vary only the model on subsamples. We next vary both the LLM and the prompt, comparing the baseline index with an alternative LLM pipeline constructed primarily using Gemini 2.5 Flash (preview) and an earlier prompt.\footnote{The alternative prompt pursues the same objective of identifying and scoring bilateral geopolitical events using the CAMEO/Goldstein framework but differs in scope, organization, and event taxonomy. Failed calls could fall back to Gemini 2.5 Pro (preview). The fallback share is not recoverable from the stored outputs.} To isolate measurement from aggregation differences, we reconstruct the alternative LLM country-year index using the same 24 major nations and time-varying GDP-share weights as the baseline.

\begin{figure}[ht]
    \centering
    \begin{subfigure}[b]{0.48\textwidth}
        \includegraphics[width=\textwidth]{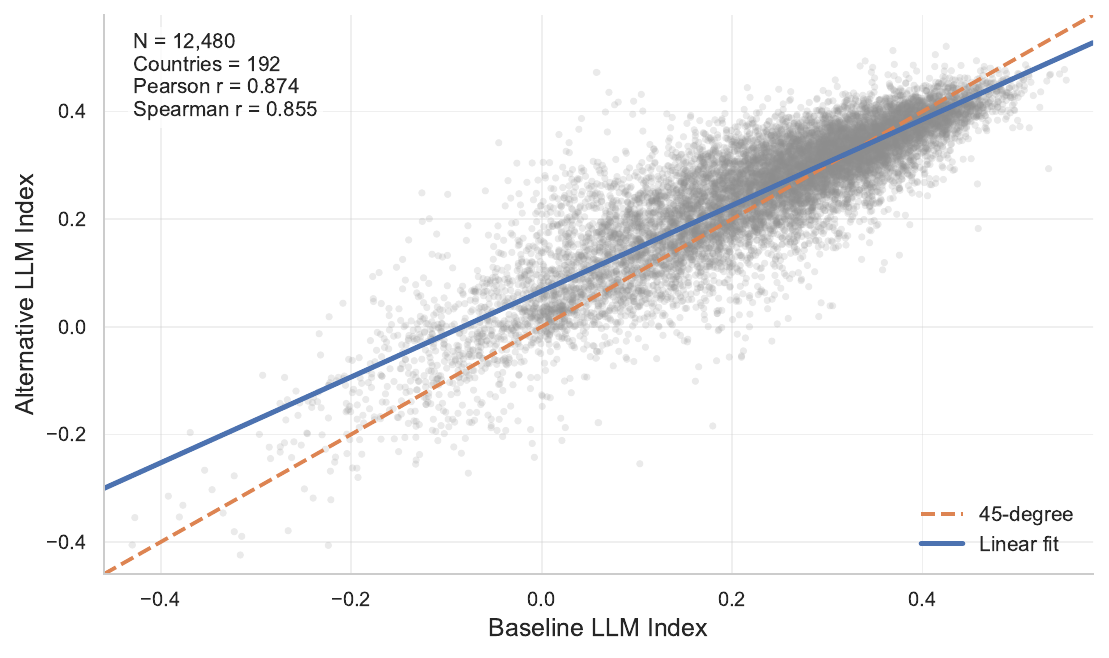}
        \caption{Country-Year Index Correlation}
        \label{fig:pipeline_correlation}
    \end{subfigure}
    \hfill
    \begin{subfigure}[b]{0.48\textwidth}
        \includegraphics[width=\textwidth]{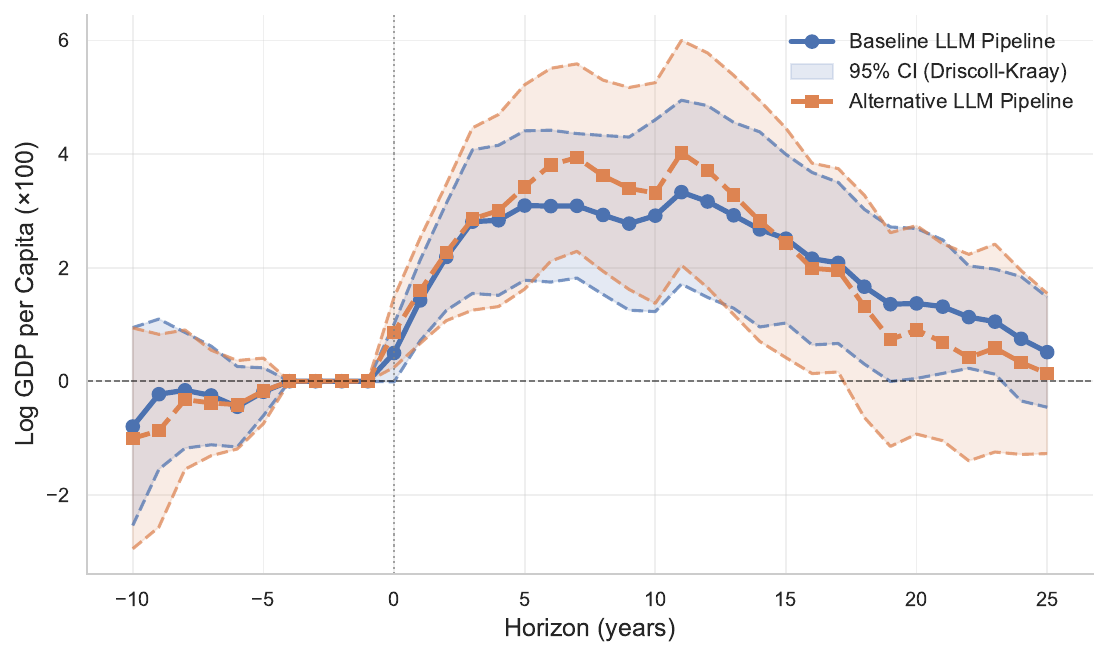}
        \caption{GDP Coefficient Paths}
        \label{fig:pipeline_irf}
    \end{subfigure}
    \caption{Full-Pipeline Robustness: Alternative LLM and Prompt Structure}
    \label{fig:pipeline_robustness}
    \note{Panel~(a) plots the country-year geopolitical alignment index from the alternative LLM pipeline, constructed primarily with Gemini 2.5 Flash (preview) and an earlier prompt, against the production Gemini 2.5 Pro baseline for all 12,480 overlapping country-year observations across 192 countries. The blue solid line shows the linear fit. The orange dashed line shows the 45-degree line. Both indices use the same 24 major nations and time-varying GDP-share weights. Panel~(b) overlays the coefficient paths of log GDP per capita ($\times 100$), per 0.141 index units (one standard deviation of the baseline index), from the baseline pipeline (blue) and the alternative pipeline (orange), both estimated on the 148-country long-coverage sample. Shaded areas are 95\% confidence intervals based on Driscoll--Kraay standard errors.}
\end{figure}

Panel~(a) of Figure~\ref{fig:pipeline_robustness} shows that the two country-year indices are strongly correlated (Pearson $\rho = 0.874$, Spearman $\rho = 0.855$), with a fitted slope below one. In panel~(b), the two pipelines produce similar coefficient paths: a hump-shaped response with its largest coefficient in year 11, followed by gradual decline. The alternative LLM pipeline's point estimates fall within the baseline 95 percent confidence interval at all 26 horizons from impact through year 25, and the correlation of point estimates across these horizons is 0.97.

\paragraph{Measurement Validity}

The exercises above establish measurement \textit{reliability}, not accuracy. Frontier LLMs draw on overlapping sources and may therefore share errors. The documented-event benchmarks in Appendix~\ref{app_a:anchoring} and the staged human audit in Appendix~\ref{app_a:human_audit} address accuracy more directly by comparing the model record with independently compiled events, blind human reconstruction, and independent recoding.

\FloatBarrier

\subsection{Human Audit of the Event Data} \label{app_a:human_audit}

The human audit covers 200 eventful dyad-years over 1960--2024 involving at least one of the 24 major nations. The stratified sample covers every decade, with 23--34 cases per decade. Based on the sign of the model's annual score, 44.5 percent of cases are conflictual and 55.5 percent cooperative. Six trained research assistants provide two retained assessments per case, giving 400 reads and 1,712 coder-event assessments of 856 model-generated events.

The audit follows a three-stage protocol. Before the audit, coders complete 15 training cases excluded from the analysis sample. In Stage~1, coders see only the country pair, year, and the codebook used by the model. They independently identify, classify, and score bilateral events. After this record is locked, Stage~2 reveals the model's event descriptions and dates while withholding its classifications and scores. Coders then assess validity and independently recode events. In Stage~3, the model's classifications and scores are revealed for diagnostic comparisons. Earlier assessments cannot be revised. Coders are prohibited from using generative AI and must document sources for each event. They are instructed to spend less than 75 minutes researching, recording, and scoring each case.

\begin{figure}[htbp]
    \centering
    \includegraphics[width=0.85\linewidth,clip]{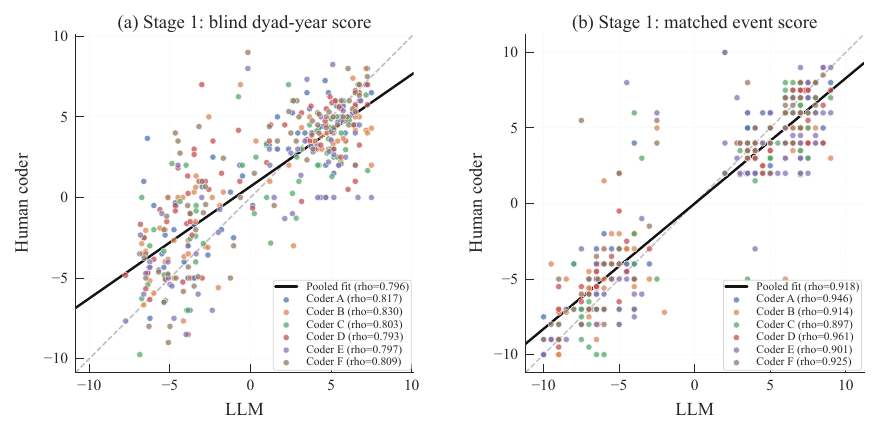}
    \caption{Independent Human Reconstruction of Bilateral Events}
    \label{fig:human_audit_agreement}
    \note{Panel~(a) compares blind human and model annual scores ($n=400$ reads). Each is the unweighted mean of its event-level Goldstein scores. For 28 reconstructions with no events, the human score is the coder's holistic rating of the dyad-year. Panel~(b) compares events with exact or near matches ($n=491$ links). Colors denote coders. Dashed and solid lines show equality and the pooled OLS fit. Both panels use 200 dyad-years and uncorrected coder scores. All scores are in Goldstein units on $[-10,10]$. $\rho$ is Pearson correlation.}
\end{figure}

Independent human reconstructions yield annual Goldstein scores that correlate at 0.80 with the model's scores (Figure~\ref{fig:human_audit_agreement}, panel~(a)), compared with 0.76 between human coders. Human reconstructions contain 2.95 events per dyad-year on average, 31.1 percent fewer than the 4.28 events in the model record.

For independently reconstructed events with exact or near matches in the model record, Goldstein scores correlate at 0.92 (panel~(b)). Coders flag these correspondences in Stage~2, but the human scores are locked in Stage~1. Strong agreement on matched events suggests that event selection is an important source of differences between the annual measures.

\begin{figure}[!t]
    \centering
    \includegraphics[width=0.85\linewidth,clip]{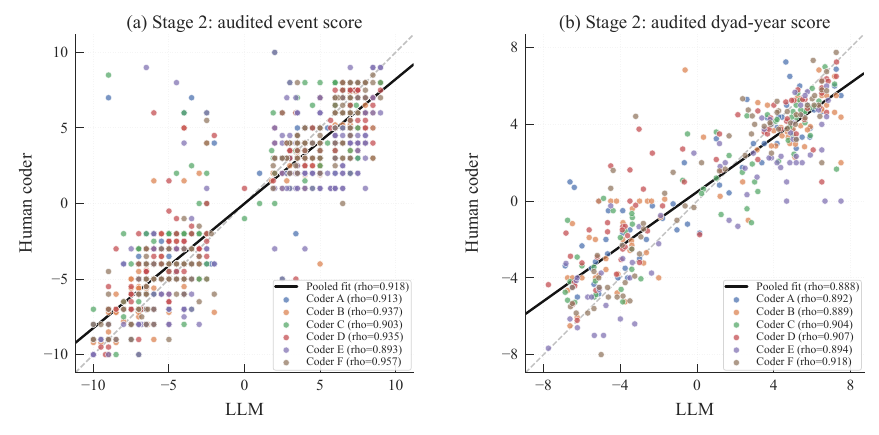}
    \caption{Human Review of Model-Generated Events}
    \label{fig:human_audit_review}
    \note{Panel~(a) uses all submitted human event recodes ($n=1{,}357$), including those for flagged records. Unscored assessments are excluded. Panel~(b) compares annual scores ($n=400$ reads). The human score is the unweighted mean over the final event set after exclusions, corrections, merges, splits, and inclusion of unmatched Stage~1 events, with a holistic fallback for an empty set. The model score averages its original events. Both panels use 200 dyad-years and uncorrected coder scores. Colors denote coders. Dashed and solid lines show equality and the pooled OLS fit. All scores are in Goldstein units on $[-10,10]$. $\rho$ is Pearson correlation.}
\end{figure}

After event review, human and model scores correlate at 0.92 at the event level and 0.89 at the annual level (Figure~\ref{fig:human_audit_review}, panels~(a) and~(b)).

The audit process identifies three fabricated records among the 856 model-generated events.\footnote{These records were classified as fabricated during adjudication: a Norwegian royal state visit to the Netherlands in 2017, a visit by US Secretary of State Condoleezza Rice to Angola in 2006, and a British condemnation attributing the 1994 AMIA bombing to Iran.} Fabrication is uncommon in the reviewed sample, accounting for less than 0.5 percent of sampled events.

Before the model's classifications are revealed, human and model assignments to the six broad event categories agree in 79.0 percent of assessments. The most common differences arise when coders classify security and defense or multilateral and global governance events as diplomatic and political relations. In a separate post-reveal comparison in Stage~3, coders prefer the model's category or regard both classifications as defensible in 87.9 percent of 1,306 comparisons. The audit thus shows agreement in annual scores and event coding within this stratified sample of eventful dyad-years. Because the sample excludes dyad-years without events and human searches are time-limited, the audit says less about the accuracy of the full event database.

\FloatBarrier

\subsection{Bilateral Outcomes} \label{app_a:bilateral_outcomes}

Table~\ref{tab:gravity_bilateral} reports gravity regressions of four bilateral outcomes on the bilateral geopolitical alignment score $S_{ij,t}$. The sample comprises directed country pairs involving at least one of the 24 major nations over 1960--2024, the pairs that enter the alignment index. Columns (1)--(3) include distance controls and origin-year and destination-year fixed effects for each sub-period. Column (4) adds dyad fixed effects, so identification comes from changes within pairs over time. Bilateral alignment is strongly associated with trade, sanctions, migration, and foreign aid. With dyad fixed effects, improvements in alignment are associated with higher trade, a lower sanction probability, and larger aid flows, consistent with the score capturing economically relevant bilateral ties.

\begin{table}[p]
\centering
\caption{Geopolitical Alignment and Bilateral Outcomes: Gravity Regressions}
\label{tab:gravity_bilateral}
\small
\resizebox{0.85\linewidth}{!}{%
\begin{tabular}{lcccc}
\toprule
 & Cold War & Globalization & Fragmentation & All Years \\
 & 1963--1990 & 1990--2010 & 2010--2024 & (Dyad FE) \\
 & (1) & (2) & (3) & (4) \\
\midrule
\multicolumn{5}{l}{\textbf{Panel A. Dependent Variable: Log Bilateral Trade}} \\
\addlinespace[0.1cm]
Baseline score & $0.995^{***}$ & $0.800^{***}$ & $0.861^{***}$ & $0.292^{***}$ \\
 & $(0.073)$ & $(0.066)$ & $(0.079)$ & $(0.027)$ \\
Excluding Economic Relations events & $0.741^{***}$ & $0.627^{***}$ & $0.718^{***}$ & $0.225^{***}$ \\
 & $(0.077)$ & $(0.069)$ & $(0.080)$ & $(0.026)$ \\
\addlinespace[0.1cm]
Observations & 142,356 & 158,306 & 123,784 & 414,391 \\
\addlinespace[0.3cm]
\multicolumn{5}{l}{\textbf{Panel B. Dependent Variable: Sanction Indicator (LPM)}} \\
\addlinespace[0.1cm]
Baseline score & $-0.028^{***}$ & $-0.052^{***}$ & $-0.146^{***}$ & $-0.044^{***}$ \\
 & $(0.004)$ & $(0.005)$ & $(0.013)$ & $(0.003)$ \\
Excluding sanction impositions and removals & $-0.025^{***}$ & $-0.046^{***}$ & $-0.124^{***}$ & $-0.037^{***}$ \\
 & $(0.005)$ & $(0.005)$ & $(0.013)$ & $(0.003)$ \\
\addlinespace[0.1cm]
Sanction Rate & 0.037 & 0.070 & 0.090 & 0.058 \\
Observations & 242,592 & 181,944 & 121,296 & 554,496 \\
\addlinespace[0.3cm]
\multicolumn{5}{l}{\textbf{Panel C. Dependent Variable: Log Bilateral Migration}} \\
\addlinespace[0.1cm]
Baseline score & $0.447^{***}$ & $0.547^{***}$ &  &  \\
 & $(0.091)$ & $(0.099)$ &  &  \\
Excluding Movement of People events & $0.416^{***}$ & $0.554^{***}$ &  &  \\
 & $(0.092)$ & $(0.099)$ &  &  \\
\addlinespace[0.1cm]
Observations & 18,464 & 13,495 &  &  \\
\addlinespace[0.3cm]
\multicolumn{5}{l}{\textbf{Panel D. Dependent Variable: Log Bilateral ODA}} \\
\addlinespace[0.1cm]
Baseline score & $2.012^{***}$ & $1.784^{***}$ & $2.046^{***}$ & $0.902^{***}$ \\
 & $(0.136)$ & $(0.116)$ & $(0.173)$ & $(0.059)$ \\
Excluding Economic Relations events & $1.421^{***}$ & $1.099^{***}$ & $1.442^{***}$ & $0.529^{***}$ \\
 & $(0.140)$ & $(0.111)$ & $(0.166)$ & $(0.056)$ \\
\addlinespace[0.1cm]
Observations & 27,884 & 36,323 & 28,973 & 90,641 \\
\midrule
Distance Controls & Yes & Yes & Yes & --- \\
Origin $\times$ Year, Dest.\ $\times$ Year FE & Yes & Yes & Yes & Yes \\
Dyad FE & --- & --- & --- & Yes \\
\bottomrule
\end{tabular}}
\note{Gravity regressions of bilateral outcomes on the geopolitical alignment score $S_{ij,t}$. The sample comprises directed country pairs involving at least one of the 24 major nations over 1960--2024. In each panel, the first row uses the baseline score, and the second row uses a score rebuilt after removing potentially outcome-linked events before dyad-year aggregation and the dynamic recursion, on an identical sample. Trade and ODA exclude all category A (Economic Relations) events, sanctions exclude sanction impositions and removals (CAMEO codes 081, 085, 163, and 172), and migration excludes category D3 (Movement of People) events. Columns (1)--(3): origin$\times$year and destination$\times$year fixed effects with distance controls. Column (4): dyad, origin$\times$year, and destination$\times$year fixed effects (distance controls absorbed by dyad FE). Period endpoints are inclusive, so 1990 enters columns (1) and (2), and 2010 enters columns (2) and (3). Panel A: log bilateral trade from COMTRADE (1962--2024). Panel B: linear probability model for bilateral sanction indicator from the Global Sanctions Database v4 (1960--2023), and Sanction Rate reports the unconditional probability. Panel C: log bilateral migration stocks from the World Bank in five decennial waves from 1960 through 2000; the reported period columns begin in 1963, so the 1960 wave is not used, and the later-period and sparse all-years dyad fixed-effects cells are omitted. Panel D: log gross ODA disbursements from the OECD DAC (1960--2024, constant prices); the sample comprises 47 donor and 156 recipient countries. Reported observation counts are before singleton fixed-effect removal. Distance controls include normalized geodesic distance, contiguity, and tree-weighted linguistic distance. Standard errors are clustered by dyad. $^{***}p<0.01$, $^{**}p<0.05$, $^{*}p<0.10$.}
\end{table}

The second row of each panel removes outcome-linked events before rebuilding the score: all Economic Relations events for trade and official development assistance (ODA), sanction impositions and removals for sanctions, and Movement of People events for migration. Signs and significance are unchanged. With dyad fixed effects, the coefficients are 0.225 for trade, $-0.037$ for sanctions, and 0.529 for ODA, versus 0.292, $-0.044$, and 0.902 in the baseline, and the period-specific migration estimates barely change. Because the exclusions alter the score's dispersion and composition, the coefficient differences are not a decomposition. The second-row estimates show that the bilateral patterns are not generated solely by coding the outcome event itself.

\FloatBarrier

\subsection{Statistics of the Geopolitical Alignment Index} \label{app_a:geo_relations}
Table~\ref{tab:geo_relation_summary_stats} presents summary statistics for the geopolitical alignment index across seven decades. The index covers all 193 countries in every year: before independence, a country's relations are recorded through the state that controlled its territory, such as the Soviet Union for the Baltic states before 1991.

\begin{table}[ht]
\centering
\caption{Summary Statistics of the Geopolitical Alignment Index by Decade, 1960--2024}
\label{tab:geo_relation_summary_stats}
\footnotesize
\resizebox{\textwidth}{!}{%
\begin{tabular}{lrrrrrrrr}
\toprule
 & 1960s & 1970s & 1980s & 1990s & 2000s & 2010s & 2020s & Overall \\
\midrule
\multicolumn{9}{l}{\textit{Summary Statistics}} \\
\addlinespace[0.1cm]
\quad Mean & 0.173 & 0.192 & 0.182 & 0.260 & 0.287 & 0.296 & 0.281 & 0.235 \\
\quad Median & 0.189 & 0.204 & 0.209 & 0.284 & 0.315 & 0.319 & 0.302 & 0.261 \\
\quad Std. Dev. & 0.141 & 0.120 & 0.138 & 0.152 & 0.132 & 0.108 & 0.114 & 0.141 \\
\quad Min & $-0.323$ & $-0.315$ & $-0.275$ & $-0.430$ & $-0.256$ & $-0.335$ & $-0.160$ & $-0.430$ \\
\quad Max & 0.499 & 0.531 & 0.466 & 0.550 & 0.548 & 0.479 & 0.482 & 0.550 \\
\quad 5th Pct. & $-0.054$ & $-0.013$ & $-0.103$ & $-0.027$ & 0.023 & 0.082 & 0.059 & $-0.029$ \\
\quad 25th Pct. & 0.060 & 0.108 & 0.108 & 0.188 & 0.226 & 0.251 & 0.223 & 0.154 \\
\quad 75th Pct. & 0.283 & 0.281 & 0.284 & 0.371 & 0.378 & 0.368 & 0.367 & 0.340 \\
\quad 95th Pct. & 0.378 & 0.368 & 0.359 & 0.451 & 0.446 & 0.425 & 0.420 & 0.422 \\
\quad N Country-Years & 1,930 & 1,930 & 1,930 & 1,930 & 1,930 & 1,930 & 965 & 12,545 \\
\addlinespace[0.2cm]
\multicolumn{9}{l}{\textit{Countries with Lowest Geopolitical Alignment Index (Bottom 5)}} \\
\addlinespace[0.1cm]
\quad 1. & S. Africa & S. Africa & Afghanistan & Libya & Myanmar & N. Korea & N. Korea & -- \\
\quad 2. & China & Albania & Libya & Iraq & Belarus & Eritrea & Nicaragua & -- \\
\quad 3. & Zimbabwe & Zimbabwe & S. Africa & Myanmar & Zimbabwe & Syria & China & -- \\
\quad 4. & Albania & Timor-L. & Latvia & Serbia & Iraq & Venezuela & Russia & -- \\
\quad 5. & Timor-L. & Chile & Lithuania & Montenegro & Gambia & Iran & Syria & -- \\
\addlinespace[0.2cm]
\multicolumn{9}{l}{\textit{Countries with Highest Geopolitical Alignment Index (Top 5)}} \\
\addlinespace[0.1cm]
\quad 1. & Ethiopia & Romania & Mozambique & Hungary & Mozambique & Peru & Mozambique & -- \\
\quad 2. & Italy & Finland & Egypt & Poland & Romania & Singapore & Qatar & -- \\
\quad 3. & Nepal & Bangladesh & Bangladesh & Mozambique & Ghana & Sierra Leone & Singapore & -- \\
\quad 4. & Afghanistan & Belgium & Senegal & Kuwait & Vietnam & Liberia & Senegal & -- \\
\quad 5. & Colombia & Colombia & Mali & Mongolia & Albania & Senegal & Laos & -- \\
\bottomrule
\end{tabular}%
}
\note{The geopolitical alignment index is the GDP-weighted aggregate of each country's bilateral geopolitical alignment scores with the 24 major nations. Statistics are calculated using all available country-year observations within each decade. Countries are ranked by their decade-average geopolitical alignment index values.}
\end{table}

The summary statistics show three patterns. First, the mean geopolitical alignment index improved from 0.173 in the 1960s to 0.296 in the 2010s---an increase of approximately 0.87 standard deviations---before declining to 0.281 in the 2020s. This trajectory matches the post-Cold War convergence and recent fragmentation documented in Section~\ref{ss:landscape_geo_relations}. Second, the standard deviation declined from 0.141 in the 1960s to 0.108 in the 2010s, indicating reduced heterogeneity as countries converged toward more cooperative average relations, though it has since risen slightly to 0.114 in the 2020s. Third, the 5th percentile rose from $-0.054$ in the 1960s to 0.082 in the 2010s, so the lower tail of the distribution also improved, before falling to 0.059 in the 2020s.

The countries at the extremes provide a face-validity check on the measure. During the Cold War decades, apartheid-era South Africa, China, and Albania ranked among the most geopolitically isolated states, reflecting ideological divisions and international ostracism. The contemporary period's most isolated countries---North Korea, Syria, and Iran in the 2010s, joined by China, Russia, and Nicaragua in the 2020s---are all subjects of extensive international sanctions or geopolitical confrontation with Western powers. Conversely, the highest-scoring countries are often small states that maintain positive relations across geopolitical divides (Singapore, Senegal, Qatar) or beneficiaries of particular historical moments (Poland and Hungary during post-Cold War democratization, Vietnam and Albania following market reforms, African states like Mozambique and Ghana engaging multiple development partners).
\FloatBarrier

\subsection{Comparison with the Geopolitical Risk Index} \label{app_a:gpr_comparison}

\begin{figure}[ht]
    \centering
    \includegraphics[width=\linewidth]{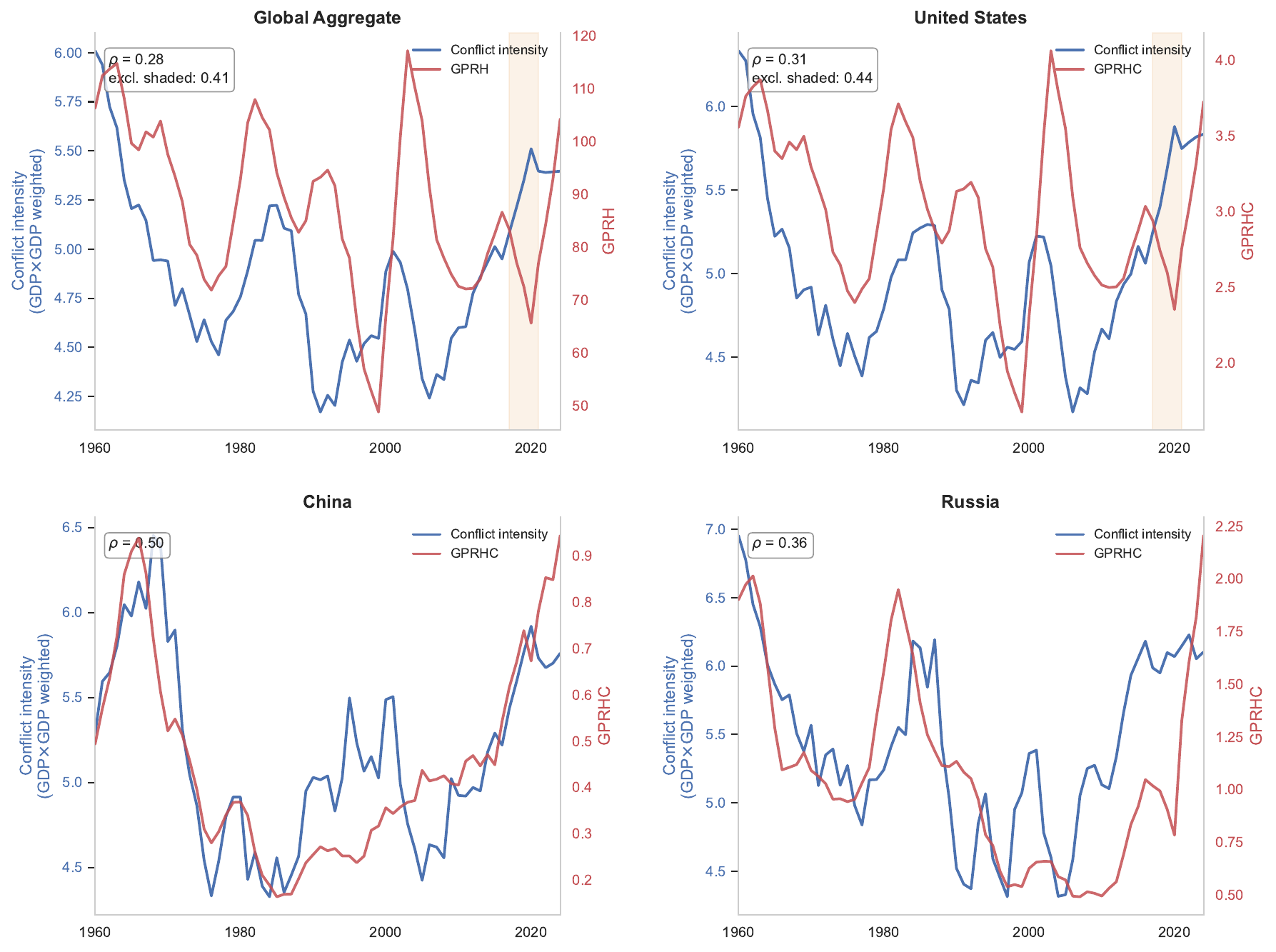}
    \caption{GDP-Weighted Conflict Intensity versus Geopolitical Risk Index}
    \label{fig:gpr_comparison}
    \note{Each panel plots our GDP$\times$GDP--weighted conflict intensity (blue, left axis) against the historical Geopolitical Risk Index (GPRH/GPRHC, red, right axis) from \citet{Caldara2022-fy}. Conflict intensity is the weighted mean severity of recorded conflict events, $-\sum_n w_n g_n/\sum_n w_n$, where the sums run over all conflict events in the year, $g_n<0$ is the Goldstein score of event $n$, and $w_n=\text{GDP}_i\times\text{GDP}_j$ for its country pair. Higher values indicate more severe conflict among economically significant country pairs. Both series are smoothed with a four-year centered moving average. Orange shading highlights the 2017--2021 geoeconomic divergence period. Boxed values are Pearson correlations of the unsmoothed annual series. For the aggregate and the United States, the second value excludes the shaded 2017--2021 period.}
\end{figure}

The Geopolitical Risk Index (GPR) of \citet{Caldara2022-fy} infers geopolitical risk from newspaper article counts discussing wars, terrorism, and military tensions---measuring media \textit{attention} rather than the events themselves. Figure~\ref{fig:gpr_comparison} compares our GDP$\times$GDP--weighted conflict intensity against the historical GPR for the global aggregate and three major nations. The two series co-move during traditional military confrontations (Vietnam War, Russia--Ukraine war). The correlations of the annual series are $\rho = 0.28$ (aggregate), $0.31$ (United States), $0.50$ (China), and $0.36$ (Russia).

Two divergences are informative. During 2001--2004, the Iraq War produces the largest post-Cold War GPR spike, yet our measure barely responds, since Iraq's GDP was about 0.1 percent of world output. During 2017--2021 (orange shading), our measure rises with the US--China trade war and technology restrictions while the GPR stays flat, as its keywords target military threats rather than geoeconomic competition. Excluding 2017--2021 raises the aggregate correlation to $\rho = 0.41$ and the US correlation to $0.44$. The GPR is also limited to countries with extensive English-language media coverage, whereas our event-based approach covers all 193 countries.

\FloatBarrier

\newpage
\section{Additional Empirical Results} \label{app_b:add_empirics}

Appendix~\ref{app_b:add_empirics} follows Sections~\ref{s:dyn_growth_effects}--\ref{s:add_robust}: data (B.1--B.2), the baseline estimates and their robustness (B.3--B.8), channels and democracy (B.9--B.10), and additional robustness and alternative measures (B.11--B.16). As in the main text, coefficients on the alignment index are per one standard deviation of the index (0.141).

\subsection{Economic Data} \label{app_b:econ_data}

Our dataset extends \citet{Acemoglu2019-bo} to cover 193 countries over 1960--2024, incorporating updated democracy measures from \citet{Acemoglu2025-lv} and economic indicators from the Penn World Table \citep{Feenstra2015-vy}.

Table~\ref{tab:data_description} organizes variables into five categories following the growth literature: enhanced Solow fundamentals (GDP, investment, capital, demographics), political institutions and governance (democracy, unrest, regime indicators), market institutions and reforms (liberalization indices, fiscal measures), human capital and labor markets (education, employment), and productivity and returns (TFP, internal rate of return, consumption, absorption). Coverage varies from 117 countries (TFP) to 192 countries (demographic variables). All monetary variables are in constant prices.

\paragraph{Domestic Unrest} Our domestic unrest indicator is a binary variable equal to one if any of eight domestic conflict components---assassinations, general strikes, guerrilla warfare, government crises, purges, riots, revolutions, and anti-government demonstrations---is strictly positive in a given country-year. We construct this variable from the 2024 edition of the Cross-National Time-Series Data Archive (CNTS), which extends coverage through 2023 (191 countries, 10,324 country-year observations). The original \citet{Acemoglu2019-bo} unrest variable, derived from an earlier CNTS edition, covers 179 countries through 2010. Our updated indicator agrees with the original in 83 percent of country-years in the overlap period, with all discrepancies reflecting events added in the 2024 edition rather than removed.

\begin{table}[htbp]
\centering
\caption{Data Description and Coverage}
\label{tab:data_description}
\footnotesize
\begin{tabular}{lccc}
\toprule
Variable & Data Source & Country Coverage & Data Period \\
\midrule
\multicolumn{4}{l}{\textit{Enhanced Solow Fundamentals}} \\
GDP per capita (Constant US Dollar) & WDI & 184 countries & 1960--2024 \\
Real GDP per capita & PWT & 172 countries & 1960--2023 \\
Trade as Share of GDP & WDI & 176 countries & 1960--2024 \\
Investment as Share of GDP & PWT & 172 countries & 1960--2023 \\
Capital Stock & PWT & 169 countries & 1960--2023 \\
Population & PWT & 172 countries & 1960--2023 \\
0--14 Population Share & WDI & 192 countries & 1960--2023 \\
15--64 Population Share & WDI & 192 countries & 1960--2023 \\
\\
\multicolumn{4}{l}{\textit{Political Institutions and Governance}} \\
Democracy (Acemoglu et al. 2019) & Acemoglu et al. (2019) & 183 countries & 1960--2010 \\
Democracy (Acemoglu et al. 2025) & Acemoglu et al. (2025) & 183 countries & 1960--2019 \\
Unrest (ANRR) & Acemoglu et al. (2019) & 179 countries & 1960--2010 \\
Unrest (CNTS 2024) & Cross-National Time-Series & 191 countries & 1960--2023 \\
Soviet Union & Acemoglu et al. (2019) & 183 countries & 1960--2024 \\
Region & Acemoglu et al. (2025) & 192 countries & 1960--2024 \\
\\
\multicolumn{4}{l}{\textit{Market Institutions and Reforms}} \\
Market Reform Index & Acemoglu et al. (2019) & 152 countries & 1960--2005 \\
Tax-to-GDP & Acemoglu et al. (2019) & 136 countries & 1960--2005 \\
\\
\multicolumn{4}{l}{\textit{Human Capital and Labor Markets}} \\
Human Capital Index & PWT & 142 countries & 1960--2023 \\
Employment & PWT & 172 countries & 1960--2023 \\
Labor Share & PWT & 132 countries & 1960--2023 \\
Primary Enrollment & Acemoglu et al. (2019) & 176 countries & 1970--2010 \\
Secondary Enrollment & Acemoglu et al. (2019) & 176 countries & 1970--2010 \\
\\
\multicolumn{4}{l}{\textit{Productivity and Returns}} \\
TFP & PWT & 117 countries & 1960--2023 \\
Internal Rate of Return & PWT & 132 countries & 1960--2023 \\
Real Consumption & PWT & 172 countries & 1960--2023 \\
Real Domestic Absorption & PWT & 172 countries & 1960--2023 \\
\bottomrule
\end{tabular}
\note{Country coverage represents the number of countries with at least one non-missing observation for each variable. Data sources: WDI = World Development Indicators; PWT = Penn World Table; CNTS = Cross-National Time-Series Data Archive (2024 edition). Unrest (ANRR) is the original variable from \citet{Acemoglu2019-bo}, retained for comparability. Unrest (CNTS 2024) is our updated indicator, used throughout the analysis.}
\end{table}

\subsection{Cross-Sectional Correlation with Economic Fundamentals} \label{app_b:fundamentals}

Table~\ref{tab:geo_vs_gdp_fundamentals} reports within-year cross-sectional correlations between economic fundamentals and (i) GDP per capita (Panel~A) and (ii) the geopolitical alignment index (Panel~B). GDP per capita correlates strongly with human capital (0.7--0.8), TFP (0.6), capital stock (0.5), and investment (0.4). Correlations with the geopolitical alignment index are generally modest and vary in sign across decades. Most are below 0.2 in absolute value, although capital stock reaches about 0.27 in the 1970s and 0.26 in the 1990s and human capital reaches 0.33 in the 1990s. The index is therefore distinct from, but not orthogonal to, standard measures of economic development.

\begin{table}[htbp]
\centering
\caption{Cross-Sectional Correlation with Economic Fundamentals}
\label{tab:geo_vs_gdp_fundamentals}
\footnotesize
\begin{threeparttable}
\begin{tabular}{lcccccccc}
\toprule
 & Overall & 1960s & 1970s & 1980s & 1990s & 2000s & 2010s & 2020s \\
\midrule

\multicolumn{9}{l}{\textit{Panel A: Correlation with GDP per Capita}} \\
\midrule
Investment Share & 0.390 & 0.404 & 0.312 & 0.376 & 0.433 & 0.446 & 0.408 & 0.441 \\
Capital Stock & 0.486 & 0.602 & 0.467 & 0.477 & 0.473 & 0.476 & 0.474 & 0.465 \\
Internal Rate of Return & $-$0.196 & $-$0.432 & $-$0.303 & $-$0.252 & $-$0.098 & $-$0.081 & $-$0.130 & $-$0.071 \\
TFP & 0.645 & 0.426 & 0.480 & 0.539 & 0.659 & 0.738 & 0.790 & 0.844 \\
Trade Openness & 0.313 & 0.094 & 0.235 & 0.295 & 0.282 & 0.343 & 0.387 & 0.411 \\
Human Capital & 0.738 & 0.764 & 0.628 & 0.732 & 0.752 & 0.766 & 0.791 & 0.775 \\
\midrule

\multicolumn{9}{l}{\textit{Panel B: Correlation with Geopolitical Alignment Index}} \\
\midrule
GDP per Capita (PWT) & 0.099 & 0.123 & 0.113 & $-$0.166 & 0.241 & 0.127 & 0.133 & 0.007 \\
Investment Share & 0.106 & $-$0.022 & 0.150 & 0.018 & 0.130 & 0.186 & 0.195 & 0.100 \\
Capital Stock & 0.137 & 0.141 & 0.266 & $-$0.006 & 0.256 & 0.163 & 0.061 & $-$0.153 \\
Internal Rate of Return & $-$0.014 & $-$0.005 & $-$0.082 & $-$0.010 & 0.001 & $-$0.041 & 0.005 & 0.137 \\
TFP & 0.007 & $-$0.025 & $-$0.031 & $-$0.128 & 0.167 & $-$0.003 & 0.005 & $-$0.072 \\
Trade Openness & 0.115 & $-$0.047 & 0.088 & 0.024 & 0.181 & 0.160 & 0.146 & 0.119 \\
Human Capital & 0.086 & 0.029 & 0.067 & $-$0.188 & 0.333 & 0.163 & 0.056 & $-$0.094 \\

\bottomrule
\end{tabular}
\end{threeparttable}
\note{Entries are pooled within-year cross-sectional correlations. Panel A reports correlations between GDP per capita and economic fundamentals. Panel B reports correlations between the geopolitical alignment index and the same variables. For each period, variables are first demeaned within year, and correlations are computed across country-year observations. Only years with at least 20 countries and nonzero cross-country variation are included. TFP is the PWT 11.0 level at current purchasing power parities (ctfp). The 2020s column covers 2020--2024.}
\end{table}

\FloatBarrier
\subsection{Additional Results for Baseline Estimates} \label{app_b:add_baseline}

Table~\ref{tab:lp_geopolitical_gdp} presents selected local-projection coefficients underlying Figure~\ref{fig:baseline_irf}. The remaining results examine index weights, inference, reverse-direction timing, and lag specifications.

\begin{table}[ht]
\centering
\caption{Local-Projection Coefficients for GDP per Capita}
\label{tab:lp_geopolitical_gdp}
\resizebox{\linewidth}{!}{
\begin{tabular}{lccccccccc}
\toprule
Horizon (years) & $-15$ & $-10$ & $-5$ & $+0$ & $+5$ & $+10$ & $+15$ & $+20$ & $+25$ \\
\midrule
Geopolitical Alignment Index & $-1.758$ & $-0.798$ & $-0.185$ & $0.497^{*}$ & $3.095^{***}$ & $2.917^{***}$ & $2.511^{***}$ & $1.371^{**}$ & $0.512$ \\
 & $(1.154)$ & $(0.889)$ & $(0.215)$ & $(0.260)$ & $(0.671)$ & $(0.862)$ & $(0.758)$ & $(0.674)$ & $(0.495)$ \\
\midrule
Within-$R^2$ & $0.553$ & $0.790$ & $0.983$ & $0.983$ & $0.786$ & $0.550$ & $0.354$ & $0.214$ & $0.120$ \\
Observations & $6{,}534$ & $7{,}274$ & $8{,}014$ & $8{,}154$ & $7{,}414$ & $6{,}674$ & $5{,}934$ & $5{,}194$ & $4{,}454$ \\
Countries & $148$ & $148$ & $148$ & $148$ & $148$ & $148$ & $148$ & $148$ & $148$ \\
\bottomrule
\end{tabular}
}
\note{This table presents the local-projection coefficients from equation~\eqref{eq:lp_spec} for log GDP per capita ($\times 100$). Each column is a separate horizon regression; negative horizons provide timing diagnostics. Coefficients are per one standard deviation of the index. All specifications include country fixed effects, region-year fixed effects, and four lags of log GDP per capita and geopolitical alignment. Driscoll--Kraay standard errors are in parentheses. A Wald test that the coefficients at $h=-10,\ldots,-5$ are jointly zero, using the covariance of 500 country-block bootstrap draws, gives a $p$-value of 0.061. The underlying 148-country long-coverage sample spans 1960--2024. Estimation support varies by horizon because of the lags and horizon-specific outcome: $h=0$ uses 1964--2024, while the displayed columns range from 1975--2024 at $h=-15$ to 1964--1999 at $h=25$. Significance levels: $^{***}$ $p<0.01$, $^{**}$ $p<0.05$, $^{*}$ $p<0.10$.}
\end{table}

\FloatBarrier
\paragraph{Alternative Index Weights and Construction}

\begin{figure}[ht]
    \centering
    \begin{subfigure}[b]{0.32\textwidth}
        \includegraphics[width=\textwidth]{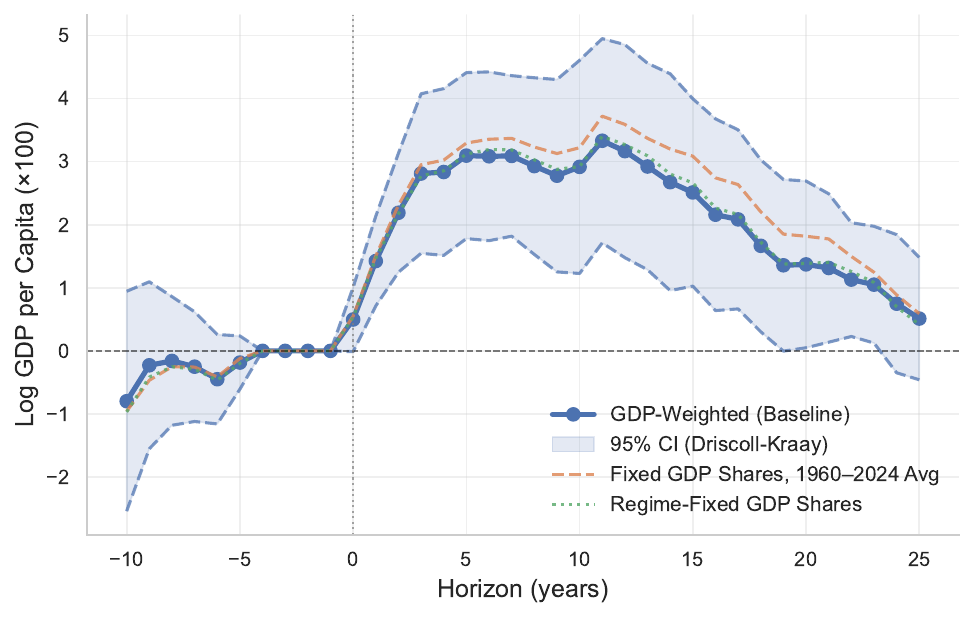}
        \caption{Fixed GDP Weights}
        \label{fig:weighting_comparison}
    \end{subfigure}
    \hfill
    \begin{subfigure}[b]{0.32\textwidth}
        \includegraphics[width=\textwidth]{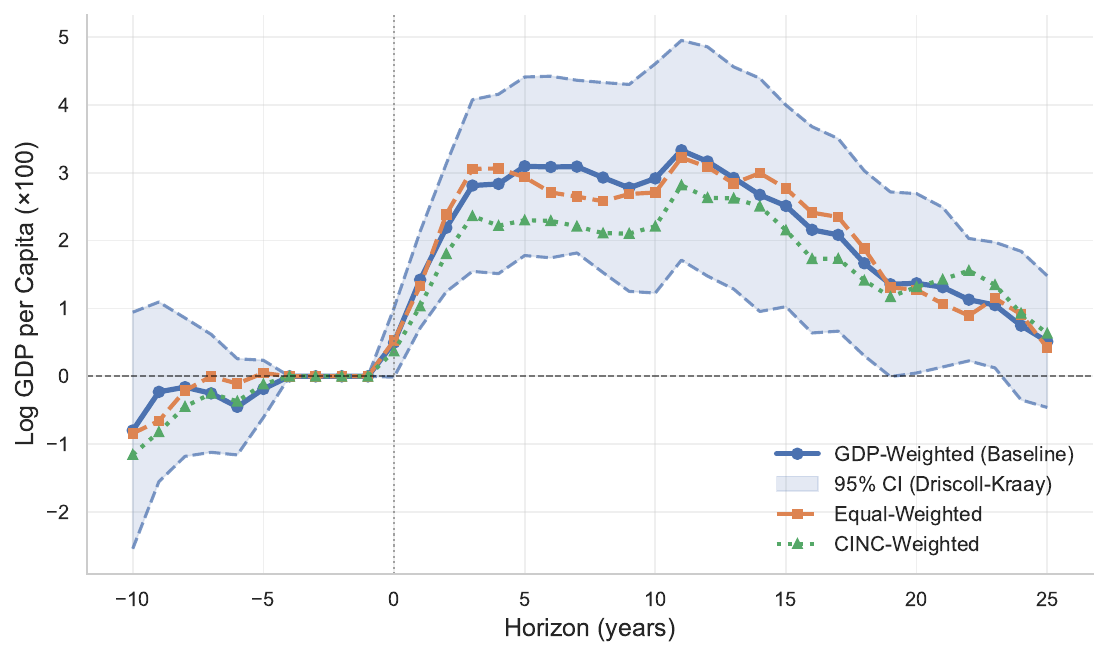}
        \caption{Equal and CINC Weights}
        \label{fig:equal_weight}
    \end{subfigure} \hfill \begin{subfigure}[b]{0.32\textwidth} \includegraphics[width=\textwidth]{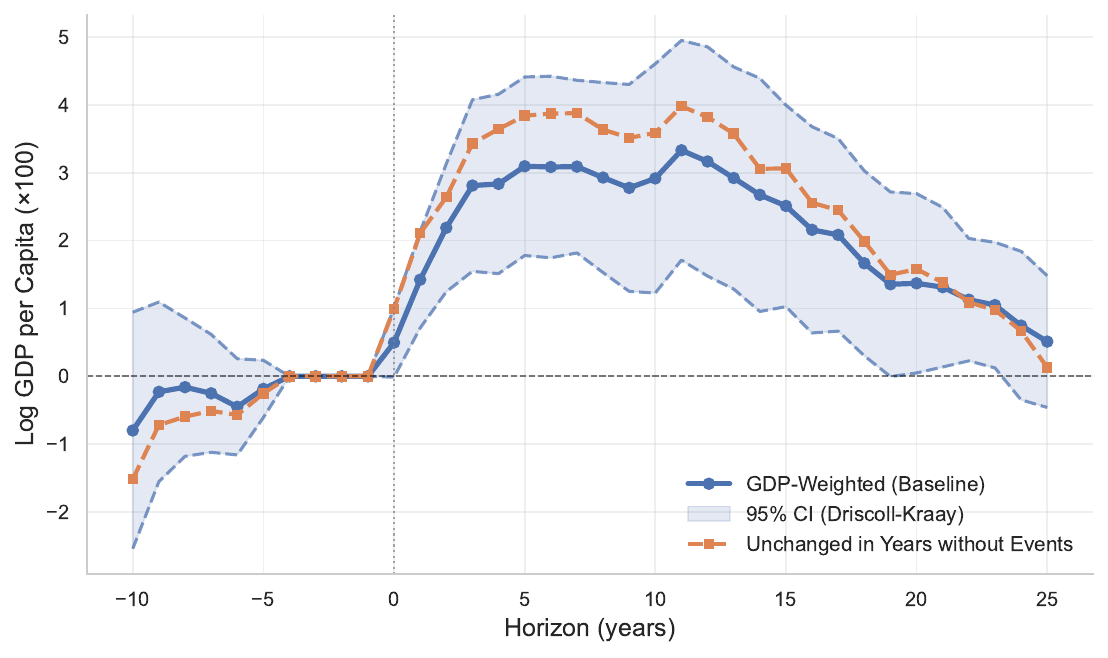} \caption{Years without Recorded Events} \label{fig:carry_forward} \end{subfigure}
    \caption{Robustness to Alternative Index Weights and Construction}
    \label{fig:weighting_robustness}
    \note{All panels report conditional local-projection coefficients for log GDP per capita ($\times 100$). Panel (a) compares the baseline using time-varying GDP shares (blue) with fixed GDP shares averaged over 1960--2024 (orange) and regime-fixed GDP shares using Cold War versus post-Cold War averages (green). Panel (b) compares the baseline GDP-weighted index (blue) with an equal-weighted index (orange) and a CINC-weighted index (green). Panel (c) compares the baseline (blue) with an index that holds bilateral scores unchanged in dyad-years without recorded events instead of moving them toward neutral (orange). All paths use the 148-country long-coverage sample. Each alternative index is rescaled to match the standard deviation of its corresponding baseline index before estimation. All specifications use the baseline controls. Shaded areas are 95\% confidence intervals for the baseline, based on Driscoll--Kraay standard errors.}
\end{figure}

Because the baseline index weights partners by time-varying GDP shares, changes in partner GDP could move the index without any change in bilateral relations. Figure~\ref{fig:weighting_robustness} replaces these weights with fixed, regime-fixed, equal, and capability-based weights.

\FloatBarrier

Panel (a) compares the baseline time-varying GDP shares with fixed shares averaged over 1960--2024 and regime-fixed shares for the Cold War and post-Cold War periods. Panel (b) instead assigns equal weight to each of the 24 major nations. The resulting index has a correlation of 0.86 with the baseline. Panel (b) also weights partners by their share of the Composite Index of National Capability \citep{singer1972capability}. This index has a correlation of 0.89 with the baseline, and its coefficient path stays within the baseline confidence interval. Panel (c) holds bilateral scores unchanged in years without recorded events instead of moving them toward neutral. The resulting index has a correlation of 0.92 with the baseline, and its 25-year sustained-path coefficient is 10.0 log points. All five alternatives produce coefficient paths that closely track the baseline, weighing against changes in GDP weights, the aggregation scheme, or the treatment of years without events as the source of the result.

\paragraph{Inference and Reverse-Direction Diagnostics}

\begin{figure}[ht]
    \centering
    \begin{subfigure}[b]{0.48\textwidth}
        \includegraphics[width=\textwidth]{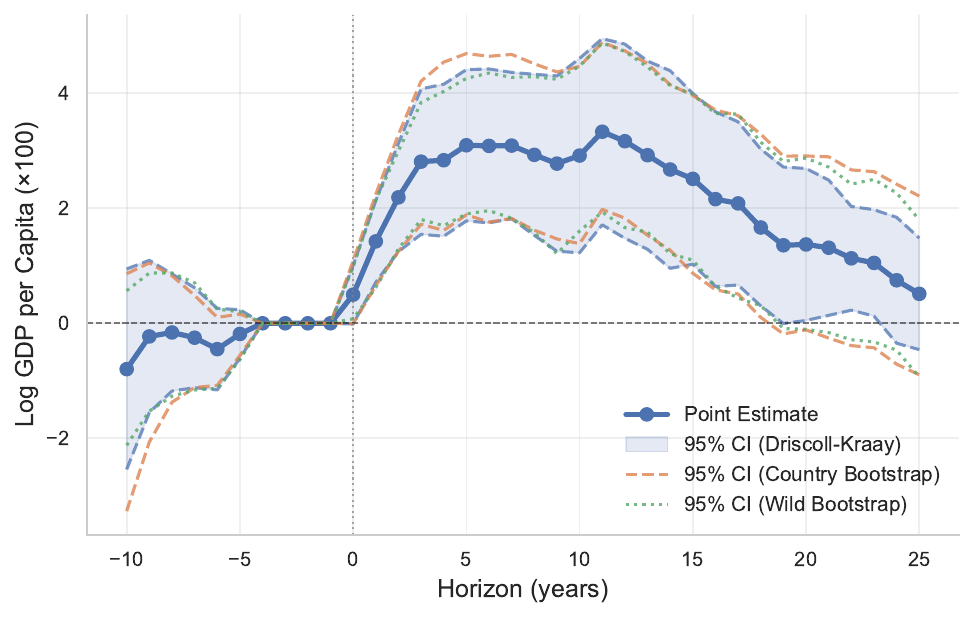}
        \caption{Alternative Inference}
        \label{fig:se_vs_bootstrap}
    \end{subfigure}
    \hfill
    \begin{subfigure}[b]{0.48\textwidth}
        \includegraphics[width=\textwidth]{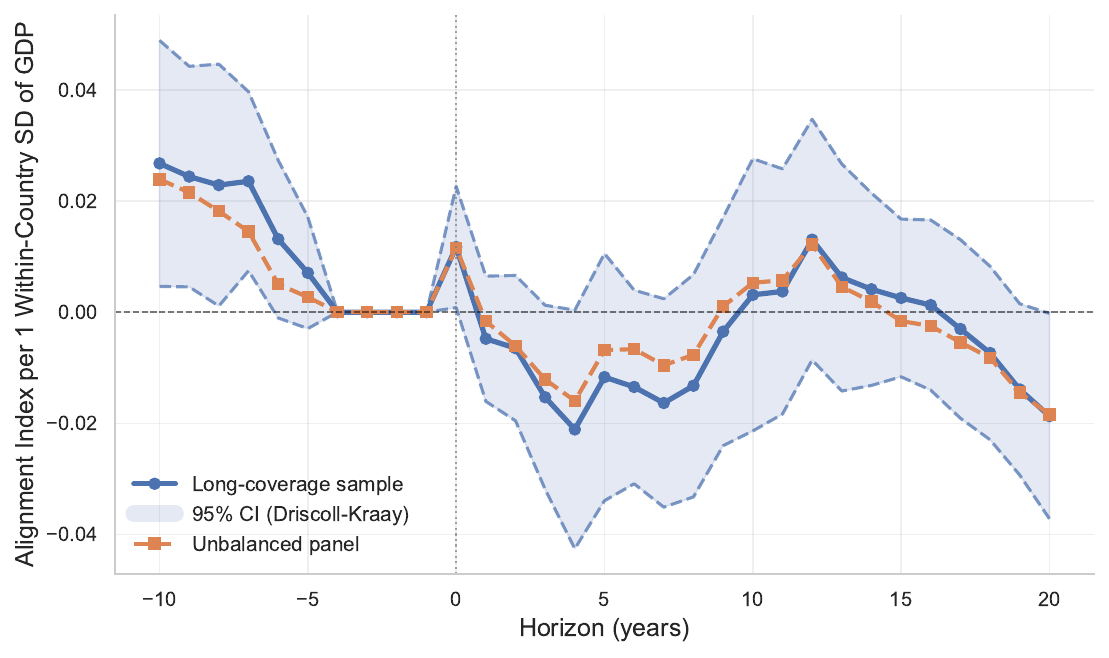}
        \caption{Reverse Direction: GDP to Alignment}
        \label{fig:reverse_irf}
    \end{subfigure}
    \caption{Inference and Reverse-Direction Diagnostics}
    \label{fig:robustness_checks_main}
    \note{Panel (a) displays the baseline GDP coefficient path with three types of confidence intervals: Driscoll--Kraay standard errors (blue shaded band), 500 country-block bootstrap iterations (orange dashed), and 500 wild bootstrap iterations using Rademacher weights (green dotted). Panel (b) reverses the projection, regressing fixed-weight geopolitical alignment at $t+h$ on log GDP per capita at $t$, conditional on four lags of both variables and the baseline fixed effects. Coefficients are per one within-country standard deviation of GDP. The blue line and shaded 95\% Driscoll--Kraay confidence interval use the long-coverage sample. The dashed orange line uses the unbalanced sample. In panel (b), $h=-4,\ldots,-1$ are mechanically zero because the corresponding lagged alignment outcomes enter the controls. Horizons $h=-10,\ldots,-5$ relate past alignment to current GDP and do not test whether GDP precedes alignment.}
\end{figure}

Figure~\ref{fig:robustness_checks_main} pairs alternative inference procedures with a reverse-direction timing diagnostic.

\FloatBarrier
Panel (a) compares the baseline Driscoll--Kraay bands with two resampling procedures. The country-block bootstrap resamples entire countries with replacement, preserving within-country dependence. The wild bootstrap applies observation-level Rademacher weights to residuals after absorbing the fixed effects, which allows for heteroskedasticity but not for serial or cross-country dependence.

The confidence intervals from 500 iterations of each bootstrap method are similar in width to the Driscoll--Kraay bands through year 20 and wider at longer horizons. The wild bootstrap intervals (green dotted) are slightly narrower than the country-block bootstrap intervals (orange dashed) at most horizons. Neither alternative changes the substantive inference.

Panel (b) asks whether current GDP predicts subsequent geopolitical alignment. Because the baseline index uses time-varying partner GDP shares, we hold each major nation's aggregation weight fixed at its sample mean, preventing changes in GDP weights from mechanically generating movement in the alignment index. In both the long-coverage and the unbalanced samples, current GDP shows no systematic or statistically meaningful association with subsequent alignment, so there is no evidence that rising GDP precedes geopolitical improvement.

\paragraph{Alternative Lag Specifications}

\begin{figure}[ht]
    \centering
    \includegraphics[width=0.9\linewidth]{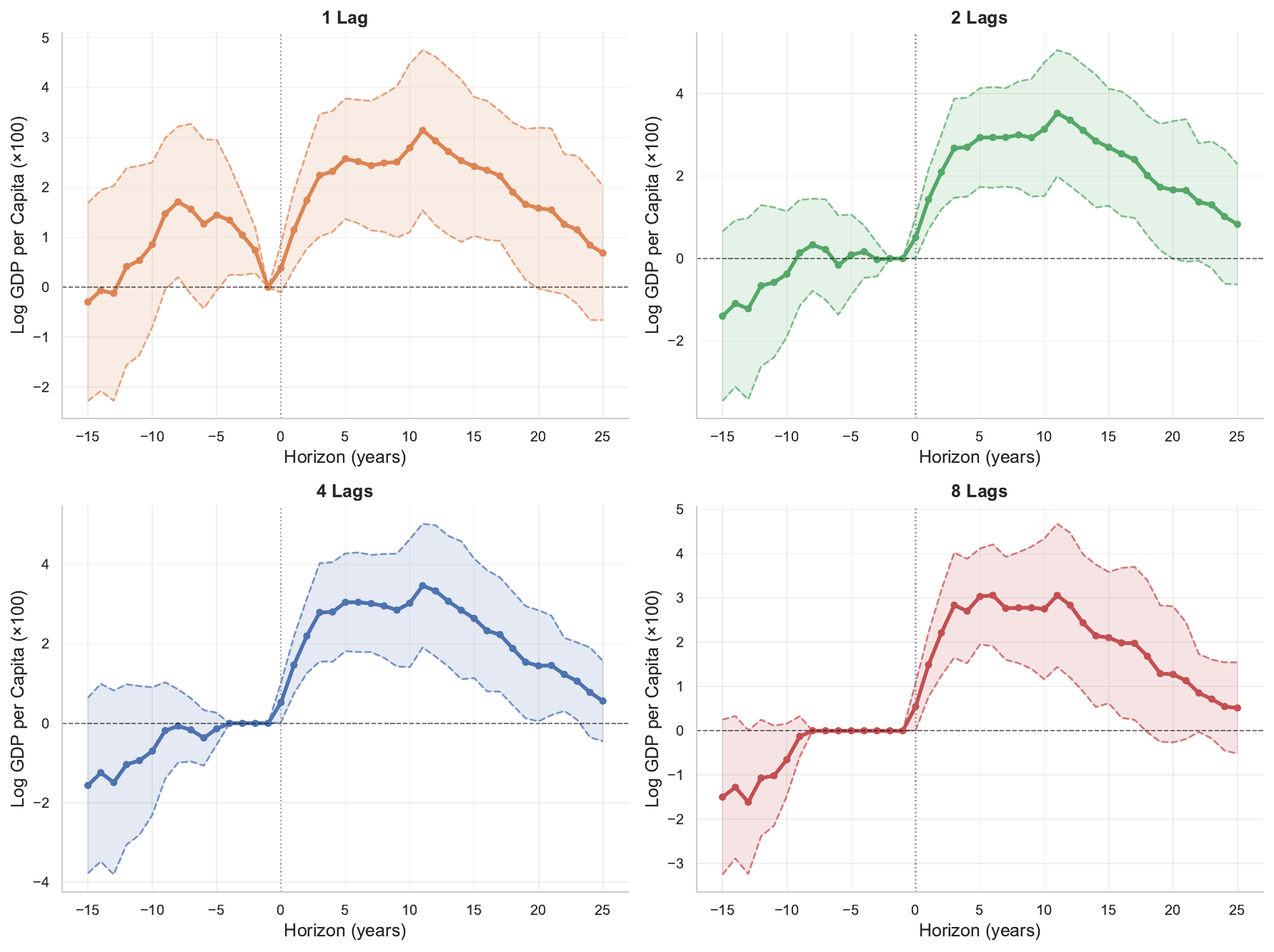}
    \caption{Conditional GDP Coefficient Paths Under Alternative Lag Specifications}
    \label{fig:lag_specifications}
    \note{Each panel reports conditional local-projection coefficients for log GDP per capita ($\times 100$) with the indicated number of lags of GDP and geopolitical alignment. All specifications include country and region-year fixed effects. Shaded areas represent 95\% confidence intervals based on Driscoll--Kraay standard errors.}
\end{figure}

Figure~\ref{fig:lag_specifications} examines sensitivity to lag length selection.

\FloatBarrier
The one-lag specification shows positive estimates at negative horizons and the two-lag specification negative ones at longer leads, indicating sensitivity to incomplete dynamic adjustment. With four and eight lags, the displayed negative-horizon coefficients are not individually distinguishable from zero and the post-zero paths are similar. We use four lags in the baseline to retain precision while allowing flexible dynamics.

\FloatBarrier
\subsection{Residualized Geopolitical Alignment and Subsequent GDP} \label{app_b:fwl}

To visualize the conditional relationship underlying the local projections, we residualize GDP and geopolitical alignment with respect to the baseline controls and fixed effects and plot their binned residual means.

\begin{figure}[ht]
    \centering
    \begin{subfigure}[b]{0.48\textwidth}
        \includegraphics[width=\textwidth]{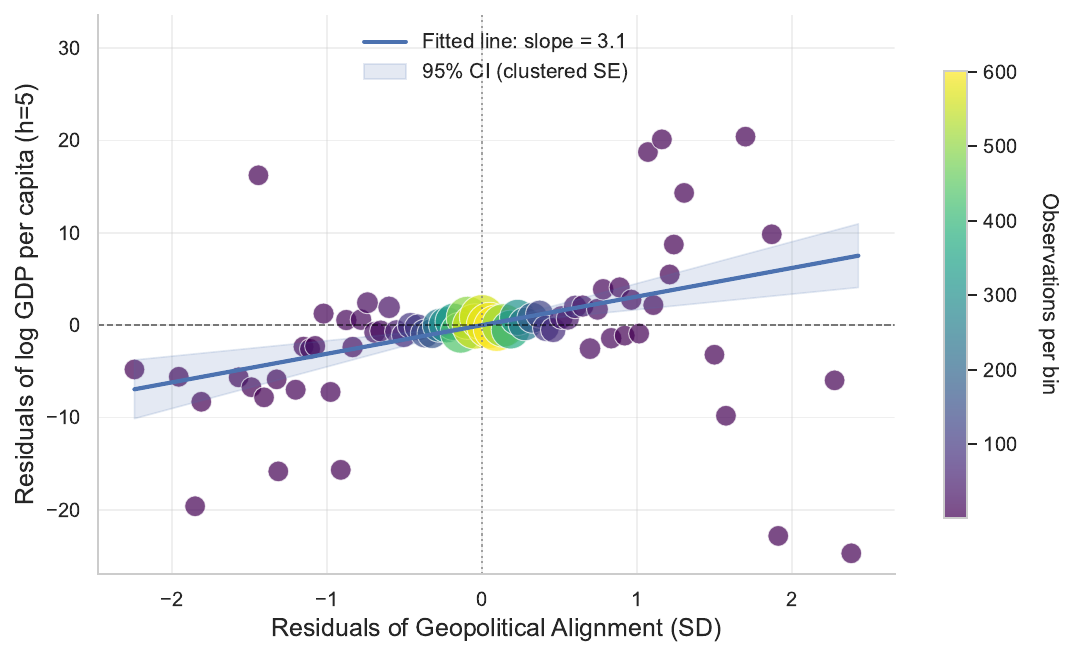}
        \caption{GDP at $t+5$ vs. Geopolitical Alignment at $t$}
        \label{fig:fwl_h5}
    \end{subfigure}
    \hfill
    \begin{subfigure}[b]{0.48\textwidth}
        \includegraphics[width=\textwidth]{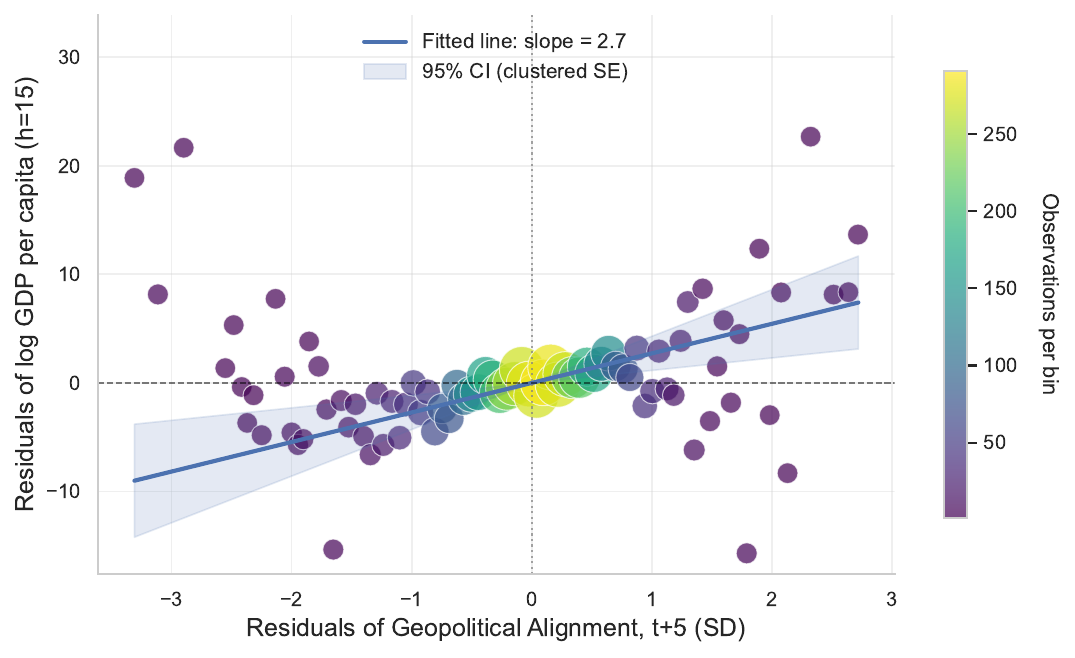}
        \caption{GDP at $t+15$ vs. Geopolitical Alignment at $t+5$}
        \label{fig:fwl_h15}
    \end{subfigure}
    \caption{Residualized Geopolitical Alignment and Subsequent GDP}
    \label{fig:fwl_binscatter}
    \note{Binned residual means of log GDP per capita against geopolitical alignment after partialling out the baseline controls and fixed effects used in Figure~\ref{fig:baseline_irf}. Panel~(a) relates GDP at $t+5$ to geopolitical alignment at $t$. Panel~(b) relates GDP at $t+15$ to forward geopolitical alignment measured at $t+5$. The plots use 100 equal-width bins. Marker size and color indicate the number of observations per bin. Shaded bands are 95\% confidence intervals for the fitted lines, with standard errors clustered by country. GDP residuals are in log points (log GDP per capita $\times 100$), and alignment residuals are divided by the standard deviation of the index (0.141).}
\end{figure}

The fitted slopes are 3.1 in panel~(a) and 2.7 in panel~(b). In both, the binned residual means follow an approximately linear positive pattern over the observed range, with no single segment driving the slope. This binscatter is a functional-form and outlier diagnostic, not evidence of causal symmetry or a basis for extrapolation.

\FloatBarrier
\subsection{Transitory and Sustained Alignment Paths} \label{app_b:irf_transitory_persistent}

The coefficient paths in Section~\ref{ss:baseline} combine the current association between geopolitical alignment and GDP with the subsequent dynamics of alignment. To isolate the component associated with a purely transitory change, we construct a counterfactual path in which the index increases by one unit on impact and returns to zero immediately thereafter. Following \citet{Sims1986-mt} and \citet{bilal2024macroeconomic}, we combine the projection coefficients for geopolitical alignment and GDP to construct this transitory path.

We begin by estimating the dynamics of the geopolitical alignment index using local projections:
\begin{equation}
    p_{c,t+h} = \rho_h^{p} p_{ct} + \gamma_h^{\prime} \mathbf{x}_{ct} + \varepsilon_{c,t+h}, \quad h = 0, 1, \ldots, H
\end{equation}
where $\{\rho_h^{p}\}_{h=0,\ldots,H}$ are the horizon-specific projection coefficients on current geopolitical alignment, and $\mathbf{x}_{ct}$ contains the baseline lags and fixed effects. To construct the purely transitory path, we introduce a series of auxiliary shocks $\{p_h^{\text{shock}}\}_{h=0}^H$ at each horizon that impose the desired response pattern $\tilde{\boldsymbol{\rho}}^p = (1, 0, \ldots, 0)'$. The required auxiliary series $\boldsymbol{p}^{\text{shock}}$ is obtained by solving:
\begin{equation}
    \underbrace{\left(\begin{array}{c}
p_0^{\text{shock}} \\
p_1^{\text{shock}} \\
\vdots \\
p_H^{\text{shock}}
\end{array}\right)}_{\boldsymbol{p}^{\text{shock}}} = \underbrace{\left(\begin{array}{cccc}
1 & 0 & \cdots & 0 \\
\rho_1^p & 1 & \cdots & 0 \\
\vdots & \vdots & \ddots & \vdots \\
\rho_H^p & \rho_{H-1}^p & \cdots & 1
\end{array}\right)^{-1}}_{\left(\mathbf{R}^p\right)^{-1}} \underbrace{\left(\begin{array}{c}
1 \\
0 \\
\vdots \\
0
\end{array}\right)}_{\tilde{\boldsymbol{\rho}}^p}
\end{equation}

Given this auxiliary series, the corresponding GDP coefficients $\tilde{\boldsymbol{\alpha}}$ under the purely transitory path are:
\begin{equation}
    \underbrace{\left(\begin{array}{c}
\tilde{\alpha}_0 \\
\tilde{\alpha}_1 \\
\vdots \\
\tilde{\alpha}_H
\end{array}\right)}_{\tilde{\boldsymbol{\alpha}}} = \underbrace{\left(\begin{array}{cccc}
p_0^{\text{shock}} & 0 & \cdots & 0 \\
p_1^{\text{shock}} & p_0^{\text{shock}} & \cdots & 0 \\
\vdots & \vdots & \ddots & \vdots \\
p_H^{\text{shock}} & p_{H-1}^{\text{shock}} & \cdots & p_0^{\text{shock}}
\end{array}\right)}_{\mathbf{P}^{\text{shock}}} \underbrace{\left(\begin{array}{c}
\alpha_0 \\
\alpha_1 \\
\vdots \\
\alpha_H
\end{array}\right)}_{\boldsymbol{\alpha}}
\end{equation}

The resulting coefficients $\tilde{\boldsymbol{\alpha}}$ trace the GDP path associated with a one-time, purely transitory improvement in geopolitical alignment. For the sustained path shown in Figure~\ref{fig:permanent_shocks}, we compute the cumulative coefficient $\sum_{s=0}^h \tilde{\alpha}_s$, which gives the GDP coefficient associated with maintaining geopolitical alignment one unit above baseline. The reported paths multiply these coefficients by 0.141, the standard deviation of the index. Confidence intervals come from a country-block bootstrap, which accounts for estimation uncertainty in both stages.

\paragraph{Methodological Caveat} This construction is an algebraic transformation of the conditional local-projection coefficients and does not add identifying variation beyond the baseline specification. A causal interpretation requires the conditions in Section~\ref{ss:interpretation}, together with linearity and the assumption that the response to a sequence of unanticipated geopolitical innovations equals the response to the same alignment path anticipated at time zero. The latter assumption abstracts from forward-looking behavior that may differ with anticipation. We use the construction to compare persistence profiles and scale magnitudes, not as a structural policy experiment.

\FloatBarrier
\subsection{Placebo Design} \label{app_b:placebo}

This section documents the placebo exercises summarized in Section~\ref{ss:placebo}.

\paragraph{Implementation} Both exercises use 500 independent placebo draws. In each draw, the full baseline local projection is re-estimated from scratch using the same specification: four lags of GDP and geopolitical alignment, country fixed effects, and region-year fixed effects, with Driscoll--Kraay standard errors. The summary statistic is the average estimated coefficient over horizons $h = 0, \ldots, 10$.

\paragraph{Placebo A: within-region-year reassignment} For each draw, we randomly permute realized alignment values across countries within each region-year cell. This preserves the region-year distribution of the index and the panel structure of outcomes and controls while breaking its country-specific assignment.

\paragraph{Placebo B: future-year timing} For each draw and each country, we replace current alignment with its value 8 to 15 years in the future, where the lead is drawn uniformly at random. To ensure that all leads are well defined, we restrict base years to 1960--2009. This preserves each country's own alignment distribution, up to a time shift, while breaking the contemporaneous timing between alignment and subsequent GDP dynamics.

\paragraph{Results}

Averaged over horizons 0 to 10, the realized coefficient is $2.51$ under Placebo~A and $2.67$ under Placebo~B, above all 500 placebo draws in each design, whose 95th percentiles are $0.29$ and $0.42$.

\FloatBarrier
\subsection{Domestic Transformation Periods} \label{app_b:transformation_episodes}

\paragraph{Definition and Boundaries} This section documents the classification of domestic transformation periods absorbed in Section~\ref{ss:bundled}. We evaluate candidate periods using four criteria: whether the period marks a recognized historical turning point, changes domestic political or economic institutions, spans multiple institutional margins (such as the constitutional order, state structure, property rights, or the economic-policy regime), and plausibly affects the country's long-run economic trajectory. A window begins when institutional flux becomes visible and ends when the new framework is operational. Transformation windows typically span two to seven years, but this is a guideline, not a hard restriction.

The classification excludes changes in geopolitical alignment, alliance membership, or bilateral relations unless they accompany a qualifying domestic transformation. It also excludes routine electoral alternation, gradual policy evolution, narrow sectoral reforms, macroeconomic crises, sanctions, and wars that do not transform the domestic institutional order. For example, Brexit is not classified because it changed external institutional membership without replacing the United Kingdom's domestic political-economic regime, and Japan's asset-price collapse is not classified because it was a macroeconomic shock within a continuous institutional framework. The rule is direction-neutral: democratizations and post-conflict reconstructions qualify alongside authoritarian consolidations and state collapses. Candidate episodes were reviewed for window placement, substantive scope, cross-country comparability, and coverage of well-established historical transformations, then normalized into a complete, nonoverlapping country-year classification.

\begin{table}[ht]
\centering
\caption{Domestic Transformation Periods: Coverage, Categories, and Examples}
\label{tab:transformation_periods}
\footnotesize
\begin{tabular}{p{0.32\textwidth}rp{0.43\textwidth}}
\toprule
\multicolumn{3}{l}{\textit{Panel A: Coverage and Duration}} \\
\midrule
Country-year grid & \multicolumn{2}{l}{12,545 observations, 193 countries over 1960--2024} \\
Coded episodes & \multicolumn{2}{l}{329 episodes across 172 countries; 21 countries with none} \\
Transformation years & \multicolumn{2}{l}{1,476 country-years, or 11.8\% of the grid} \\
Episode duration & \multicolumn{2}{l}{Mean 4.5 years; median 4 years; range 1--13 years} \\
Duration distribution & \multicolumn{2}{l}{79.6\% last at most 5 years; 91.8\% at most 7 years} \\
\midrule
\multicolumn{3}{l}{\textit{Panel B: Categories and Representative Examples}} \\
\midrule
Category & \multicolumn{1}{c}{Episodes} & Representative example \\
\midrule
Regime change & 91 & Chile, military coup and economic restructuring, 1973--1976 \\
Democratization & 67 & South Africa, end of apartheid, 1990--1994 \\
Independence & 57 & Botswana, independence and state construction, 1966--1969 \\
Authoritarian consolidation & 30 & Pakistan, Zia coup and Islamization, 1977--1980 \\
Post-conflict transition & 25 & Rwanda, genocide and Rwandan Patriotic Front regime establishment, 1994--1999 \\
Post-socialist transition & 23 & Poland, transition to market democracy, 1989--1992 \\
Economic transition & 18 & China, reform and opening, 1978--1984 \\
State collapse & 14 & Somalia, state collapse and civil war, 1988--1992 \\
Reconstruction & 4 & Germany, reunification and institutional integration, 1990--1994 \\
\bottomrule
\end{tabular}
\note{Panel A summarizes the complete country-year classification. Panel B reports the mutually exclusive category assigned to each episode and one representative case. The replication package contains the complete episode and country-year panels.}
\end{table}

\paragraph{Use in Estimation} The transformation-year specification adds a separate country-year fixed effect for every observation classified as part of a domestic transformation. This specification is algebraically equivalent to excluding these observations. We do not absorb fixed effects for all country-era cells, including stable-era means. Because geopolitical reorientation is persistent, full country-era fixed effects would remove the low-frequency component of alignment across long stable periods and change the estimand relative to the baseline.

As an alternative classification, we use V-Dem's \textit{Regimes of the World} measure and absorb observations within three years before or after a change in regime type. V-Dem covers 8,385 of the 9,620 country-years in the long-coverage sample. It omits 16 countries, primarily microstates, and has shorter historical coverage for eight others. Observations without a V-Dem classification are excluded from this exercise rather than treated as stable. Among covered country-years, both rules flag 7.9 percent, the domestic-transformation classification alone flags 4.7 percent, V-Dem alone flags 19.6 percent, and neither flags 67.8 percent. The domestic-transformation classification covers broader episode types, whereas V-Dem applies a mechanical seven-year window around each regime-type change. At year 5, the transformation-year and V-Dem exclusions yield coefficients of 2.0 and 2.4 log points, respectively. On the 124 countries with complete V-Dem coverage, the year-5 coefficient is 3.6 log points (95 percent confidence interval 2.3 to 4.9) in the corresponding baseline and 2.5 log points (0.8 to 4.1) under the V-Dem exclusion. The V-Dem sustained-path magnitude is slightly smaller than under the transformation-year exclusion despite its larger year-5 coefficient, partly because the mapping re-estimates alignment persistence within each exclusion sample. We therefore interpret V-Dem as an alternative exclusion rule, not as validation of individual historical episodes.

\FloatBarrier
\subsection{Leadership Transitions in Major Nations: Robustness} \label{app_b:leader_iv}

This section documents the leadership-change IV based on bilateral shifts measured around 67 leadership changes across 21 of 24 major nations: 25 deaths in office (from the Archigos database, 1950--2015) and 42 close election turnovers with vote margins below 10 percentage points (from \citet{marx2024elections}, 1960--2018). Switzerland is excluded due to its annually rotating presidency.\footnote{For France we use presidential elections only. We drop elections flagged in the source or lacking a reported margin, as well as elections within two years of a death in office in the same country. We also correct two Archigos codings. Deaths in office are Archigos exits by natural death, assassination, suicide, or irregular removal followed by death.} For each event in major nation $j$ at time $t$, the instrument is $z_{ct}^{\text{LC}} = (S_{cj,t+1} - S_{cj,t-1}) \times \text{GDP share}_{j,t-1}$. Because the instrument uses the realized bilateral shift rather than the transition alone, it may also capture other bilateral developments in the event window.

\paragraph{Window Robustness} Table~\ref{tab:window_robustness} reports reduced-form and LP-IV estimates at horizon $h=5$ under six bilateral-shift windows. The reduced-form point estimates range from 3.4 to 4.4 log points, while the IV estimates are larger for wider post-transition windows as the first stage weakens. The main-text window ($S_{t+1} - S_{t-1}$) is relatively short and has a first-stage $F$-statistic of 16.9.

\begin{table}[ht]
\centering
\caption{Leadership-Change IV: Window Robustness}
\label{tab:window_robustness}
\footnotesize
\begin{tabular}{lccc}
\toprule
Window & First-stage $F$ & RF$(h=5)$ & IV$(h=5)$ \\
\midrule
$S_t - S_{t-1}$       & 53.7 & 4.4 (2.3) & 3.1 (1.4) \\
$S_{t+1} - S_{t-1}$   & 16.9 & 3.4 (2.4) & 6.0 (2.5) \\
$S_{t+2} - S_{t-1}$   & 17.8 & 3.5 (2.8) & 8.1 (3.7) \\
$S_t - S_{t-2}$       & 35.4 & 4.1 (2.6) & 3.0 (1.8) \\
$S_{t+1} - S_{t-2}$   & 15.7 & 3.4 (2.8) & 5.8 (3.1) \\
$S_{t+2} - S_{t-2}$   & 14.7 & 3.5 (3.2) & 8.0 (4.4) \\
\bottomrule
\end{tabular}
\note{Each row uses a different bilateral-shift window to construct the leadership-change instrument. First-stage $F$: Wald $F$-statistic for the instrument in the first-stage projection of alignment. RF$(h=5)$ and IV$(h=5)$: reduced-form and LP-IV coefficients for log GDP per capita ($\times 100$) at $h=5$ with Driscoll--Kraay standard errors in parentheses. Both are multiplied by 0.141, so the IV coefficient is per one standard deviation of the index and the reduced form is per 0.141 units of the instrument. All specifications use 67 events (25 deaths + 42 close elections), country and region-year fixed effects, and four lags of GDP and geopolitical alignment. Sample: 148-country long-coverage sample.}
\end{table}

\begin{figure}[ht]
    \centering
    \begin{subfigure}[b]{0.48\textwidth}
        \includegraphics[width=\textwidth]{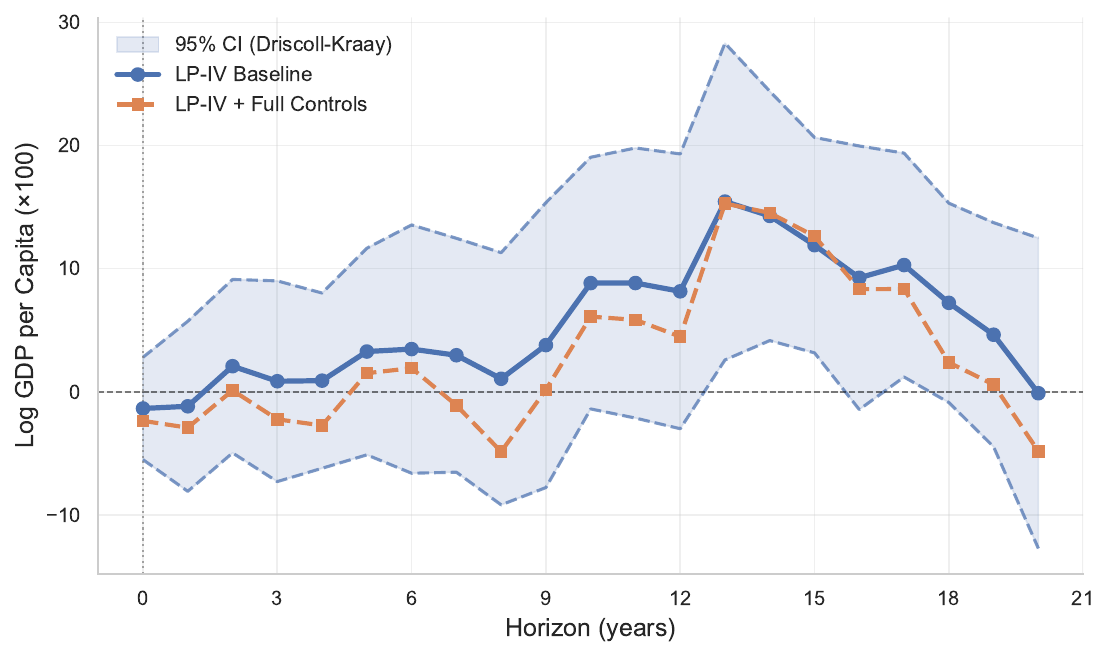}
        \caption{Deaths in Office (25 events)}
        \label{fig:leader_iv_death}
    \end{subfigure}
    \hfill
    \begin{subfigure}[b]{0.48\textwidth}
        \includegraphics[width=\textwidth]{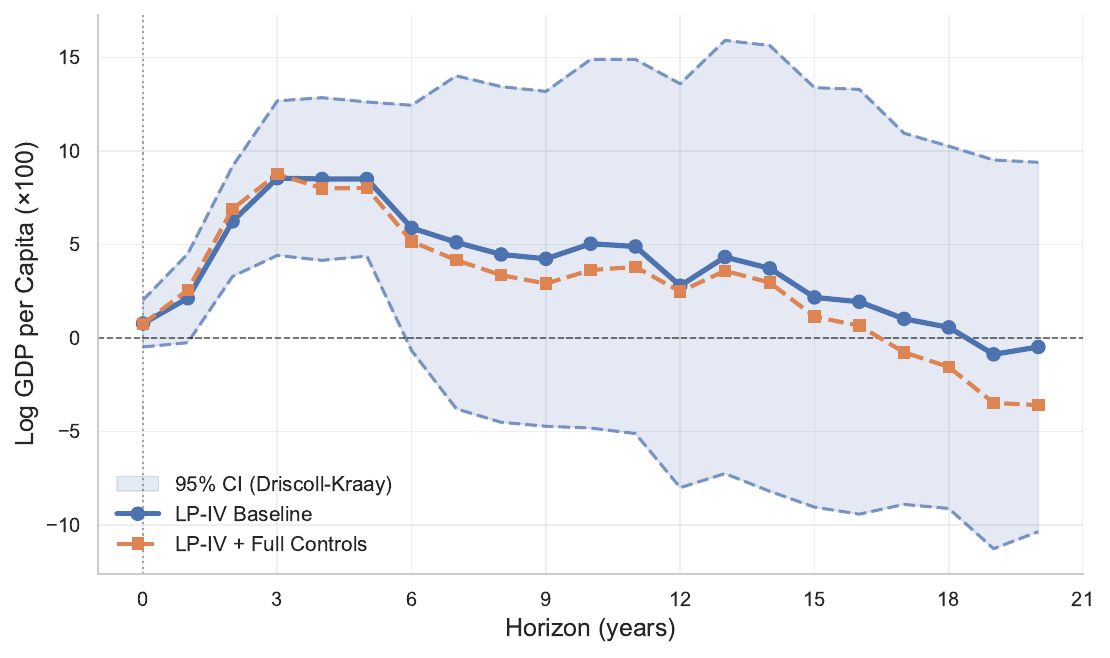}
        \caption{Close Election Turnovers (42 events)}
        \label{fig:leader_iv_election}
    \end{subfigure}
    \caption{Leadership-Change IV by Event Type}
    \label{fig:leader_iv_split}
    \note{LP-IV coefficient paths for log GDP per capita ($\times 100$), estimated separately using deaths in office (panel~(a)) and close election turnovers with margins below 10 percentage points (panel~(b)). Blue solid: baseline specification. Orange dashed: adding four lags of trade openness, domestic unrest, war exposure, and V-Dem institutional indices (on the 109-country common sample, which the plotted baseline also uses). Shaded bands show 95\% confidence intervals for the baseline, based on Driscoll--Kraay standard errors.}
\end{figure}

\paragraph{Deaths versus Close Elections} Figure~\ref{fig:leader_iv_split} splits the leadership-change IV by event type. The close-election component (panel~(b)) yields a smoother hump-shaped path, with coefficients distinguishable from zero over horizons 2--5. The deaths-in-office component (panel~(a)) is more volatile: its first-decade estimates are imprecise, and large positive coefficients appear at several horizons from 13 to 17. The death events provide more plausibly external timing, but their first stage is weak ($F=7.9$, against 26.5 for close elections). Pooling the two event types does not remove the caveat created by using realized bilateral shifts.

\FloatBarrier
\subsection{Other Correlates of Growth} \label{app_b:other_growth_correlates}

\begin{figure}[ht]
    \centering
    \begin{subfigure}[b]{0.48\textwidth}
        \includegraphics[width=\textwidth]{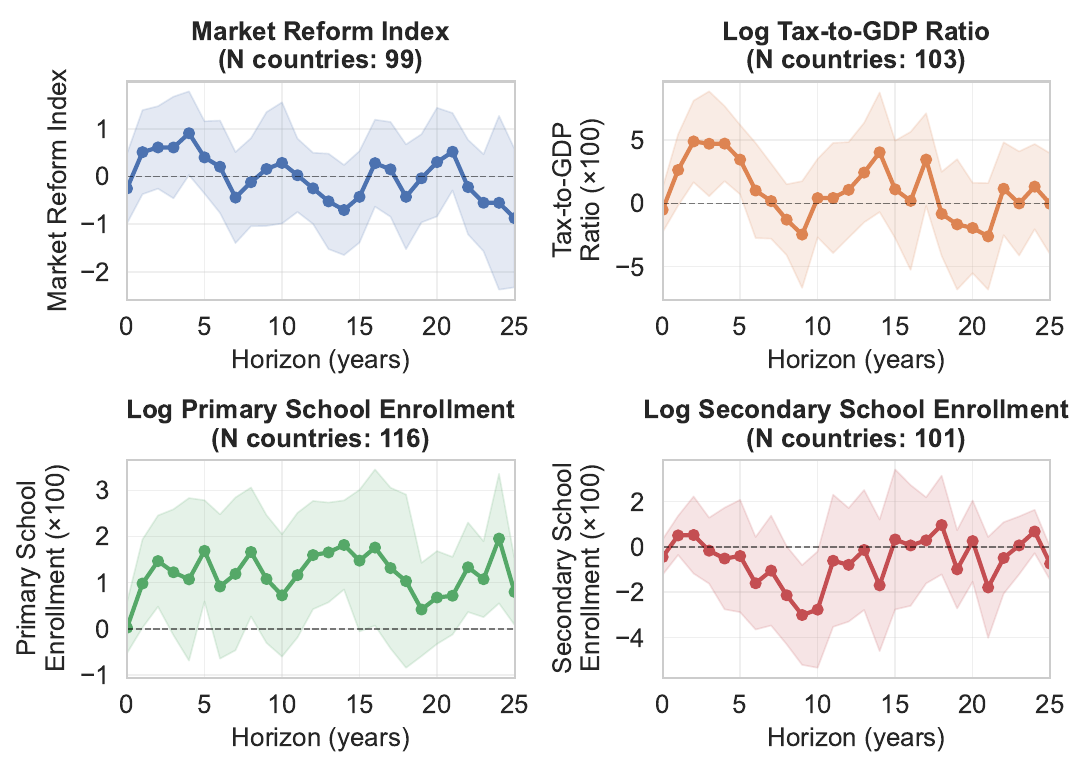}
        \caption{Market Reforms and Education}
        \label{fig:panel_A_market_education}
    \end{subfigure}
    \hfill
    \begin{subfigure}[b]{0.48\textwidth}
        \includegraphics[width=\textwidth]{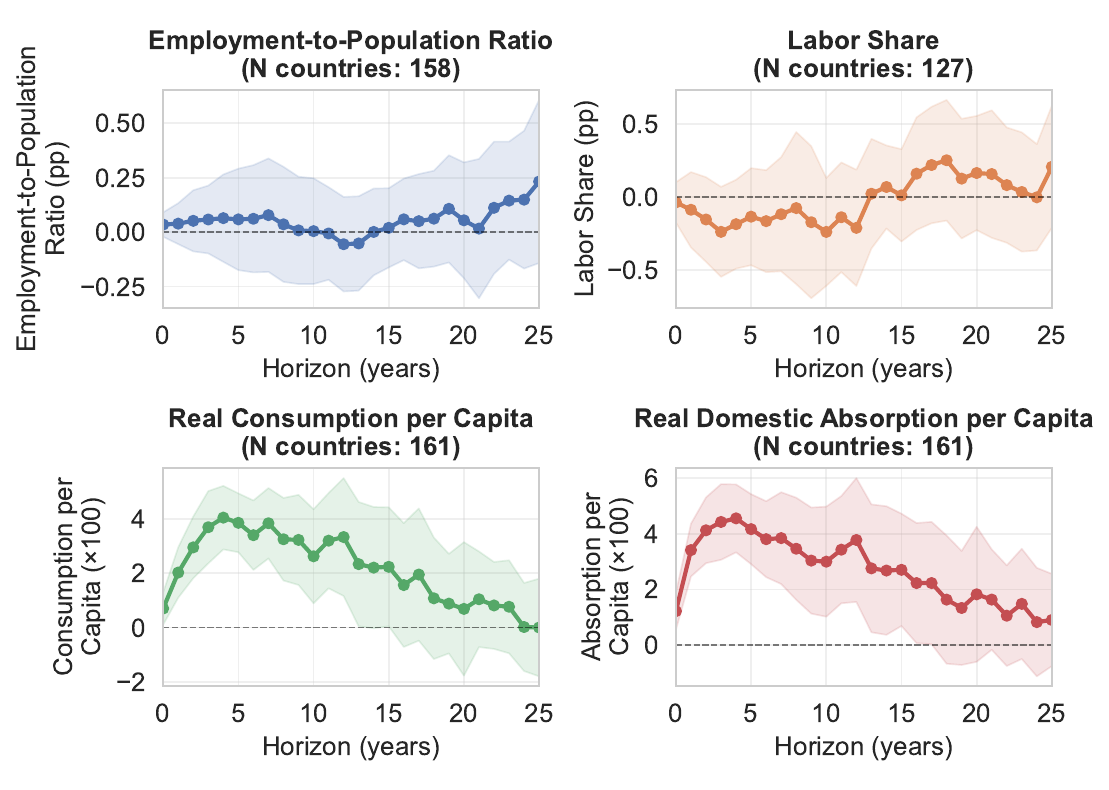}
        \caption{Employment and Expenditure}
        \label{fig:panel_B_employment_consumption}
    \end{subfigure}
    \caption{Dynamic Responses of Additional Growth Correlates to Geopolitical Alignment}
    \label{fig:additional_correlates}
    \note{This figure reports horizon-specific coefficients on the country-level geopolitical alignment index from local projections of the listed outcomes. Panel~(a) shows the market reform index, tax-to-GDP ratio, and primary and secondary enrollment. Panel~(b) shows the employment-to-population ratio, labor share, real consumption, and domestic absorption per capita. Log outcomes are multiplied by 100. All specifications follow equation~\eqref{eq:correlates}. Each outcome uses an outcome-specific country set, with counts shown in parentheses. Shaded areas represent 95\% confidence intervals based on Driscoll--Kraay standard errors.}
\end{figure}

Figure~\ref{fig:additional_correlates} extends the growth-fundamentals analysis in Figure~\ref{fig:growth_correlates} to institutional, educational, labor-market, and expenditure outcomes.

\paragraph{Market Reforms and Human Capital Formation}

Panel~(a) shows a short-run positive response in the tax-to-GDP ratio and a more gradual increase in primary enrollment. Market reform and secondary enrollment show no systematic response.

\paragraph{Labor Markets and Domestic Absorption}

Panel~(b) shows positive, GDP-like paths for real consumption and domestic absorption per capita. The employment-to-population ratio and labor share show no systematic response. As in Figure~\ref{fig:growth_correlates}, these outcomes are estimated separately and should be read as accompanying responses rather than identified mediating channels.

\FloatBarrier
\subsection{Additional Results for Democracy and Geopolitics} \label{app_b:geo_democracy}

This section complements Section~\ref{ss:geo_democracy} by mapping the separate and joint democracy coefficients into transitory and sustained paths and by examining democratic status, aggregate alignment, and GDP under alternative fixed effects.

\subsubsection{Transitory and Sustained Democracy Paths}

We apply the construction in Appendix~\ref{app_b:irf_transitory_persistent} to the separate and joint specifications in Figure~\ref{fig:horse_race}, extending them through year~25. For each specification, the estimated response of democratic status to itself determines the sequence of innovations that produces a one-period democracy path. Convolving this sequence with the GDP coefficients yields the transitory path, and its cumulative sum yields the sustained path.

\begin{figure}[ht]
    \centering
    \begin{subfigure}[b]{0.48\textwidth}
        \includegraphics[width=\textwidth]{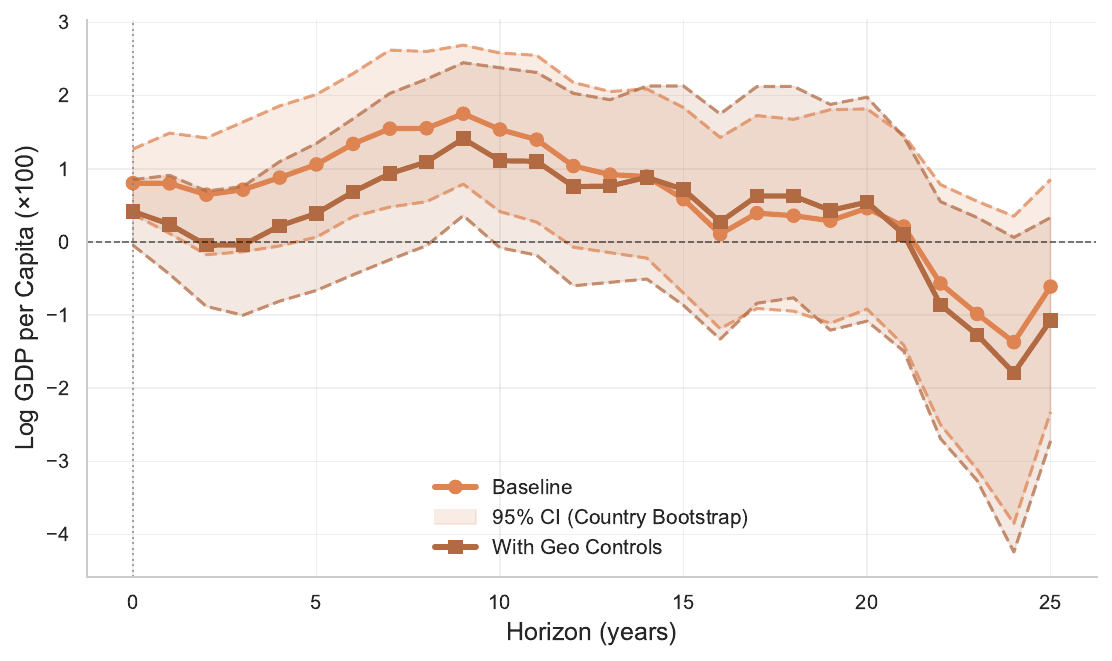}
        \caption{Transitory Democracy Path}
        \label{fig:dem_transitory}
    \end{subfigure}
    \hfill
    \begin{subfigure}[b]{0.48\textwidth}
        \includegraphics[width=\textwidth]{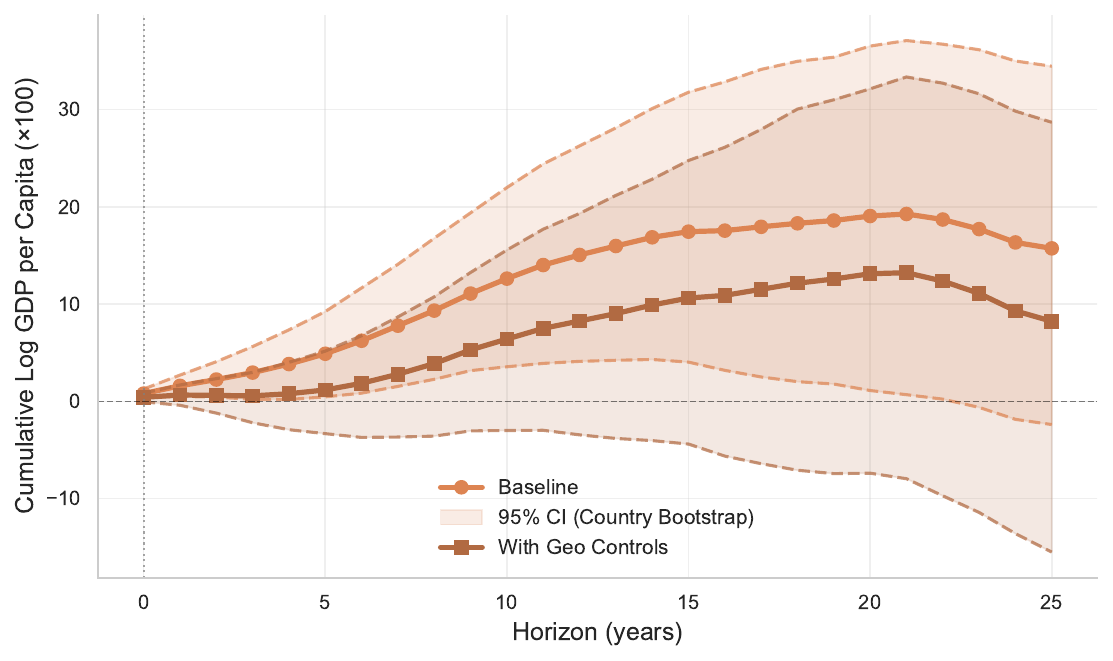}
        \caption{Sustained Democracy Path}
        \label{fig:dem_permanent}
    \end{subfigure}
    \caption{GDP Mappings for Transitory and Sustained Democracy Paths}
    \label{fig:democracy_shock_decomposition}
    \note{Panel~(a) maps the democracy coefficients into the GDP path associated with a one-period increase in democratic status. Panel~(b) cumulates that path to construct a sustained increase. The separate specification includes four GDP lags. The joint specification also includes current geopolitical alignment and its four lags. Both use the original ANRR GDP series and include country and year fixed effects. Shaded bands are pointwise 95\% confidence intervals from 500 country-block bootstrap draws. The legends label the separate specification ``Baseline'' and the joint specification ``With Geo Controls''. The outcome is log GDP per capita ($\times 100$).}
\end{figure}

Panel~(a) shows small transitory paths that peak at 1.8 log points in the separate specification and 1.4 log points when alignment is included, before declining at longer horizons. Panel~(b) shows the corresponding sustained paths. At year~20, the separate and joint estimates reach 19.1 and 13.1 log points. By year~25, they are 15.7 and 8.2 log points. These differences summarize conditional overlap between democracy and geopolitical alignment, not a mediation share.

\subsubsection{Aggregate Alignment and Region-Year Fixed Effects}

\begin{figure}[ht]
    \centering
    \begin{subfigure}[b]{0.48\textwidth}
        \includegraphics[width=\textwidth]{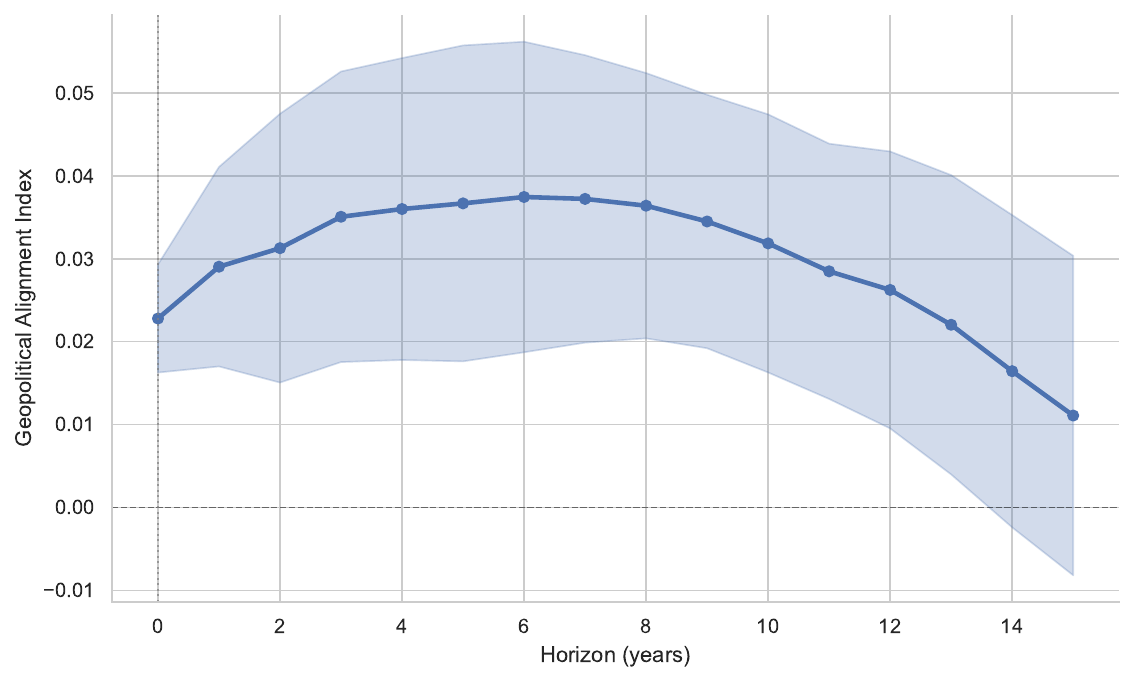}
        \caption{Democratic Status and Aggregate Alignment}
    \end{subfigure}
    \hfill
    \begin{subfigure}[b]{0.48\textwidth}
        \includegraphics[width=\textwidth]{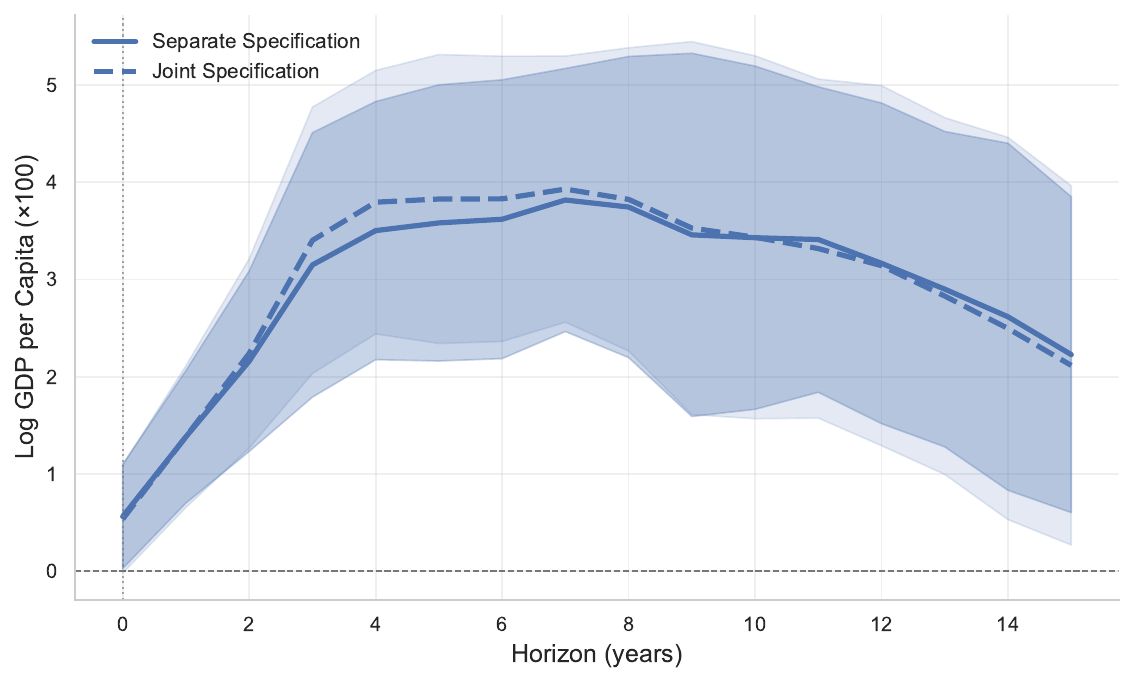}
        \caption{Alignment and GDP, Region-Year FE}
    \end{subfigure}
    \caption{Democracy and Geopolitical Alignment: Additional Results}
    \label{fig:democracy_geo_appendix}
    \note{Panel~(a) reports horizon-specific coefficients from local projections of aggregate geopolitical alignment on democratic status, including four alignment lags and country and year fixed effects. Panel~(b) reports coefficients on geopolitical alignment from local projections of log GDP per capita ($\times 100$), with and without current democratic status, using the original ANRR GDP series. Both specifications in panel~(b) include four GDP and alignment lags and country and region-year fixed effects. Panel~(a) coefficients are in index units, and panel~(b) coefficients are per one standard deviation of the index. Shaded bands report 95\% confidence intervals based on Driscoll--Kraay standard errors.}
\end{figure}

Figure~\ref{fig:democracy_geo_appendix} provides two additional checks. Panel~(a) relates democratic status to aggregate geopolitical alignment, and panel~(b) estimates the alignment--GDP relationship under the baseline region-year fixed effects.

Panel~(a) shows a positive, persistent conditional relationship between democratic status and aggregate alignment. The coefficient is 0.023 on impact, peaks at 0.037 in year~6, and declines to 0.011 by year~15. Together with Figure~\ref{fig:dem_bilateral_effects}, this pattern shows that democratic status is associated with both partner-specific geopolitical reorientation and a higher aggregate index.

Panel~(b) shows that adding democratic status leaves the alignment--GDP coefficient path nearly unchanged under country and region-year fixed effects. Both paths peak near 3.8--3.9 log points, and their horizon-by-horizon correlation is 0.99.

\subsection{Dynamic Panel Estimates} \label{app_b:dynamic_panel}

This section reports the dynamic panel estimates summarized in Section~\ref{ss:dynamic_panel}. Following the preferred specification in \citet{Acemoglu2019-bo}, we estimate
\begin{align*}
    y_{ct} &= \alpha p_{ct} + \sum_{\ell = 1}^{J}\beta_{\ell}y_{c,t-\ell} +\sum_{\ell=1}^J \gamma_\ell p_{c,t-\ell} + \eta_c + \eta_{r(c)t} + \varepsilon_{ct}
\end{align*}

where $y_{ct}$ denotes log GDP per capita ($\times 100$), $p_{ct}$ is the geopolitical alignment index, and $\eta_c$ and $\eta_{r(c)t}$ denote country and region-year fixed effects. We set $J=4$. A causal interpretation of the resulting responses rests on Assumption~\ref{assumption_1} (Section~\ref{ss:interpretation}).\footnote{For a one-period change in alignment, the response satisfies $\psi_0=\alpha$ and $\psi_k=\sum_{\ell=1}^{\min(k,J)}\beta_\ell\psi_{k-\ell}+\gamma_k\mathbf{1}[k\leq J]$ for $k\geq1$. The response to a sustained unit path at horizon $h$ is $\Psi_h=\sum_{s=0}^{h}\psi_s$, with long-run multiplier $\Psi_\infty=(\alpha+\sum_{\ell=1}^{J}\gamma_\ell)/(1-\sum_{\ell=1}^{J}\beta_\ell)$.}

\begin{figure}[!t]
    \centering
    \begin{subfigure}[b]{0.48\textwidth}
        \includegraphics[width=\textwidth]{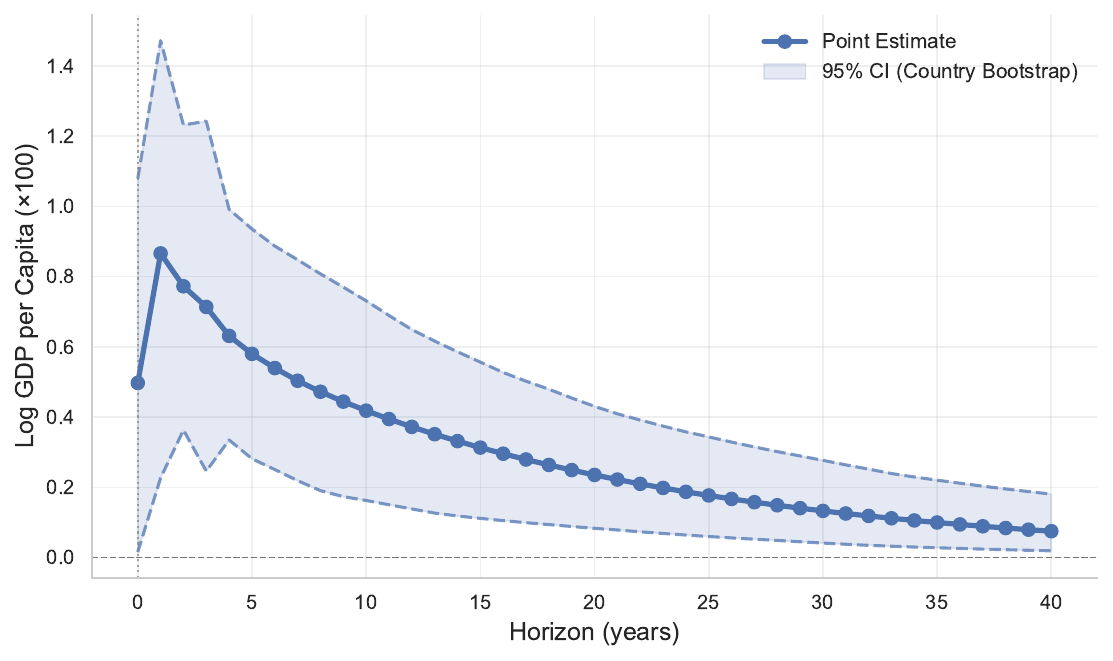}
        \caption{Transitory Alignment Change}
        \label{fig:ar_transitory}
    \end{subfigure}
    \hfill
    \begin{subfigure}[b]{0.48\textwidth}
        \includegraphics[width=\textwidth]{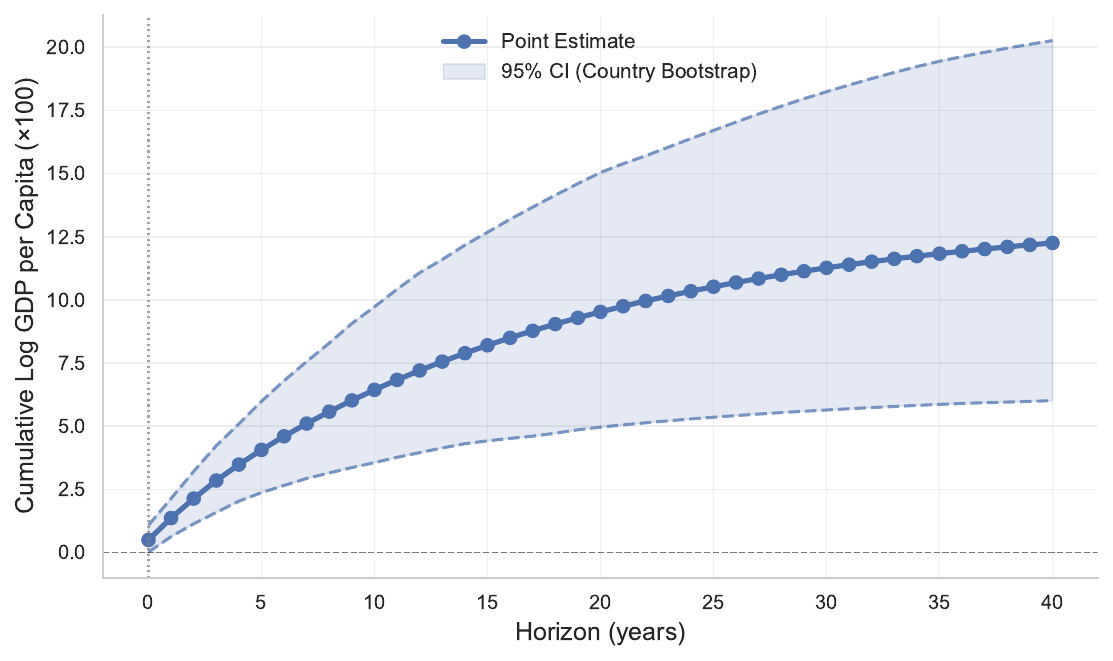}
        \caption{Sustained Alignment Path}
        \label{fig:ar_permanent}
    \end{subfigure}
    \caption{Dynamic Panel Estimates: GDP Responses to Transitory and Sustained Alignment Paths}
    \label{fig:ar_irf}
    \note{Panel (a) shows the response of log GDP per capita ($\times 100$) to a one-period, one-standard-deviation change in geopolitical alignment. Panel (b) cumulates these responses to represent a sustained one-standard-deviation alignment path. Both panels include country and region-year fixed effects and four lags of GDP and alignment. Shaded areas report 95\% confidence intervals from 500 country-block bootstrap iterations.}
\end{figure}

Panel (a) of Figure~\ref{fig:ar_irf} shows the response to a one-period alignment change. It rises from 0.5 log points on impact to 0.9 after one year, then declines to 0.3 at year 15 and 0.2 at year 25. The point estimates remain positive, and their confidence intervals exclude zero throughout the displayed 40-year horizon.

In panel (b), the sustained-path coefficient reaches 4.1 log points at year 5, 8.2 at year 15, 10.5 at year 25, and 12.3 at year 40. At year 25, the 95 percent confidence interval is 5.4 to 16.7 log points, close to the local-projection estimate of 10.1 and its interval of 5.3 to 16.8.

\FloatBarrier
\subsection{Dynamics of the Unsmoothed Index} \label{app_b:irf_events}

\begin{figure}[ht]
    \centering
    \begin{subfigure}[b]{0.48\textwidth}
        \includegraphics[width=\textwidth]{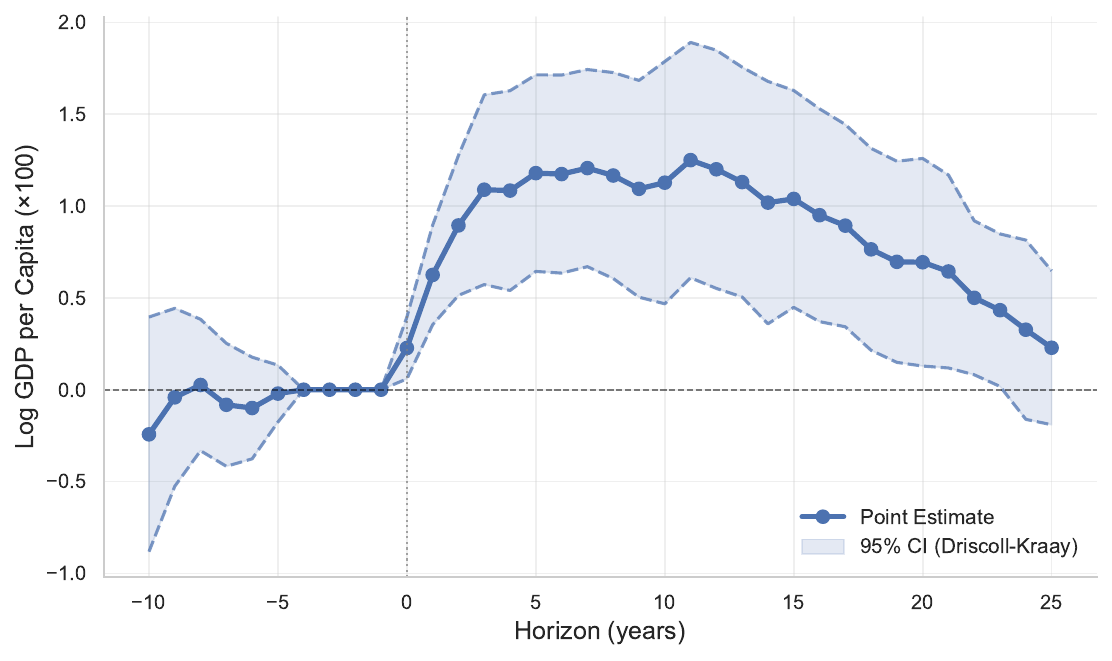}
        \caption{Unsmoothed Index}
        \label{fig:irf_events}
    \end{subfigure}
    \hfill
    \begin{subfigure}[b]{0.48\textwidth}
        \includegraphics[width=\textwidth]{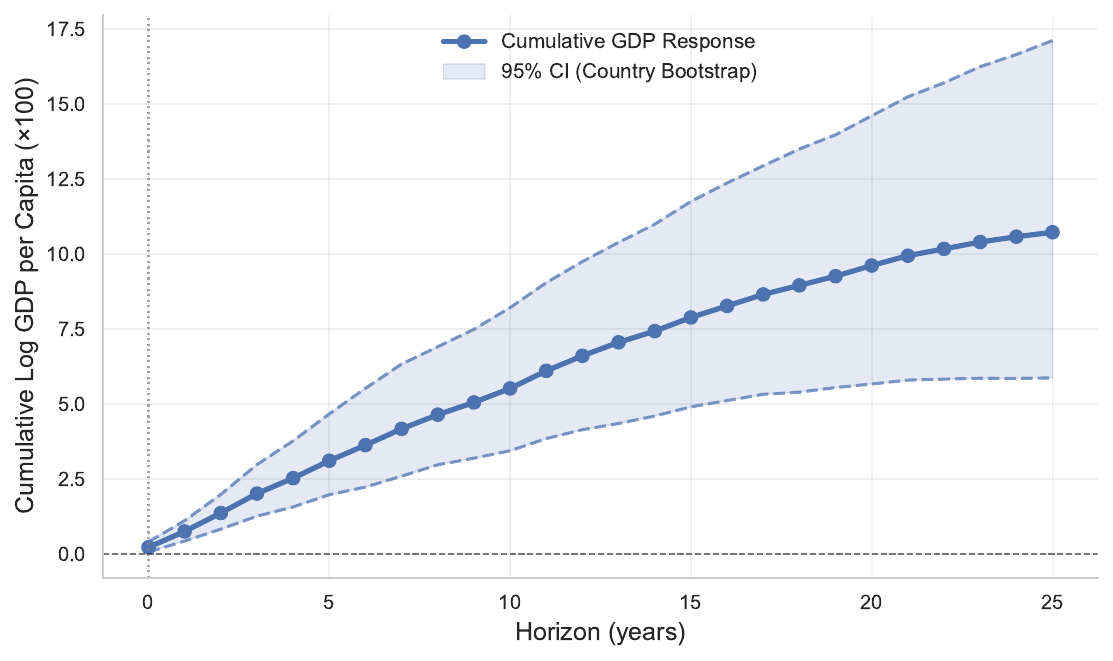}
        \caption{Sustained Path, Unsmoothed Index}
        \label{fig:irf_events_permanent}
    \end{subfigure}
    \caption{GDP Responses to the Unsmoothed Index}
    \label{fig:event_based_irf}
    \note{Panel (a) reports conditional local-projection coefficients for log GDP per capita ($\times 100$) on the unsmoothed index, per 0.141 index units (one standard deviation of the baseline index). Panel (b) reports the sustained-path mapping obtained by cumulating the constructed transitory coefficients. Both panels include four lags of GDP and of the unsmoothed index, country fixed effects, and region-year fixed effects. Panel (a) reports 95\% Driscoll--Kraay confidence intervals. Panel (b) reports 95\% confidence intervals from 500 country-block bootstrap iterations.}
\end{figure}

This section reports persistence and transitory-path results for the unsmoothed index $\tilde{S}_{ct}$ examined in Section~\ref{ss:irf_geo_shocks}. Figure~\ref{fig:event_based_irf} reports its GDP responses.

\begin{figure}[ht]
    \centering
    \begin{subfigure}[b]{0.48\textwidth}
        \includegraphics[width=\textwidth]{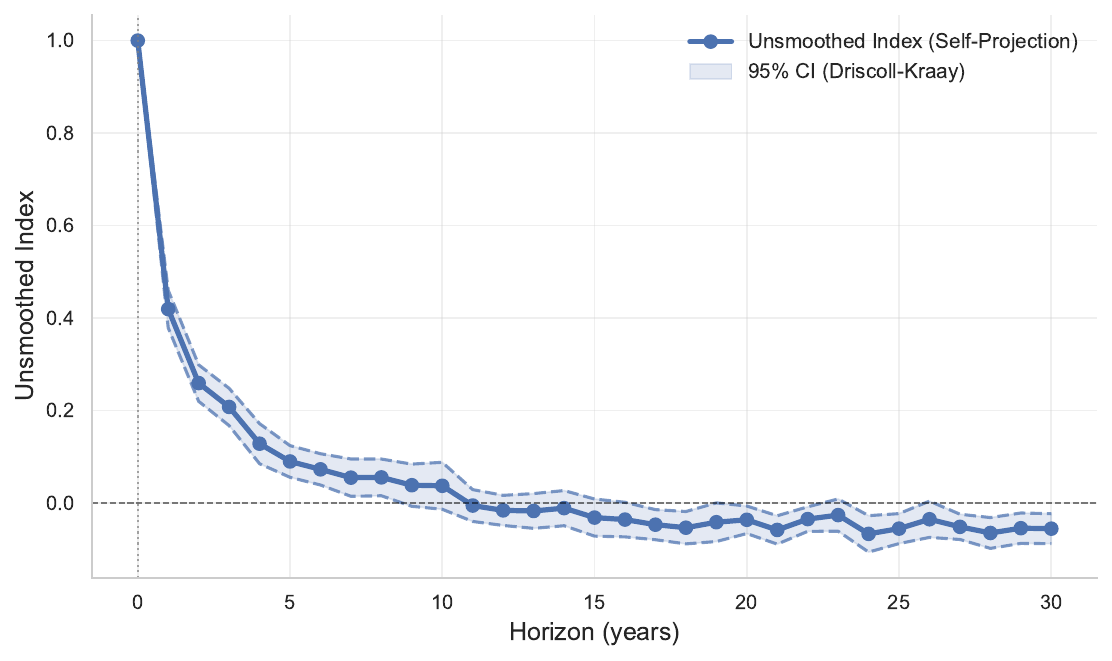}
        \caption{Persistence of the Unsmoothed Index}
        \label{fig:event_persistence}
    \end{subfigure}
    \hfill
    \begin{subfigure}[b]{0.48\textwidth}
        \includegraphics[width=\textwidth]{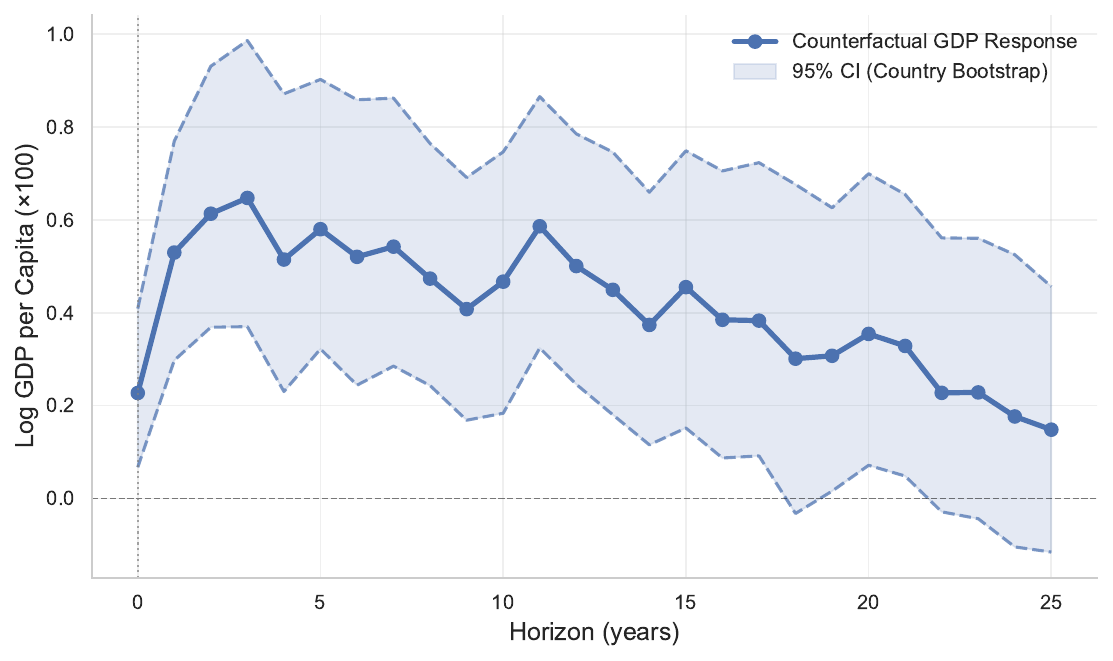}
        \caption{Transitory Path, Unsmoothed Index}
        \label{fig:event_transitory}
    \end{subfigure}
    \caption{Dynamics of the Unsmoothed Index}
    \label{fig:event_dynamics}
    \note{Panel (a) reports coefficients from projecting the unsmoothed index at $t+h$ on its current value, conditional on the baseline controls. Panel (b) reports the constructed GDP path associated with a 0.141 change in the unsmoothed index at $h=0$ that returns to zero thereafter. Appendix~\ref{app_b:irf_transitory_persistent} details the construction. Panel (a) reports 95\% Driscoll--Kraay confidence intervals. Panel (b) reports 95\% confidence intervals from 500 country-block bootstrap iterations.}
\end{figure}

Panel (a) of Figure~\ref{fig:event_dynamics} shows that the unsmoothed index mean-reverts substantially faster than the baseline index (panel~(a) of Figure~\ref{fig:baseline_irf}). The coefficient falls from one on impact to 0.419 after one year, so 58.1 percent of the initial change dissipates. It is 0.055 at year 8, and its confidence interval first includes zero at year 9. Later point estimates are small and sometimes negative.

Panel (b) applies the construction in Appendix~\ref{app_b:irf_transitory_persistent} to remove this persistence. The constructed GDP coefficient peaks at 0.6 log points in year 3 and declines thereafter. The later estimates are less precise.

\FloatBarrier
\subsection{Machine-Coded Event Databases} \label{app_b:measurement_alternatives}

\begin{figure}[ht]
    \centering
    \includegraphics[width=\linewidth]{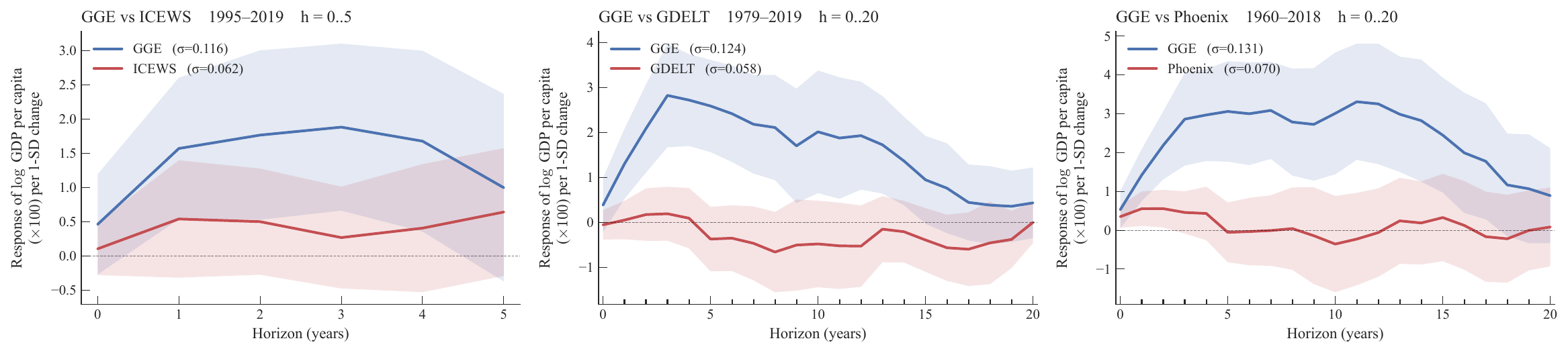}
    \caption{Growth Responses Using Baseline and Machine-Coded Event Databases}
    \label{fig:corpus_irf}
    \note{Local-projection coefficients for log GDP per capita ($\times 100$) per one-standard-deviation improvement in geopolitical alignment. In the legends, GGE denotes the baseline index, and $\sigma$ is each index's standard deviation in the regression sample, which scales the coefficients. Each panel compares the baseline index with one machine-coded database over the longest common estimation window: ICEWS, 1995--2019 and horizons 0--5; GDELT, 1979--2019 and horizons 0--20; Phoenix, 1960--2018 and horizons 0--20. All specifications include country and region-year fixed effects and four lags of GDP and geopolitical alignment. Shaded areas are 95\% confidence intervals based on Driscoll--Kraay standard errors.}
\end{figure}

This section reports the downstream GDP comparison for the machine-coded event databases introduced in Appendix~\ref{app_a:event_comparison}. Each panel of Figure~\ref{fig:corpus_irf} estimates the baseline specification with the baseline index and with an index built from one alternative database, over their longest common window. Both indices use the same aggregation and dynamic recursion and are standardized by their in-sample standard deviations. The baseline coefficient path peaks at 1.88 log points after three years in the ICEWS window, 2.83 after three years in the GDELT window, and 3.31 after eleven years in the Phoenix window. The ICEWS and GDELT estimates remain statistically indistinguishable from zero. Phoenix yields a small positive response at horizons 0--2, peaking near 0.56 log points, but it fades by year 5. Standardization rules out pure scale differences as an explanation for the weaker and less persistent machine-coded paths. This exercise compares the persistence of the GDP responses. Appendix~\ref{app_a:event_comparison} evaluates measurement accuracy and calibration directly.

\FloatBarrier
\subsection{UNGA Voting and Economic Growth} \label{app_b:UNGA_votes}

\begin{figure}[ht]
    \centering
    \begin{subfigure}[b]{0.48\textwidth}
        \includegraphics[width=\textwidth]{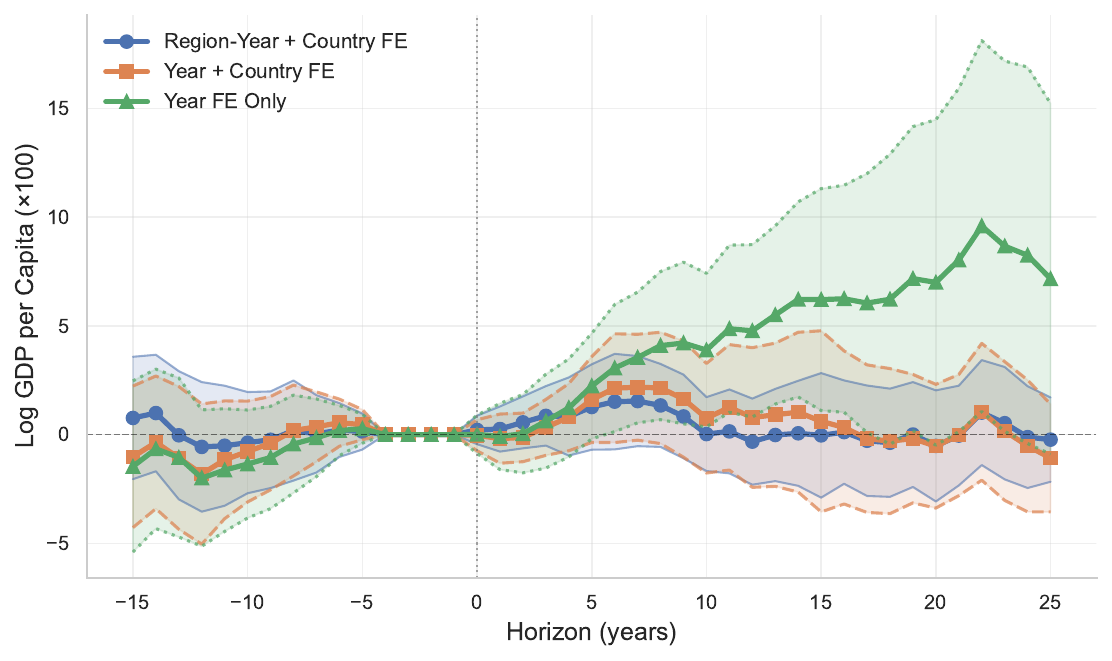}
        \caption{Voting Proximity to United States}
        \label{fig:ipd_usa}
    \end{subfigure}
    \hfill
    \begin{subfigure}[b]{0.48\textwidth}
        \includegraphics[width=\textwidth]{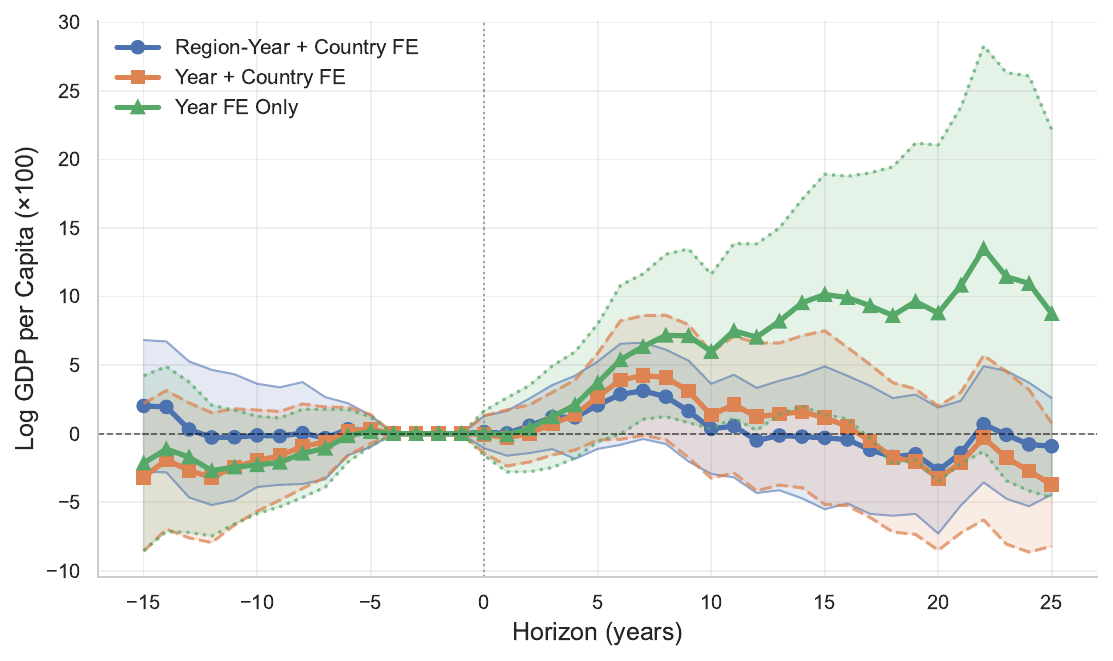}
        \caption{GDP-Weighted Voting Proximity}
        \label{fig:ipd_weighted}
    \end{subfigure}
    \caption{GDP Coefficient Paths Using UNGA Voting Proximity}
    \label{fig:unga_results}
    \note{Local-projection coefficients for log GDP per capita ($\times 100$) on UNGA voting proximity. Panel (a) uses negative IPD to the United States. Panel (b) uses the GDP-weighted average of negative IPD to the 24 major nations. The specifications include country and region-year fixed effects (blue, our baseline), country and year fixed effects (orange), or year fixed effects (green), together with four lags of GDP and the relevant IPD measure. Coefficients are per unit of negative IPD. Shaded areas report 95\% confidence intervals based on Driscoll--Kraay standard errors.}
\end{figure}

UNGA ideal points locate countries in a common multilateral voting space. We use the negative Ideal Point Distance (IPD) from \citet{Bailey2017-po}, which ranges from about $-5.4$ (complete disagreement) to 0 (perfect alignment), so higher values indicate closer voting positions. Panel~(a) of Figure~\ref{fig:unga_results} uses the United States as the reference point and is our primary comparison. Panel~(b) instead takes the GDP-weighted average of negative IPD to the 24 major nations. This aggregation broadens the reference set, but it is not an exact analogue of the geopolitical alignment index because ideal-point distances describe relative positions in a common multilateral space rather than the intensity of bilateral relationships. We therefore interpret panel~(b) as a check on whether the United States reference drives the comparison.

Panel~(a) shows that the coefficient path is statistically indistinguishable from zero under the baseline country and region-year fixed effects throughout the horizon. Replacing region-year fixed effects with year fixed effects produces a similar path. A positive coefficient emerges only when country fixed effects are removed, reaching approximately 7 log points after 20 years. This specification combines cross-country covariance between income and United States voting proximity with within-country changes, whereas the baseline design uses within-country variation.

Panel~(b) reports the GDP-weighted check. With country fixed effects, the coefficient paths are statistically indistinguishable from zero. A positive path appears only with year fixed effects alone, at horizons 7 to 16. Broadening the reference set therefore does not alter the result from the United States comparison.

These results do not imply that UNGA voting is uninformative. The two measures describe different objects: multilateral voting positions can remain stable while bilateral relations change, as the historical cases in Section~\ref{ss:measure_geo_relations} illustrate. Within-country variation in IPD therefore does not reproduce the same bilateral dynamics in the growth application.

\FloatBarrier
\subsection{Sanctions and Geopolitical Alignment} \label{app_b:sanctions}

This section asks whether active sanctions account for the alignment--GDP relationship. We use the Global Sanctions Database (GSDB) \citep{felbermayr2020global,yalcin2025global} to identify active sanctions imposed by major nation $j$ on country $c$ from 1960 through 2023. We aggregate the dyadic indicators using the same 24 major nations as in the alignment index:
\[
p^{\text{Sanction}}_{ct} = \sum_{j \in \mathcal{N}\setminus\{c\}} \mathbf{1}_{cjt} \times \text{GDP share}_{jt},
\]
where $\mathbf{1}_{cjt}$ equals one if major nation $j$ has an active sanction on country $c$ in year $t$. All countries in the baseline panel are retained. Within the database's coverage, dyad-years with no active sanction are coded zero. Appendix~\ref{app_a:anchoring} separately benchmarks the event database against GSDB sanction onsets and removals.

\paragraph{Empirical Specification}

We estimate separate and joint specifications:

\begin{align}
\text{Separate:} \quad y_{c,t+h} &= \alpha_h^{k} p_{ct}^{k} + \sum_{\ell=1}^{4} \beta_{h\ell} y_{c,t-\ell} + \sum_{\ell=1}^{4} \gamma_{h\ell}^{k} p_{c,t-\ell}^{k} + \eta_c + \eta_{r(c)t} + \varepsilon_{c,t+h} \label{eq:sanctions_univ}\\
\text{Joint:} \quad y_{c,t+h} &= \alpha_h^{\text{Geo}} p_{ct} + \alpha_h^{\text{Sanc}} p_{ct}^{\text{Sanction}} + \sum_{\ell=1}^{4} \beta_{h\ell} y_{c,t-\ell} + \sum_{\ell=1}^{4} \gamma_{h\ell}^{\text{Geo}} p_{c,t-\ell} \nonumber\\*
&\quad + \sum_{\ell=1}^{4} \gamma_{h\ell}^{\text{Sanc}} p_{c,t-\ell}^{\text{Sanction}} + \eta_c + \eta_{r(c)t} + \varepsilon_{c,t+h} \label{eq:sanctions_joint}
\end{align}

where $k \in \{\text{Geo}, \text{Sanction}\}$ indexes the measure. The outcome and sample follow the baseline: log GDP per capita, the 148-country long-coverage sample, country and region-year fixed effects, and four lags of GDP and each included geopolitical measure. At every horizon, the separate and joint specifications use identical country-year observations.

\begin{figure}[ht]
    \centering
    \begin{subfigure}[b]{0.48\textwidth}
        \includegraphics[width=\textwidth]{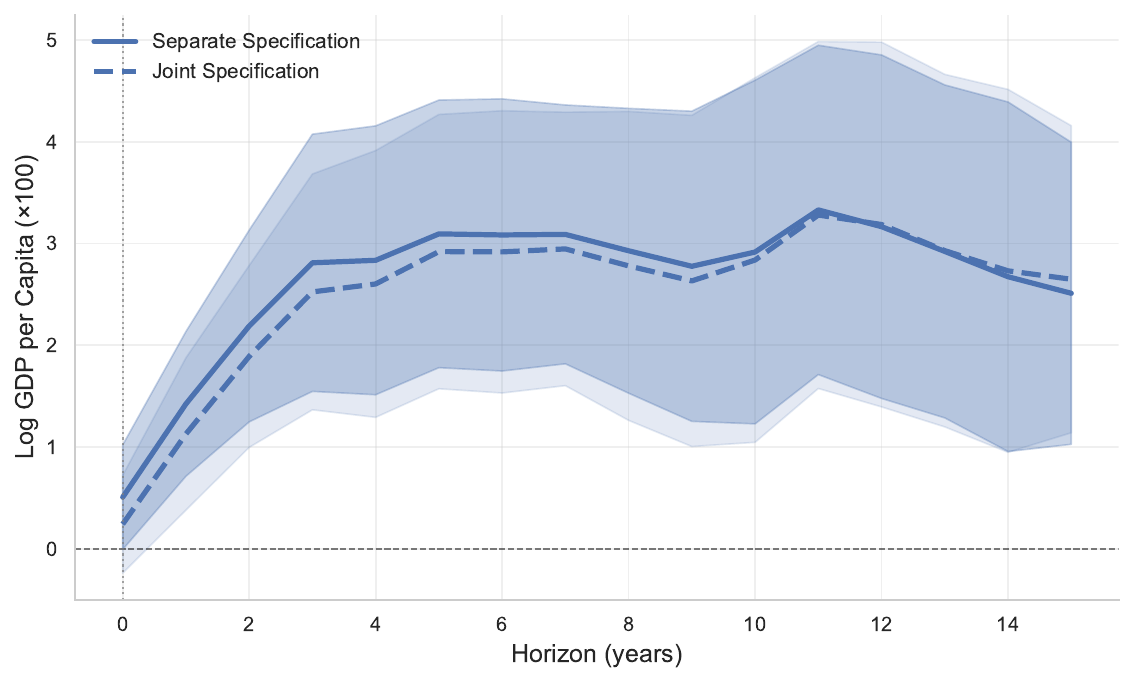}
        \caption{Geopolitical Alignment Index}
        \label{fig:geo_horse_race}
    \end{subfigure}
    \hfill
    \begin{subfigure}[b]{0.48\textwidth}
        \includegraphics[width=\textwidth]{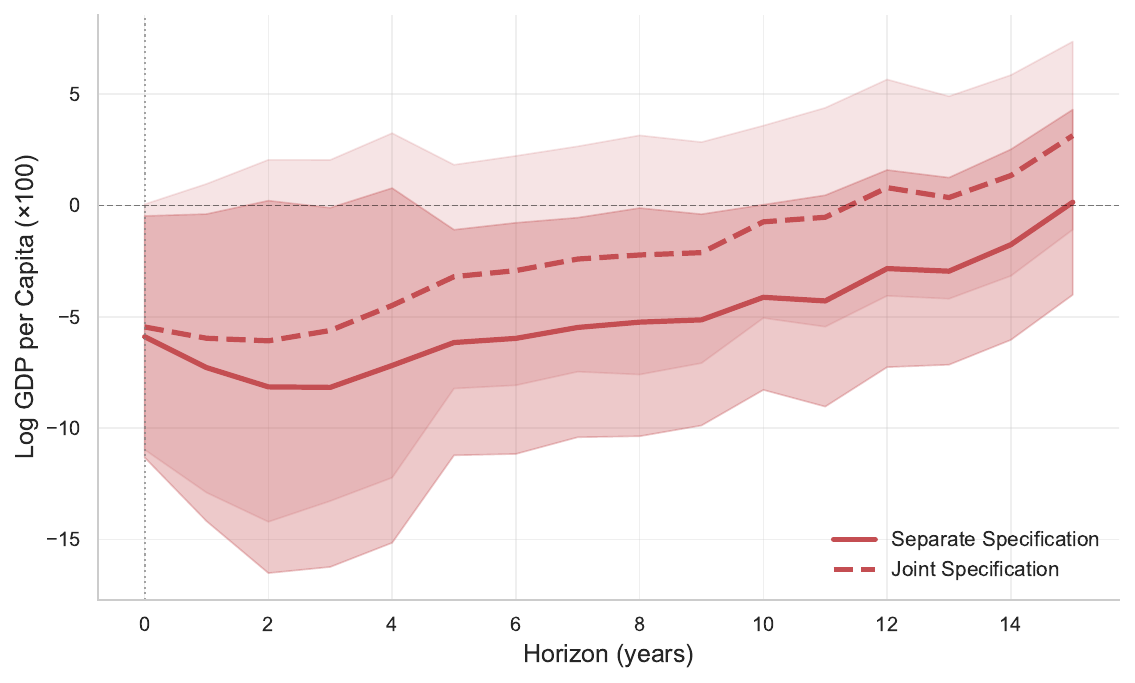}
        \caption{Sanctions}
        \label{fig:sanc_horse_race}
    \end{subfigure}
    \caption{Geopolitical Alignment and Sanctions}
    \label{fig:sanctions_horse_race}
    \note{Local-projection coefficient paths for log GDP per capita ($\times 100$). Panel (a) compares separate and joint specifications for geopolitical alignment. Panel (b) does the same for sanctions exposure. All specifications use identical observations at each horizon and include four lags of GDP and the included geopolitical measures, country fixed effects, and region-year fixed effects. Shaded areas report 95\% confidence intervals based on Driscoll--Kraay standard errors. Alignment coefficients are per one standard deviation of the index. Sanctions exposure is measured in raw GDP-share units, and the vertical scales differ across panels.}
\end{figure}

\paragraph{Results and Interpretation}

In panel~(a) of Figure~\ref{fig:sanctions_horse_race}, the alignment coefficient paths track one another closely and peak in year~11 at 3.3 log points in both the separate and the joint specification. In panel~(b), the separate sanctions coefficient reaches $-8.2$ log points in year~3. The corresponding joint estimate is $-5.6$, a 31 percent attenuation, and no joint sanctions coefficient is statistically distinguishable from zero. By year~10, the joint estimate is close to zero.

The two measures differ in timing. The alignment index records diplomatic deterioration and the announcement or lifting of sanctions, while the GSDB indicator records active enforcement. The stable alignment path shows that the narrower enforcement measure does not account for the broader alignment--GDP relationship.

\FloatBarrier
\subsection{Formal Alliances and Geopolitical Alignment} \label{app_b:alliances}

\paragraph{Data and Variable Construction}

This section asks whether formal military alliances account for the alignment--GDP relationship. We use the Alliance Treaty Obligations and Provisions (ATOP) dataset, version 5.1 \citep{leeds2002alliance}, which codes formal military alliances through 2018.\footnote{The original ATOP dataset \citep{leeds2002alliance} covered 1815--1944. We use version 5.1, which extends coverage through 2018.} For each country $c$ and year $t$, we construct a GDP-weighted alliance exposure measure:
\[
p^{\text{Alliance}}_{ct} = \sum_{j \in \mathcal{N}\setminus\{c\}} \mathbf{1}^{A}_{cjt} \times \text{GDP share}_{jt},
\]
where $\mathbf{1}^{A}_{cjt}$ equals one if country $c$ maintains any active formal alliance (a defense pact, offensive pact, neutrality agreement, non-aggression pact, or consultation agreement) with major nation $j$ in year $t$. We restrict the sample to years with observed ATOP coverage and retain all countries in the baseline panel. Within that period, dyad-years with no active alliance are coded zero. The dyadic indicators are binary, but the aggregate is continuous and can change with the major nations' GDP shares even when treaty status is unchanged. The measure ranges from zero to approximately 0.80.

\paragraph{Results}

We estimate the same separate and joint local projections, on identical observations, as in equations~\eqref{eq:sanctions_univ}--\eqref{eq:sanctions_joint}, replacing $p^{\text{Sanction}}_{ct}$ with $p^{\text{Alliance}}_{ct}$.

\begin{figure}[ht]
    \centering
    \begin{subfigure}[b]{0.48\textwidth}
        \includegraphics[width=\textwidth]{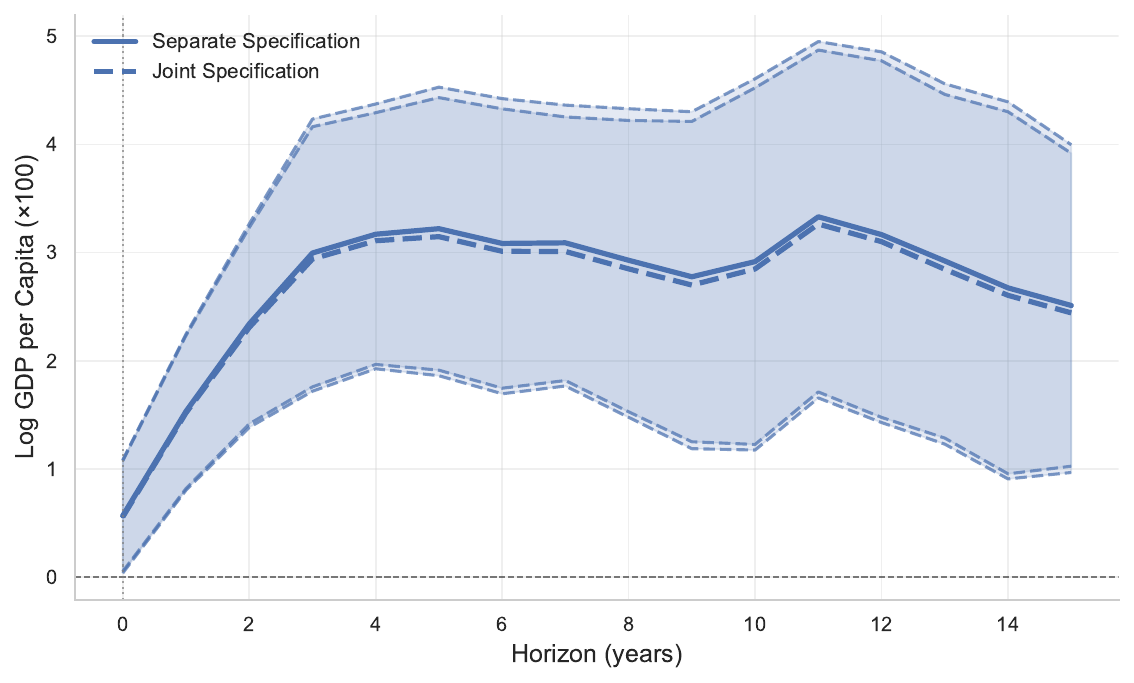}
        \caption{Geopolitical Alignment Index}
        \label{fig:geo_ally_horse_race}
    \end{subfigure}
    \hfill
    \begin{subfigure}[b]{0.48\textwidth}
        \includegraphics[width=\textwidth]{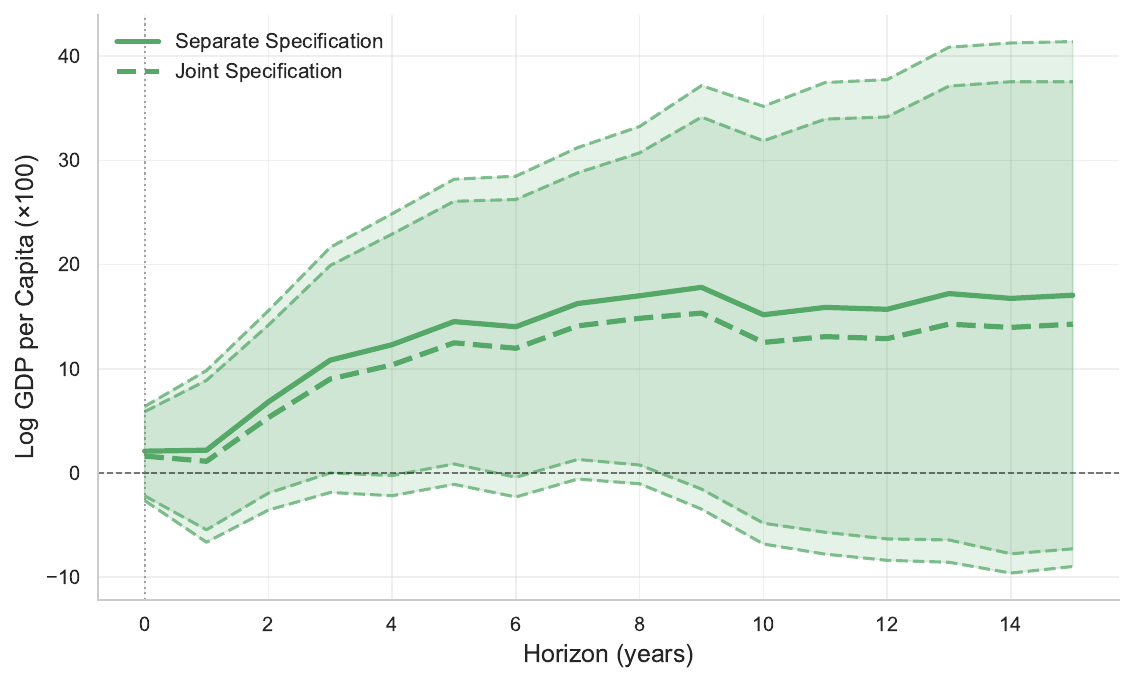}
        \caption{Formal Alliances}
        \label{fig:ally_horse_race}
    \end{subfigure}
    \caption{Geopolitical Alignment and Formal Alliances}
    \label{fig:alliance_horse_race}
    \note{Local-projection coefficient paths for log GDP per capita ($\times 100$). Panel (a) compares separate and joint specifications for geopolitical alignment. Panel (b) does the same for formal alliance exposure. All specifications use identical observations at each horizon and include four lags of GDP and the included geopolitical measures, country fixed effects, and region-year fixed effects. Shaded areas report 95\% confidence intervals based on Driscoll--Kraay standard errors. Alignment coefficients are per one standard deviation of the index. Alliance exposure is measured in raw GDP-share units, and the vertical scales differ across panels.}
\end{figure}

In panel~(a) of Figure~\ref{fig:alliance_horse_race}, the alignment coefficient paths are nearly identical and peak in year~11 at 3.3 log points in both the separate and the joint specification. In panel~(b), the alliance coefficients are positive and build gradually through the first decade. In year~9, the estimate declines from 17.8 to 15.3 log points when alignment is included, a 14 percent attenuation. The joint alliance coefficients are statistically indistinguishable from zero throughout the horizon.

The two measures differ in timing: the alignment index records diplomatic cooperation surrounding alliance formation and maintenance, while the ATOP indicator records a persistent treaty stock. The alignment coefficient path remains stable after this stock is included.

\paragraph{Defense Pacts}

As a robustness check, we restrict the alliance measure to defense pacts, the strongest form of alliance commitment. The alignment coefficient path remains nearly unchanged, while the defense-pact coefficients are imprecise in both the separate and joint specifications.

\newpage
\section{LLM Prompt: Event Category, CAMEO, and Goldstein Score} \label{app_c:llm}

Appendix~\ref{app_c:llm} describes the prompt structure (C.1), the event-selection criteria (C.2), and the CAMEO and Goldstein coding rules (C.3).

\subsection{Prompt Structure} \label{app_c:prompt}

This section describes the LLM prompt structure for compiling and analyzing major political events shaping the bilateral relationship between two countries, \{country\_1\} (code: \{country\_1\_code\}) and \{country\_2\} (code: \{country\_2\_code\}), or their historical predecessors, during \{target\_year\}. The analysis employs the Conflict and Mediation Event Observations (CAMEO) framework and the Goldstein scale, both detailed in Appendix~\ref{app_c:cameo}, to classify events and assess their intensity. If no major events are identified, a historical context-based relationship assessment is provided. The output is a single JSON object.

\paragraph{Relationship Assessment Framework} 
The bilateral relationship for \{target\_year\} is classified into one of the following mutually exclusive categories, reflecting interaction intensity and nature:

\begin{itemize}
    \item \textbf{State of War / Active Conflict}: Sustained, large-scale armed conflict.
    \item \textbf{Crisis / Intense Confrontation}: High tension with disputes or limited clashes, short of war.
    \item \textbf{Hostile / Antagonistic Relationship}: Animosity marked by sanctions or diplomatic friction.
    \item \textbf{Competitive / Rivalrous Relationship}: Strategic competition with limited cooperation.
    \item \textbf{Limited Contact / Cool Relationship}: Minimal, neutral interaction.
    \item \textbf{Selective Cooperation / Transactional Relationship}: Cooperation on specific interests amid competition.
    \item \textbf{Broad Cooperation / Partnership}: Extensive sectoral cooperation with regular dialogue.
    \item \textbf{Strategic Partnership}: Deep coordination on strategic issues with high trust.
    \item \textbf{Alliance}: Formal treaty-based mutual support, often military.
\end{itemize}

\paragraph{Analytical Steps}
The prompt specifies five steps for event identification, classification, and assessment:

\begin{enumerate}
    \item \textbf{Verify Political Entities}: Confirm \{country\_1\} (code: \{country\_1\_code\}) and \{country\_2\} (code: \{country\_2\_code\}) existed in \{target\_year\}. If not, identify the primary political entity controlling the relevant territory (e.g., Soviet Union for Russia before December 26, 1991; Russian Federation thereafter). Ambiguities are noted in the evaluation summary, with analysis using the best-identified entities, reflected in JSON fields \texttt{country1} and \texttt{country2}.
    
    \item \textbf{Data Collection}: Use search tools to compile interactions between verified entities during \{target\_year\}.
    
    \item \textbf{Identify Major Political Events}: Identify events in \{target\_year\} significantly influencing the bilateral relationship, verified by reliable sources. Events include economic diplomacy, diplomatic actions, high-level interactions, and security measures, with details in Appendix~\ref{app_c:events}.
    
    \item \textbf{Event Analysis}: For each event:
        \begin{enumerate}
            \item Assign \texttt{country1} and \texttt{country2} as initiator/target or participants.
            \item Provide \texttt{event\_name} and \texttt{event\_description}.
            \item Classify using CAMEO (see Appendix~\ref{app_c:cameo}): \texttt{CAMEO\_quad\_class} (Verbal/Material Cooperation/Conflict), \texttt{CAMEO\_root\_code} (e.g., 04), \texttt{CAMEO\_event\_code} (e.g., 043), emphasizing economic actions (e.g., 163 for sanctions) and mediation (e.g., 045).
            \item Assign \texttt{Goldstein\_Scale} ($-10.0$ to $+10.0$, see Appendix~\ref{app_c:cameo}), reflecting intensity, adjusted for bilateral context but consistent with CAMEO.
            \item Assign \texttt{event\_category}, the primary subcategory listed in Table~\ref{tab:event_categories}. Record the event date at the finest verifiable resolution and any involvement of international organizations.
            \item Provide \texttt{evaluation\_summary} justifying classifications and scores.
        \end{enumerate}
    
    \item \textbf{Overall Relationship Assessment}: Select one \texttt{relationship} category for \{target\_year\}, integrating event patterns (via CAMEO/Goldstein) and historical context. If no events are found, assess based on interaction absence and historical trends.
\end{enumerate}

\paragraph{JSON Output} 
The output is a JSON object with key \texttt{historical\_political\_events}:

\begin{itemize}
    \item If events found: List of objects with fields:
        \begin{itemize}
            \item \texttt{year}, \texttt{country1}, \texttt{country2}, \texttt{event\_name}, \texttt{event\_description},
            \item \texttt{event\_category}, \texttt{event\_exact\_date}, \texttt{event\_month}, \texttt{event\_quarter}, \texttt{event\_year},
            \item \texttt{CAMEO\_quad\_class}, \texttt{CAMEO\_root\_code}, \texttt{CAMEO\_event\_code}, \texttt{Goldstein\_Scale},
            \item \texttt{international\_org\_involvement}, \texttt{relationship}, \texttt{evaluation\_summary}.
        \end{itemize}
    \item If no events found: Single object with \texttt{event\_name} = ``No Significant Bilateral Events Found,'' null category, CAMEO, and Goldstein fields, and context-based \texttt{relationship}.
\end{itemize}

The scores in this paper use only the event-level \texttt{Goldstein\_Scale} values. The \texttt{relationship} field is not used.

\FloatBarrier
\subsection{Event Identification Details} \label{app_c:events}

This section elaborates on the criteria for identifying major political events that significantly influence the bilateral relationship during \{target\_year\} (Step 3 of Appendix~\ref{app_c:prompt}).

\paragraph{Event Verification} 
The prompt instructs the model to check potential events against reliable sources and to exclude unverified or fabricated events. When events span multiple categories, the primary classification reflects the dominant mechanism or domain of impact, with secondary dimensions noted in the event description.

\paragraph{Event Dimensions} 
Events are identified across six major dimensions of bilateral relations. Each event is assigned one primary subcategory (\texttt{event\_category}). Table~\ref{tab:event_categories} lists all 25 subcategories, and the descriptions below summarize the main ones.

\subparagraph{Economic Relations.}
This dimension covers economically mediated bilateral interactions, both cooperative and coercive.

\begin{itemize}
    \item \textit{Trade Policy and Market Access}: Imposition, adjustment, or removal of tariffs; non-tariff barriers such as technical standards, sanitary measures, and import licensing; trade agreement negotiations, signings, ratifications, withdrawals, or dispute settlement proceedings.
    
    \item \textit{Financial and Monetary Relations}: Financial sanctions including asset freezes, transaction bans, and SWIFT exclusions; currency swap agreements and bilateral payment arrangements; foreign investment restrictions or liberalizations; bilateral investment treaty developments.
    
    \item \textit{Economic Coercion and Inducements}: Comprehensive trade embargoes and sectoral sanctions; export controls and entity list designations; technology transfer restrictions; foreign aid packages, development finance, and debt relief programs.
    
    \item \textit{Strategic Economic Sectors}: Energy supply agreements and pipeline projects; telecommunications and digital economy restrictions (e.g., 5G bans); critical resource arrangements covering rare earth elements, strategic minerals, and food security.
    
    \item \textit{Economic Integration and Infrastructure}: Bilateral infrastructure initiatives; supply chain arrangements including friend-shoring and critical supply agreements; regional economic arrangement participation.
\end{itemize}

\subparagraph{Diplomatic and Political Relations.}
This dimension captures formal state-to-state interactions and official communications that shape the bilateral relationship.

\begin{itemize}
    \item \textit{Formal Diplomatic Engagement}: Embassy or consulate openings and closures; ambassador appointments and recalls; diplomatic staff expulsions; formal protests, d\'{e}marches, and official condemnations or commendations.
    
    \item \textit{High-Level Political Interactions}: Presidential or prime ministerial visits and bilateral summits; ministerial meetings across foreign affairs, defense, and trade portfolios; joint commission sessions and strategic dialogues.
    
    \item \textit{Public Diplomacy and Rhetoric}: Major policy speeches affecting bilateral relations; parliamentary resolutions; government white papers on the bilateral relationship; significant public campaigns or propaganda efforts.
\end{itemize}

\subparagraph{Security and Defense.}
This dimension encompasses military cooperation, competition, and security-related incidents between the two countries.

\begin{itemize}
    \item \textit{Military Cooperation and Competition}: Defense cooperation agreements and military alliances; status of forces agreements; major weapons sales or cancellations; arms embargoes; joint military exercises and military-to-military exchanges.
    
    \item \textit{Security Incidents}: Border skirmishes and territorial claim assertions; airspace or maritime violations; naval encounters and air intercepts; demarcation agreements.
    
    \item \textit{Intelligence and Cyber Operations}: Publicly revealed espionage scandals; intelligence officer expulsions; intelligence sharing agreements or suspensions; state-sponsored cyber attacks and cyber security cooperation.
\end{itemize}

\subparagraph{Legal, Territorial, and Movement.}
This dimension covers international legal proceedings, sovereignty disputes, and policies governing the movement of people.

\begin{itemize}
    \item \textit{International Legal Actions}: Bilateral disputes before international courts such as the International Court of Justice or International Tribunal for the Law of the Sea; WTO disputes with significant political dimensions; contentious extradition cases.
    
    \item \textit{Territorial and Maritime Issues}: Exclusive economic zone disputes; continental shelf claims; fisheries agreements or conflicts; freedom of navigation operations; border demarcation and sovereignty recognition.
    
    \item \textit{Movement of People}: Visa regime changes and travel bans; visa-free agreements; guest worker programs; readmission agreements; refugee and asylum policy changes.
\end{itemize}

\subparagraph{Multilateral and Global Governance.}
This dimension captures bilateral dynamics manifested through international organizations and global issue areas.

\begin{itemize}
    \item \textit{International Organizations}: United Nations Security Council confrontations; General Assembly coalition building; specialized agency disputes; regional organization membership changes.
    
    \item \textit{Global Issues}: Climate agreement positions and environmental cooperation; pandemic response coordination and vaccine diplomacy; human rights criticism or defense in international fora.
\end{itemize}

\subparagraph{Other Significant Events.}
This residual category encompasses additional politically consequential interactions.

\begin{itemize}
    \item \textit{Historical and Symbolic}: Apologies for historical wrongs; memorial visits; monument disputes; anniversary commemorations with bilateral significance.
    
    \item \textit{Humanitarian and Disaster Response}: Aid offers or rejections following natural disasters; joint rescue operations; humanitarian access disputes.
    
    \item \textit{Technology and Space}: Joint space missions; satellite cooperation; technology theft accusations; research collaboration terminations.
    
    \item \textit{Communications and Media}: Journalist expulsions; broadcasting restrictions; undersea cable disputes.
\end{itemize}

\paragraph{Selection Criteria}
Events are prioritized based on their significant effect on, or strong indication of, the bilateral relationship's trajectory. When economic tools are employed for political purposes, events are classified under economic categories.

\FloatBarrier
\subsection{Conflict and Mediation Event Observations and Goldstein Score} \label{app_c:cameo}

Our analysis employs the Conflict and Mediation Event Observations (CAMEO) framework \citep{schrodt2012cameo} to classify bilateral political events. CAMEO categorizes international political actions along two primary dimensions: the nature of the interaction (cooperation versus conflict) and the form of action (verbal versus material).

\paragraph{CAMEO Classification Structure} 
The two dimensions define four quadrant classes:

\begin{itemize}
\item \textbf{Verbal Cooperation}: Diplomatic statements, consultations, expressions of intent to cooperate, and formal diplomatic cooperation including treaty signing and public endorsements.
\item \textbf{Material Cooperation}: Tangible cooperative actions such as economic aid provision, military cooperation, intelligence sharing, and policy concessions.
\item \textbf{Verbal Conflict}: Critical statements, accusations, demands, rejections, threats, and public protests that express disagreement or hostility.
\item \textbf{Material Conflict}: Concrete hostile actions including economic sanctions, military demonstrations, coercive measures, and various forms of violence.
\end{itemize}

Within each quadrant, CAMEO provides a hierarchical coding system with root codes (two-digit) representing general action categories and event codes (three-digit) specifying precise actions. For example, root code 16 (REDUCE RELATIONS) includes event codes 163 (impose economic sanctions) and 161 (reduce diplomatic relations).

\paragraph{Implementation in Our Analysis} 
The prompt walks the model through CAMEO classification. The process begins with identifying the core bilateral action in each event description, determining its cooperative or conflictual nature, and assessing whether the action is primarily verbal or material. The LLM then selects the most appropriate root code within the identified quadrant class and chooses the specific event code that best fits the action.

We pay particular attention to economic statecraft. For instance, we distinguish broad economic sanctions (code 163) from targeted administrative sanctions (code 172). We also distinguish routine consultations (code 040) from formal agreement signing (code 057).

\paragraph{Integration with Goldstein Scale Scoring} 
CAMEO classifications provide the first reference point for Goldstein scale scoring \citep{goldstein1992}, which assigns numerical values from $-10.0$ (maximum conflict) to $+10.0$ (maximum cooperation). After classifying an event, the LLM evaluates its conflict--cooperation intensity against the scale's coding conventions, intensity bands, and action-based anchor cases, then allows limited adjustments for the event's specific bilateral context and severity. The anchors are the same for all dyads. Event selection reflects an event's significance, and the Goldstein score reflects its intensity of conflict or cooperation.

\stopappendixcontents

\end{document}